\documentclass[showpacs, nofootinbib, aps, prd, superscriptaddress, 
10pt, showkeys, notitlepage,onecolumn]{revtex4-1}

\usepackage{graphicx,amssymb,amsmath,amsthm,amsfonts,epsfig,natbib}
\usepackage[linktocpage]{hyperref}
\usepackage[utf8]{inputenc}
\usepackage[usenames,dvipsnames]{color}
\usepackage{epstopdf}
\usepackage{aas_macros}
\usepackage{tensor}
\usepackage{mathtools}
\usepackage{amsbsy}
\usepackage{bm}

\definecolor{darkred}{rgb}{0.5,0,0}
\definecolor{darkgreen}{rgb}{0,0.5,0}
\definecolor{darkblue}{rgb}{0,0,0.5}
\definecolor{prussian}{rgb}{0.0, 0.19, 0.33}

\hypersetup{colorlinks=true, citecolor=darkblue, linkcolor=darkblue,
urlcolor = darkblue}

\newcommand{\be}{\begin{equation}}
\newcommand{\ee}{\end{equation}}
\newcommand{\bear}{\begin{eqnarray}}
\newcommand{\eear}{\end{eqnarray}}
\newcommand{\ba}{\begin{aligned}}
\newcommand{\ea}{\end{aligned}}

\newcommand{\xp}{x_{\rm p}}
\newcommand{\bnabla}{\boldsymbol{\nabla}}

\newcommand{\cE}{{\cal E}}
\newcommand{\cN}{{\cal N}}

\newcommand{\cS}{{\cal S}}
\newcommand{\cR}{{\cal R}}

\newcommand{\cM}{{\cal M}}

\newcommand{\rA}{{\rm A}}
\newcommand{\rE}{{\rm E}}

\newcommand{\rK}{{\rm K}}

\begin{document}
\title{Gravitational waves from single neutron stars: an advanced detector era survey}

\begin{abstract}

With the doors beginning to swing open on the new gravitational wave astronomy, this review provides an up-to-date survey of the most important physical mechanisms  that could lead to emission of potentially  detectable gravitational radiation from isolated and accreting neutron stars. In particular we discuss the gravitational  wave-driven instability and asteroseismology formalism of the $f$- and $r$-modes, the different ways that a neutron star could form and sustain a non-axisymmetric quadrupolar ``mountain'' deformation, the excitation of oscillations during magnetar flares and the possible gravitational wave signature of pulsar glitches. We focus on progress made in the recent years in each topic, make a fresh assessment of the gravitational wave detectability of each mechanism and, finally, highlight key problems and desiderata for future work.

\end{abstract}

\author{Kostas Glampedakis}
\email{kostas@um.es}
\affiliation{Departamento de F\'isica, Universidad de Murcia,
Murcia, E-30100, Spain}
\affiliation{Theoretical Astrophysics, University of T\"ubingen, Auf der Morgenstelle 10, T\"ubingen, D-72076, Germany}

\author{Leonardo Gualtieri}
\email{leonardo.gualtieri@roma1.infn.it}
\affiliation{Dipartimento di Fisica, ``Sapienza'' Universit\`a di Roma, Piazzale Aldo Moro 5, 00185, Roma, Italy}
\affiliation{Sezione INFN Roma1, Piazzale Aldo Moro 5, 00185, Roma, Italy}

\date{{\today}}

\maketitle




\section{Introduction}
\label{sec:intro}

September 14, 2015 marked a milestone for gravitational physics with the first direct detection of gravitational waves
(GWs) by the twin advanced LIGO observatories \cite{GW150914}.  Since then, and with the additional participation 
of the advanced Virgo detector, more detections have been announced \cite{GW151226,GW170104,GW170608}, with all 
observed signals having been identified as mergers of binary black hole systems.
 Finally, it was the turn of neutron stars to be ``discovered'' by the interferometric telescopes of the new GW
astronomy: on August 17, 2017, the first GW signal from neutron stars has been detected \cite{GW170817}, together with
transient counterparts across the entire electromagnetic spectrum \cite{EM170817}.  This signal came, as expected, from
the same systems that indirectly established for the first time the very existence of gravitational radiation: binary
neutron star systems, and in particular their final GW-driven merger \cite{GW170817}.  In fact, these catastrophic
events had already been routinely observed by photon astronomy in the form of short gamma-ray bursts (GRBs). 

\emph{Single} neutron stars, the subject of this chapter, await their turn to be added to the catalogue of observed GW
sources -- what is still uncertain is if this goal will be reached by Advanced LIGO/Virgo or by the next
generation instruments such as the Einstein Telescope (ET)~\cite{Punturo:2010zza}. The term ``single'' comprises (i)
isolated neutron stars, that is, the bulk of the known neutron star population which includes normal radio pulsars, old
recycled millisecond pulsars (MSPs), magnetars, etcetera, and (ii) accreting systems, from which the neutron stars in
low mass X-ray binaries (LMXBs) are the ones relevant to this review.

The purpose of this chapter is to provide a 2017 status report of our understanding of the 
small mosaic of known mechanisms that could turn single neutron stars into potentially detectable sources of GWs. 
More specifically, we discuss (i) the GW-driven instability of the fundamental $f$-mode and of the inertial $r$-mode  
(including a basic introduction to the CFS theory of unstable oscillations) 
(ii) the GW asteroseismology formalism  for stable and unstable $f$-modes (iii) the different ways that non-axisymmetric 
quadrupolar deformations could be induced and sustained in neutron stars  thereby making them continuous 
sources of GWs (iv) the excitation of magneto-elastic oscillations and GW emission produced by magnetar flares, 
and finally (v) the GW detectability of pulsar glitches. 

Although being mostly interested in providing a sober assessment of the prospects for detecting GWs from
these various single neutron star sources, we also present in some detail key astrophysical ``side effects'' 
of GW emission such as the $r$-mode driven spin-temperature evolution in accreting neutron stars and 
the antagonism between the $f$-mode and magnetic dipole torque for controlling the dynamics
of neutron star merger remnants.

Given the space limitations, our discussion is, of course, far from being comprehensive -- the objective is to 
present and discuss key recent developments in each research topic and give pointers for future work rather than 
explaining things from scratch. References to textbooks, earlier review articles and selected research papers are 
plentifully given for the reader less  familiar with the topics discussed here. 


\subsection{Summary, conventions and plan of this review}
\label{sec:plan}

Table~\ref{tab:summary} is a bird's eye view of the chapter: it provides a list of the various topics that are discussed 
in the following sections and of what we consider to be the most important recent progress and open problems in 
each area. 

\begin{table}[tbh]
    \begin{center}
     \begin{tabular}{p{3.5cm}  p{6cm} p{0.2cm}  p{5cm} p{0.2cm} p{1cm}}	
	\hline
	Topic & Recent progress &  & Key future goals    & Section\\
	\hline\hline
	Gravitational wave asteroseismology 
        & Relativistic-Cowling formulae, alternative parametrisations based on moment of inertia/compactness   instead of radius. 
        & & Formalism with dynamical spacetime. Find  ``optimal'' set of asteroseismology parameters.  
	                                    & &\ref{sec:seismo}\\
	\hline
	$f$-mode instability    
        & Relativistic-Cowling instability windows with realistic equation of state. Saturation amplitude.
        Application to supramassive post-merger remnants and constraints from short gamma-ray bursts. 
        Spin-temperature evolutions in Newtonian gravity. 
	& & Models with dynamical spacetime. Obtain saturation amplitude using relativistic gravity and fast rotation.  
	                                      & & \ref{sec:fmode}\\  
	\hline 
	$r$-mode instability   
        & Several $r$-mode unstable systems in minimum damping models. Upper limits on mode amplitude 
	from low-mass X-ray binaries and millisecond pulsar data and implications. $r$-mode puzzle: small amplitude
        or enhanced damping?
	Multifluid $r$-mode modelling. Modified instability window due to mode resonances. 
	& & Improve understanding of ``conventional'' damping: Ekman layer with elastic crust, magnetic field and 
	paired matter. Vortex-fluxtube interactions. New saturation mechanisms.
	                                  & & \ref{sec:rmode} \\
	\hline
	Magnetar oscillations
        & Models of magneto-elastic oscillations including superfluidity, crust phases, complex magnetic field structures.
        General relativistic magnetohydrodynamical simulations of magnetic field instability.
        & & Analytical expressions for magnetar asteroseismology with quasi-periodic oscillations.
        Models of giant flares.
        & & \ref{sec:magnetars}\\
	\hline
	Mountains
        & Deformations of neutron stars with exotic matter and/or 
        pinned superfluidity. Constraints from short gamma-ray bursts.
        Observational evidence suggesting the existence of
        thermal mountains in accreting neutron stars. 
	& & Understanding stability of magnetic field configurations in neutron stars.
        Rigorous magnetic field modelling in the presence of exotic matter. 
        Estimates of the actual size of thermal and 
        core mountains. 
        & &\ref{sec:mountains}\\
	\hline 
	Pulsar glitches
        & Numerical simulations of pulsar glitches. Models of the post-glitch relaxation.
        Models of collective vortex motion using the Gross-Pitaevskii equation; extraction of the gravitational
        wave signal.
        & & Interface between Gross-Pitaevskii vortex simulations and hydrodynamical
        models, including magnetic field and crust-core coupling.
        Magnetohydrodynamical simulations of pulsar glitches.
        Modelling of post-glitch relaxation including magnetic fields and mutual friction.
        & &\ref{sec:glitches}\\
	 \hline\hline
	 \end{tabular}
\end{center}
\caption{Summary of topics discussed in this review. }
\label{tab:summary}
\end{table}

Several equations presented below feature physical parameters normalised to some canonical value. 
The physical meaning of these parameters and their normalisations are summarised in Table~\ref{tab:norms}.

Oscillation modes are assumed to be proportional to $e^{i\omega t + im\varphi}$.
We also make frequent use of the term ``canonical neutron star parameters''. This refers to
a non-rotating neutron star with a typical mass $M = 1.4 M_\odot$ and typical radius $R = 12\,\mbox{km}$. 
Although we discuss stellar models constructed with various equations of state (EOS) for matter,
our frequently used canonical model will be that of a simple $n=1$ Newtonian polytrope.

\begin{table}[tbh]
\begin{center}
\begin{tabular}{p{4cm}  p{1.5cm}  p{2.0cm} }
\hline\hline
Stellar mass    &  $M_{1.4}$          & $M/1.4 M_\odot$            \\
Stellar radius  & $ R_6 $               &  $R/10^6\,\mbox{cm}$    \\
Spin period     & $ P_{-3} $            &  $P/1\,\mbox{ms} $        \\
Spin period  derivative& $ \dot{P}_{-10} $   &  $\dot{P}/10^{-10} $        \\
Proton fraction & $ x_{{\rm p} 1}$  &  $ x_{\rm p}/0.01  $     \\
Magnetic field  &  $B_{\rm n} $       & $ B/10^{\rm n} \,\mbox{G} $  \\
Distance        & $D_1$   & $D/1\,\mbox{kpc}$   \\
Moment of inertia    &   $I_{45}$   &    $I/10^{45}\mbox{ g cm}^2$   \\
\hline\hline
\end{tabular}
\end{center}
\caption{Table of frequently used normalised parameters.}
\label{tab:norms}
\end{table}

The remainder of the chapter is structured as follows: in Section~\ref{sec:seismo} we discuss neutron star GW 
asteroseismology. Section~\ref{sec:cfs} provides a basic ``primer'' introduction to the theory of unstable oscillation 
modes in rotating neutron stars. The two most important cases of unstable modes, the $f$-mode and the $r$-mode
are discussed in detail in Sections ~\ref{sec:fmode} \& \ref{sec:rmode} respectively. Section~\ref{sec:magnetars} 
is devoted to magnetars and reviews the properties (including the emission of GWs) of the oscillations that are believed to 
be excited in these objects by their magnetic field activity. The physics and GW detectability of neutron stars with quadrupolar 
deformations (i.e. neutron star ``mountains'') is discussed in Section~\ref{sec:mountains}.  GW emission associated
with pulsar glitches is discussed in Section~\ref{sec:glitches}. Finally, our concluding remarks can be found in 
Section~\ref{sec:conclusions}.


\section{Neutron star asteroseismology}
\label{sec:seismo}

The premise behind the notion of neutron star GW ``asteroseismology'' is more or less the same as the one of the more
traditional helioseismology: use GW observations of pulsating neutron stars as probes of their interiors. In fact,
neutron star asteroseismology is a bit more specialised: the key idea is to parametrise an oscillation mode's angular
frequency $\omega$ and GW damping timescale $\tau_{\rm gw}$ in terms of the neutron star's three bulk parameters, that
is, the mass $M$, the radius $R$ and the angular frequency $\Omega$.  A clever combination of these parameters could
then allow the construction of ``universal'' (i.e. EOS-insensitive) parametrised relations. Then, an actual observation
of a pulsating neutron star, through the measurement of the $\omega, \tau_{\rm gw}$ pair for one or more modes and the
inversion of the parametric relations, would in principle allow the inference of the three basic stellar
parameters. Moreover, if the mass of the neutron star is known, a measurement of its oscillation modes would give direct
information on the neutron star EOS: its stiffness, the presence of hyperons/quarks,
etc.~\cite{Benhar:2004xg}.

The basic idea of neutron star asteroseismology was first put forward almost two decades ago\, \cite{NAKK98}.  This
initial work was focused on the asteroseismology of the fundamental $f$-mode and that was done for a good reason: the
$f$-mode is the oscillation most likely to be excited in violent processes such as neutron star mergers or neutron star
formation by supernova core-collapse. The $f$-mode has the extra advantage of being a copious emitter of gravitational
radiation -- the same property is responsible for the mode's rapid damping.

The main drawback of the early astreroseismology models was the assumption of non-rotating systems -- astrophysical
neutron stars always have some degree of rotation. In fact, the most relevant scenarios from the point of view of GW
detectability of pulsation modes are likely to involve rapidly rotating systems.
 It was thus imperative that the asteroseismology scheme should become ``fast'' so that it could be applied to rapidly
spinning neutron stars.  At the same time it should be able to adapt to the possibility of having modes growing in
amplitude rather than decaying under the emission of GWs (this is the so-called CFS instability which is discussed in
more detail in Section~\ref{sec:cfs}).


\subsection{Fast relativistic asteroseismology}
\label{sec:fast}
 
The problem of extending the asteroseismology scheme to rotating neutron stars was tackled just a few years 
ago \cite{gaertig_KK2011, doneva_etal2013}. This new effort was based on relativistic stellar models within the 
so-called Cowling approximation (where only the fluid is perturbed while the spacetime metric remains a fixed background) 
and was again focused on the $f$-mode. The resulting parametrised formulae for $\omega, \tau_{\rm gw}$  are constructed 
with a  two-step procedure and are polynomial-type fits to the numerical data. The typical accuracy of these fits lies in the range 
between few percent and few tens of percent, the larger error typically associated with high spin rates. 
 
In the first step we have an expression for the mode's frequency $\omega_0$ of a non-rotating star of the same
gravitational mass as the rotating one \cite{NAKK98, gaertig_KK2011, doneva_etal2013} ,
\be
\frac{\omega_0 (\mbox{kHz})}{2\pi} =  \kappa_\ell + \mu_\ell  \left ( \frac{M_{1.4}}{R_6^3} \right )^{1/2}.
\label{om_seismo1}
\ee
This fit is clearly inspired by the simple Newtonian relation between the $f$-mode's frequency and the mean stellar density,  
$\omega \sim \rho^{1/2}$. For the most interesting case of a quadrupolar ($\ell=2$) mode the numerical coefficients are 
$ (\kappa_2,\mu_2) =  (0.498, 2.418)$.

In the second step, the mode frequency $\omega_r$ as measured in the stellar rotating frame is expressed 
in terms of  $\Omega$ normalised by the mass shedding angular frequency  $\Omega_\rK$ (i.e. the Kepler limit).
The resulting parametric formula is the quadratic expression 
\cite{gaertig_KK2011, doneva_etal2013},
\be
\frac{\omega_r^{u,s}}{\omega_0} = 1 + a_\ell^{u,s} \frac{\Omega}{\Omega_\rK} + b_\ell^{u,s}\left (  \frac{\Omega}{\Omega_\rK} \right )^2,
\label{om_seismo2}
\ee
where the labels $u, s$ stand for the $m>0$ (retrograde and potentially unstable) and $m<0$ (prograde and stable) $f$-mode 
branches, respectively. The stable branch turns out to be the one admitting the easiest parametrisation since \emph{all} relevant $m=-\ell$ 
modes are well approximated by the single pair of coefficients $(\alpha_\ell^s,b_\ell^s) = (-0.235,-0.358)$. In contrast, the unstable
branch requires different coefficients for each $m=\ell$ multipole. For example, for the lowest multipole we have $(a^u_2,b^u_2) =(0.402, - 0.406)$.   
In all cases the mode frequency in the inertial frame can be recovered with the help of the general formula,
\be
\label{eqomegair}
\omega_i = \omega_r -m \Omega.
\ee
The parametrisation of the $f$-mode's damping (or growth) timescale proceeds along the same lines. First, the damping timescale $\tau_0$
of the non-rotating system's mode is parametrised as:
\be
\frac{1}{\tau_0 (\mbox{s})} = \frac{M_{1.4}^{\ell+1}}{R_6^{\ell+2}}\left[ \nu_\ell  + \xi_\ell \left(\frac{M_{1.4}}{R_6}\right)\right].
\label{tau_seismo1}
\ee
This expression makes direct contact with the  $f$-mode timescale of a Newtonian uniform density star 
$\tau_{\rm gw} \sim R (R/M)^{\ell+1}$ \cite{detweiler1975}. For the particular case of the $\ell=2$ multipole 
the numerical coefficients are $(\nu_2, \xi_2) = (78.55, -46.71 )$  \cite{doneva_etal2013}.

The timescale $\tau_{\rm gw}$ for the modes of the rotating system are polynomial expansions with respect to 
the frequency ratio $\omega_r/\omega_0$ (or $\omega_i/\omega_0$). 
For the stable $f$-mode branch we have \cite{gaertig_KK2011, doneva_etal2013}
\be
\frac{\tau_{\rm gw}}{\tau_0} = \left [\, c_\ell + d_\ell \left ( \frac{\omega^s_r}{\omega_0} \right )  
+ e_\ell \left ( \frac{\omega_r^s}{\omega_0} \right )^2 + f_\ell \left ( \frac{\omega_r^s}{\omega_0} \right )^3 \, \right ]^{2\ell},
\label{tau_seismo2}
\ee
where, unlike the previous frequency fit, different multipoles require different sets of coefficients. 
For example, for the $m=-2$ mode these take the values $(c_2,d_2,e_2,f_2) = ( -0.127,  3.264,  -5.486, 3.349)$.

This time it is the unstable branch that is well approximated by a single formula for all $\ell=m$ multipoles 
\cite{gaertig_KK2011, doneva_etal2013}
\be
\frac{\tau_{\rm gw}}{\tau_0} = \mbox{sgn} (\omega_i^u)  \left ( \frac{\omega^u_i}{\omega_0} \right )^{-2\ell}   
\left [\, 0.9 - 0.057 \left ( \frac{\omega_i^u}{\omega_0} \right ) +  0.157\left ( \frac{\omega_i^u}{\omega_0} \right )^2 \, \right ]^{-2\ell},
\label{tau_seismo3}
\ee
where we can notice the appearance of the inertial frame frequency $\omega_i$. This expression has the CFS instability of the 
$f$-mode (see Section~\ref{sec:cfs}) hardwired in it since the change of sign in $\tau_{\rm gw}$ is synced with the  change of sign in $\omega_i$ 
(or, equivalently, with the transition of the mode's pattern speed from retrograde to prograde with respect to the stellar rotation, see
Eq. (\ref{OmCFS}) below).  


\subsection{$R$ or $I$ asteroseismology?}
\label{sec:Iseismo}

The previous $f$-mode formulae are constructed from the basic three parameters $M, R$ and $\Omega$.  However, this may not 
be, after all, the best choice of parameters for building ``universal'' asteroseismology expressions. Seeking an optimal
parametrisation with a higher degree of EOS-independence than the previous results, recent work has followed an alternative approach 
that relies on the use of  the stellar moment of inertia $I$ instead of the radius $R$. The original $f$-mode model for non-rotating 
stars \cite{lau_etal2010} was subsequently generalised to rapidly rotating systems by Doneva \& Kokkotas \cite{doneva_KK2015}. 
The final outcome of that work comprises  expressions that directly relate the $f$-mode's inertial frame frequency and damping/growth 
time to $M,\Omega,I$ without the need to involve the mode properties of a non-rotating star. We thus have for the stable and unstable $f$-mode 
branches,
\begin{align}
M_1 \omega_i^u (\mbox{kHz})&= a_\ell^u + b_\ell^u M_1 \Omega_1 + c_\ell^u (M_1 \Omega_1)^2 
+  \left  [\, d_\ell^u + e_\ell^u M_1 \Omega + f_\ell^u (M_1 \Omega)^2 \, \right ] \eta,
\label{om_seismo3}
\\
M_1 \omega_i^s (\mbox{kHz}) &= a_\ell^s + b_\ell^s M_1 \Omega_1 +   \left  (\, d_\ell^s + e_\ell^s M_1 \Omega_1  \, \right ) \eta,
\label{om_seismo4}
\end{align}
where $M_1 = M/M_\odot, \,\Omega_1 = \Omega/\mbox{kHz}$ and the effective compactness parameter
\be
\eta^2 =  M^3_1 I_{45}^{-1},
\ee
serves as a proxy for the moment of inertia\footnote{As pointed out in \cite{doneva_KK2015}, 
this asteroseismology formalism could equally well have been based on the compactness $M/R$ 
rather than $\eta$.}.
The numerical coefficients appearing in (\ref{om_seismo3}), (\ref{om_seismo4}) 
are  tabulated in Ref. \cite{doneva_KK2015} and of course should not be confused with the ones of the preceding section. 
For the $m=\pm 2$ modes these coefficients take the values:
\begin{align}
(a_2^u,b_2^u,c_2^u,d_2^u,e_2^u,f_2^u) &= (-1.76, -0.143, -0.00665, 3.64, -0.0436, 0.002),
\\
(a_2^s,b_2^s, d_2^s, e_2^s ) &= (-1.66, -0.249, 3.66, 0.0633).
\end{align}

Moving on to the $\tau_{\rm gw} $ timescale, only the formula for the potentially unstable branch has been constructed 
\cite{doneva_KK2015}:
\be
\frac{1}{\tau_{\rm gw} (s)} = \frac{\eta^{2(1-\ell)}}{M_1} \left [\, g_\ell^u M_1 \left (\frac{\omega_i^u}{\mbox{kHz}}\right ) 
+ h_\ell^u M_1^2 \left ( \frac{\omega^u_i}{\mbox{kHz}} \right)^2 \, \right ]^{2\ell},
\label{tau_seismo4}
\ee
where for the $\ell=m=2$ mode the coefficients take the value $(g_2^u, h_2^u) = (0.644, 0.0207)$.

Clearly, these ``$I$-asteroseismology'' relations are algebraically simpler than their ``$R$-asteroseismology'' counterparts
of the preceding section. This is combined with an enhanced universality with respect to the EOS of matter, although
at the price of a lower overall accuracy caused by the adoption of the Cowling approximation \cite{doneva_KK2015}.


\subsection{Future directions}

As described in the preceding sections, the last few years have seen an impressive advance in the modelling
of $f$-mode asteroseismology in neutron stars. There is, however, much room for further improvement or extension. 
The main simplification affecting the accuracy of the entire asteroseismology structure is the adoption of the Cowling 
approximation in the numerical computation of the $f$-modes of relativistic stars. The error caused by this
approximation can be easily gauged by calculating the $f$-mode frequency of a uniform non-rotating star in Newtonian
gravity. The outcome of this straightforward exercise is $\omega^2 =  8\pi G\rho f_\ell/3$ with $f_\ell = \ell/2$
and  $f_\ell = \ell(\ell-1)/(2\ell+1)$  when the Cowling approximation is used or not used, respectively. Although this simple
calculation slightly overestimates the Cowling-induced error (which is typically $\sim 10-30\,\%$ ) it does tell us that the 
approximate result systematically lies above the exact one and that higher multipoles are less affected. 
These same trends are indeed  found in more rigorous general relativistic (GR) $f$-mode calculations with or without the Cowling
approximation \cite{zink_etal2010}. 
For a given error in the mode frequency the corresponding error in the GW timescale $\tau_{\rm gw}$ is much
more pronounced because of the steep dependence of this parameter with respect to $\omega$.  This is clearly
exemplified by comparing the expressions for the $\ell=2$ $f$-mode damping timescale in non-rotating stars. 
The coefficients in the Cowling-approximated formula (\ref{tau_seismo1}) are about a factor three larger than the
ones appearing in the non-Cowling expressions of \cite{NAKK98, gaertig_KK2011}. This large mismatch is caused by 
the $\sim \omega^{2\ell+2}$ dependence of the GW damping rate (see Eq.~(\ref{dEgw}) below) and the aforementioned
error in the mode frequency. 

The upshot of this discussion is that a further improvement of the asteroseismology scheme should be based on the 
removal of the Cowling approximation and the use of full GR computations of  $f$-modes in rapidly rotating stars. 
This effort could be combined with further experimentation with the variables used in the asteroseismology 
formulae along the lines discussed in this section. For example, is $\Omega$ the ``best'' parameter for representing
rotation or is there a more suitable parametrisation (e.g. the Kerr parameter $cJ/GM^2$, where $J$ is the stellar angular
momentum)?

Neutron star asteroseismology should also be extended beyond the $f$-mode. Indeed, as the modelling of neutron stars
improves including more and more physics, new classes of modes appear. Although the $f$-mode is likely to be the most
relevant for GW detection at the birth and at the death of a neutron star, other modes could be relevant at different
stages of its life. Magneto-elastic oscillations - a class of modes associated to the elastic strain of the crust and to
the magnetic field in the crust and in the core - can be excited in giant flares of strongly magnetized neutron stars;
they are discussed in detail in Sec.~\ref{subsec:magnqpo}. Moreover, when a superfluid phase is present in the neutron
star core, the multiplicity of the oscillation modes doubles (see the discussion at the end of
Sec.~\ref{sec:extra_damping}). At characteristic values of the temperature (of the order of $\sim10^8$ K), the damping
times of these ``superfluid'' modes can become small enough to make them - at least in principle - efficient sources of
GWs~\cite{Gualtieri:2014lsa}. Although the superfluid phase is not present in the most violent stages of the neutron
star life, superfluid $g$-modes~\cite{Gusakov:2013eoa,Kantor:2014lja,Passamonti:2015oia,Dommes:2015wul} can be unstable
by convection in the first year of neutron star's life, or - if hyperons are present - can be excited in neutron star's
coalescences; superfluid $r$-modes~\cite{gusakov_etal2014a,gusakov_etal2014b} will be discussed in
Sec.~\ref{sec:extra_damping}.

Neutron star asteroseismology requires the knowledge of analytic expressions of the frequencies and damping times of the
modes in terms of the fundamental parameters of the star, such as those in Eqns.~(\ref{om_seismo1}), (\ref{om_seismo2}),
(\ref{tau_seismo1}) -
(\ref{om_seismo4}), (\ref{tau_seismo4}). In recent years, these quantities have been computed in specific models for
magneto-elastic modes and superfluid modes, but we still need a better understanding of the underlying physics, in order
to derive analytic expressions useful for asteroseismology.


\section{Unstable oscillation modes in rotating stars: a CFS theory primer}
\label{sec:cfs}

The theory of instabilities in rotating self-gravitating fluid systems came of age in the 1970s with the seminal work of 
Chandrasekhar, Friedman and Schutz \cite{chandra1970, friedman_schutz1978a, friedman_schutz1978b, friedman1978} 
but its seeds had been sown much earlier with the development and completion 
of the Newtonian theory of classical ellipsoids \cite{chandra_book}. 
Nowadays this framework is widely known as the CFS theory or CFS instability mechanism and 
includes two types of instabilities, the secular and the dynamical. In reality they all are
rotational ``frame-dragging'' instabilities that are based on the notion of perturbations with a 
conserved canonical energy that can become zero (dynamical case) or even negative (secular case) 
above a certain rotational frequency threshold. Apart from this similarity the two instability types 
are formally very different: the secular instability requires the fluid to be coupled to some dissipative mechanism
such as GWs or viscosity while the dynamical instability can take place in an entirely dissipationless system.

Our own discussion of the CFS theory is focussed on the GW-driven secular instability of normal modes which is the most relevant 
mechanism for ``normal'' neutron stars. In contrast, the dynamical instability, which is briefly discussed at the end of this section, 
is known to require the presence of differential rotation in the system (unless we consider less realistic, uniform density stellar models) 
and therefore can take place for very brief periods of time and in very special environments such as neutron star mergers. 
Given the ``primer'' character of this section  our presentation of the CFS theory is very far from being comprehensive; the reader 
can find more detailed reviews of the subject in Refs. \cite{NAKK01,NA03, FSbook, NAGC2007}.

The rotational threshold for the onset of the secular CFS instability marks the transition of an initially counter-moving mode 
to a co-moving one as perceived by an inertial observer. In slightly more technical terms, a mode becomes CFS-unstable
when its pattern speed (that is, the azimuthal propagation speed of a constant phase wavefront),
\be
\sigma_p = \frac{d\varphi}{dt} =  - \frac{\omega_i}{m},
\ee
changes sign, going from negative to positive. This is achieved when the stellar angular frequency $\Omega$
exceeds the mode's pattern speed in the rotating frame (see Eq.~(\ref{eqomegair})): 
\be
\Omega >   \frac{\omega_r}{m}\,.
\label{OmCFS}
\ee
This condition is the first of the two prerequisites for triggering the secular instability. The second one is the coupling of
the oscillation to a dissipative mechanism which would allow the mode's negative canonical energy to grow (in absolute value)
while more and more positive energy is removed from the system.  Assuming coupling to GWs, the radiated power $\dot{E}_{\rm gw}$  
is equal and opposite to the change in the mode energy $E_{\rm mode}$ and  can be calculated with the standard multipole moment 
formula \cite{Thorne80}:
\be
\dot{E}_{\rm gw} = -\dot{E}_{\rm mode} = \omega_r \sum_{\ell \geq 2} \omega_i^{2\ell+1} N_\ell \left [\, | \cM_{\ell m} |^2 + | \cS_{\ell m} |^2 \, \right ],
\label{dEgw}
\ee
where $\cM_{\ell m}, \cS_{\ell m}$ are the mass and current multipoles respectively and $N_\ell$ a positive constant. 
In fact, this same formula allows a quick ``rediscovery'' of the CFS instability since it predicts $\dot{E}_{\rm mode} >0$
for $\omega_i < 0$, i.e. the criterion (\ref{OmCFS}). The GW timescale (for both stable and unstable modes) is defined as
(see, e.g. \cite{IL1991})
\be
\tau_{\rm gw} = \frac{2E_{\rm mode}}{\dot{E}_{\rm gw}}.
\label{tgw_def}
\ee

It should be emphasised that the CFS instability is not strictly a GW-driven effect; electromagnetic radiation and/or
normal shear viscosity can also drive unstable oscillations\,\footnote{We note with some amusement that the literature
  contains a mechanical pendulum analogue of the viscosity-driven CFS instability in a 1908 paper by Lamb
  \cite{lamb1908}. Another occurrence of a CFS-like instability in the context of wave dynamics can be found in the 1974
  textbook by Pierce (Chapter 11) \cite{pierce_book}.}
but as it turns out
gravitational radiation is almost always the dominant mechanism. In a realistic scenario, an oscillation mode can become
CFS-unstable provided the mode-dragging condition (\ref{OmCFS}) is satisfied and the growth rate due to GW emission is
faster than the viscous damping rate.  The damping timescale associated with a given viscosity mechanism is defined as
\be \tau_{\rm visc} = \frac{2E_{\rm mode}}{\dot{E}_{\rm visc}}.
\label{tvisc_def}
\ee
where $\dot{E}_{\rm visc}$ is the damping rate. In the most common case where several dissipative mechanisms
are operating simultaneously the viscous timescales are combined according to the ``parallel resistors'' rule. 

The resulting instability parameter space is commonly depicted as a ``window'' in the $\Omega-T$ plane and is typically
a V-shaped region limited by bulk/shear viscosity at high/low temperatures and by the spin threshold (\ref{OmCFS}) 
from below. 
Note that in this review we shall always consider (unless otherwise specified) the standard form for these viscosities, 
namely, shear viscosity due to electron-electron scattering and bulk viscosity due to the modified URCA reactions
(the corresponding viscosity coefficients can be found in \cite{sawyer1989, andersson_etal2005b}). The boundary of the 
instability window (also known as the critical curve) is then defined by, 
\be
\tau_{\rm gw}^{-1} = \tau_{\rm sv}^{-1} + \tau_{\rm bv}^{-1},
\label{inst_window}
\ee
where $\tau_{\rm sv}$ and $\tau_{\rm bv}$ are respectively the shear and bulk viscosity damping timescales and
the growth timescale $\tau_{\rm gw}$ is hereafter taken to be a positive parameter. The extension of (\ref{inst_window})
to cases with additional dissipation is straightforward.

Among the various neutron star oscillation modes, the ones most promising to become potential sources of GWs and
play an important role in the dynamics of neutron stars via the CFS mechanism are the fundamental $f$-mode and
the inertial $r$-mode. Appropriately, these have been the subject of most work in this area and we discuss them in 
some detail in Sections~\ref{sec:fmode} \& \ref{sec:rmode}. 

The $f$-mode is also the one associated with the dynamical CFS instability\footnote{Remarkably, a similar CFS dynamical
  instability is known to exist in an entirely different context, namely, that of a rotating liquid drop with surface
  tension playing the role of gravity. Laboratory experiments have probed the impact of the instability on the shape of
  the drop and have revealed a rich phenomenology, see e.g. \cite{brown_scriven1980, hill_eaves2008}.}
(for that reason it is nicknamed the ``bar-mode'' instability from the dominant $\ell=|m|=2$ multipole).  The instability
erupts once the system reaches a critical value $\beta_d$ of the rotational to gravitational binding energy ratio $\beta
= T/|W|$. Mathematically this amounts to the merger of the $m= \pm \ell$ mode frequencies.  The $\beta_d$ threshold is
significantly higher than the corresponding $\beta_s$ for secular mode instabilities and for realistic models of rigidly
rotating neutron stars it lies beyond the Kepler break-up limit. What this means in practice is that the dynamical
$f$-mode instability is likely to appear in systems that are formed from the outset with fast and differential rotation
so that $\beta > \beta_d$. This could happen in the immediate aftermath of binary neutron star merger -- a scenario that
is corroborated by numerical simulations, for some relatively recent work see e.g.  \cite{baiotti_etal2007,
  manca_etal2007,franci_etal2013}.  The ensuing bar mode instability acts as a potent source of gravitational radiation
but its duration is severely limited by the resulting rapid spin-down of the merger remnant and the quenching of
differential rotation by the magnetic field \cite{shapiro2000}.

The neutron star arsenal of CFS instabilities can be more extensive than what has been described so far if we allow 
for a high degree of differential rotation or a superfluid component. In the former case a low-$\beta$ (i.e. $\beta \ll \beta_d$) 
dynamical instability may arise via the system's quadrupole $f$-modes. This shear instability was first stumbled upon 
in numerical studies of the dynamics of differentially rotating neutron stars \cite{shibata_etal2002} and is akin to shear 
instabilities in accretion disks. Since then low-$\beta$ instability has been seen to be excited in core-collapse numerical 
simulations (see e.g. \cite{kuroda_umeda2010}). Follow up work \cite{watts_etal2005, passamonti_andersson2015} 
established the instability's intimate connection with the presence of a corotation band (that is, a region where the mode's 
pattern speed matches the local $\Omega$) in the stellar interior.

The presence of a superfluid opens the possibility for another type of $r$-mode CFS instability. 
This  newly discovered ``two-stream'' instability \cite{KGNA2009, andersson_etal2013} requires a spin lag between the 
neutron and proton fluids and is driven by vortex mutual friction instead of gravitational radiation. Although this instability 
could be a viable mechanism for triggering pulsar glitches \cite{KGNA2009} it remains to be seen if it is of any relevance 
as a source of GWs.


\section{The $f$-mode instability}
\label{sec:fmode}

\subsection{Newtonian origins and recent developments}

The $f$-mode was chronologically  the first to be studied in relation with the CFS instability, with initial work dating
back to the 1980s \cite{friedman1983, IL1989, IL1990, IL1991}.  The main conclusion of those early papers 
was rather disappointing, predicting that  in Newtonian stellar models and under the dissipative action of standard 
shear and bulk viscosity the instability sets in at a rotation rate very close to the Kepler limit, typically 
$\Omega \approx 0.95 \Omega_\rK$. 
Subsequent work \cite{lindblom_mendell1995} showed that the instability is essentially wiped out in neutron stars 
with a superfluid core as a result of the dissipative coupling between the superfluid's vortex array and the electrons 
(this is what we call ``standard mutual friction'' \cite{als1984, andersson_etal2006}).  This result has been considered 
as a showstopper for the $f$-mode instability since neutron stars are expected to become superfluid very soon after 
they are formed. This leaves as the only realistic candidates for harbouring unstable $f$-modes newly formed neutron 
stars with hot non-superfluid matter (and obviously with fast rotation). 

The traditional astrophysical scenario where these requirements could be met is that of neutron star formation in the
aftermath of a core-collapse supernova. The $f$-mode GW detectability of such an event was first analysed in detail in
\cite{lai_shapiro1994}. This scenario is somewhat less favoured nowadays due to the uncertainty in forming near-Kepler
limit spinning neutron stars (see \cite{stergioulas2003} and references therein).  The alternative scenario that has
attracted much attention in the recent years is that of binary neutron star mergers -- a posterchild source of GWs.
These violent events may provide the only suitable arena for the occurrence of the $f$-mode instability by playing the
role of cosmic factories of metastable \emph{supramassive} neutron stars\footnote{The term ``supramassive'' refers to a
  rapidly rotating neutron star with mass above the maximum allowed mass for spherical non-rotating neutron stars. The
  excess mass is supported by rotation which means that a supramassive system is dynamically stable only above a certain
  spin rate.}.  As we are about to see in the following section it is this property of high mass, in combination with
fast rotation and relativistic gravity, that makes these systems prime candidates for harbouring unstable $f$-modes.

\begin{figure*}[htb!]
\includegraphics[width=0.45\textwidth]{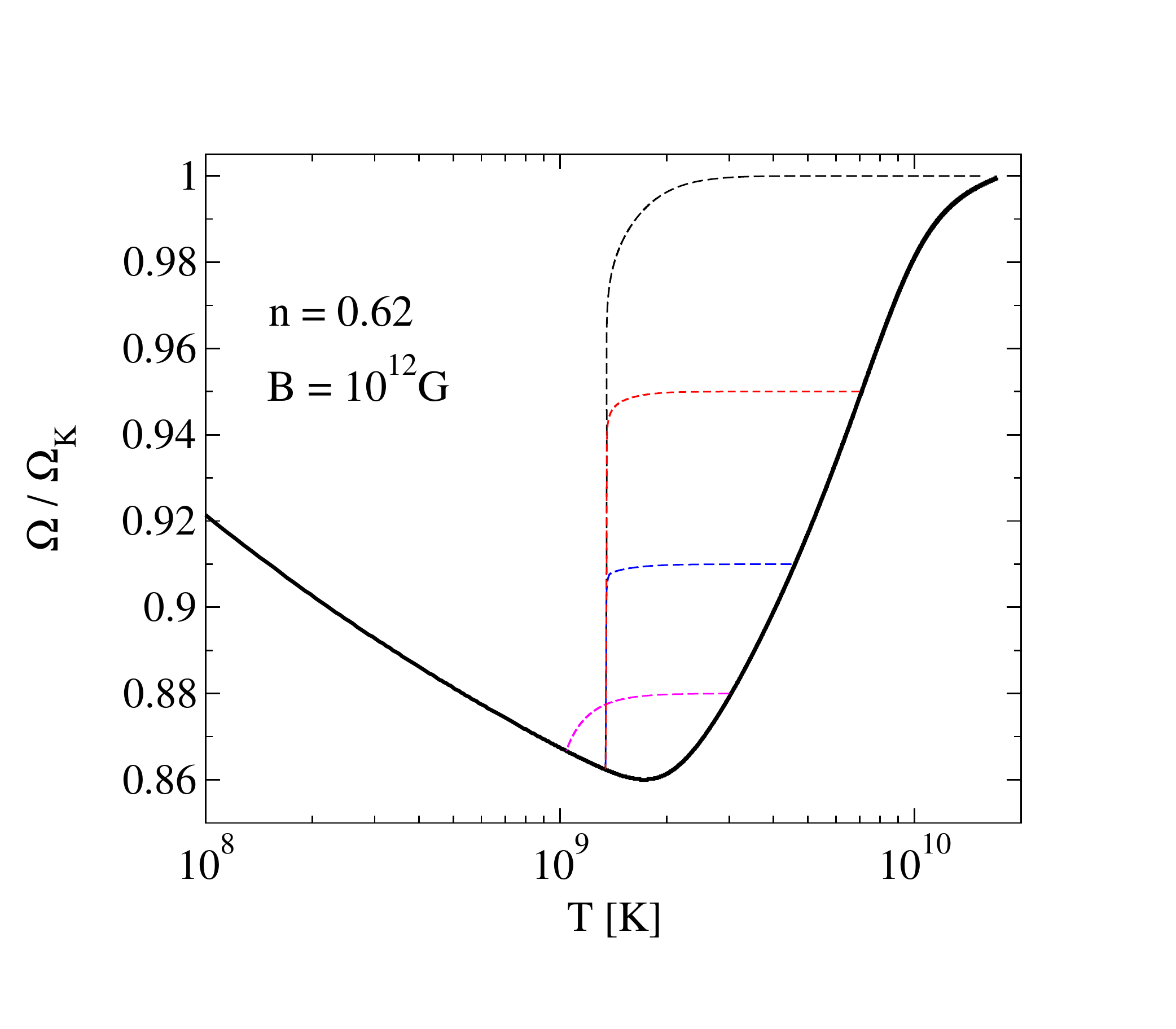}
\includegraphics[width=0.45\textwidth]{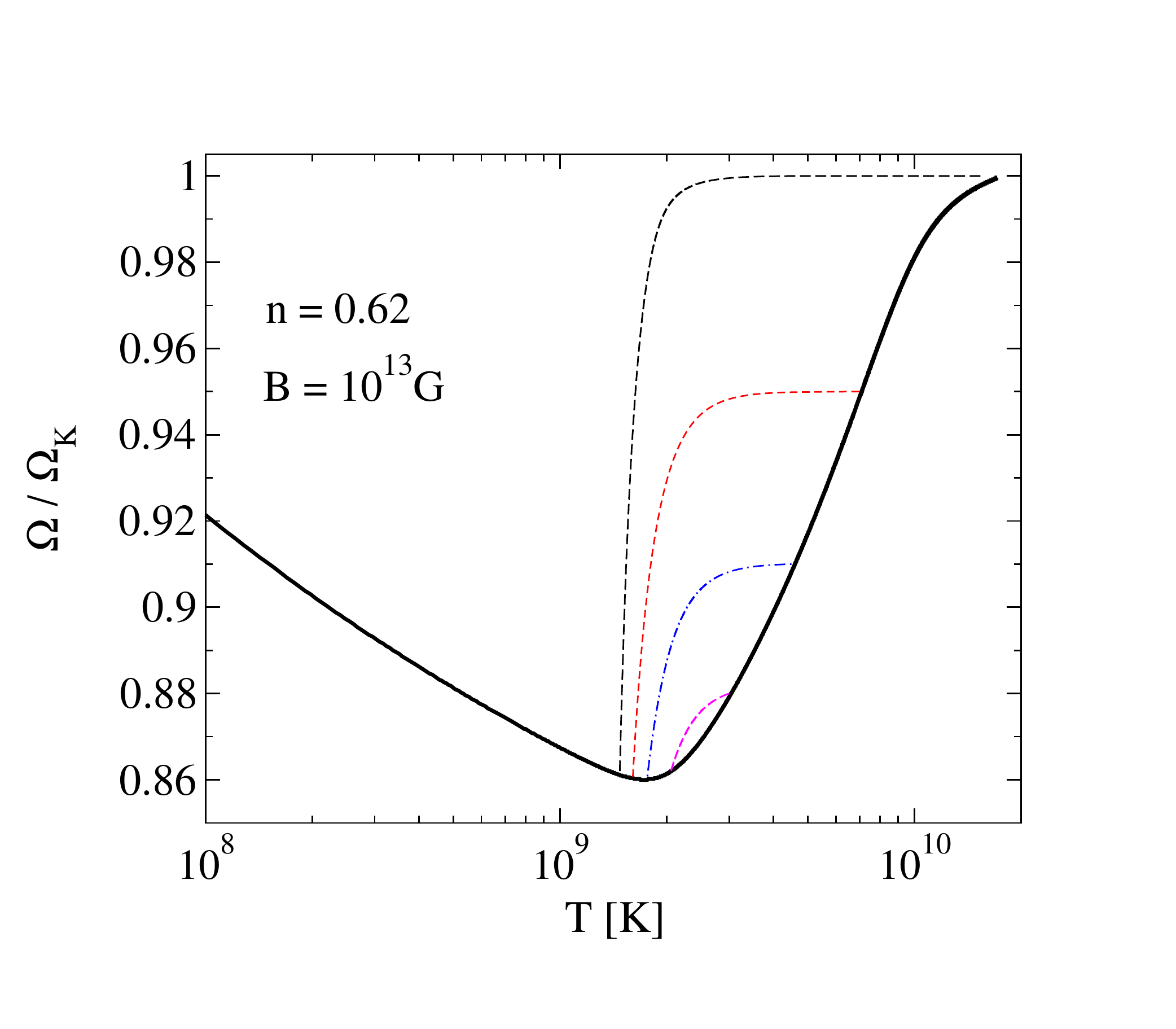}
\caption{$f$-\emph{mode spin-temperature evolution}:  adapted from Ref.~\cite{passamonti_etal2013},  this figure
displays the coupled $\Omega-T$ evolution of a Newtonian $n=0.62$ polytropic model with gravitational mass
(in the non-rotating limit) $M=1.98\,M_\odot$ under the combined influence of an 
unstable $\ell=m=4$ $f$-mode and magnetic dipole torque. The thick solid curve represents the instability window 
(accounting for standard shear and bulk viscosity but not for superfluid mutual friction). Left panel: the various evolution 
trajectories correspond to different initial $\Omega$ and a surface dipole field $B = 10^{12}\,\mbox{G}$. 
Right panel: the same initial $\Omega$ values as before for a surface dipole field  $B = 10^{13}\,\mbox{G}$.}
\label{fig:fmode_evolution}
\end{figure*}

But before opening the discussion on the $f$-mode instability in relativistic stars it is worth reviewing the recent progress
achieved with the use of the ``outdated'' Newtonian framework. The first problem to be addressed was that of the
coupled spin-temperature evolution of a neutron star undergoing an $f$-mode instability \cite{passamonti_etal2013}.
Using a formalism similar to that previously employed for the $r$-mode instability \cite{owen_etal1998}, this work
considered hot and rapidly rotating newly formed systems entering the instability window as the temperature drops below 
$\sim 10^{10}\,\mbox{K}$ -- this value marks the regime where bulk viscosity becomes negligibly weak. 
The unstable $f$-mode is promptly saturated, at which point the star begins to spin down with emission of gravitational 
radiation at almost constant temperature until it exits the instability window. 

A representative example of this evolution is shown in Fig.~\ref{fig:fmode_evolution} for the most unstable 
$\ell=m=4$ $f$-mode of a massive $M=1.98\,M_\odot$ stellar model with a $n=0.62$ polytropic EOS.
A first noteworthy feature of these $f$-mode trajectories is that the system is unlikely to ever 
enter the region where neutrons pair to form a superfluid phase (this is expected to happen at a temperature 
$T \lesssim 10^9\,\mbox{K}$) and the instability is suppressed by vortex mutual friction.
Even more important is the interplay between the stellar magnetic field and the unstable $f$-mode. 
A neutron star with surface field $B \gg 10^{12}\,\mbox{G}$ will evolve along a shorter spin-temperature trajectory (see right 
panel of Fig.~\ref{fig:fmode_evolution}) with its spin evolution mostly driven by the magnetic dipole radiation rather 
than gravitational radiation, hence leading to a deteriorated GW detectability.  A similar situation may arise if in parallel with the 
$f$-mode there is also an unstable $r$-mode present in the system -- a not unlikely scenario given the
much wider instability window of the latter, see Ref. \cite{passamonti_etal2013} for more details. 

A second key recent development concerns saturation amplitude (or equivalently saturation energy) of the instability. 
This arduous calculation was undertaken in \cite{pnigouras_KK2015, pnigouras_KK2016} by means of a nonlinear mode-coupling model 
similar to the one that had been employed before for the $r$-modes (see Section~\ref{sec:rmode}) but with the 
added property of non-uniform, stratified matter. According to the aforementioned work the mode's energy is primarily drained by the 
non-linear coupling to $g$-modes; this leads to a saturation energy that may fluctuate considerably across the instability's 
$\Omega-T$ parameter space. A representative maximum value for the saturation energy can be taken to be 
$E_{\rm sat} \sim 10^{-6} M c^2$ \cite{pnigouras_KK2015, pnigouras_KK2016}. This result has obvious implications for the 
GW detectability of unstable $f$-modes and will be discussed in more detail below.


\subsection{The $f$-mode instability in relativistic stars: a story of revival?}
\label{sec:revival}

\begin{figure*}[htb!]
\includegraphics[width=0.35\textwidth]{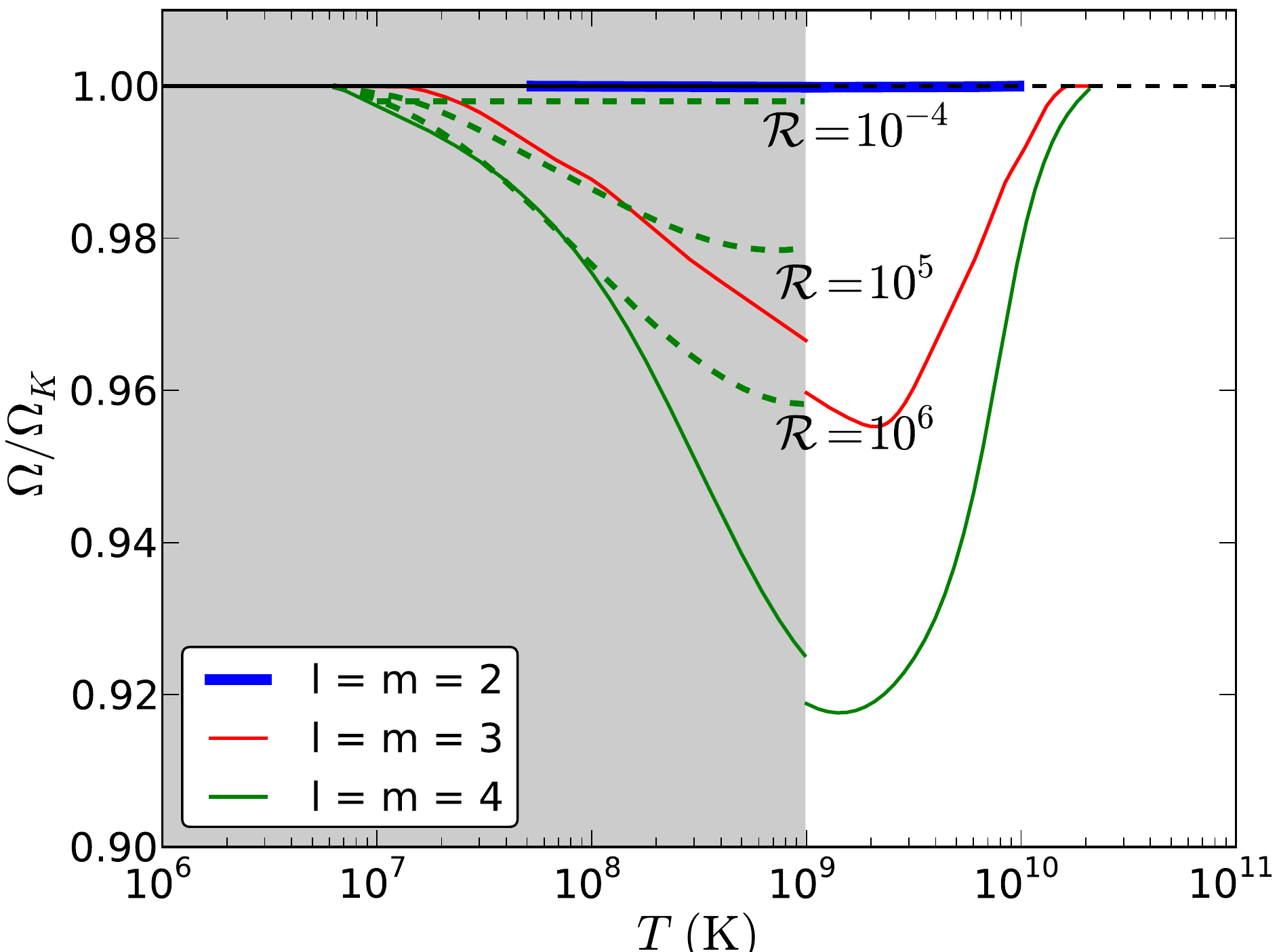}
\includegraphics[width=0.35\textwidth]{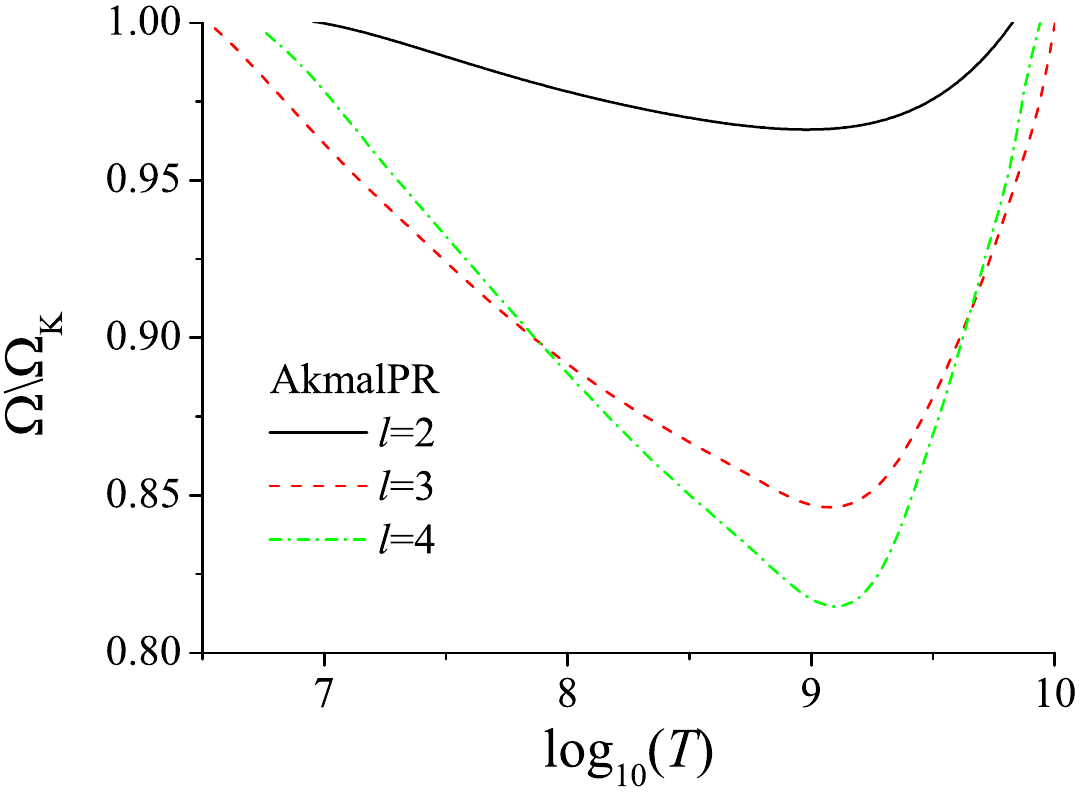}
\caption{\emph{Relativistic $f$-mode instability window}. Left panel: this figure, taken from Ref. \cite{gaertig_etal2011}, shows 
the $f$-mode instability window (solid curves) corresponding to the first three multipoles $\ell=m=2,3,4$ and for a $n=0.73$ 
relativistic polytrope with parameters $M= 1.48 M_\odot$ and $R= 10.47 \,\mbox{km}$ (in the $\Omega=0$ limit). 
The shaded area indicates the parameter space where neutron superfluidity is present (with a fiducial onset temperature 
$T_{\rm cn}=10^9\,\mbox{K}$ ). The critical curves are discontinuous at $T_{\rm cn}$ 
as a result of the use of different shear viscosity coefficients in the superfluid and normal region. 
The dashed curves show the instability window of the $\ell=4$ multipole with vortex mutual friction accounted for
and the drag coefficient shifted from its canonical value $\cR \approx 10^{-4}$ (which leads to a fully suppressed instability). 
Right panel: the instability window for the same $f$-mode multipoles (this time assuming non-superfluid matter) 
and for a neutron star model with a realistic EOS (in this particular example APR) and parameters 
$M = 2 M_\odot$, $R= 10.88\,\mbox{km} $, see Ref. \cite{doneva_etal2013}.}
\label{fig:fmode_window1}
\end{figure*}

The last few years have seen a renewed interest in the $f$-mode instability with the objective of revising
the earlier Newtonian results using  relativistic and rapidly rotating neutron star models. This effort was spearheaded 
by Ref. \cite{gaertig_etal2011} which considered polytropic models and relativistic gravity in the Cowling approximation. 
These first relativistic results were promising, predicting a revised $f$-mode growth timescale about an order
of magnitude shorter than the Newtonian value for the same canonical stellar parameters. Accordingly, the instability window
was found to be larger than its Newtonian counterpart, with the $\ell=m=4$ being the most unstable multipole, 
see Fig.~\ref{fig:fmode_window1} (left panel). Follow-up work \cite{doneva_etal2013} established that the combination 
of a realistic EOS model with a higher stellar mass ($\approx 2 M_\odot$) can support an even wider instability window
-- this is shown on the right panel of Fig.~\ref{fig:fmode_window1} and can be directly compared against the 
Newtonian window of Fig.~\ref{fig:fmode_evolution} (which corresponds to a Newtonian polytrope of the same mass). 

These calculations allow us to draw a clear
conclusion: for a given rotation $\Omega$, massive relativistic systems (which are also the most compact ones) have 
significantly enhanced $f$-mode instability properties as compared to their Newtonian and/or less massive counterparts.

\begin{figure*}[htb!]
\includegraphics[width=0.3\textwidth]{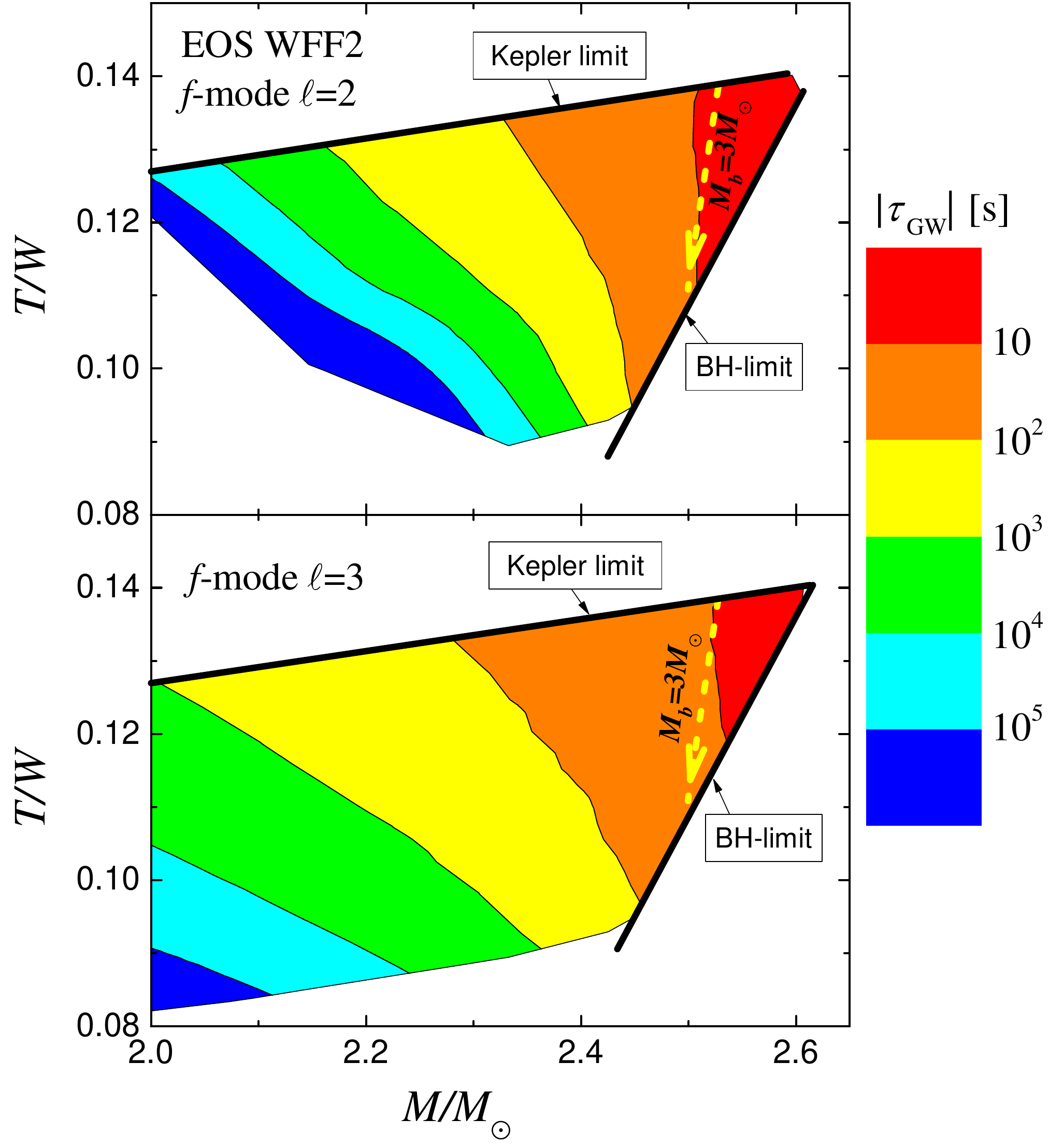}
\includegraphics[width=0.5\textwidth]{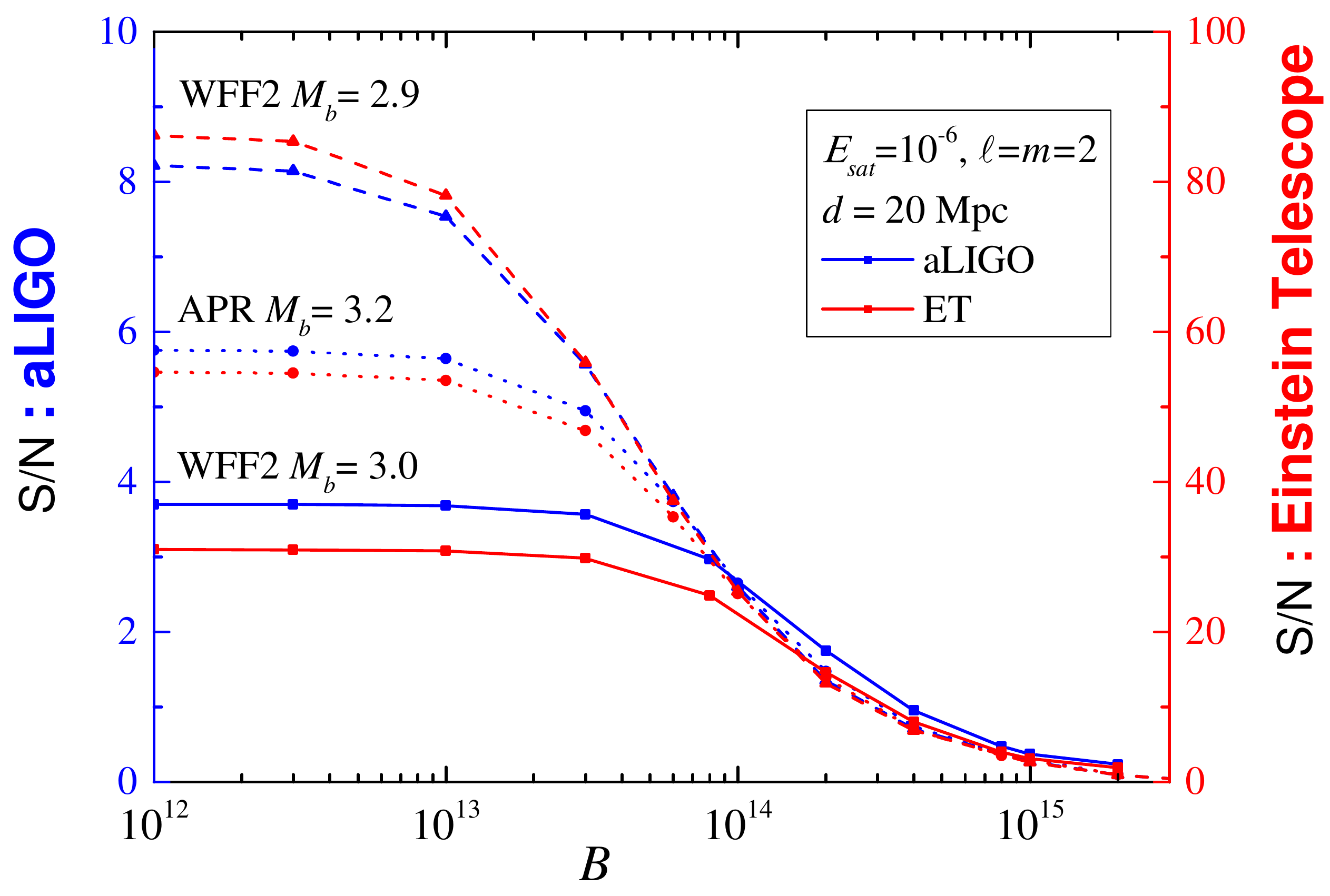}
\caption{$f$-\emph{mode instability in supramassive neutron stars}. Left panel: this figure is adapted from Ref. \cite{doneva_etal2015} 
and shows the $f$-mode growth timescale (for the first two multipoles and assuming a WFF2 EOS stellar model) as a function
of the supramassive neutron star's mass.  Right panel: the GW signal-to-noise ratio as a function of the surface dipole field
for a supramassive neutron star located at a fiducial distance $20\,\mbox{Mpc}$, spinning down under the combined influence of 
the $f$-mode instability (saturated at $E_{\rm sat} = 10^{-6} M c^2$) and magnetic dipole radiation, for different choices of 
baryonic mass and EOS (figure adapted from \cite{doneva_etal2015}).}
\label{fig:fmode_afterglow}
\end{figure*}

The most dramatic manifestation of this conclusion may take place in the immediate aftermath of the merger 
of a binary neutron star system: for the expected range of initial masses the merger may not immediately produce a black hole but, 
instead, lead to a transient phase of a supramassive ($M \sim 2.5 M_\odot$) and $\Omega \approx \Omega_\rK$ 
neutron star remnant \footnote{A relatively low mass remnant may settle down to a normal neutron star existence
without ever collapsing to a black hole.}. 
Besides their central role in GW astronomy, these mergers have come to be seen as the leading theoretical model for the central 
engine powering short GRBs (see e.g. \cite{dai_lu1998, zhang_meszaros2001,rowlinson_etal2013}). 
The supramassive remnant, which now takes the form of a proto-magnetar as a result of
strong magnetic field amplification (see e.g. \cite{rezzolla_etal2011, kiuchi_etal2014, giacomazzo_etal2015}), 
is believed to power the burst's late-time emission and is associated with the  X-ray plateau and power-law tail seen 
in the light curves of several of these events \cite{rowlinson_etal2013, metzger_piro2014}.  According to the GRB data, 
the remnant's  lifetime spans a range $10^2-10^5\, \mbox{s}$ which is determined by the spin-down timescale
due to magnetic dipole radiation (the same mechanism is responsible for powering the system's X-ray emission).

It is during this X-ray ``afterglow''/supramassive phase where the $f$-mode instability is most likely to take place and become
a potentially strong source of GWs. Viscosity is not likely to be an impeding factor in these circumstances since the system is 
expected to cool below $10^{10}\,\mbox{K}$ very shortly after the supramassive remnant has been formed, see e.g. \cite{PLKG2016}. 
This scenario has been put forward in Ref. \cite{doneva_etal2015} and is backed up by $f$-mode calculations 
that suggest surprisingly short growth timescales, $\tau_{\rm gw} \sim 10-100\,\mbox{s}$, see Fig.~\ref{fig:fmode_afterglow} (left panel).
However, a short $\tau_{\rm gw}$ does not necessarily translate into a conspicuous $f$-mode GW signal. To what extent post-merger
supramassive remnants could be realistic targets for present and next generation GW detectors is discussed in more
detail in the following section.  


\subsection{The observability of the $f$-mode instability}

The overall amplitude of the $f$-mode signal is limited by the distance of the source and the mode's maximum saturation
amplitude.  As we have already seen, the latter parameter was recently obtained and expressed as a saturation energy,
$E_{\rm sat} \sim 10^{-6} M c^2$ \cite{pnigouras_KK2015, pnigouras_KK2016}.  A perhaps more intuitive way to quantify
this result is via the $f$-mode-induced ellipticity in the stellar shape. A back-of-the-envelope calculation leads
to~\cite{PLKG2016}, \be \frac{\delta R}{R} \sim \left ( \frac{E_{\rm sat}}{M c^2} \right )^{1/2} \left ( \frac{cR}{GM}
\right )^{1/2}, \ee which for a typical neutron star compactness returns $ \delta R/R \sim 10^{-3}$. In other words, the
mode gets saturated at a ``linear'' level. The GW observability of an unstable $f$-mode saturated at this maximum
amplitude is shown in Fig.~\ref{fig:fmode_afterglow} (right panel) in the form of a signal-to-noise ratio~(SNR) for the
Advanced LIGO/Virgo and ET detectors as a function of the surface dipole field
\cite{doneva_etal2015}. Detectability is strongly diminished in systems with magnetar-like fields for the simple reason
that the spin-down timescale is controlled by magnetic dipole radiation and is much shorter than the duration of a
GW-driven spin-down (see also Fig.~\ref{fig:fmode_evolution}). Less magnetised systems with
$B \lesssim 10^{14}\,\mbox{G}$ are likely to be marginally detectable by Advanced LIGO but should be ``in the bag'' for ET.

A more empirical assessment of the $f$-mode's GW observability can be made with the help of the short GRB X-ray 
data \cite{PLKG2016}. The $ \sim t^{-2}$ late time decay profile seen in several light curves of these events can be 
taken as evidence of an electromagnetic radiation dominated spin-down, in accordance with the GRB proto-magnetar 
model. This information, in combination with the observed duration of the X-ray plateaus, can be used to set upper 
limits in the saturation amplitude of unstable $f$-modes. Interestingly, the resulting limit is similar to the theoretically 
predicted maximum amplitude -- this result suggests that the $f$-mode instability could in principle play an important 
role in the dynamics of the post-merger remnant. Unfortunately, the predicted $f$-mode detectability is rather pessimistic 
even for ET, limited by the shortness of the spin-down timescale and the large distances ($\gtrsim 500\,\mbox{Mpc}$)  
associated with short GRBs.


\subsection{Future directions}
\label{sec:fmode_future}

The $f$-mode calculations discussed in the preceding sections were carried out using the Cowling approximation. 
We have already pointed out in a previous section how this approximation affects the $f$-mode asteroseismology formalism.
As far as the $f$-mode instability is concerned, it is known \cite{zink_etal2010, yoshida2012} that the Cowling approximation 
makes relativistic stars less prone to the CFS instability by increasing the rotation threshold (\ref{OmCFS}). It is therefore expected 
that the $f$-mode instability in neutron star models with fully relativistic dynamical spacetime will be enhanced but a detailed 
quantitative calculation of this modification is still lacking.

Another desideratum in this area should be the further improvement in the modelling of the $f$-mode's non-linear saturation 
physics. The very recent state-of-art calculation of \cite{pnigouras_KK2015, pnigouras_KK2016} is ``primitive'' in the sense that
it is based on a Newtonian framework and a slow rotation approximation. Taking this calculation to the next level (with
relativistic gravity and/or fast rotation) is likely to prove a very challenging -- but necessary -- endeavour.

Finally, it should be borne in mind that the $f$-mode instability window could be significantly modified by viscosity
due to the presence of exotic phases of matter in the interior of neutron stars, such as hyperons and quarks.  
Although this scenario has been exhaustively explored in the context of the $r$-mode instability (see Section~\ref{sec:rmode}) 
very little is known about its impact on the  $f$-mode instability.


\section{The $r$-mode instability}
\label{sec:rmode}

The discovery of the inertial $r$-mode CFS instability in 1998 came as something of a surprise to the neutron star 
community \cite{NA98, FM98}. Since then this instability has received the lion's share of the published work 
on the subject of neutron star oscillations as a consequence of its potentially key role in the spin evolution of neutron stars 
and as a promising source of GWs (for early comprehensive reviews on the subject  see \cite{NAKK01, FSbook}; 
a more specialised recent review can be found in \cite{haskell15}). Accordingly, this $r$-mode section occupies 
a central place in this review.

Two key characteristics are associated with the $r$-mode instability: 

(i) the mode frequency in the rotating frame is 
\be
\omega_r = \frac{2m \Omega}{\ell(\ell+1)},
\ee
so that with dissipation switched off, the mode becomes CFS-unstable as soon as the star acquires rotation 
(that is, the condition (\ref{OmCFS}) is automatically satisfied for any $\Omega$). 
With dissipation restored, the resulting $\Omega-T$ instability window is in general quite large. 

(ii) the mode's predominant axial geometry implies a nearly horizontal  fluid flow pattern and leads to GW emission 
dominated by the current multipole $\cS_{22}$ rather than the mass multipoles. The resulting  growth timescale
$\tau_{\rm gw}$ exhibits a characteristic $\sim \Omega^{-6}$ dependence and is very short.
For a canonical $n=1$ Newtonian polytropic star this is \cite{lindblom_etal1998, NAKK01},
\be
\tau_{\rm gw} \approx 50 \, M_{1.4}^{-1} R_6^{-4} P_{-3}^6\,\mbox{s}. 
\label{tgw_rmode}
\ee

The two above properties alone are sufficient to guarantee the astrophysical relevance of the $r$-mode instability. 
In addition, practitioners of neutron star dynamics enjoy the luxury of being able to do most of $r$-mode physics
within a Newtonian/slow rotation framework rather than having to struggle with the complexity 
of rapidly rotating GR stars (as it was the case for the $f$-mode instability). 
Nonetheless, relativistic $r$-mode calculations have been performed 
\cite{lockitch_etal2001, ruoff_KK2001, yoshida_lee2002, lockitch_etal2003, idrisy_etal2015}
and have demonstrated the corrections to the Newtonian mode eigenfunction GW growth timescale to be typically very small. 
At a qualitative level, GR changes the purely axial slow-rotation Newtonian $r$-mode into an axial-led inertial mode -- this is 
similar to the modification caused by fast rotation in Newtonian theory. More pronounced is the relativistic correction to the mode 
frequency which is of the order of $\sim 20\%$. This is large enough to be accounted for in searches for GW signals from 
unstable $r$-modes \cite{owen2010}. 
The $r$-mode's insensitiveness extends to variations in the EOS of matter as well. This has been demonstrated by
the very recent analysis of \cite{idrisy_etal2015} and is in agreement with earlier investigations that made use of 
polytropic models \cite{lockitch_etal2003}. The upshot is that $r$-mode calculations based on a canonical polytropic Newtonian 
model (or even a uniform density model) are sufficiently robust and accurate for most practical applications. This will
be our benchmark model for the reminder of our discussion of the $r$-mode instability unless otherwise specified.

\subsection{$r$-mode phenomenology}
\label{sec:rmode_pheno}

\begin{figure*}[htb!]
\includegraphics[width=0.5\textwidth]{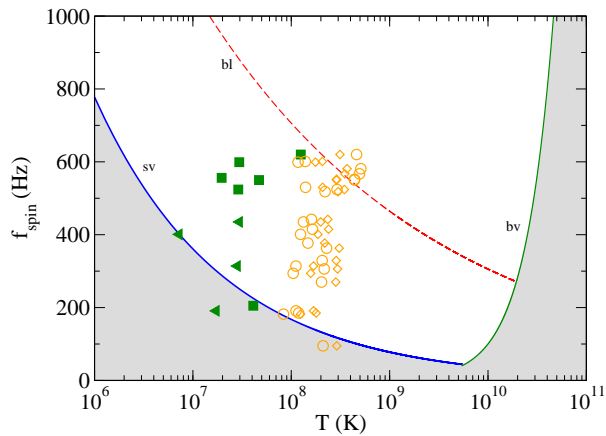}
\caption{\emph{$r$-mode instability window}. This figure shows the $r$-mode spin-temperature 
instability window (in terms of the spin frequency $f_{\rm spin} = \Omega/2\pi$) assuming a canonical neutron 
star model and accounting for standard shear (blue curve) and bulk (green curve) viscosity. The corresponding 
viscous damping timescales were taken from \cite{lindblom_etal1998} and \cite{NAKK01} while Eq.~(\ref{tgw_rmode}) 
was used for the growth timescale. The shaded area represents the region of $r$-mode stability. 
Besides the minimum damping window, the figure also includes the critical curve due to Ekman boundary layer 
damping (red dashed curve) which is discussed in Section~\ref{sec:crust}. 
The data points represent LMXBs with known spin frequencies. The core temperature is inferred by flux measurements 
during phases of quiescence \cite{MS2013} (filled squares and triangles for upper limits) or is 
theoretically predicted assuming a combined $r$-mode spin and thermal equilibrium during phases of accretion 
\cite{Ho_etal2011} (open diamonds and circles for cooling dominated by mURCA reactions 
and Cooper pair processes respectively) see Section~\ref{sec:small} for details.}
\label{fig:rmode_window1}
\end{figure*}

Most of the uncertainty related to the $r$-mode instability has to do with its damping. 
In a ``minimum physics'' (although not necessarily realistic!) model that assumes dissipation only 
due to standard shear and bulk viscosity the resulting instability window overlaps with the 
box-shaped region occupied by the known population of rapidly rotating neutron stars in LMXBs 
and MSPs. This $r$-mode window is shown in Fig.~\ref{fig:rmode_window1}, where we have also included
LMXB data tabulated in \cite{MS2013, Ho_etal2011} (to be discussed below). 
The striking feature of  Fig.~\ref{fig:rmode_window1} is that several known neutron stars
reside well inside the minimum damping window and therefore should harbour unstable 
$r$-modes. This observation forms the basis of what we shall call the ``$r$-mode puzzle'' in the following section. 

\begin{figure*}[htb!]
\includegraphics[width=0.6\textwidth]{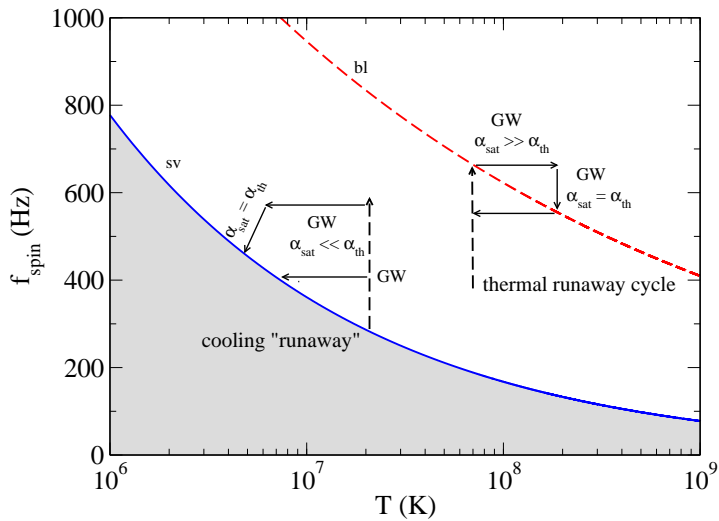}
\caption{\emph{$r$-mode spin-temperature evolution}. The $r$-mode-driven $f_{\rm spin}-T$ evolution of an
accreting neutron star is determined by (i) the instability curve's slope at the point where the system,
spun up by accretion, first enters the instability window (dashed vertical lines) and (ii) the maximum saturation amplitude
$\alpha_{\rm sat}$. In this figure we show the canonical scenario of a negative slopping instability curve due to
shear viscosity (blue curve) or an Ekman layer (red curve). 
If the saturation amplitude is not too small ($ \alpha_{\rm sat} \gg \alpha_{\rm th}$ initially), the system is 
forced into a cyclic thermal runaway loop in the vicinity of the instability curve where the GW emitting portions are 
indicated in the figure, see Sections~\ref{sec:rmode_pheno} \& \ref{sec:small} for more details. The cycle's nearly 
horizontal path ($r$-mode heating) continues until thermal equilibrium is established, $\alpha_{\rm sat} = \alpha_{\rm th}$,  
followed by $r$-mode spin-down towards the instability curve. In the opposite scenario of a very small saturation amplitude 
(i.e. $ \alpha_{\rm sat} \ll \alpha_{\rm th}$ initially) a weak $r$-mode instability is active during accretion
and, once the latter has ended, cannot prevent the system from cooling (at nearly constant spin) 
towards a state of thermal equilibrium or until the instability curve is crossed again (this scenario is shown
here with respect to the minimum damping curve).}
\label{fig:rmode_window2}
\end{figure*}

The spin distribution of these neutron stars has been something of a mystery since under the 
unhindered action of accretion, LMXBs (and their MSP descendants) should have been expected to straddle the Kepler 
frequency limit, $f_\rK \gtrsim  1\,\mbox{kHz}$  (see also e.g. the discussion in \cite{Haskell:2012vg}).
The apparent spin cut-off at a much lower frequency has been taken as evidence of the presence of a spin-down torque 
that counteracts accretion. A suggestion that has attracted much attention since its conception is that of a GW torque supplied 
either by a deformation in the stellar shape (i.e. a neutron star ``mountain'') or an unstable oscillation mode 
\cite{PP78, bildsten1998, andersson_etal1999}. This idea obviously combines well with the minimum dissipation $r$-mode 
model but we should not be too hasty in drawing conclusions.  Spin equilibrium in LMXBs could be achieved by an alternative 
\emph{non}-GW mechanism, namely, the coupling of the stellar magnetic field with the accretion disk 
\cite{ghosh_lamb1979, wang1995, rappaport_etal2004, andersson_etal2005} and it is here fitting to open a parenthesis and discuss it.

In the disk coupling model the global stellar magnetic field threads the material of the disk and the field lines, being 
simultaneously anchored in the disk and on the stellar surface, provide a very efficient braking mechanism. 
With the input of canonical surface dipole fields, $B \sim 10^8-10^9\,\mbox{G}$, and reasonable physical assumptions
the predictions of the available phenomenological disk coupling models compare fairly well with LMXB spin data
\cite{andersson_etal2005, haskell_patruno2011, patruno_watts2012} thus  dispelling much of the mystery behind 
their spin distribution cut off . Although this is compelling evidence in favour of these models it is certainly premature to 
shelve the alternative GW-based mechanisms. In fact, recent work \cite{bhattacharyya_chakrabarty2017, bhattacharyya2017} 
suggests that the transient nature of accretion could seriously weaken the efficiency of the disk coupling model
thus calling for an additional spindown torque. What can be said with certainty is that the existing data do not exclude 
the realistic possibility of having some $r$-mode activity (or indeed a neutron star mountain) in LMXB systems that could 
otherwise be dominated by magnetic disk coupling or by  magnetic dipole spin-down when in quiescence. 
The disk coupling model does, however, remove the need to cling to solely GW-based spin equilibrium
models. With this observation in mind we can resume our main $r$-mode discussion.

Once inside the instability window, the $r$-mode drives a coupled spin-temperature evolution the details of
which are largely determined by the maximum (saturation) value of the mode amplitude $\alpha_r$. 
This dimensionless parameter is commonly defined in  the literature via the velocity field of the dominant $\ell=m$ 
mode \cite{owen_etal1998}
\be
\delta \mathbf{v} = \alpha_r  \left (\frac{r}{R} \right )^\ell \Omega R \mathbf{Y}_{\ell\ell}^B e^{i\omega_r t},
\ee 
where $\mathbf{Y}_{\ell m}^B$
is a vector spherical harmonic 
of the magnetic type. More simply, we can think of the amplitude as the ratio $\alpha_r \approx \delta v/\Omega R $.

Considering first accreting neutrons stars, once the system moves across a $d\Omega / dT <0$ segment of 
the instability curve during spin up it will undergo a cyclic thermal runaway 
\cite{levin99, andersson_etal2000, bondarescu_etal2007}, see Fig.~\ref{fig:rmode_window2}.

The cycle begins with the mode growing under the emission of gravitational radiation, rapidly reaching its
maximum saturation value $\alpha_{\rm sat}$. When this happens, and assuming $\alpha_{\rm sat} \ll 1$, 
the viscous timescale becomes comparable to the growth timescale,  $\tau_{\rm visc} \approx \tau_{\rm gw}$, 
(see the $r$-mode evolution equations in \cite{owen_etal1998, levin99}). 
At the same time the heat deposited by shear viscosity lifts the stellar core temperature. This happens at 
almost constant spin frequency since the GW spin-down timescale, 
$ \tau_{\rm sd}^{\rm gw} \equiv \Omega / |\dot{\Omega}| $, is always much longer than the heating timescale 
$\tau_{\rm heat} \equiv T/|\dot{T}|$. These timescales are (see e.g. \cite{levin99}):
\be
\tau_{\rm sd}^{\rm gw}  \approx \frac{5.3}{\alpha_{\rm sat}^2} \tau_{\rm gw}, 
\qquad 
\tau_{\rm heat} = \frac{C_{\rm v} T}{\dot{E}_{\rm visc}} 
\approx 7.8 \times 10^{-6}\, \frac{P_{-3}^2\, T_8^2}{\alpha_{\rm sat}^2 M_{1.4}   R_6^{2} } \,   \tau_{\rm gw}.
\label{tau_sd+heat}
\ee
where in the last equation the damping rate $\dot{E}_{\rm visc}$ is typically fixed by shear viscosity or a crust-core boundary 
layer (see below) and the heat capacity $C_{\rm v} \approx 1.4\times 10^{38}\, T_8 \,\mbox{erg}/\mbox{K}$ is that of ordinary 
matter \cite{STbook} (superfluidity would reduce $\tau_{\rm heat}$ even further by decreasing $C_{\rm v}$).  
Also note that we have explicitly used the timescale equality $\tau_{\rm visc} \approx \tau_{\rm gw}$ for a saturated mode.

The thermal runaway continues up to the point where stellar cooling can efficiently balance viscous heating. 
Assuming neutrino cooling due to the  standard modified URCA process (with the corresponding emissivity 
$L_\nu \approx 7 \times 10^{31}\, T_8^8 \,\mbox{erg}/\mbox{s}$) \cite{STbook} and a minimum damping 
instability curve, we can convert thermal equilibrium into a ``thermal'' $r$-mode amplitude,
\be
\dot{E}_{\rm visc} =  \dot{E}_{\rm sv} = L_\nu ~ \Rightarrow ~\alpha_{\rm th} 
\approx 1.4 \times 10^{-9} M_{1.4}^{-1} R_6^{-3} P_{-3}^4 T_8^4.
\label{alpha_th}
\ee
Note that this result is indicative as it obviously depends on the assumed cooling physics.
For the temperature range of LMXBs the cooling could be dominated by superfluid Cooper-pair processes
\cite{Ho_etal2011} or even surface photon emission \cite{MS2013}.

At the stage of thermal equilibrium the temperature remains essentially constant and the $r$-mode-driven spin-down 
steers the system towards the instability curve; once the curve is crossed, the star becomes stable again and rapidly cools. 
Eventually, accretion will once again spin up the star preparing the ground for the next cycle. 
The duration of the GW-emitting portion of this cycle depends on  $\alpha_{\rm sat}$. 
A large amplitude (the first papers to explore the implications of this $r$-mode evolution were somewhat optimistically
assuming $\alpha_{\rm sat} \sim 0.1-1$, see e.g. \cite{andersson_etal2000}]) translates to fast GW-driven evolution but 
a tiny GW duty cycle, making LMXBs uninteresting sources of GWs. If $\alpha_{\rm sat}$ is small, the system does 
not wander off much from the instability curve and the cycle's GW efficiency can improve dramatically. 
We can make this argument quantitative by approximating the cycle's GW efficiency as the ratio between the 
time the system spends emitting radiation and the typical LMXB lifetime \cite{heyl2002},
\be
D_{\rm cycle} \approx \frac{t_{\rm cycle}}{10^7\,\mbox{yr}} \approx \frac{10^{-11}}{\alpha_{\rm sat}^2}.
\label{Dcyc1}
\ee
Combining this with the estimated LMXB birth rate $\sim 10^{-5} /\mbox{yr} /\mbox{galaxy}$, it is not too difficult 
to see that in order to have a system always switched on in our Galaxy,
\be
D_{\rm cycle} \lesssim 10^{-2} ~ \Rightarrow ~ \alpha_{\rm sat} \lesssim 10^{-4}.
\label{Dcyc2}
\ee
Thus, somewhat counterintuitively, a relatively small amplitude is likely to improve the $r$-mode's GW detectability in
LMXBs. Of course, the amplitude should not be too small for otherwise these sources would be too faint. 
As discussed below, the upper limit (\ref{Dcyc2}) is compatible with the theoretically predicted $r$-mode 
saturation amplitude. 

In the thermal runaway scenario described above the cut off in the spin distribution of LMXBs is set by the spin frequency
at which the systems enter the instability window. Taking into account that the expected temperature range for LMXBs
is $T \sim 10^7- 5\times 10^8\,\mbox{K}$ it becomes immediately clear that the previously defined minimum damping 
window is in disagreement with the observed $\sim 600\,\mbox{Hz}$ cut off (see Fig.~\ref{fig:rmode_window1}).
The model does much better if we invoke a  ``canonical'' $r$-mode instability window which, in addition to
shear and bulk viscosity, accounts for dissipation due to a viscous Ekman boundary layer at the crust-core
boundary. The implications of this Ekman layer could be crucial for the survivability of the $r$-mode instability 
and are discussed in detail below in Sections~\ref{sec:crust} \& \ref{sec:requiem}. 

It should also be emphasised that the cyclic  evolution is \emph{not} an unavoidable outcome of the $r$-mode evolution
in LMXBs. For instance, the presence of exotic neutron star matter in the form of hyperons or quarks could lead to
an instability curve with positive $d\Omega/dT$ slope in the temperature range relevant to LMXBs. As discussed 
by several authors,  this configuration could effectively trap accreting systems near the critical curve and turn them 
into persistent sources of GWs \cite{andersson_etal2002, nayyar_owen2006, haskell_andersson2010}. 
Another way to prevent the cyclic evolution from happening is by invoking a saturation $\alpha_{\rm sat}$ sufficiently 
small so that $\tau_{\rm heat}$ (and consequently $\tau_{\rm sd}^{\rm gw}$) exceeds the accretion timescale 
($\sim 10^7-10^8\, \mbox{yr}$). In this scenario cooling dominates over $r$-mode heating, i.e. 
$\alpha_{\rm sat} \ll \alpha_{\rm th}$, and once accretion comes to an end the system undergoes  a ``cooling runaway'', 
moving towards the low $T$ part of the instability window until thermal equilibrium is established 
or the instability curve is crossed \cite{alford_schwenzer2015}, see Fig.~\ref{fig:rmode_window2}. 
We elaborate more on this small amplitude scenario below in Section~\ref{sec:small}.

The $r$-mode-driven evolution of rapidly rotating non-accreting neutron stars is somewhat simpler than the
cyclic scenario of the preceding paragraphs. These systems can be either old MSPs or, more speculatively, 
very young neutron stars such as the central compact objects (CCOs) associated with supernova remnants. 
The latter objects have already been the target of broad band GW searches by LIGO  and, in spite of their 
unknown spin frequencies (which are likely to be low), have led to direct upper limits on the $r$-mode amplitude 
\cite{aasi_etal2015}. The key  parameter here is the spin-down timescale (\ref{tau_sd+heat}) which becomes 
\be
\tau_{\rm sd}^{\rm gw} \approx 800\, \left ( \frac{10^{-4}}{\alpha_{\rm sat}} \right )^2 M_{1.4}^{-1} R_6^{-4} P_{-3}^6\, \mbox{yr}.
\label{tau_sd2}
\ee
According to this expression, a large amplitude $r$-mode drives a very rapid spin-down thus seriously diminishing 
the system's GW observability. In order to have a $\tau_{\rm sd}^{\rm gw}$ that is compatible with the estimated 
ages of the CCOs (i.e. $\sim 10^2 - 10^3\,\mbox{yr}$) we would need to invoke $\alpha_{\rm sat} \sim 10^{-3}$. 
A much smaller amplitude is required if $\tau_{\rm sd}^{\rm gw}$ is associated with the observed spin-down age of MSPs. 
We will return to this point later, in Section~\ref{sec:small}.


\subsection{An $r$-mode puzzle?}
\label{sec:puzzle}

How do actual observations compare against the basic $r$-mode phenomenology described in the preceding 
section? As we have seen, the minimum damping model clearly predicts that the $r$-mode instability 
should be operating in a large portion of the LMXB and MSP populations (and perhaps in some young sources). 
The fact that these objects do \emph{not} appear to show any evidence of strong $r$-mode activity should have 
important implications for the physics in their interior. For example, and as already pointed out, 
the data points shown in Fig.~\ref{fig:rmode_window1} are clearly incompatible with a thermal runaway cycle taking 
place near the standard shear viscosity instability curve. Similarly, the MSP timing data are incompatible with the 
$r$-mode instability unless $\alpha_{\rm sat}$ is very small. 

A clue is provided by the observed long-term spin-down of accreting millisecond X-ray pulsars (AMXPs) in quiescence (see
\cite{patruno_watts2012} for a review). For example, two of these objects, SAX J1808-3658 and IGRJ00291+5934,
are likely to sit inside the instability window (given the uncertainties on the pulsar parameters),
and therefore could 
experience $r$-mode-driven spin-down. However, the
measured spin-down rate is consistent with that caused by a canonical LMXB magnetic field ($\sim 10^8\,\mbox{G}$), hence
suggesting that the $r$-mode instability is not the dominant effect here \cite{haskell_patruno2011,papitto_etal2011}.

This apparent tension between the minimum damping $r$-mode model and the spin-temperature data 
of known rapidly rotating neutron stars may be dubbed the ``$r$-mode puzzle''.  

There are two complementary ways to make theory and observations mutually compatible. 
The first one relies on the presence of additional damping mechanisms that could modify the instability window and render 
$r$-mode-stable the systems in question. The required extra damping could be provided, for example, by exotic matter in the 
neutron star core, strong superfluid vortex mutual friction or an Ekman-type viscous boundary layer at the crust-core interface. 
The second possibility is that of a small saturation amplitude $r$-mode. In this scenario the $r$-mode instability does operate 
in (at least) some rapidly rotating neutron stars but  $\alpha_{\rm sat}$ is sufficiently small so that the ensuing sluggish $r$-mode 
evolution is compatible with observations. 
We now can take a closer look at these two resolutions of the $r$-mode puzzle.


\subsection{Small amplitude $r$-modes}
\label{sec:small}

Theoretical calculations already provide constraints on $\alpha_{\rm sat}$ and obviously need to be incorporated in 
any realistic $r$-mode model.  The most robust saturation mechanism is provided by non-linear couplings between
the $r$-mode and other (primarily inertial) modes 
\cite{schenk_etal2002, arras_etal2003, bondarescu_etal2007, bondarescu_etal2009}.
A series of impressive \emph{tour de force} calculations have revealed a complicated spin-temperature evolution 
pattern for $\alpha_{\rm sat}$ but as a rule of the thumb estimate for the time-averaged amplitude we can take 
$\alpha_{\rm sat} \sim 10^{-4} - 10^{-3}$ \cite{schenk_etal2002, arras_etal2003, bondarescu_etal2007, bondarescu_etal2009}.

This level of saturation is obviously low  but not low enough for the purposes of the small amplitude scenario! To illustrate this,
we consider MSPs with measured spin-down rates that reside inside the minimum damping instability window. 
These data can be used to set upper limits on the $r$-mode amplitude (obviously, these limits make sense
provided the systems in question are $r$-mode-unstable in the first place -- this may not be the case in a more realistic 
enhanced damping scenario). 
Equating the total rate of change of the stellar rotational energy to the GW  torque leads to a spin-down $r$-mode amplitude 
(see e.g. \cite{owen2010}) : 
\be
\alpha_{\rm sd}  \approx 5.7\times 10^{-3} P^{5/2}_{-3} \dot{P}_{-10}^{1/2} M_{1.4}^{-1/2} R_6^{-2}.
\ee
The strongest constraints from the available MSP data imply a very small amplitude, $\alpha_{\rm sd} \lesssim 10^{-7}$ 
\cite{alford_schwenzer2014b, alford_schwenzer2015}.

Given that MSPs are almost certainly spinning down via standard magnetic dipole radiation, it is meaningful to  
compare the relative $\dot{P}$ contribution of GW emission. It is an easy exercise to derive the following formula 
for the spin-down ratio $\dot{P}_{\rm gw}/ \dot{P}_{\rm em}$ which is also equal to the spin-down age ratio 
$\tau_{\rm sd}^{\rm em}/\tau^{\rm gw}_{\rm sd}$:
\be
\frac{\dot{P}_{\rm gw}}{\dot{P}_{\rm em}} = \frac{\tau_{\rm sd}^{\rm em}}{\tau^{\rm gw}_{\rm sd}} 
\approx 0.23 \left (\frac{\alpha_r}{10^{-7}} \right )^2 M_{1.4}^2 B_8^{-2} P_{-3}^{-4}.
\label{Pdot_ratio}
\ee
 This expression indeed verifies that for canonical MSP parameters the $r$-mode torque is much weaker 
than the electromagnetic one.

Two more key $r$-mode amplitudes can be calculated by making contact with LMXB observations. 
The first one comes from the assumption of spin equilibrium (discussed earlier).  Using a simple fiducial
spin up torque that ignores the effect of the magnetic field on the accretion dynamics leads to \cite{BU2000}
\be
 \alpha_{\rm acc} \approx 1.2 \times 10^{-8} \left (\frac{L_{\rm acc}}{10^{35}\,\mbox{erg}\, \mbox{s}^{-1}} \right )^{1/2} 
 P_{-3}^{7/2}
\ee
where $L_{\rm acc}$ is the accretion luminosity. This estimate is very close to the spin equilibrium amplitude 
obtained in Ref. \cite{MS2013} using a spin-up torque extracted observationally from the time-averaging
over a succession of accretion episodes.  

The second  ``handle'' for estimating $r$-mode amplitudes is provided by the consideration of thermal equilibrium
in LMXBs \cite{Ho_etal2011, MS2013}. Measuring the luminosity of these objects in quiescence allows the inference 
of their surface and core temperature and in turn of their cooling rate. If heating is attributed to the dissipation of a 
steady-state unstable $r$-mode, the mode amplitude can be calculated  by invoking thermal equilibrium. 
This is the approach taken in \cite{MS2013} and is of the same logic that led to the amplitude (\ref{alpha_th}). 

The quiescent LMXBs considered in \cite{MS2013} have $T \lesssim 10^8\,\mbox{K}$ (see data points in 
Fig.~\ref{fig:rmode_window1}) -- this implies a cooling dominated by surface photon emission rather that neutrinos. 
The resulting thermal equilibrium amplitudes lie in the range $\alpha_{\rm th} \sim 10^{-8} - 10^{-7}$. A closer look
at the tabulated data of \cite{MS2013} reveals that for sources with both spin and thermal equilibrium data
$\alpha_{\rm th} \approx 0.1 \alpha_{\rm acc}$.
This means within the minimum damping model,  and treating the inferred $\alpha_{\rm th}$ as
an empirical saturation amplitude,  $r$-modes \emph{cannot} balance the long-term accretion torque in LMXBs. 
Furthermore, using $\alpha_{\rm th}$ in Eq.~(\ref{Pdot_ratio}) we  find $\dot{P}_{\rm gw} \ll \dot{P}_{\rm em}$. 
In other words, quiescent LMXBs are expected to predominantly spin-down via magnetic dipole radiation, 
in agreement with observations \cite{haskell_patruno2011}.

A different angle of approach that makes more contact with the accreting phase of LMXBs rather that
their quiescence assumes \emph{both} $r$-mode thermal and spin equilibrium to take place at the same time. 
Then, the equality $\alpha_{\rm acc} = \alpha_{\rm th}$ relates $T$ to the observables $L_{\rm acc}$ and $P$.
As shown in \cite{Ho_etal2011} the LMXB core temperatures calculated in this way lie in the range 
$T\sim 10^8-5\times 10^8\,\mbox{K}$ and are consistently higher than those in \cite{MS2013}
(see also Fig. ~\ref{fig:rmode_window1}). Of course this model would run into difficulties in explaining the
quiescence data (at least for some systems).  

What is noteworthy from the preceding discussion is that the three ``observable'' amplitudes 
$\alpha_{\rm sd}, \alpha_{\rm acc}, \alpha_{\rm th}$ lie well below the predicted $\alpha_{\rm sat}$ due 
to non-linear couplings. This is clearly problematic from a theoretical point of view and therefore other 
saturation mechanisms must be sought in order to fill the gap. A recently suggested alternative mechanism 
is based on the dissipative coupling between the superfluid vortex array and the quantised magnetic fluxtubes 
in regions of the star where a neutron superfluid co-exists with a proton superconductor \cite{haskell_etal2014}. 
The resulting saturation amplitude could be as small as $\alpha_{\rm sat} \sim 10^{-5}-10^{-6}$. Although this is 
much smaller than the mode-coupling $\alpha_{\rm sat}$, there is still some significant difference with the 
observable amplitudes.  

The small $r$-mode amplitude scenario has been further explored in a recent series of papers by Alford \& Schwenzer
\cite{alford_schwenzer2014a, alford_schwenzer2014b, alford_schwenzer2015}. Following an analysis similar to that of
\cite{bondarescu_etal2007, bondarescu_etal2009} but allowing for a very small $\alpha_{\rm sat} (\Omega,T)$, they derive
steady-state $r$-mode evolution trajectories for LMXBs/MSPs and young neutron stars. As mentioned earlier, in their model the 
cyclic thermal runaway in accreting systems does not take place since the $r$-mode is too feeble to heat up (or spin down)
the star efficiently.  After the end of accretion, the system simply cools until it reaches a steady state at a lower temperature
(see Fig.~\ref{fig:rmode_window2}). This scenario, however, cannot explain the cut off in the LMXB spin distribution since, 
in the absence of any other spin-down mechanism, these systems would be free to reach the Kepler limit.  

At this point it is worth pausing to consider the actual GW detectability of $r$-mode-active neutron stars as a function 
of the amplitude and the spin frequency.

\subsection{$r$-mode detectability}
 \label{sec:rmode_detect}
 
\begin{figure*}[htb!]
\includegraphics[width=0.55\textwidth]{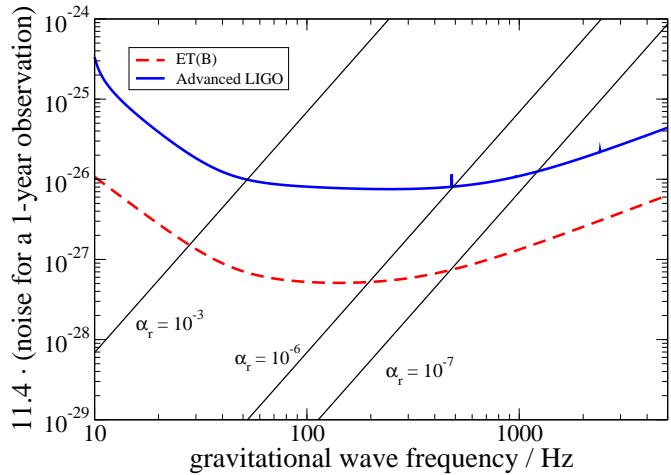}
\caption{\emph{$r$-mode GW detectability}. We show the $r$-mode strain (\ref{h0_rmode}) assuming a one-year
Advanced LIGO/ET phase-coherent observation. For the source parameters we have set $M_{1.4} = 1, R_6 = 1.2, D_1=1$.
The detector noise curves were taken from \cite{LIGOcurve, ETcurve}. The three selected mode amplitudes
represent approximate upper limits (discussed in the main text) for saturation due to (i) non-linear mode couplings 
($\alpha_r = 10^{-3}$), (ii) spin equilibrium in LMXBs  ($\alpha_r= 10^{-6}$) and (iii) MSP spin-down and thermal 
equilibrium in LMXBs ($\alpha_r = 10^{-7}$). }
\label{fig:h0_rmode}
\end{figure*}

Given the broad scope of this review, the discussion of the $r$-mode's detectability will necessarily be brief  -- 
a more detailed recent analysis can be found in \cite{KK_schwenzer2016}. The intrinsic GW strain associated with an 
$r$-mode can be computed with the help of Thorne's multipole moment formula \cite{Thorne80}. The contribution of the 
dominant $\ell=m=2$  current multipole is (see e.g. \cite{owen2010}):
\be
h_0 \approx  4 \times 10^{-23} \,\alpha_r M_{1.4} R_6^3 \left ( \frac{1\,\mbox{kpc}}{D} \right ) 
\left (\frac{f_{\rm gw}}{100\,\mbox{Hz}} \right )^3
\label{h0_rmode}
\ee
where we have normalised the distance $D$ to a galactic source and used the equality 
between the GW frequency and the mode's inertial frame frequency,
\be
f_{\rm gw} = | f_{\rm mode} | = \frac{4}{3} f_{\rm spin}.
\label{fgw_frmode}
\ee
The detectability of the strain (\ref{h0_rmode}) is shown in Fig.~\ref{fig:h0_rmode} assuming 
a one year phase-coherent observation and a source at $D= 1\,\mbox{kpc}$. 
First of all, the smallness of the $r$-mode amplitude essentially eliminates extragalactic sources from being
candidate targets for detection unless the source is located within our local group and the amplitude is
much higher than the upper limits suggested by LMXB/MSP spin-down and thermal data. 
This could be a realistic possibility if the systems from which the data came from are not $r$-mode unstable
 -- in that case it would make sense to use a fiducial amplitude  $\alpha_r = 10^{-3}$, corresponding to a 
 maximum saturation amplitude due to non-linear mode couplings.  The GW strain for this optimistic scenario 
could indeed be detectable by Advanced LIGO/Virgo (for sources located anywhere in the Galaxy) and of course by ET
for sources located further way or rotating at a lower frequency, see Fig.~\ref{fig:h0_rmode}. 
Such $r$-mode signals could be associated with very young neutron stars with still undepleted fast rotation
(i.e. fast spinning CCOs), see discussion at the end of  Section~\ref{sec:rmode_pheno}.

Taking at face value the aforementioned  LMXB/MSP upper limits, the amplitude is constrained to be
$\alpha_r \lesssim 10^{-7}$  and we can see that the prospects for detection of any $r$-mode signal look rather bleak. 
Only a next generation instrument like ET could score a detection provided the source is rapidly rotating and relatively 
close ($D \sim 1\,\mbox{kpc}$). 

Assuming a neutron star source with known spin frequency, the identification of an $r$-mode signal would be
unmistakable due to the  $f_{\rm gw}-f_{\rm spin}$ relation (\ref{fgw_frmode})  which is unique among 
the various neutron star GW emission mechanisms. Making a source parameter estimation through the
GW measured $f_{\rm mode}$ requires the use of the fully relativistic $r$-mode frequency.  
To leading post-Newtonian order, the correction to the Newtonian frequency comes from the combined effect of gravitational 
redshift and frame dragging and is of the order of the stellar compactness $M/R$; for a neutron star this translates to an
appreciable  $\sim 20\%$ frequency shift. With relativistic corrections accounted for, an $r$-mode GW detection would thus
lead to a measurement of the compactness. 

The importance of using the fully relativistic $r$-mode frequency was recently exemplified by the oscillation discovered
in the light curve of a burst from the AMXP XTE 1751-305 \cite{sm2014}.  The observed frequency is close to the expected 
Newtonian $r$-mode frequency for this neutron star's known spin frequency but an exact match for reasonable stellar mass 
and radius parameters is only possible if relativistic corrections are taken into account \cite{andersson_etal2014}. 
Unfortunately, the inferred $r$-mode amplitude is too large to be reconciled with the system's spin evolution, a result that 
hints at a different interpretation of the observed oscillation.

\subsection{Beyond the minimum damping model}
\label{sec:extra_damping}

The alternative scenario of enhanced damping can be seen as a pessimistic standpoint as it attempts to resolve 
the $r$-mode puzzle by invoking a much reduced instability parameter space that prevents much of the known
rapidly rotating neutron stars from becoming $r$-mode active. With regard to the realism of this scenario,  
it can be safely stated that none of the additional damping mechanisms mentioned earlier  
(exotic matter, mutual friction, Ekman layer) can be ruled out given our present level of understanding (or ignorance!). 

The situation is particularly murky when it comes to exotic matter. For example, the presence of hyperons in 
neutron star cores and their strong bulk viscosity was once thought to be lethal for the $r$-mode instability 
\cite{jones2001, lindblom_owen2002}. The day was saved by the likely superfluidity of these particles
that effectively shuts down their viscosity below a temperature $\sim 10^9\,\mbox{K}$, see e.g.  
\cite{nayyar_owen2006, haskell_andersson2010}.

The degree of uncertainty is even higher when it comes to strange quark matter with the resulting instability windows 
being extremely sensitive to the details of quark pairing, see e.g. \cite{madsen1998, madsen2000}. More recent work  
\cite{mannarelli_etal2008, andersson_etal2010} focused on the multifluid aspect of strange stars with 
colour-flavor-locked paired quark matter and established that the $r$-mode instability suffers very little
damping in this type of stars. Unfortunately, given that the very existence (let alone key properties such as 
viscosity and pairing) of such exotic phases of matter in neutron stars is still a matter of debate, we presently 
cannot make any reliable prediction.  

\begin{figure*}[htb!]
\includegraphics[width=0.24\textwidth]{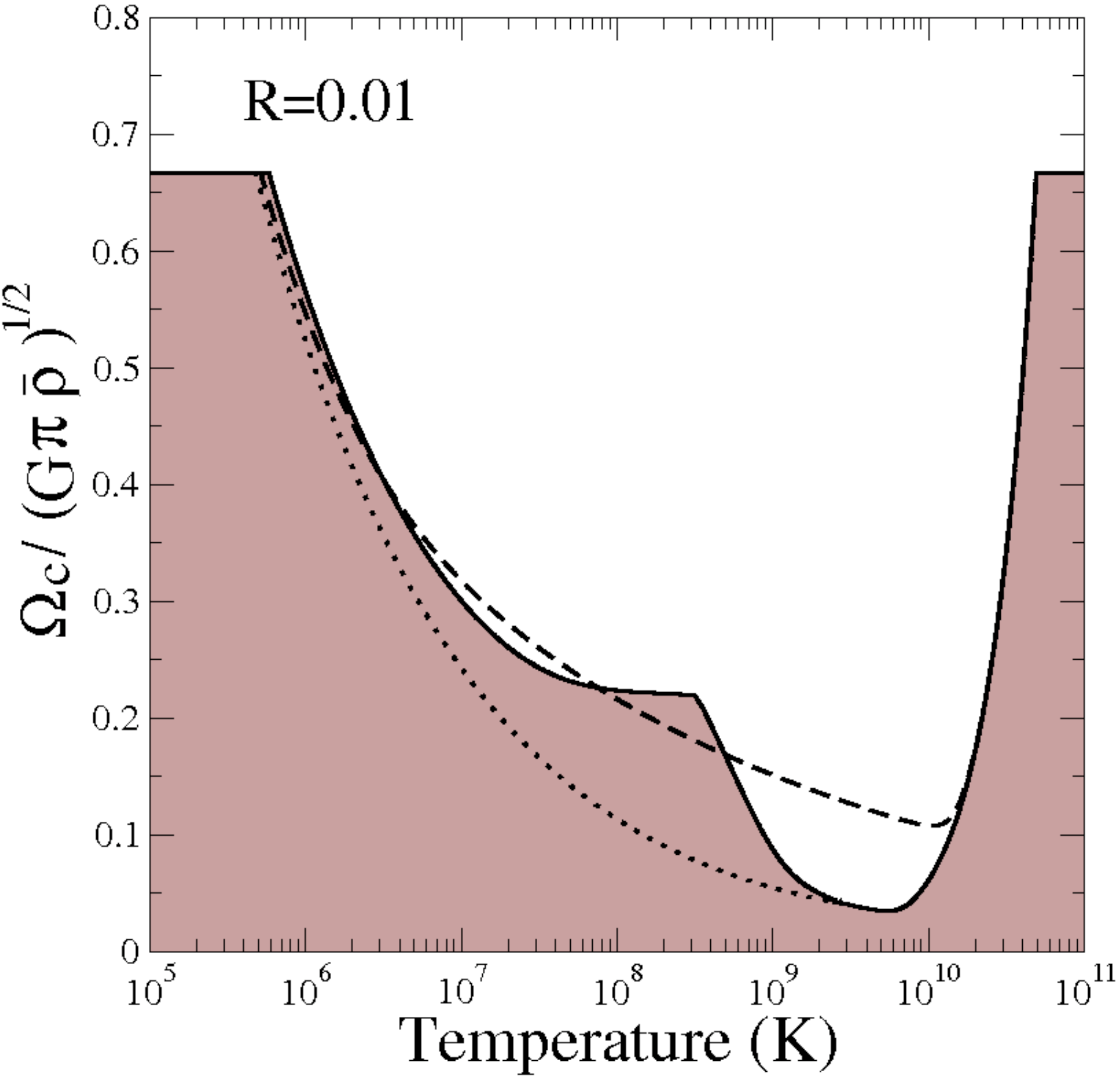}
\includegraphics[width=0.24\textwidth]{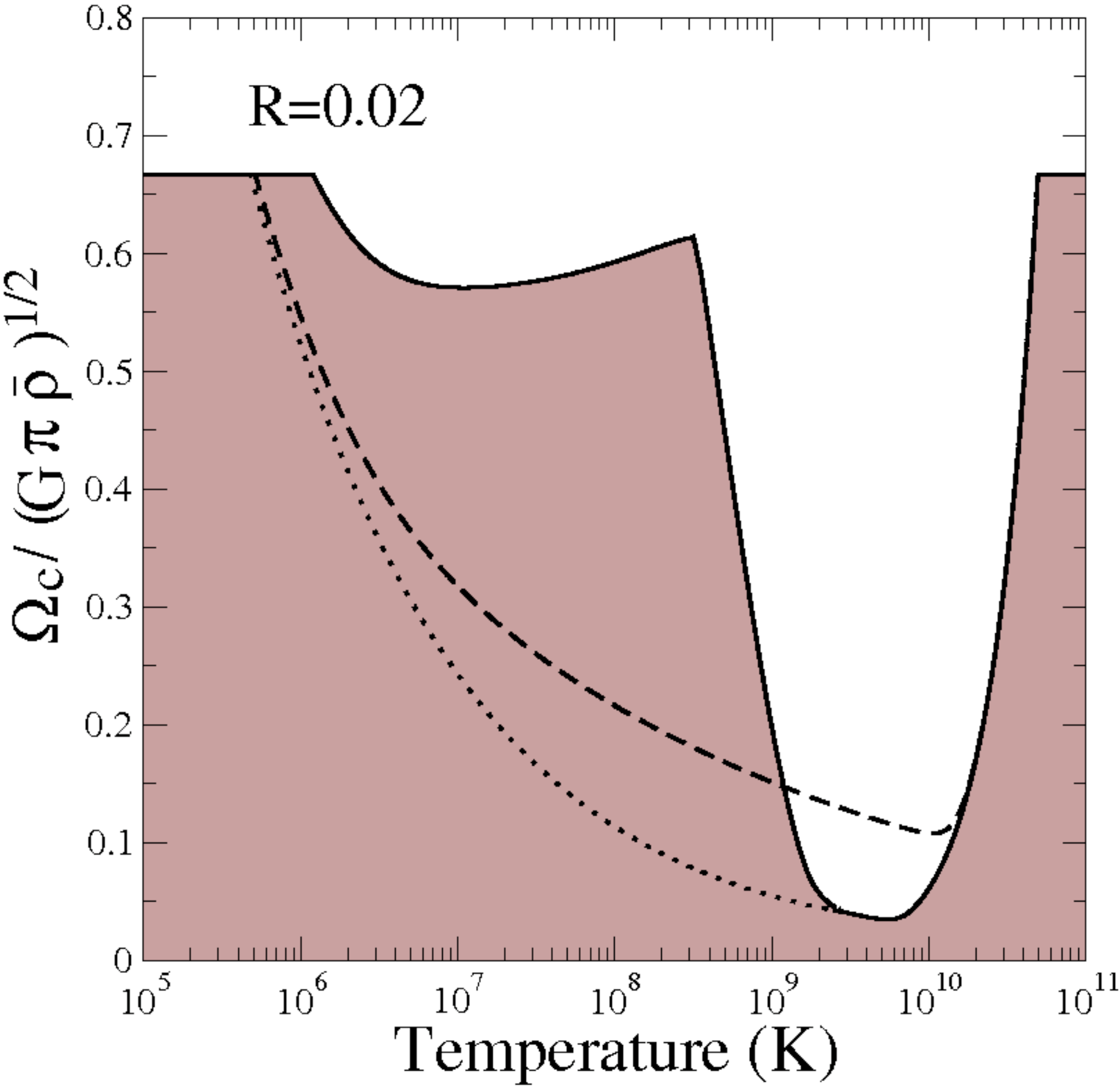}
\includegraphics[width=0.24\textwidth]{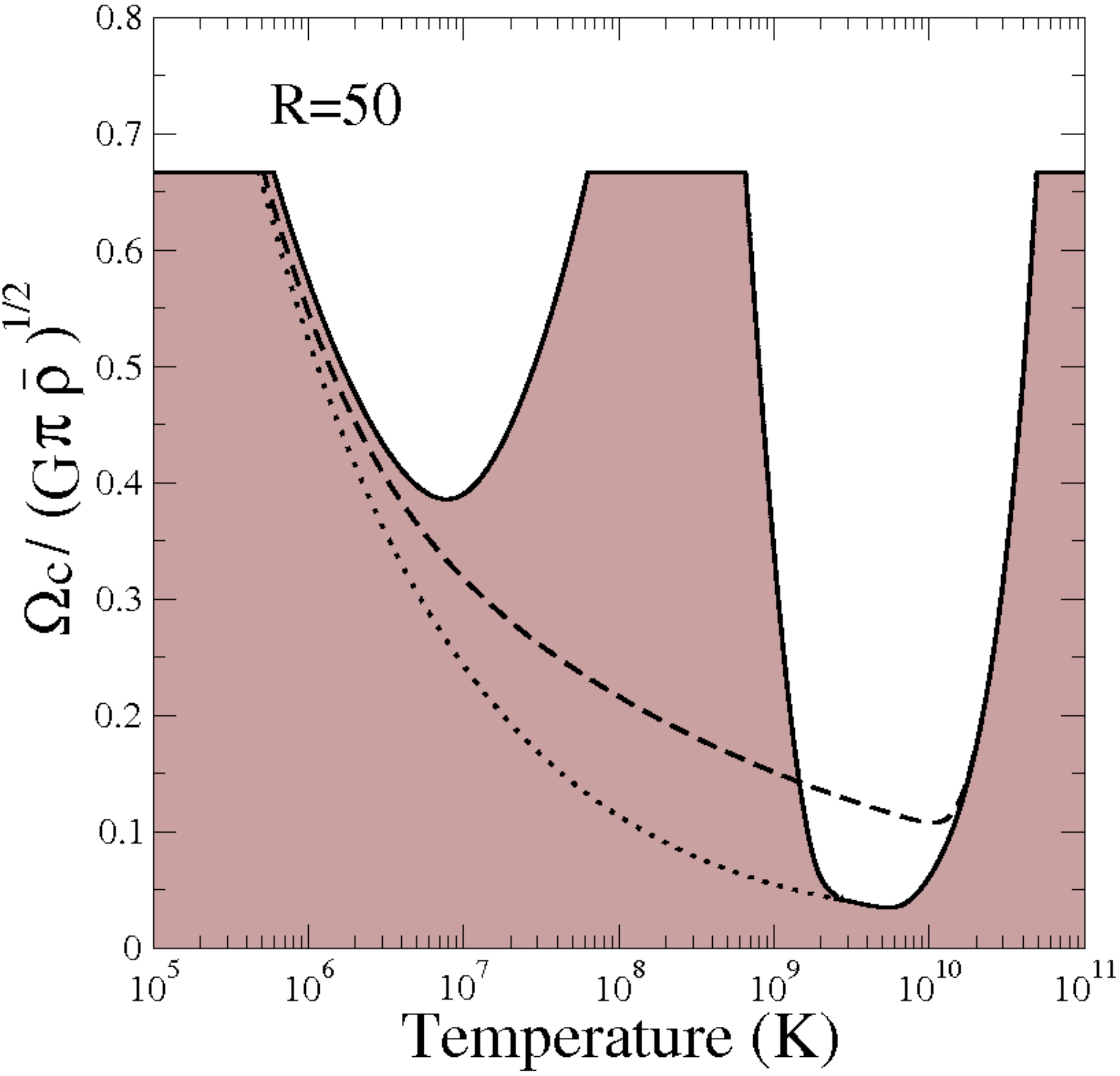}
\includegraphics[width=0.24\textwidth]{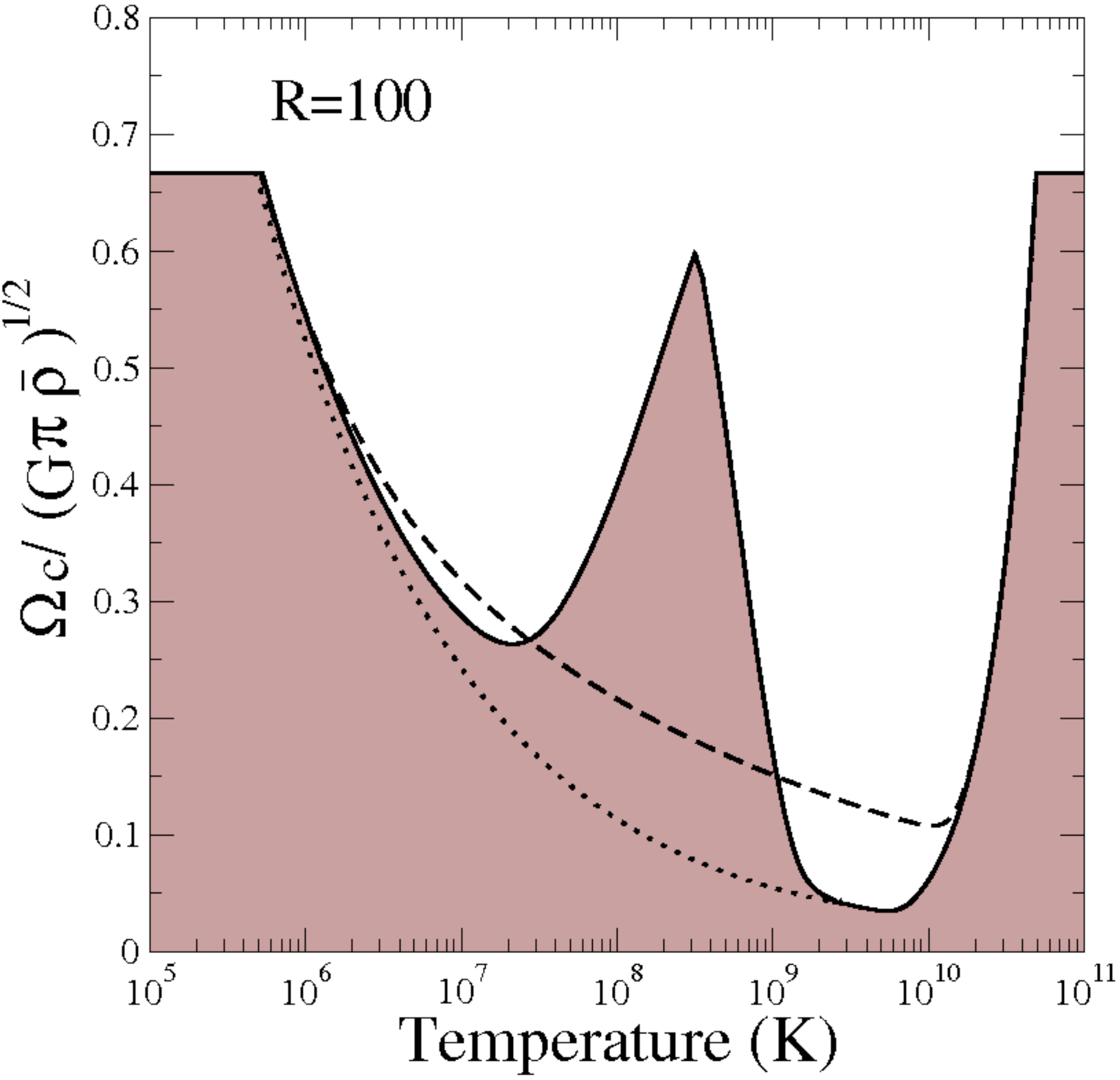}
\caption{\emph{$r$-mode instability window with phenomenological mutual friction}. This figure represents
the ``strong superfluidity'' model of Ref.~\cite{haskell_etal2009} and shows the impact that an increasingly 
amplified vortex mutual friction force would have on the $r$-mode instability window. Note that standard mutual friction 
(electron scattering by vortices) has a drag coefficient $\cR \sim 10^{-4}$ and leads to negligibly small 
$r$-mode damping. A much higher $\cR$ could result from vortex-fluxtube interactions. 
The figure also shows instability curves (for a canonical stellar model) due to standard shear and bulk viscosity as well 
as due to a slippage-modified Ekman layer (dashed curve).}
\label{fig:mutual_friction}
\end{figure*}

In our view, it makes more sense to focus on damping mechanisms that rely on less exotic physics. 
For instance, taking the commonly accepted view that the outer core of mature neutron stars contains a mixture
of superfluid neutrons and superconducting protons, the presence of vortex mutual friction is unavoidable.
The standard (that is, best understood) form of this type of friction originates from the scattering of electrons by the 
superfluid's magnetised vortices and it has been shown to have a negligible effect on $r$-modes \cite{lindblom_mendell2000}. 
However, additional mutual friction may originate from the direct interaction between the vortices and the magnetic fluxtubes 
threading the superconductor, see e.g. \cite{ruderman_etal1998, link2003}. Our understanding of this mechanism is (at best) 
rudimentary, so far having been used in $r$-mode saturation amplitude calculations \cite{haskell_etal2014}. 
A more phenomenological approach to this problem is to consider the standard form for the mutual friction force 
and explore its impact on the $r$-mode by artificially increasing its strength \cite{haskell_etal2009}, 
see Fig.~\ref{fig:mutual_friction}. It is then found that a factor $\sim 100$  increase in the mutual
friction drag parameter is sufficient for suppressing the $r$-mode instability in a large portion of the parameter space. 
Vortex-fluxtube  interactions could well lead to friction of this magnitude but more work is required in order to make a safe prediction.

An entirely different kind of effect, taking place in superfluid neutron stars, could modify the $r$-mode instability's
parameter space \cite{gusakov_etal2014a, gusakov_etal2014b}.  In general superfluid matter with two fluid degrees of
freedom supports twice as many oscillation modes as compared to ordinary matter. For $r$-modes in particular, this
doubling leads to ``ordinary'' and ``superfluid'' modes, the real physical distinction between them being the relative
amplitude of the two fluids' co-moving and counter-moving degrees of freedom. The ordinary (superfluid) $r$-mode is
mostly co-moving (counter-moving).  The mechanism proposed in \cite{gusakov_etal2014a, gusakov_etal2014b}
(see also \cite{Gusakov:2016drh,Kantor:2017xuo}) is based on the observation that these two mode branches
can experience resonant avoided crossings in a temperature range relevant for LMXBs. Close to these resonances the
ordinary $r$-mode is mixed with its counter-moving counterpart and suffers strong damping from vortex mutual
friction. The resulting $r$-mode instability window exhibits resonant ``stability spikes'' in the temperature range $T
\sim 10^7-10^8\,\mbox{K}$. The LMXB evolution model proposed in \cite{gusakov_etal2014a, gusakov_etal2014b} represents
an interesting modification of the earlier discussed runaway cycle, envisaging $r$-mode unstable systems climbing up
these spikes while emitting GWs with the peak of a given spike setting the spin frequency upper limit. This
mode-resonance mechanism is based on more or less conventional physics and, given its impact on the standard $r$-mode
window, it deserves to be explored further.

The global interaction of a growing $r$-mode with the stellar magnetic field is another type of enhanced damping 
mechanism that needs to be discussed  (there is also a local interaction with the field at the location of the crust-core 
Ekman layer, see Section~\ref{sec:requiem} below). Early work  put forward the scenario of a magnetic field ``wind-up'' 
by an unstable $r$-mode \cite{rezzolla_etal2000, rezzolla_etal2001a, rezzolla_etal2001b}. This is a non-linear effect 
associated with the mode's differential rotation and the resulting  Stokes drift experienced by the oscillating fluid elements. 
In the absence of any back-reaction from the field itself, this could potentially become a mechanism for generating 
a strong azimuthal component from an initial weaker poloidal field while sapping the mode's energy in the process 
\cite{rezzolla_etal2000, rezzolla_etal2001a, rezzolla_etal2001b,cuofano_etal2012}. 
In practice, however, one would expect that at some stage the back-reaction of the perturbed field should kick in and
self-regulate the process. This becomes obvious from the fact that once the magnetic energy becomes comparable 
to the mode energy one cannot even speak of an $r$-mode. Recent detailed work \cite{chugunov2015, friedman_etal2016} 
suggests that, with back-reaction included, this mechanism is unlikely to suppress the $r$-mode instability and produce 
strongly magnetised neutron stars (in particular, Ref. \cite{chugunov2015} shows that  $r$-mode activity in weakly 
magnetised systems such as LMXBs cannot amplify the field beyond  $\sim 10^8 (\alpha_{\rm sat}/10^{-4} )^2\,\mbox{G}$).

\subsection{The role of the crust}
\label{sec:crust}

The remaining mechanism of enhanced $r$-mode dissipation, the viscous Ekman layer, may be considered as
the most robust one since it relies on more or less conventional physics \cite{BU2000b}. 
In its most basic form, the damping originates from the ``rubbing'' of the mode's flow against the solid crust and 
the thin viscous boundary layer formed at that region. The layer thickness is roughly given by 
$\delta_{\rm E} \sim \sqrt{\nu/\Omega} $, where $\nu$ is the shear viscosity coefficient, and is of the
order of a few centimetres. The Ekman damping timescale $\tau_{\rm E}$ is related to that of shear viscosity
by $\tau_{\rm E} \sim \tau_{\rm sv} / \sqrt{Re}$ where $Re \sim R^2 \Omega/\nu$ is the characteristic
Reynolds number. For neutron star matter $Re \gg 1$, suggesting that the Ekman layer can be 
strongly dissipative. Indeed, early calculations showed that the Ekman layer could dominate $r$-mode 
damping for any temperature $T \lesssim 10^{10}\,\mbox{K}$ \cite{BU2000b, lindblom_etal2000, rieutord2001}.

This basic model was soon refined to account for the fluid's stratification and compressibility \cite{KG_NA2006b} 
and the crust's elasticity \cite{LU2001, KG_NA2006a}. 
This latter property is crucial as it allows the crust to participate in the global $r$-mode oscillation 
albeit with a velocity jump at the crust-core interface. The resulting damping rate is weakened with 
respect to that of a solid crust by a factor $\cS^2$ , where $\cS $ is the dimensionless crust-core ``slippage'' 
parameter (a typical value of which is $\cS \approx 0.05$ \cite{LU2001, KG_NA2006a}).

The $r$-mode instability window produced by this ``jelly'' crust model with its slippage-modified Ekman layer 
can be considered as the canonical one. The Ekman layer curve shown in Figs.~\ref{fig:rmode_window1} \& 
\ref{fig:rmode_window2} assumes no slippage and was calculated using the formalism of Ref.~\cite{lindblom_etal2000} 
for a canonical neutron star model with the crust-core boundary assumed at $R_c = 0.9 R$ and the shear viscosity 
coefficient taken from \cite{andersson_etal2005} (with the proton fraction set to $x_{\rm p} =0.03 $). 
The resulting Ekman timescale is, 
\be
\tau_{\rm E} \approx 154\, P_{-3}^{1/2} T_8 \,\mbox{s}.
\ee
The addition of slippage lowers the Ekman critical curve (i.e. $\tau_{\rm E} \to \tau_{\rm E} /\cS^2$) and makes 
the instability window larger. Similar but much less pronounced would be the modification due to a more 
accurate shear viscosity coefficient \cite{shternin_yakovlev2008} (which is about a factor three lower than the one used here). 
The curve can also move up or down as a result of changing the location of the crust-core curve and the matter's
symmetry energy \cite{wen_etal2012}.

Comparison of the Ekman layer-modified instability window against the LMXB quiescence data 
\cite{MS2013} reveals that essentially all systems should be $r$-mode stable provided there is no 
crust-core slippage (i.e. $\cS=1$), see Fig.~\ref{fig:rmode_window1}. In contrast, LMXBs with 
assumed $r$-mode spin and thermal equilibrium \cite{Ho_etal2011} are significantly hotter and some of 
them spill out of the instability window. If the slippage-modified Ekman layer damping is instead used, both 
sets of data would imply unstable $r$-modes in several sources. In practice, therefore, the slippage-modified 
$\tau_{\rm E}$ leads to a ``minimum damping'' window and therefore belongs to the previous small amplitude scenario.

Slippage aside, the instability window may not be what shown in Fig.~\ref{fig:rmode_window1} because
of the possibility of having \emph{resonances} between the $r$-mode and the various crustal shear modes 
\cite{LU2001, KG_NA2006a}. These resonances, by selectively amplifying damping near the resonant 
spin frequency, can lead to a ``spiky'' instability window with a large $\Omega-T$ instability swathe carved out in the 
region where LMXB and MSP may reside, see e.g. \cite{Ho_etal2011}.

\subsection{Requiem for the $r$-mode instability?}
\label{sec:requiem}

The situation could change even more drastically by a more realistic modelling of the crust-core interface that 
takes into account the presence of a magnetic field threading the two regions. The discontinuity at the interface 
leads to a kink in the oscillating magnetic field lines and the launching of short wavelength Alfv\'en waves which 
are subsequently damped by viscosity.  The physics of this  \emph{magnetised} Ekman layer was first explored 
in an early paper by Mendell \cite{mendell2001} albeit with the assumption of a solid crust. The main result of that 
work was that dissipation is significantly enhanced for field strengths $B \sim 10^{11} - 10^{13}\,\mbox{G}$, hence 
leaving little or no room for the $r$-mode instability in systems like normal radio pulsars. On the other hand, weaker 
fields $B \lesssim 10^{10}\,\mbox{G}$ have a negligible impact on the Ekman layer suggesting that the magnetic field 
is not a factor for the $r$-mode instability of LMXBs or MSPs whose magnetic fields are typically
concentrated around $\sim 10^8\,\mbox{G}$.

This conclusion, however, could be premature. The outer core of these neutron stars
is expected to be in a superconducting state which, among other things, means that the (squared) Alfv\'en speed 
is boosted by a factor $H_c/ \xp B$ with respect to its ordinary value $v_\rA^2= B^2/4\pi \rho$, where 
$H_c \approx 10^{15}\,\mbox{G}$ is the critical field for superconductivity, and at the same time the shear viscosity 
coefficient is rescaled as $\nu \to \nu/\xp$ \cite{KG_etal2011}. 
These modifications can be incorporated in Mendell's result \cite{mendell2001} for the relation between 
the damping timescale $\tau_{\rm mag}$ of the magnetised layer and the timescale $\tau_{\rm E}$ 
of the ordinary Ekman layer:
\be
\tau_{\rm mag} \approx  \tau_\rE \left (\frac{v_\rA^2}{\Omega \nu} \right )^{-1/2}  
\approx  \tau_\rE\left ( \frac{B_{\rm cr}}{B} \right )^{1/2}.
\label{tmag}
\ee
The threshold $B_{\rm cr} \approx 10^6 \rho_{14}^{3/2} x_{\rm p1}^{3/2} T_8^{-2} P_{-3}^{-1}\,\mbox{G}$
marks the transition between a magnetic field-dominated Ekman layer ($B \gg B_{\rm cr}$) and
a non-magnetic one ($B\ll B_{\rm cr}$). The above formula assumes the former limit and
leads to a markedly shorter damping timescale for systems like LMXBs and MSPs. 
Crucially, this result remains accurate in a more realistic model which combines the 
magnetic field with an elastic crust \cite{KGetal_inprep}.
The dramatic increase in the local Alfv\'en speed stretches the boundary layer, increasing its thickness 
by a factor  $\delta_{\rm mag} \approx  (B/B_{\rm crit})^{3/2} \delta_\rE$.

The magnetic field has another local effect that has not been sufficiently stressed in the literature although it is 
commonly adopted in the modelling of magnetar oscillations 
(see e.g. \cite{Glampedakis:2006apa, Gabler:2010rp, Colaiuda:2010pc, Gabler:2011am, Gabler:2016rth}).
As a consequence of the junction conditions satisfied by the Maxwell equations, the crust-core velocity slippage
is suppressed\footnote{Starting from the perturbed Faraday's law,
$\bnabla \times \delta \mathbf{E}  = -i\omega \delta \mathbf{B}/c $, and applying the usual ``circuit'' argument 
across the crust-core boundary leads to $\hat{\mathbf{n}} \times \langle \delta \mathbf{E} \rangle =0$, where
$\hat{\mathbf{n}}$ is the unit normal vector to the boundary and $\langle  ... \rangle$ stands for 
the jump across the boundary. Using $ c \delta \mathbf{E} = - \delta \mathbf{v} \times \mathbf{B}$ for the 
$r$-mode-induced electric field, we find $ ( \hat{\mathbf{n}} \cdot \mathbf{B} )   \langle \delta v \rangle = 0$ which
for a general magnetic field implies a vanishing slippage.} and the discontinuity now first appears 
in the radial velocity derivative. 

The  superconductivity-rescaled magnetic Ekman timescale (\ref{tmag}), in combination with the suppression 
of the crust-core slippage, paints a very pessimistic picture for the viability of the $r$-mode instability in LMXBs and MSPs. 
The instability curve becomes a horizontal temperature-independent line located at the critical frequency 
(here we assume canonical stellar parameters),
\be
f_{\rm mag} \approx  886\, B_8^{1/12}\, \mbox{Hz}.
\ee
This amounts to an angular frequency $ \Omega_{\rm mag} \approx 0.9 B_8^{1/12} \Omega_\rK$ and
obviously implies the complete suppression of the $r$-mode instability in \emph{all} known rapidly rotating neutron stars. 

The significance of this result is obvious but we are still very far from reaching definitive conclusions. Much of the analysis
presented here is preliminary and has been based on a simplified toy model calculation that does not fully take into account 
the rich superfluid/superconducting physics of the system \cite{kinney_mendell2003}. Nor does it include the possibility 
of resonances between an $r$-mode and crustal modes or a finite thickness ``pasta''-phase crust-core interface 
\cite{gearheart_etal2011}. All these effects could be important and ought to be included in a more realistic modelling of 
$r$-mode Ekman layer damping.

\section{Magnetar oscillations}
\label{sec:magnetars}
Magnetars are strongly magnetized, isolated neutron stars, with dipole mangetic fields $\sim10^{14}-10^{15}$ G at the
surface, and possibly even larger in the interior~\cite{Duncan:1992dt}. They are observed as soft $\gamma$-ray repeaters
(SGRs) and anomalous X-ray pulsars (AXPs)~\cite{Woods:2004kb}. These extreme magnetic fields power an intense
electromagnetic activity, with very energetic ``giant flares'', having peak luminosities as large as $10^{44}-10^{47}$
erg/s. For a recent review on the subject, we refer the reader to~\cite{Turolla:2015mwa}.

\subsection{Magneto-elastic oscillations}
\label{subsec:magnqpo}

Quasi-periodic oscillations (QPOs) have been observed in the tails of giant flares from two magnetars, SGR
$1804-20$~\cite{Israel:2005av,Strohmayer:2006py} and SGR$ 1900+14$~\cite{Strohmayer:2005ks}. Similar QPOs have later
been observed in less intense bursts from the same objects~\cite{Huppenkothen:2014cla} and from SGR
J$1550-5418$~\cite{Huppenkothen:2014pba}. Most of these oscillations have frequencies between $18$ Hz and $200$ Hz, but
two high-frequency QPOs (with frequencies of $625$ Hz and $1840$ Hz) have also been observed in SGR $1804-20$.
 
QPOs have been interpreted as stellar oscillations. This means that we have already observed neutron star oscillations,
and neutron star asteroseismology is in principle possible even before the first detection of GWs from neutron stars! In
particular, once the QPO mechanism will be fully understood, it will be possible to extract information, from the
frequencies of these oscillations, on the EOS of nuclear matter (see
e.g.~\cite{Samuelsson:2006tt,Watts:2006hk,Sotani:2012qc,Sotani:2013jya}). An alternative explanation of QPOs is that
they are oscillations of the neutron star magnetosphere, which is expected to be filled of plasma during a giant
flare. However, the ``standard model'' based on neutron star oscillations is much more promising, since it allows to
explain the details of QPO phenomenology.

After the first observations of magnetar QPOs a great effort has been devoted to the theoretical modelling of these
oscillations (see e.g.~\cite{Watts:2016uzu} and references therein).  Magnetar QPOs have initially been identified with
torsional (i.e. axial parity) elastic modes of the crust~\cite{Israel:2005av,Samuelsson:2006tt}, but it was soon
understood that the crustal oscillations are strongly coupled with Alfv\'en modes, i.e. magnetic modes of the neutron
star core~\cite{Levin:2006ck,Glampedakis:2006apa}.  QPOs are now believed to be magneto-elastic oscillations, which
involve both the core and the crust.  The first models of torsional magneto-elastic oscillations of neutron stars have
shown that Alfv\'en modes are not discrete, but instead have a continuous spectrum. However, the edges of this continuum
can yield long-lived QPOs~\cite{Levin:2006qd,Sotani:2007pp}. Subsequently, several groups developed GR
magnetohydrodynamics (MHD) numerical codes which allowed to model torsional magneto-elastic oscillations of magnetars,
improving our understanding of this phenomenom. In~\cite{vanHoven:2010gy,Colaiuda:2010pc,vanHoven:2011it} it was shown
that torsional magneto-elastic oscillations of a poloidal magnetic field have continuous bands. The crustal modes
falling in the bands are absorbed by the continuum (see also~\cite{Gabler:2011am}), while those falling in the gaps
survive as discrete modes. Further discrete modes are given by the edges of the bands. Similar results were obtained
in~\cite{Asai:2014al,Asai:2015qha} using finite series expansions.

More recently, the models of magnetar oscillations - using GR MHD simulations - have been extended in two
directions. Some works kept studying axial parity oscillations, including more and more physical contributions:
superconductivity in the neutron star
interior~\cite{Passamonti:2012ma,Sotani:2012xd,Passamonti:2013zra,Jones:2013yea,Gabler:2013ova,Gabler:2016rth}, the
neutron star magnetosphere~\cite{Gabler:2014bza,Jones:2013yea},``pasta phases'' in the neutron star
crust~\cite{Passamonti:2016jfo}. The oscillation spectrum found in these works had the structure discussed above:
continuous bands and discrete modes in the gaps.  Other works, instead, studied axial-polar coupled oscillations. As
shown in~\cite{Colaiuda:2011aa}, if the background magnetic field is not purely poloidal (they considered the so-called
``twisted torus'' configuration, which is discussed in Section~\ref{subsec:magmountains}), oscillations with axial and
polar parities are coupled. This coupling seems to destroy, or at least reduce, the Alfv\'en continuous spectrum, leaving
a set of discrete modes. Similar results were found in~\cite{Sotani:2015rla,Link:2015cza,Link:2016dxt}, where a
different choice of background magnetic field (the so-called ``tilted torus'') also removed the continuous spectrum.

We can now describe, at least qualitatively, low-frequency QPOs as magneto-elastic oscillations of magnetars.
High-frequency QPOs are more difficult to model: if they are magneto-elastic oscillations they are expected to be damped
in less than one second, but the observations show long-living oscillations. It has recently been
suggested~\cite{Huppenkothen:2014ufa} that the reason of this discrepancy could be the interpretation of the
observations: high-frequency QPOs would die out in a short timescale, consistently with the theoretical model, but then
they would be excited again, several times during the tail of the giant flare.

Our qualitative understanding of magnetar QPOs is not sufficient to carry on neutron star asteroseismology with these
oscillations. Indeed, in order to extract information on the inner structure of the star from QPOs, we need
(approximate) analytical expressions of the QPO frequencies in terms of the main features of the neutron star (mass,
radius, magnetic field, etc.). We are very far away from deriving such expressions, for two reasons: the scarcity of
observational data (we only detected few QPOs from three giant flares) and the limitation of the theoretical modelling,
which still does not include all relevant physical processes and mechanisms. Hopefully, we will soon have more
observational data (especially when new large-area X-ray detectors~\cite{arzoumanian2014neutron,feroci2012large,eXTP}
will be operating) and a deeper theoretical understanding of these phenomena.

\subsection{Gravitational waves from giant flares}
\begin{figure*}[htb!]
\includegraphics[width=0.48\textwidth]{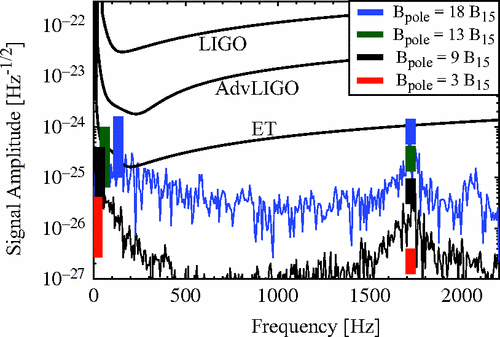}
\caption{{\it GW detectability of magnetar oscillations in giant flares.} We show the signal amplitude $\sqrt{T}|h|$,
  where $h$ is the estimated dimensionless strain from magnetar oscillations in giant flares, and $T$ is the estimated
  damping time of the oscillation.  The giant flare is modeled as the instability of the neutron star magnetic field
  (see text). The colored boxes on the left represent the $f$-mode signal, assuming $50$ ms$\le T\le200$ ms. The colored
  boxes on the right represent the signal of a lower-frequency mode (Alfv\'en and/or $g$-modes), assuming $10$ ms$\le
  T\le1$ s.  Different colors correspond to different values of the magnetic field at the poles. We also show the signal
  for the entire spectrum, assuming a damping time $T=100$ ms (from~\cite{Zink:2011kq}).}
\label{fig:zinkestimates}
\end{figure*}
Giant flares of magnetars are among the most luminous events in the universe. They are believed to be due to
large-scale rearrangements of the magnetic fields, either in the stellar core~\cite{Thompson:1995gw,Thompson:2001ie} or
in the magnetosphere~\cite{Lyutikov:2005un,Gill:2010qx}.  Even a tiny part of their energy could excite non-radial
oscillations of the neutron star - in particular, the $f$-mode - and thus be emitted through GWs. Preliminary
estimates~\cite{Corsi:2011zi} were very optimistic on the detectability of the $f$-mode excited by giant flares, but
they were based on the assumption that the energy emitted in GWs by $f$-mode excitation is comparable to the total
magnetic energy of the star. More accurate estimates~\cite{Levin:2011vh} showed that only a small part of the magnetic
energy is converted in GWs. Assuming that the released electromagnetic energy is of the order of the magnetic energy of
the star, a giant flare at $D=10$ kpc would excite an $f$-mode GW signal detectable by Advanced LIGO/Virgo with a
SNR~\cite{Levin:2011vh}
\begin{equation}
\frac{S}{N}\lesssim10^{-2}B_{15}^2\,,\label{levinest}
\end{equation}
where $B_{15}$ is the (normalized) magnetic field-strength at the poles. This result, based on a perturbative analysis,
implies that the $f$-mode would definitely not be detectable by second-generation interferometers if $B\sim10^{15}$ G,
but it may be marginally detectable for $B\sim10^{16}$ G.

Subsequently, giant flares and their coupling with the neutron star oscillations have been studied using
general-relativistic MHD numerical simulations of the magnetic field instability in
magnetars~\cite{Ciolfi:2012en,Zink:2011kq}. This is a toy-model which mimics the catastrophic magnetic field
rearrangments during a giant flare, and allows to estimate the amplitude and the features of the GW emission.  The
numerical simulations show a power-law relation between the emitted gravitational signal and the surface magnetic
field-strength (in these models the magnetic field-strength at the surface and in the interior are
comparable)~\cite{Zink:2011kq}:
\begin{equation}
h\sim7.6\times10^{-27}\frac{B_{15}^{3.3}}{D_1}\,,
\label{fitzink}  
\end{equation}
where $D_1$ is the distance of the source normalized to $1$ kpc. Most of the energy in the signal is in the $f$-mode,
but also lower frequency modes (likely Alfv\'en and/or composition $g$-modes) are excited (see
Fig.~\ref{fig:zinkestimates}).

The results of numerical simulations~\cite{Ciolfi:2012en,Zink:2011kq} are even more pessimistic than those of
perturbation theory~\cite{Levin:2011vh} shown in Eq.~(\ref{levinest}). The SNR of the GW signal from an $f$-mode excitd
by a giant flare has been estimated to be~\cite{Ciolfi:2012en} $S/N\sim10^{-4}B_{15}^2$ for Advanced LIGO/Virgo, and
$S/N\sim10^{-2}B_{15}^2$ for ET.

The numerical simulations of~\cite{Zink:2011kq} give similar results, see Fig.~\ref{fig:zinkestimates}. The SNR can be
read off the figure, as the ratio between the signal and the noise curve, for a given $B_{pole}$ and for a given
detector. As mentioned above, Fig.~\ref{fig:zinkestimates} also shows a low-frequency mode, whose SNR is comparable (but
sligthly larger) than that of the $f$-mode\footnote{It is worth noting 
that the authors of~\cite{Zink:2011kq} fit the numerical data with a power-law relation, finding that the GW amplitude
is proportional to $B^{3.3}$. The authors of~\cite{Ciolfi:2012en}, instead, have $h\sim B^2$, because they fit the data
with a quadratic function; this choice is correct as a first approximation, since in the perturbative results
of~\cite{Levin:2011vh} the GW amplitude is a quadratic function of the magnetic field strength.}.
  
These studies confirm that the gravitational emission from mode excitation in magnetar flares can not be detected by
advanced GW detectors. This signal is expected to be too weak to be detected even by third-generation GW experiments,
such as ET, unless the surface magnetic field is $\gtrsim 10^{16}$ G~\cite{Zink:2011kq,Ciolfi:2012en}, i.e. an order of
magnitude larger than the observed magnetic fields.

\section{Neutron star ``mountains''}
\label{sec:mountains}
Rotating neutron stars symmetric with respect to the rotation axis do not emit gravitational radiation. However, if a
neutron star is not perfectly axisymmetric it emits GWs, mostly at frequencies $f_{\rm spin}$ and $2f_{\rm spin}$ (where
$f_{\rm spin}$ is the rotation frequency). Most of the observed neutron stars
have rotation frequencies in the range between $\sim10$ Hz and $1$ kHz, which is the range where ground-based
interferometers such as Advanced LIGO/Virgo are most sensitive. Therefore, if the deviation from axisymmetry is large enough,
rotating neutron stars can be promising sources of GWs.

The deviation from axisymmetry is described by the {\it quadrupole ellipticity} 
\begin{equation}
  \varepsilon=\frac{Q}{I}\,,
\end{equation}
where $Q$ is the mass quadrupole moment associated to the distortion (i.e. excluding the contribution of rotation,
which is necessarily axisymmetric and is not associated to GW emission), and $I$ is the moment of inertia of the
rotation axis. Such distortions are called {\it mountains}. When $\varepsilon>0$ the star is oblate, while when
$\varepsilon<0$ it is prolate. 

If the distortion has a symmetry axis forming an angle $\alpha$ (called {\it wobble angle}) with the rotation axis,
the amplitude of the GW emission is~\cite{Zimmermann:1979ip,Bonazzola:1995rb}
\begin{equation}
  h_0\simeq\frac{4\pi^2G}{Dc^4}f_{\rm gw}^2I\varepsilon\sin\alpha\simeq 4\times10^{-24}
  P_{-3}^{-2}\,D_1^{-1}I_{45}
  \left(\frac{\varepsilon}{10^{-6}}\right)\,.
  \label{h0mount}
\end{equation}
The deformation can be due to different mechanisms: from the magnetic field of the
star~\cite{Bonazzola:1995rb,Haskell:2007bh,Ciolfi:2010td} (``magnetic mountains''), to
temperature gradients in the crust leading to local deformation sustained by the elastic strain
~\cite{Bildsten:1998ey,Ushomirsky:2000ax}
(``thermal mountains''). The former can arise both in isolated neutron stars (magnetars, but also in ``ordinary''
neutron stars), or in accreting neutron stars. The latter, instead, can only be formed in accreting neutron stars,
because the temperature gradients are created by asymmetries in the matter accreting by a companion object.

Current searches for GWs from rotating pulsars using first generation LIGO/Virgo have shown no sign of such
emission (see e.g.~\cite{Aasi:2013sia} and references therein). However, this negative result allows us to place upper
limits on the quadrupole ellipticity of the pulsars under study; in the case of the Crab pulsar,
$\varepsilon\lesssim8.6\times10^{-4}$; in the case of other pulsars, the upper limits are significantly weaker. These
limits are below the theoretical bounds (see below), but (in some cases) they exceed the spin-down limit on
$\varepsilon$, which has been obtained with the unrealistic assumption that the observed pulsar spin-down is only due to
GW emission. 

In the following we shall first discuss current models of neutron stars deformed by a magnetic field, and then the
different astrophyiscal scenarios for magnetic mountains. We shall then discuss thermal mountains, and other possible
neutron star deformations in the context of neutron star models with exotic matter and/or pinned superfluidity.

\subsection{Modelling magnetic mountains}
\label{subsec:magmountains}

Chandrasekhar and Fermi~\cite{Chandrasekhar:1953ef} (see also~\cite{Ferraro:1954ff}) first pointed out that magnetic
fields induce quadrupolar deformations on spherically symmetric stars, with ellipticities (which we hereafter denote by
$\varepsilon_{\rm B}$) of the order of the ratio between the magnetic energy $E_{\rm B}$ and the gravitational energy $E_G$, i.e.
\begin{equation}
\varepsilon_{\rm B}\sim\frac{E_{\rm B}}{E_{\rm G}}\sim\frac{R^3\langle
  B^2\rangle}{GM^2/R}\sim(10^{-6}-10^{-5})\langle B_{15}^2\rangle\,,\label{energyest}
\end{equation}
where $\langle B^2\rangle$ is the volume-averaged square field strength.  It was later
noted~\cite{Wentzel:1960ww,Ostriker:1969og} that the magnetic-induced quadrupole ellipticity $\varepsilon_{\rm B}$ can
be positive or negative, i.e. the shape of the star can be oblate or prolate, depending on the magnetic field
structure. Indeed, neutron star magnetic fields can have poloidal or toroidal structure\footnote{In polar coordinates,
  the field-strength components $B_r$, $B_\vartheta$ are poloidal, while $B_\varphi$ is toroidal.}, and poloidal fields
$B_{\rm pol}$ tend to deform the star to an oblate shape, while toroidal fields $B_{\rm tor}$ tend to deform it to a
prolate shape.  When poloidal and toroidal components are both present, the deformation has both positive and negative
contributions.  For instance, in a specific example of magnetic field geometry the deformation has been estimated to
be~\cite{Ostriker:1969og}
\begin{equation}
\varepsilon_{\rm B}\sim E_G^{-1}R^3\left(3\langle B^2_{\rm pol}\rangle-\langle B^2_{\rm tor}\rangle\right)\,.\label{estog}
\end{equation}
The relative contributions of poloidal and toroidal components can change if the magnetic field geometry is different.

Similar estimates have been derived for a neutron star with a superconducting phase (see the discussion below), in the
case of a purely toroidal field~\cite{Cutler:2002nw}:
\begin{equation}
\varepsilon_{\rm B}\sim10^{-6}\langle B_{15}\rangle\frac{H_{\rm c1}}{10^{15}\,G}\,,
  \label{cutlest2}
\end{equation}
where $H_{\rm c1}$ (which is believed to be $\sim10^{15}$ G in neutron star cores~\cite{Glampedakis:2010sk}) is the
critical field strength characterizing superconductivity.  These expressions have been computed for a constant density
star, and assuming a dipole field configuration. However, as we discuss below, more sophisticated numerical analyses
show that Eqns. (\ref{energyest}), (\ref{estog}), (\ref{cutlest2}) give good order-of-magnitude estimates for magnetic
deformations of neutron stars.

When a neutron star is prolate (which requires a prevailing toroidal field), a spin-flip mechanism can take
place~\cite{Jones:1975jj,Cutler:2002nw}, in which the wobble angle grows until the symmetry axis and the rotation axis
become orthogonal: an optimal geometry for GW emission (see Eq.~\eqref{h0mount}). 

Presently, we do not have direct information on the internal structure of neutron star's magnetic fields. However, it is
generally believed that they should have both a poloidal and a toroidal component. Indeed, analytical
computations~\cite{Prendergast:1956kh,Tayler:1973ty} and numerical
simulations~\cite{Braithwaite:2005su,Braithwaite:2007fy} show that purely poloidal and purely toroidal configuration are
unstable in an Alfv\'en timescale\footnote{The instability of purely toroidal configurations is also shown
  in~\cite{Akgun:2007ph} for a superconducting neutron star, by studying the variations of an energy
  functional.}. Conversely, poloidal-toroidal magnetic fields can develop from the evolution of arbitrary initial
fields~\cite{Braithwaite:2005ps}. The fields found in~\cite{Braithwaite:2005ps,Braithwaite:2005xi} have a so-called
``twisted-torus'' structure: a poloidal magnetic field extending throughout the star and in the exterior, and a toroidal
field confined to a torus-shaped region inside the star.

GR and Newtonian models of stationary neutron star configurations with a twisted-torus magnetic field have been carried
out in~\cite{Ciolfi:2010td,Lander:2010lj}. They are based on the numerical integration of the GR MHD equations for an
axisymmetric ideal fluid (which can be reduced to a single partial differential equation, the so-called Grad-Shafranov
(GS) equation), and predict that the magnetic field induces a quadrupole ellipticity $\sim(10^{-6}-10^{-5})~B_{15}^2$
(where $B_{15}$ is the surface magnetic field)~\footnote{In these models the magnetic field-strength has as the same
  order of magnitude on the surface and in the interior.}, consistent with the estimates (\ref{energyest}),
(\ref{estog}).  Moreover, the results of~\cite{Ciolfi:2010td,Lander:2010lj} show that in a twisted-torus configuration,
the ratio of toroidal field energy over poloidal field energy is $\langle B_{\rm tor}^2\rangle/\langle B_{\rm
  pol}^2\rangle\lesssim0.15$. Therefore, in these configurations the poloidal field prevails, the star is oblate, and
the spin-flip mechanism does not occur.  More recently, twisted-torus configurations (with $\langle B_{\rm
  tor}^2\rangle/\langle B_{\rm pol}^2\rangle\lesssim0.1$) have also been described by solving the fully non-linear
Einstein-Maxwell's equations, see e.g.~\cite{Pili:2014npa,Uryu:2014tda}.

A different class  of twisted-torus configurations has later been found
in~\cite{Ciolfi:2013dta}, where - with an appropriate prescription of the azimuthal currents - the ratio
$\langle B_{\rm tor}^2\rangle/\langle B_{\rm pol}^2\rangle$ can be as large as $\sim0.9$, i.e. the toroidal field can
prevail. In these models the deformation of the star is larger than the prediction of Eqns.~(\ref{energyest}),
(\ref{estog}) by a factor $\sim100$
\begin{equation}
\varepsilon_{\rm B}\sim10^{-4}B_{15}^2\label{estimatec13}\,.
\end{equation}
The equilibrium properties of the magnetized star depend on whether the EOS is barotropic or non-barotropic. The models
in~\cite{Ciolfi:2010td,Lander:2010lj,Ciolfi:2013dta} assume a barotropic EOS. As shown
in~\cite{Mastrano:2011tf,Glampedakis:2016gl}, if the EOS is non-barotropic one is free to prescribe the magnetic field,
and the ratio between toroidal and poloidal energy becomes a free parameter, even though it is likely to be constrained
by the requirement of stability. In the model studied in~\cite{Mastrano:2011tf},
\begin{equation}
  \varepsilon_{\rm B}\sim10^{-5}B_{15}^2
  \left(1-0.64\frac{\langle B_{\rm tor}^2\rangle}{\langle B_{\rm pol}^2\rangle}
  \right)\,.
\end{equation}
If the ratio $\langle B_{\rm tor}^2\rangle/\langle B_{\rm pol}^2\rangle$ is sufficiently large, i.e. the toroidal field
prevails, the deformation can be as large as that described by Eq.~(\ref{estimatec13}). In these configurations (such as
in those studied in~\cite{Ciolfi:2013dta}) the neutron star is prolate, and the spin-flip mechanism can take place.

If higher-order multipoles of the magnetic field are present, the (prolate) deformation can be even
larger~\cite{Mastrano:2015rfa}. Alternative geometries with poloidal-toroidal fields have also been studied, such as the
``tilted torus'' configuration, in which the poloidal and the toroidal fields have misaligned symmetry
axes~\cite{Lasky:2013bpa}.

The models discussed above do not take into account the fact that outer cores of neutron star are expected to
contain superconducting protons (and superfluid neutrons). In a (type II) superconducting phase, the magnetic field is
quantized into fluxtubes surrounded by non-magnetised matter.  The force exerted
by the quantized field is not the Lorentz force, but a fluxtube tension force~\cite{Mendell:1991mn,Glampedakis:2010sk};
therefore, the description based on the GS equation is not adequate to model the superconducting phase (even though they
still obey a GS-type equation). In recent years, the deformation of magnetized superconducting neutron stars has been
computed by numerical integrations of the equations for the fluxtube magnetic force for toroidal
fields~\cite{Akgun:2007ph,Lander:2011yr}, for purely poloidal fields~\cite{Henriksson:2012qz}, and for twisted-torus
fields~\cite{Lander:2012zz}. All these computations give deformations consistent with the analytical
estimates~\cite{Cutler:2002nw} of Eq.~(\ref{cutlest2}).

We remark that $\varepsilon_{\rm B}\sim B^2$ in non-superconducting neutron stars, while
$\varepsilon_{\rm B}\sim B$ in neutron stars with a superconducting phase. This is due to the different magnetic force
acting in superconducting matter. Comparing Eqns.~(\ref{energyest}), (\ref{estimatec13}), we can see that in magnetars
(with $B\sim10^{15}$ G) the magnetic deformations have the same order of magnitude regardless of whether a
superconducting phase is present; conversely, in ordinary neutron stars - such as the observed pulsars - the magnetic
field strength is $\sim10^{12}$ G or smaller, and magnetic deformations are much larger if a (type II) superconducting
phase is present.

\subsection{Magnetic mountains in newly-born magnetars and in known pulsars}
\label{subsec:astroscen}
\begin{figure*}[htb!]
  \includegraphics[width=0.55\textwidth]{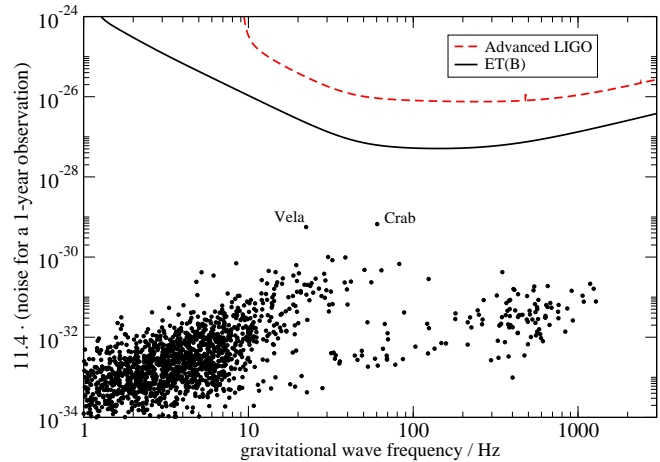}
  \caption{{\it GW detectability of magnetic mountains in known pulsars.} GW strain from the population of known
    pulsars, assuming a $1$-year phase coherent observation and quadrupole deformations given by Eq.~(\ref{cutlest2}).
    The detector noise curves were taken from \cite{LIGOcurve, ETcurve}. All pulsar data were taken from the ATNF
    database~\cite{ATNF}.}\label{magmondetect}
\end{figure*}

As we discussed in Sec.~\ref{sec:magnetars}, observations of SGRs and AXPs, with their large spin-down rates and intense
burst activity, suggest that these objects are magnetars~\cite{Duncan:1992dt}. Due to their very large magnetic fields,
magnetars are
expected to have the largest magnetic mountains. However, since the observed SGRs and AXPs have low spin frequencies, 
they cannot be GW sources for ground-based interferometers.

The standard magnetar model~\cite{Duncan:1992dt,Goldreich:1992gr} predicts that the strong magnetic fields form at the
birth of the neutron star, through convection and dynamo effects, and can last for $\sim10^4-10^5$
years~\cite{Goldreich:1992gr,Pons:2007vf}.  If this scenario is accurate, a non-negligible fraction ($\sim10\%$) of
newly-formed neutron stars can be born as magnetars~\cite{Kouveliotou:1998ze,Woods:2004kb}, and, if their rotation rates
are large enough, they can be potential sources for GW detectors (see e.g.~\cite{DallOsso:2008kll}).  Presently, it is
not clear what the actual rotation rate of newly-born neutron stars is; current estimates suggest that it should not be
larger than few hundreds of Hz~(see e.g.~\cite{Camelio:2016fan} and references therein). A newly-born magnetar with
$B\sim10^{15}$ G could have a quadrupole deformation $\varepsilon_{\rm B}\sim10^{-6}-10^{-5}$~(\ref{energyest}) (even
larger if the toroidal field prevails, see Eq.~(\ref{estimatec13})). If $D\sim10$ kpc and $P\sim10$ ms, it may yield a
GW strain~(\ref{h0mount}) $h_0\sim10^{-26}$ , potentially detectable by ET. However, the event rate of galactic
newly-born magnetars (comparable with that of galactic supernovae) is too low to make them a promising GW source.

We have observed, instead, several pulsars with magnetic field-strengths $\lesssim10^{12}$ G, and rotation rates in the
bandwidth of ground-based GW detectors. The GW emission from these objects is negligible, unless a type II
superconducting phase is present. Indeed, when $B\sim10^{12}$ the estimate for (type II) superconducting
stars~(\ref{cutlest2}) is $\varepsilon_{\rm B}\sim10^{-9}$, while the estimate for non-superconducting
stars~(\ref{energyest}) is a factor $\sim1000$ smaller. As noted in~\cite{Andersson:2009yt}, a
(superconducting) MSP with $B\sim10^{12}$ G at $D=1$ kpc would then emit a GW strain $h_0\sim10^{-27}$,
marginally detectable by ET. This estimate, however, is probably too optimistic: all known MSPs have
magnetic fields much smaller than $10^{12}$ G, while pulsars with $B\sim10^{12}$ have periods much larger than $\sim1$
ms. In Fig.~\ref{magmondetect} we show the GW strain for the population of known pulsars, assuming that the quadrupole
deformation is given by the estimate~(\ref{cutlest2}), corresponding to the presence of a type-II superconducting phase
in the core. We can see that, under these assumptions, the expected signal from known pulsars would be well below the
sensitivity curves of second- and third-generation interferometers.

\subsection{Magnetic mountains in millisecond magnetars from compact binary mergers}
\begin{figure*}[htb!]
\includegraphics[width=.44\textwidth]{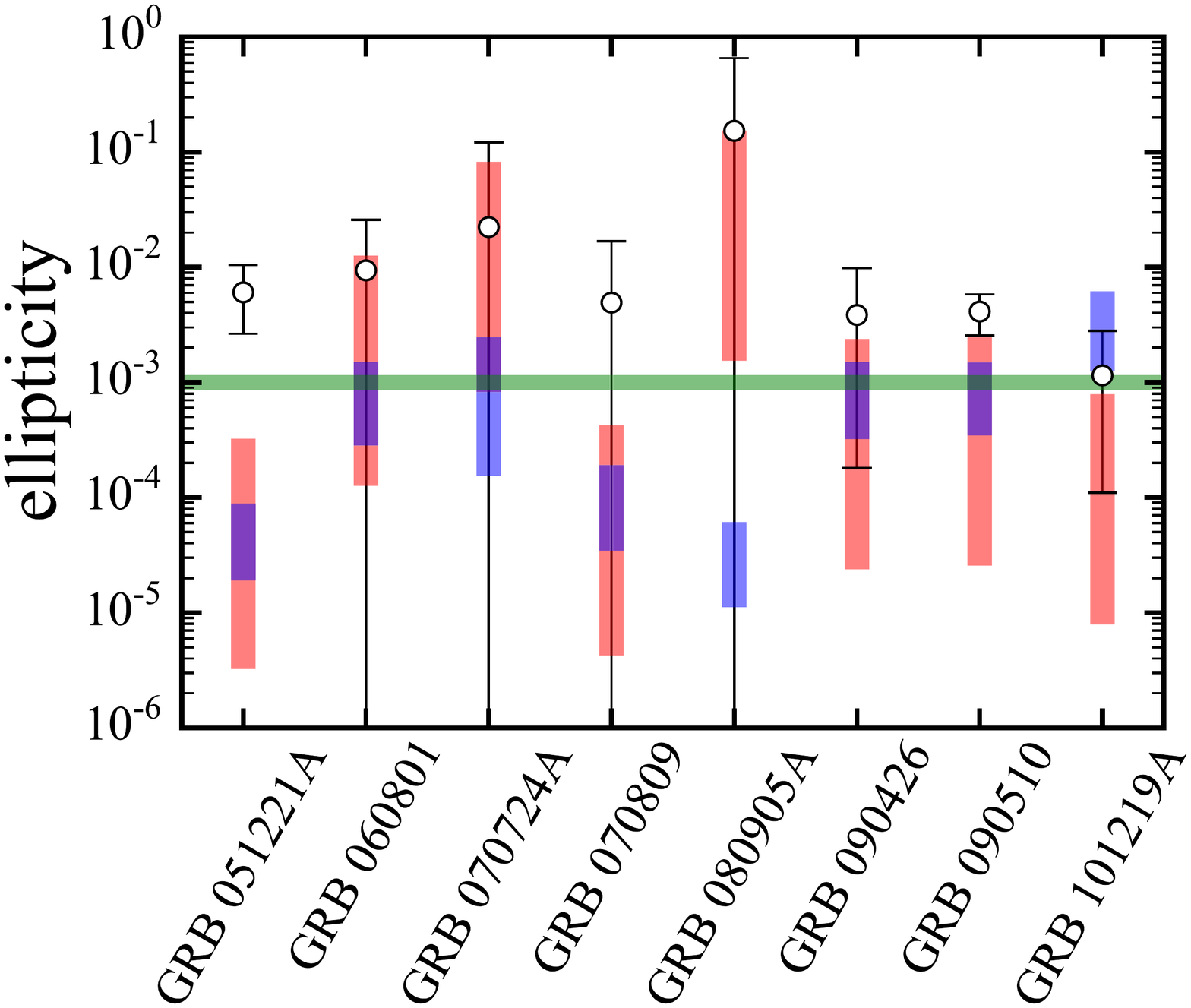}
\includegraphics[width=.462\textwidth]{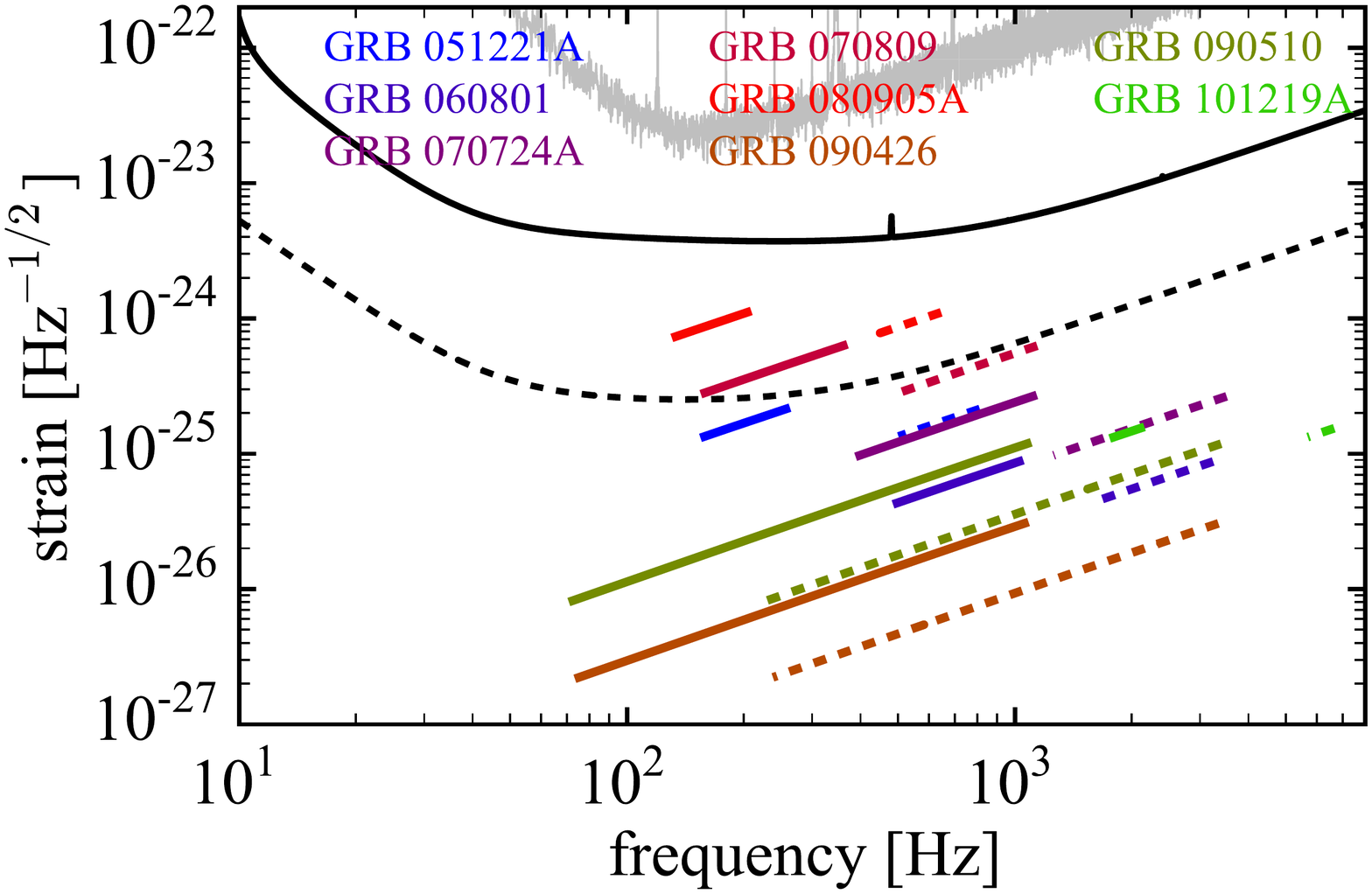}
\caption{\emph{Observational upper limits on magnetar deformation and GW emission.} Left panel: Upper limits on the
  neutron star ellipticity from eight short GRBs, given by Eq.~(\ref{lgllimit}) (black point with error bars). The red bars
  show the theoretical estimate from Eq.~(\ref{energyest}). The blue bars show the range of maximal ellipticities for
  which the spin flip can take place. The green line represents the ellipticity which can be induced by $f$-mode
  excitation (see section~\ref{sec:fmode}).
  Right panel: Upper limits on the GW strain amplitude evolution, compared with the sensitivity curves of Advanced LIGO and ET,
  for $\eta=0.1$ (dotted lines) and for $\eta=1$ (solid lines). The observation time is assumed to be $10^5$ s for GRBs
  with power-law decay, while it coincides with the GRB emission time for those with steeper decay (figure
  from~\cite{PLKG2016}).
}
\label{fig:lgl}
\end{figure*}

Binary neutron star mergers can result in strongly magnetized neutron stars~\cite{Zhang:2012wt,Giacomazzo:2013uua}. They
can be meta-stable supramassive neutron stars living from few seconds to hours~\cite{Ravi:2014gxa}, but can also be
stable neutron stars~\cite{Giacomazzo:2013uua}. These objects have been proposed to be the central engine of short GRB,
in the so-called {\it millisecond magnetar model}~\cite{dai_lu1998,zhang_meszaros2001,rowlinson_etal2013}. In this
scenario, the energy injection in the GRB is due to dipole radiation from a spinning-down magnetar. Strong evidence
supporting the millisecond magnetar model is its ability to explain the shape of the observed GRB X-ray spectra, which
are characterized by a plateau of $\sim10-10^2$ s followed, in most cases, by a power-law decay; in a subset of short
GRBs, instead, the decay is much steeper. The X-ray spectrum from a spinning-down magnetar resulting from a compact
binary coalescence would have the same structure: a plateau followed by a power-law decay if the neutron star is stable,
by a steeper decay if it is a supramassive star eventually collapsing to a black hole.

The millisecond magnetar model is probably the most promising astrophysical scenario for GW detection from magnetic
mountains.  Indeed, numerical simulations show that the magnetars resulting from a binary neutron star coalescence would
have periods of the order of milliseconds - significantly smaller than the expected periods of newly-born neutron
stars. Moreover, in this scenario we have astrophysical data from these objects, since they would be the engine of short
GRBs.

Preliminary estimates~\cite{DallOsso:2014hpa} show that the GW emission from these sources could be marginally
detectable by second-generation interferometers, and detectable by third-generation interferometers such as ET (even
more optimistic estimates have been derived in~\cite{Corsi:2009jt}).  However, these computations are based on the
assumption that GW emission gives a significant contribution to the magnetar spin-down. Since the millisecond magnetars
predicted in this model are progenitors of short GRBs, the assumptions of the model can be tested against GRB observations.

In~\cite{PLKG2016}, observations of the X-ray light curve of short GRBs have been used to constrain the ellipticity of the
(stable) neutron stars produced in the merger, and then to set a bound on the GW emission from these objects.  Indeed,
the late time tail of the observed X-ray spectrum is $\sim t^{-2}$, suggesting that the spin-down is mainly due to dipole
electromagnetic emission rather than to GW emission. Therefore, the comparison between the observed X-ray spectrum and
the predictions of the model allows to set
an observational upper limit on the neutron star ellipticities:
\begin{equation}
  \varepsilon_{\rm B}\lesssim0.33\,\eta\, I_{45}^{1/2}\,\left(\frac{L_{\rm em}}{10^{49}\,{\rm erg/s}}\right)^{-1}\,
    \left(\frac{t_{\rm b}}{100\,{\rm s}}\right)^{-3/2}\,,
\label{lgllimit}
\end{equation}
where $\eta\le1$ is the conversion efficiency of spin-down energy into X-ray luminosity, $L_{{\rm em}}$ is the short GRB
luminosity in the inital plaeau phase, and $t_{\rm b}$ is the plateau duration. This upper limit (through
Eq.~(\ref{h0mount})) corresponds to an upper limit on the expected GW signal. The analysis of~\cite{PLKG2016} shows that
the agreement between the observed X-ray spectrum and the predictions of the model requires that the magnetar spin-down
is mainly due to dipole emission, with a very small contribution from GW emission. Therefore, these signals are not
expected to be detected by Advanced LIGO/Virgo, but are potentially detectable by third-generation interferometers such as
ET. This can be seen in Fig.~\ref{fig:lgl}. In the left panel we show the upper limit (\ref{lgllimit}) on the neutron
star ellipticities, assuming $\eta=0.1$ (black points with error bars) for eight observed short GRBs. In the right panel
we show the corresponding upper limits on the GW strain amplitude evolution, for $\eta=0.1$ (dotted lines) and for
$\eta=1$ (solid lines), compared with the sensitivity curves of Advanced LIGO and ET.

\subsection{Magnetic mountains in accreting systems}

Magnetic mountains can also form in LMXBs. In this case, the mountain - a localized deformation - is formed by the
accreting matter, but it is sustained by the magnetic field. Since the mountain is not sustained by crustal rigidity,
its quadrupole ellipticity can be larger than the upper bound associated to the maximum crustal
strain~\cite{Ushomirsky:2000ax}. In this scenario~\cite{Payne:2004vt,Melatos:2005ez,Vigelius:2008fv}, accreted matter
accumulates in a column over the polar cap, distorting the magnetic field and reducing the magnetic moment of the star.
As the stars accretes, the exterior magnetic field evolves approximately as $B=B_*(1+M_{\rm acc}/M_{\rm c})^{-1}$ where
$B_*$ the field when accretion starts, $M_{\rm acc}$ the accreted matter and $M_{\rm c}$ the critical accreted mass
above which the magnetic moment starts to change~\cite{Shibazaki:1989sm}. The quadrupole
ellipticity associated to magnetic mountains in LMXB is~\cite{Priymak:2011zv,Haskell:2015psa}
\begin{equation}
  \varepsilon\sim\frac{M_{\rm acc}}{M_\odot}\left(1+\frac{M_{\rm acc}}{M_{\rm c}}\right)^{-1}\,,\label{epslmxb}
\end{equation}
where $M_{\rm c}\sim10^{-7}(B_*/10^{12}G)^{4/3}$.  This formula is only accurate for $M_{\rm acc}\lesssim M_{\rm c}$:
for larger values of the accreted mass, the quadrupole ellipticity saturates.  Dynamical MHD simulations suggest that
these deformations can persist for timescales of the order of $\sim10^5-10^8$ years~\cite{Vigelius:2009eg}.

Although magnetic fields of LMXBs are much weaker than those of magnetars, the quadrupole ellipticities of local
mountains can be much larger than those produced by a global magnetic deformation of the star. Moreover, LMXBs have
rotation rates much larger than those of magnetars, well within the sensitivity band of ground-based
interferometers. Fast rotation also enhances the GW signal from a deformed star. Therefore, as shown
in~\cite{Priymak:2011zv,Haskell:2015psa}, a buried field $B_*\gtrsim10^{12}$ G could generate a deformation, and then a
GW emission, strong enough to be detected by Advanced LIGO/Virgo~\cite{Priymak:2011zv,Haskell:2015psa}. 

\subsection{Thermal mountains}
\label{subsec:thermal}
Accreting neutron stars in LMXBs can have quadrupolar deformations due to temperature gradients in the accreted
crust~\cite{Bildsten:1998ey,Ushomirsky:2000ax}. Indeed, as the matter accretes, it is buried and compressed until
nuclear reactions occur (electron capture, neutron emission, etc.)~\cite{Haensel:1990hz}. These reactions heat the
crust, and since the accreted matter is expected to be asymmetric, the temperature gradient produced by nuclear reactions
is also asymmetric, giving rise to quadrupolar deformations in the neutron star. These deformations are called ``thermal
mountains''; they also belong to the broader class of ``elastic mountains'', i.e. deformations sustained by the elastic
strain of the crust. Presently, thermal gradients are the only known viable mechanism to produce elastic mountains.

\begin{figure*}[htb!]
\includegraphics[width=.53\textwidth]{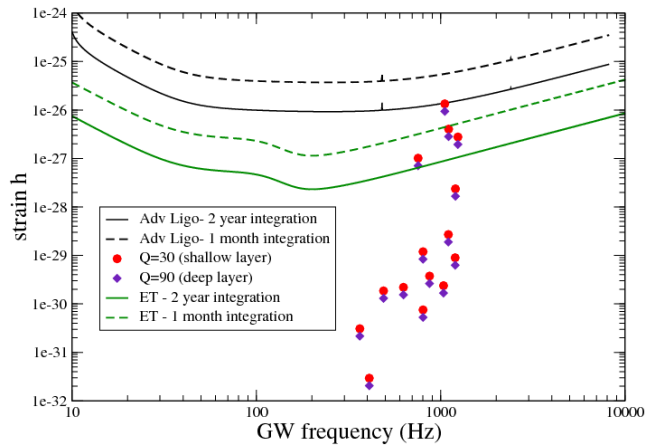}
\caption{\emph{Detectability of thermal mountains.} We show the estimated GW strain from thermal mountains formed in a
  set of observed LMXB outbursts, for different values of the reaction threshold energy $Q$, assuming $\delta T_{\rm
    q}\simeq10^5$ K. The detector noise curves assume one month (dashed) or two years (solid) of phase-coherent
  observation (from~\cite{Haskell:2015psa}).
}
\label{fig:therm}
\end{figure*}

The deformation due to a thermal gradient with quadrupolar component $\delta T_{\rm q}$ (which we denote by
$\varepsilon_{\rm th}$) is~\cite{Ushomirsky:2000ax,Haskell:2015psa}
\begin{equation}
\varepsilon_{\rm th}\sim10^{-10}R_6^4\left(\frac{\delta T_{\rm q}}{10^5\,{\rm K}}\right)\left(\frac{Q}{30\,{\rm MeV}}\right)^3
\,,\label{epstherm}
\end{equation}
where $Q$ is the threshold energy of electron capture by nuclei. The quadrupolar thermal gradient is just a fraction of
the total thermal gradient $\delta T$, which is expected to be, at most, of the order of $10^6$
K~\cite{Ushomirsky:2001pd}, if produced by an outburst. In the most optimistic scenario $\delta T_{\rm
  q}\lesssim0.1\,\delta T\lesssim10^5$ K , yielding a thermal mountain of the order of $\varepsilon_{\rm
  th}\sim10^{-10}$~\cite{Haskell:2015psa}.  The GW strain~(\ref{h0mount}) from this deformation would be
$h_0\sim10^{-27}-10^{-28}$, too weak to be detected by Advanced LIGO/Virgo, but potentially detectable by third-generation GW
intereferometers such as ET. This can be seen in Fig.~\ref{fig:therm}, where we show the estimated GW strain from a set
of observed LMXBs, computed from Eqns.~(\ref{epstherm}) and (\ref{h0mount}), for different values of the electron
capture threshold energy $Q$ and assuming $\delta T_{\rm q}\simeq10^5$ K. In the figure, the Advanced LIGO and ET sensitivity
curves are shown for different values of the integration time (one month, which is a typical outburst duration, and two
years) because the persistence timescale of the mountain, and then the typical duration of the signal, is presently
poorly constrained. In the most favourable case (the deformation persists after the outburst, $\delta T_{\rm q}/\delta
T\simeq0.1$) some of the systems could emit a signal detectable by ET. We note, however, that this value of $\delta
T_{\rm q}/\delta T$ is an upper limit with no solid physical motivation; if the actual value of this ratio is smaller
than $\sim0.1$, even third-generation GW interferometers will not be able to detect thermal mountains.

The GW emission from thermal mountains, regardless of its direct detectability, may have a role in the spin evolution of
LMXBs.  Twenty years ago, thermal mountains have been proposed to explain the apparent spin cut-off in
LMXBs~\cite{bildsten1998}. However (as discussed in detail in Sec.~\ref{sec:rmode_pheno}), it was later understood that
this feature can probably be explained in terms of the coupling of the stellar magnetic field with the accretion disk.
More recently, thermal mountains have been revived by the observations of the pulsar PSR J$1023+0038$, which shows a
transition between a radio MSP state and a LMXB state~\cite{Archibald:2009zb}. Remarkably, the pulsar spins down faster
(by $\sim20\%$) during the LMXB state. It has been suggested~\cite{Haskell:2017ajb} that this enhanced spin-down is due
to GW emission by a thermal mountain. To explain the additional spin-down, a deformation $\varepsilon_{\rm
  th}\sim5\times10^{-10}$ would be required, which - as we have discussed above - can indeed be a thermal mountain, and
would emit gravitational radiation detectable by third-generation interferometers such as ET. If these results are
confirmed, they will provide the first observational indirect evidence supporting the existence of GW emission from
neutron star mountains.

Thermal mountains (and more generally, elastic mountains) are limited by the maximum stress that the crust can sustain
before breaking~\cite{Ushomirsky:2000ax,Haskell:2006sv,JohnsonMcDaniel:2012wg}:
\begin{equation}
\varepsilon_{\rm th}\lesssim\frac{\mu_{\rm cr}\sigma_{\rm br}V_{\rm cr}}{GM^2/R}\sim10^{-5}\left(\frac{\sigma_{\rm br}}{0.1}\right)\,,
\label{limitelast}
\end{equation}
where $\mu_{\rm cr}$ is the shear modulus of the crust, $V_{\rm cr}$ is the volume of the crust, and $\sigma_{\rm br}$
is the crustal breaking strain, which could be as large as $\sim0.1$~\cite{Horowitz:2009ya}~\footnote{It should be
  mentioned that the breaking strain found in~\cite{Horowitz:2009ya} is the result of numerical simulations with
  duration much shorter than the timescale associated with crust straining/relaxation.}.

We remark that this is just an upper bound: there is no reason to believe that neutron star have deformations close to
this value. For instance, the thermal mountain which has been suggested in~\cite{Haskell:2017ajb} to explain the
observations of PSR J$1023+0038$ is much smaller than the deformation in Eq.~\eqref{limitelast}. We also remark that
this bound applies to thermal mountains (more generally, to elastic mountains), but it does not necessarily apply to
magnetic mountains. Indeed, when the crust forms in a newly-born neutron star the magnetic deformation may already be
present; in this case, the equilibrium shape of the crust would be non-spherical, and the crust would not have to
sustain an elastic strain, at least while the magnetic field is present.

\subsection{Exotic mountains}
\label{sec:exotic}

The ellipticities associated with the neutron star deformations discussed so far are clearly pessimistic from the
perspective of GW observability. A somewhat more promising situation may arise if we consider neutron stars with quark
matter cores. If present, quarks are most likely to find themselves in a color-superconducting state, the exotic
properties of which could accommodate significantly larger quadrupolar deformations (of both magnetic and elastic
nature) as compared to conventional hadronic matter.

\begin{figure*}[htb!]
\includegraphics[width=0.55\textwidth]{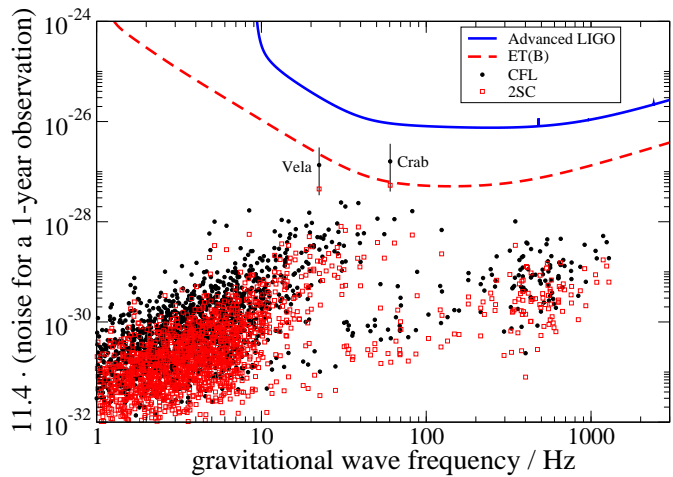}
\caption{\emph{GW detectability of color-magnetic mountains}. In this figure, adapted from \cite{KG_etal2012}, we show the
estimated GW strain (assuming a 1-year phase-coherent observation) from the population of known pulsars, assuming they
have CFL (or 2SC) quark cores. We have used the fiducial parameters $\mu_q = 400\,\mbox{MeV}$, $V_q = 0.5 V_{\rm star}$,
and $\langle B\rangle = 2 B_{\rm surf}$. }
\label{fig:color_mountain}
\end{figure*}

Among the various possible quark matter incarnations (for a review see \cite{alford_etal2008}) the color-flavor-locked (CFL) 
phase, where all three quark species $(u,d,s)$ are paired, appears to be the most favoured one. Another well-studied phase 
is the two-flavor superconducting (2SC) state where just the $(u,d)$ quarks pair. Both of these phases come with remarkable 
magnetic permeability properties. The stellar magnetic field threading a CFL/2SC core is likely to do so by forming an array of 
quantised  vortices in the same way that fluxtubes are formed in the more familiar type II protonic superconductor 
\cite{iida_baym2002, alford_sedrakian2010}. These vortices, however, are \emph{color-magnetic} (rather than just magnetic) 
in the sense that they are carriers of an admixture of magnetic and gluonic field degrees of freedom. The latter component 
is the dominant one and, as a consequence, the energy per unit length $\cE_{\rm X}$ (where X labels the phase) of a color-magnetic 
vortex can be 2-3 orders of magnitude higher than  that of a conventional protonic fluxtube \cite{iida_baym2002, alford_sedrakian2010}. 
This property entails an amplified vortex array tension and opens the possibility of creating a large deformation in a neutron star's 
CFL/2SC core \cite{KG_etal2012}. The ellipticity of this \emph{internal} color-magnetic mountain can be estimated 
by means of the ratio of the vortex array tension energy to the stellar gravitational energy: 
\be
\varepsilon_{\rm X} \approx \frac{\cN_{\rm X} \cE_{\rm X} V_q}{GM^2/R},
\label{epsilon_X}
\ee
where $\cN_{\rm X}$ is the vortex surface density and $V_q$ is the volume of the quark core.  
For a CFL core (and assuming canonical stellar parameters) this expression leads to \cite{KG_etal2012},
\be
\varepsilon_{\rm cfl} \approx 10^{-7} \langle B_{12} \rangle\left (\frac{V_q}{V_{\rm star}} \right )
\left (\frac{\mu_q}{400\,\mbox{MeV}} \right )^2,
\label{epsilon_cfl}
\ee
where $\mu_q$ is the strange quark chemical potential (normalised to a canonical value) and, as before, 
$\langle B\rangle$ represents the volume-averaged interior magnetic field.

The GW strain associated with this deformation can be estimated with the help of Eq.~(\ref{h0mount}), using reasonable
values for the various parameters, i.e. $\mu_q = 400\,\mbox{MeV}$, $V_q = 0.5 V_{\rm star}$ and $\langle B\rangle = 2 B_{\rm
  surf}$ (as suggested by models of MHD equilibria in neutron stars, see e.g. \cite{ciolfi_etal2009, lander_jones2009}).
The resulting GW detectability by Advanced LIGO and ET of the pulsar population with known surface dipole fields $B_{\rm surf}$
is shown in Fig.~\ref{fig:color_mountain}. When compared to Fig.~\ref{magmondetect}, it is evident that the presence of
a color-magnetic deformation in neutron stars leads to a marked improvement in the strength of their GW signature. The
results presented here suggest that young pulsars such as the Crab or the Vela could be detectable by ET. A much larger
fraction of the pulsar population could become detectable (even by Advanced LIGO/Virgo) if we assume a higher magnetic field ratio
$\langle B\rangle/B_{\rm surf}$. 

A drastically different phase of quark matter could be that of a \emph{crystalline} color-superconductor (CCS)
\cite{rajagopal_sharma2006, mannarelli_etal2007}, effectively leading to a neutron star with a solid
core. Given that the shear modulus $\mu_{\rm ccs}$ of CCS matter is estimated to be much higher than that of the crustal 
bcc lattice \cite{mannarelli_etal2007}, the elastic deformation that could be sustained by a solid quark core
could be significantly greater than that of normal hadronic neutron stars.  For obtaining the ellipticity of the CCS mountain 
we can use a formula similar to Eq.~(\ref{energyest}). We find,
\be
\varepsilon_{\rm ccs} \approx \frac{\mu_{\rm ccs} V_q \sigma_{\rm br}}{G M^2/R} \approx  6 \times 10^{-4} 
\left (\frac{V_q}{V_{\rm star}} \right )  \left ( \frac{\mu_q}{400\, \mbox{MeV}} \right )^2
\left ( \frac{\Delta_q}{10\, \mbox{MeV}} \right )^2 \left (\frac{\sigma_{\rm br}}{0.01} \right ),
\label{epsilon_ccs}
\ee where $\Delta_q$ is a quark gap parameter (normalised to a canonical value) and the (highly uncertain) breaking
strain of a crystalline quark core is normalized to the fiducial value $\sigma_{\rm br}\sim0.01$. This simple estimate
is in good agreement with the results of more rigorous calculations \cite{owen2005, haskell_etal2007} and is about a
factor $\mu_{\rm ccs}/\mu_{\rm cr} \sim 10^3$ higher than the maximum elastic deformation of the crust
(Eq.~(\ref{limitelast})).  As discussed in \cite{haskell_etal2007}, the presence of such a large internal deformation in
neutron cores could in principle be tested by the upper limits set by GW searches. According to the recently published
Advanced LIGO results \cite{abbott_etal2017} the most stringent upper limit is $\varepsilon \approx 10^{-7}$ which is well below
the result (\ref{epsilon_ccs}). However, the non-detection of GWs can only be used as a constraint for the
(unidentified) non-axisymmetric straining mechanism or the poorly known $\sigma_{\rm br}$ and not as evidence against
the existence of CCS matter.

From our brief exposition of deformations related to the presence of quark matter in neutron star cores
it should have become clear that observation of GWs from such sources could provide strong
evidence in favour of the existence of  these very exotic states of matter. But it should also be 
remembered that the properties of quark matter in neutron stars are largely uncertain and therefore
any theoretical predictions based on them should be viewed in a similar way. 


\subsection{Mountains and pinned superfluidity}
\label{sec:pinnedSF}

The presence of a pinned neutron superfluid component in the interior of neutron stars may lead to glitches (see next section) 
but could also have a number of interesting implications related to mountains. Vortex pinning could take place in the crust 
(with the pinning sites provided by the crustal lattice) and/or in the fluid core as a result of the interaction of the 
neutron vortices with the fluxtubes of the proton superconductor. 

As discussed in \cite{jones2010}, an elastically or magnetically deformed neutron star with a pinned superfluid component 
that has its angular momentum axis slightly misaligned with the stellar symmetry axes can modify the harmonic
structure of the emitted GW signal by allowing emission at both frequencies $2\Omega$ and $\Omega$ (whereas only the
former should be present in the absence of pinning and assuming a non-precessing state). 

Incidentally, it is worth mentioning that the same pinned superfluid acts as a fixed gyroscope and could easily stabilise the previously 
discussed spin-flip instability of a strong toroidal magnetic field component \cite{KG_DIJ2010}. 

Finally, a pinned superfluid could itself act as a source of deformation and lead to GW mountain emission \cite{jones2002}. 
Since this ``Magnus mountain''  is the only deformation mechanism that has not been considered  so far  in detail it is 
worth discussing it a bit more. As the name suggests, the straining mechanism responsible for this mountain is the Magnus force
acting on the pinned superfluid vortices due to the spin lag between the superfluid and the  normal stellar  component onto which
the vortices are attached (see \cite{ruderman1991a, ruderman1991b} for a more detailed discussion). 
Assuming both spin angular frequencies $\mathbf{\Omega}_{\rm n}, \mathbf{\Omega}$ to be aligned 
along the same axis (say the $z$-axis) the resulting Magnus force is purely radial in cylindrical coordinates and has a per unit volume 
magnitude,
\be
f_{\rm m} = 2 \varpi \rho_n  \Omega_{\rm lag} \Omega, 
\label{magnus1}
\ee where $\Omega_{\rm lag} = \Omega_{\rm n} - \Omega$ is the spin lag and $\varpi$ is the standard cylindrical radius.
In order to produce a non-axisymmetric quadrupolar deformation, the force (\ref{magnus1}) itself has to have a similar
non-axisymmetry. In turn, this requires a non-uniform distribution of pinned vortices which would entail a non-rigid
body rotational profile for the superfluid. Although detailed calculations for this non-axisymmetric Magnus force are
still lacking it is not too difficult to see how this property could come about in a realistic neutron star model. For
example, if pinning is provided by the magnetic fluxtubes then the generic non-axisymmetry of the stellar magnetic field
would imply a similar property for the spatial distribution of the pinning sites. In our present back-of
the-envelope-analysis the non-axisymmetric character of the Magnus force can be accounted for by multiplication of
(\ref{magnus1}) with a phenomenological dimensionless factor $\lambda \lesssim 1$.

The ellipticity of the resulting Magnus mountain can be estimated by means of the ratio between the volume-integrated Magnus 
force $F_{\rm m} = \int dV f_{\rm m}$ and the gravitational binding force \cite{jones2002}:
\be
\varepsilon_{\rm m} \sim \frac{F_{\rm m}}{F_{\rm grav}} \sim \frac{2 \lambda\Omega_{\rm lag} 
\Omega R^3 \rho V_{\rm pin} }{GM^2} \sim  \frac{2 \lambda \Omega_{\rm lag} \Omega R^3 M_{\rm pin}}{GM^2},
\ee
where $V_{\rm pin}$ and $M_{\rm pin} $ represent the volume and mass of the pinned superfluid. Parametrising this
result and assuming canonical stellar parameters we find,
\be
\varepsilon_{\rm m} \sim 5 \times 10^{-7}  \lambda P_{-3}^{-1} \left ( \frac{\Omega_{\rm lag}}{0.01\,\mbox{Hz}} \right ) 
\left ( \frac{M_{\rm pin}}{M_\odot}\right ).
\ee
For the spin-lag we have used a representative value so that the Magnus force is at most comparable to the pinning force 
in the crust or in the core, see \cite{link_cutler2002, link2003}. Assuming $\lambda \sim 1$, we can see that a 
reasonable range for ellipticity could be $\varepsilon_{\rm m} \sim 10^{-7} - 10^{-10}$,
with the most favourable case coming from the scenario of vortex pinning in the core ($ M_{\rm pin} \sim M_\odot$).
As also concluded in \cite{jones2002}, the expected Magnus mountain could be comparable to that
due to magnetic or elastic stresses and therefore could be of interest. 

\subsection{Future directions}

In recent years our understanding of the different processes which can lead to neutron star mountains greatly
improved. However, we still do not know which mountains are actually present in the different kinds of neutron stars
(magnetars, radio pulsars, LMXB, etc.). Indeed, past and present astrophysical observations mainly probe the surface and
the exterior of the star, while we have very limited information from its interior.

For instance, spin-down measurements give us accurate estimates of the exterior magnetic field of magnetars (and of
ordinary neutron stars), but if the field inside the star is much larger than in the exterior, we will need know the
interior field to find the size and shape of magnetic mountains.  Theoretical modelling can be improved, for instance to
better understand the stability properties of magnetic field configurations (which can tell us how much of the field is
toroidal and how much is poloidal). MHD calculations of neutron stars with exotic matter would also be very important,
because these objects can accomodate larger deformations than stars with conventional hadronic matter; as discussed
above, existing calculations of these systems are only back-of-the-envelope estimates. In any case, we need observational data to know the actual (volume averaged) strength of the interior magnetic fields.

The same holds for thermal mountains. Theoretical modelling can clarify the mechanism leading
to temperature anisotropies and can give us estimates of their actual shape and magnitude. In particular, we need to know the size of quadrupole contribution to temperature anisotropies, $\delta T_{\rm q}$. For instance, numerical simulations of
the post-burst thermal evolution would help us understanding how large these mountains can be.
However, even for thermal mountains, the power of theoretical modelling alone is limited:
only observational data can tell us definitely how large actual mountains are.

The most promising observational probe to study neutron star mountains is the GW signal they emit as the star
rotates. Once observed, this signal would provide a direct measurement of the mountain, shedding light on the formation
mechanisms. Even an indirect GW observation, i.e., the observation in the electromagnetic spectrum of a process due to
GW emission (such as the spin-down increase discussed in~\cite{Haskell:2017ajb}) would give us valuable information to
understand magnetic and thermal mountains and their formation.

\section{Glitches} 
\label{sec:glitches}
Radio pulsars have extremely stable rotation rates - they are the most precise natural clocks in the universe - but some
of them exhibit sudden increases in the rotation rate, called {\it glitches}. After the first glitch observations in the
Vela and Crab pulsars~\cite{Radjakrishnan:1969rm,Reichley:1969rd,boynton:1969pr,richards:1969gl}, more than three hundred
glitches have been observed in $\sim100$ pulsars~\cite{Espinosa:2011els}. More recently, glitches have also been
observed in gamma-ray pulsars~\cite{Ray:2011rk,Pletsch:2012ct} and in magnetars~\cite{Dib:2008dk} (MSPs
can also have small glitches, but they are very rare~\cite{Cognard:2004av}).

The glitch occurs in a very short timescale, still unresolved by observations (the best upper limit is $\sim40$
s~\cite{Dodson:2002gy}).  The relative increase in the rotation rate $\Delta\Omega/\Omega$
ranges from $\sim10^{-11}$ to $\sim10^{-5}$~\cite{Espinosa:2011els}, and is generally followed by an increase
in the spin-down rate.  Most glitches have $\Delta\Omega/\Omega\lesssim10^{-7}$, but few of them (such as those of the
Vela pulsar) are significantly larger, with $\Delta\Omega/\Omega>10^{-6}$.

The standard model of glitches is the superfluid model, in which glitches are due to superfluid neutrons in the neutron
star interior, which store the angular momentum in a quantized array of vortices.  The star spins down but its
superfluid component does not, because the vortices are ``pinned'', i.e. their positions are fixed.  Periodically, some
of the vortices ``unpin'', move outwards and in this way the superfluid component removes the excess angular momentum,
which is released in glitches.

An alternative model is the starquake model, in which glitches are due to the rigidity of the crust, which maintains its
shape as the star spins down and reduces its quadrupole deformation; the strain on the crust increases and,
periodically, there is a starquake, with a rearrangement of the moment of inertia and a glitch. This model can not
explain large Vela-like glitches, but it is in principle possible that some small glitches are due to this mechanism.
For a detailed discussion of the different glitch models, we refer the reader to~\cite{Haskell:2015jra} and references
therein.

When a glitch occurs, there is a sudden rearrangement of the neutron star structure (in both the superfluid and the
starquake models); therefore, glitches are expected to emit GWs.  In 2006, during the fifth Science Run of LIGO, the
Vela pulsar underwent a glitch. An analysis of the first-generation LIGO data looking for the excitation of stellar
oscillations did not find a GW signal, thus setting an upper limit on the GW strain
$h\le1.4\times10^{-20}$~\cite{Abadie:2010sf}. However, second-generation detectors are expected to be about one order
of magnitude more sensitive to this kind of signal. 

Different mechanisms have been suggested for GW emission by pulsar glitches. They belong to two classes: the signals
emitted during the glitch itself and those emitted during post-glitch relaxation.
\begin{itemize}
\item The energy released during the glitch can excite stellar oscillations~\cite{Sidery:2009at,Keer:2014uva} (see
  also~\cite{Sedrakian:2003sb,KGNA2009,SantiagoPrieto:2012ew}). The GW strain can be expressed in terms of the total
  energy $\Delta E$ deposited into the mode as~\cite{Kokkotas:1999mn}
  \begin{equation}
    h\sim2\times 10^{-23}D_1^{-1}\left(\frac{\Delta E}{10^{-12}M_\odot c^2}\right)^{1/2}
    \left(\frac{1\,{\rm kHz}}{f_{\rm gw}}\right)\left(\frac{0.1\,{\rm s}}{\tau_{\rm gw}}\right)^{1/2}\,,
    \label{hdeltae}
  \end{equation}
  where $f_{\rm gw}$, $\tau_{\rm gw}$ are the frequency and damping time of the oscillation (and of the GW emission).  A
  back-of-the-envelope estimate of the deposited energy, assuming that the rotational energy change associated to a
  glitch is entirely converted into non-radial oscillation, is
  $\Delta E\sim (2\pi)^2I\Delta\Omega/\Omega\lesssim10^{-12}M_\odot\,c^2$~\cite{Andersson:2001kx}. Then,
  assuming that most of the energy is deposited into the fundamental mode (as suggested by hydrodynamical
  simulations~\cite{Sidery:2009at,Keer:2014uva}), $f_{\rm gw}\sim 1$ kHz, $\tau_{\rm gw}\sim0.1$ s and the GW
  strain~(\ref{hdeltae}) is too weak to be detected by Advanced LIGO/Virgo, but potentially detectable by an ET-class detector.

  These estimates, however, are probably too optimistic, at least in the two-fluid scenario. Indeed, a detailed
  modelling~\cite{Sidery:2009at} shows that non-superfluid matter (ions in the crust and protons in the core) spin up,
  while superfluid neutrons spin down; therefore, the available energy is smaller by a factor $\sim10^7$. This leads to
  a GW strain $\sim10^{-26}$ or smaller, too weak to be detected even by third-generation interferometers. In the
  starquake scenario, instead, a careful analysis of the interplay between deformation energy and rotational kinetic
  energy shows that the GW strain can be as large as $\sim10^{-23}$~\cite{Keer:2014uva}; however, as mentioned above,
  this scenario fails to describe Vela-like glitches.

\item During a glitch, a large number 
  of vortices move outwards, in a sort of ``avalanche'', transferring angular momentum to the crust. This process has
  been studied using a very sophisticated toy-model in which the superfluid neutron star is represented by a
  zero-temperature Bose condensate with dissipation described by a non-linear Schr\"odinger equation, the
  Gross-Pitaevskii (GP)
  equation~\cite{Warszawski:2011vy,Warszawski:2012zq,Warszawski:2012mb,Warszawski:2013wm}. Numerical simulations of the
  GP equation model describe the collective vortex migration associated to a glitch. The simulations of the GP equations
  have then been extended using Monte-Carlo techniques, to include a larger number of vortices, compute the GW emission,
  and find the distribution of glitch parameters to be compared with observational data.

  In the GP model the star is described as an infinite, rotating cylinder of fluid, with a rectangular grid of pinning
  sites. Applying an external spin-down torque, the vortices unpin and repin, with an average displacement of $\Delta r$
  (which is a key parameter of the model).  The GP equation gives the evolution of the vortex distribution, as they
  unpin and repin.  Each vortex generates a solenoidal velocity field, which is affected by the displacement
  process. Therefore, the velocity of the fluid acquires a non-axisymmetric component, which can be reconstructed
  by the vortex distribution. The non-axisymmetric velocity generates a time-varying current
  quadrupole moment, and then a GW emission with strain amplitude:
  \begin{equation}
    h\sim10^{-23}D_1^{-1}\left(\frac{\Delta r}{1\, {\rm cm}}\right)^{-1}\left(\frac{\Delta\Omega/\Omega}{10^{-5}}\right)
    \left(\frac{f_{\rm spin}}{100\,{\rm Hz}}\right)^3\,.
    \label{aval}
  \end{equation}
  The gravitational signal decreases as the travel distance $\Delta r$ increases. This is due to the fact that, for
  larger values of $\Delta r$ (keeping fixed $\Delta\Omega/\Omega$), the number of vortices involved is
  smaller and the current quadrupole is also smaller.
Recently, it has been suggested~\cite{Melatos:2015oca} that part of the large scale non-axisymmetries in
the velocity field produced by a vortex avalanche can persist in the inter-glitch recovery phase. If this is true, the
GW signal can be as large as $h\sim10^{-21}$, potentially detectable even by second-generation interferometers.

  We remark that the GP model is based on simplifying assumptions, which may affect the GW detectability estimates: it
  is not an hydrodynamical model, it is assumed that the system is weakly interacting, the grid of pinning sites is
  defined {\it a priori}, and the effect of magnetic field, which can be relevant for vortex pinning in the core, is
  neglected. However, this model is very powerful, since it captures the collective behaviour of vortices, which is
  likely to have a fundamental role in pulsar glitches. 

\item After the glitch, the neutron star resumes its spin-down, entering in a recovery phase which can last from months
  to years. However, since - just after the glitch - the crust has a rotation rate larger than the superfluid interior,
  during the first part of the glitch recovery the superfluid interior spins-up due to viscous interactions, restoring
  corotation with the crust, and erasing the non-axisymmetries in the velocity field. In this ``relaxation phase''
  viscous interactions act through the process of Ekman pumping~\cite{benton1974spin}, which operates on a timescale
  (the ``Ekman time'') which has been estimated to range from days to weeks.  As the core spins up, the time-dependent
  mass and current quadrupole moments due to non-axisymmetric meridional circulation emit a continuous GW.  Numerical
  simulations of the relaxation phase (assuming initial data in which the components with different azimuthal numbers
  have comparable amplitude) yield a GW strain~\cite{vanEysden:2008pd,Bennett:2010tm,Singh:2016ilt}
\begin{equation}
h\sim6\times10^{-27}D_1^{-1}\left(\frac{\Delta\Omega/\Omega}{10^{-5}}\right)\left(\frac{f_{\rm spin}}{100\,{\rm Hz}}\right)^3\,.
\end{equation}
Although the GW signal emitted during the relaxation phase has a lower dimensionless strain amplitude than the burst
emitted during the glitch, its duration is larger, and this increases the measured strain $\sim
h\sqrt{T}$~\cite{Prix:2011qv}. Therefore, the detectability of this signal crucially depends on its actual duration.
Current models of the relaxation phase~\cite{vanEysden:2008pd,Bennett:2010tm,Singh:2016ilt} find that the signal can
last up to some weeks, and - if emitted by a large nearby neutron star after a large glitch - it may be marginally
detectable even by second-generation GW interferometers. However, these models neglect the contribution of magnetic
field and mutual friction, which couple the crust with the core faster than Ekman pumping, reducing the duration of the
signal. These contributions could significantly reduce the actual detectability of the GW emission in the post-glitch
recovery phase.
\end{itemize}
The estimates of the GW signal from a glitch have recently been extended, for most of the mechanisms discussed above, to
the observed glitches in gamma-ray pulsars (most of which are radio-quiet), finding that the detectability of the GW
signal from gamma-ray pulsar glitches is comparable with that of the signal from radio pulsar
glitches~\cite{Stopnitzky:2013wza}.  

Summarizing, GW emission from glitches is still not fully understood, and the predictions can be very different, from
pessimistic (signal too weak even for third-generation interferometers, as in~\cite{Sidery:2009at}), to moderately
optimistic (signal potentially detectable by ET, as in~\cite{Keer:2014uva,Warszawski:2012zq}), to very optimistic
(signal marginally detectable by Advanced LIGO/Virgo, as
in~\cite{vanEysden:2008pd,Bennett:2010tm,Prix:2011qv,Melatos:2015oca}). We remark that the models predicting
detectability by second-generation interferometers require a persistent signal after the glitch - either due to
crust-core viscous interaction, or to a persistent current quadrupole moment produced by a vortex avalanche - but the
actual duration of the signal is still matter of debate.

\subsection{Future directions}

Neutron star glitches are a very complex process, which we are just starting to understand. Our theoretical models
probably are able to capture the main features of this process, but they are often qualitative, and based on simplifying
assumptions. We still need to build a general model which captures the different aspects of pulsar glitches, and to
perform MHD simulations based on this model. 

In order to understand the glitch phase, we need a reliable description of the collective motion of vortices. To this
aim, the GP model is a good starting point, but it needs to be extended and interfaced to an hydrodynamical model,
including magnetic fields and crust-core couplings.

Concerning the post-glitch relaxation phase, our understanding of the mechanisms leading to density and current asymmetries
(crust-core differential rotation, two-stream instabilities, cracks and tilts in the crust,
etc.~\cite{vanEysden:2008pd}) is only qualitative. We need an accurate model of these processes, including magnetic
field and mutual friction, in order to understand the actual duration of the relaxation phase and then of the GW signal.

Only a detailed and quantitative description of the entire glitch and post-glitch relaxation phases, will allow us to
determine the GW emission associated to this process, and to definitely assess its detectability by GW interferometers.


\section{Concluding remarks}
\label{sec:conclusions}

In this Chapter we have surveyed a number of physical processes in isolated and accreting neutron stars that could be
interesting sources for the newborn field of GW astronomy. A general conclusion that can be drawn from the results
summarised here is that the GW signals from these single neutron stars are not expected to be as loud as the ones
produced by binary systems of black holes or neutron stars.  This is mostly due to the basic fact that there is less
gravitational mass ``sloshing around'' in single systems than in binary ones.
This handicap, however,
can be partially offset by the continuous character of the GW signal associated with some of the emission mechanisms
(e.g. mountains) and/or the closeness to the source (e.g. known neutron stars in our Galaxy).  All relevant factors
accounted for, GWs from single neutron stars are more likely to be detected by future ET-class observatories.

It should be emphasized that in some cases our present level of understanding of a particular mechanism
does not even allow us to make a safe prediction as to whether GW emission will operate in the first place. 
An example of this is provided by the $r$-mode instability where the mode's maximum amplitude and
instability window are still largely uncertain factors. This situation is not surprising given the multi-faceted 
physics that has bearing on the problem.   

There is one more key point worth considering here, namely, the possibility of GW emission mechanisms having an impact
on photon astronomy observations of neutron stars. This has, of course, already happened when the orbital evolution of
binary pulsar systems was found to be consistent with GW emission, thus providing evidence of the existence of GWs
several decades before their first direct detection. A more recent remarkable example of this synergy may have been
provided by the observed spin-down profile of PSR J1203+0038 which is a member of a LMXB system
\cite{Haskell:2017ajb}. The enhanced spin-down rate of this pulsar immediately after an accretion phase has been
attributed to the GW emission by a thermal mountain, but the required deformation is too small to be directly detected
by Advanced LIGO. A similar situation of GWs ``seen'' in the electromagnetic channel may arise if, for example, a small amplitude
$r$-mode drives the spin evolution of a MSP.


\acknowledgements

We thank Nils Andersson, Daniela Doneva, Brynmor Haskell, Wynn Ho,  Ian Jones, Kostas Kokkotas, Cristiano Palomba, George Pappas, Andrea Passamonti and Kai Schwenzer for useful discussions during the course of this work and for providing data and figures.

This work was supported by the H2020-MSCA-RISE-2015 Grant No. StronGrHEP-690904 and by the COST actions MP1304 and CA16104.

\bibliography{biblio.bib}

\begin{thebibliography}{303}%
\makeatletter
\providecommand \@ifxundefined [1]{%
 \@ifx{#1\undefined}
}%
\providecommand \@ifnum [1]{%
 \ifnum #1\expandafter \@firstoftwo
 \else \expandafter \@secondoftwo
 \fi
}%
\providecommand \@ifx [1]{%
 \ifx #1\expandafter \@firstoftwo
 \else \expandafter \@secondoftwo
 \fi
}%
\providecommand \natexlab [1]{#1}%
\providecommand \enquote  [1]{``#1''}%
\providecommand \bibnamefont  [1]{#1}%
\providecommand \bibfnamefont [1]{#1}%
\providecommand \citenamefont [1]{#1}%
\providecommand \href@noop [0]{\@secondoftwo}%
\providecommand \href [0]{\begingroup \@sanitize@url \@href}%
\providecommand \@href[1]{\@@startlink{#1}\@@href}%
\providecommand \@@href[1]{\endgroup#1\@@endlink}%
\providecommand \@sanitize@url [0]{\catcode `\\12\catcode `\$12\catcode
  `\&12\catcode `\#12\catcode `\^12\catcode `\_12\catcode `\%12\relax}%
\providecommand \@@startlink[1]{}%
\providecommand \@@endlink[0]{}%
\providecommand \url  [0]{\begingroup\@sanitize@url \@url }%
\providecommand \@url [1]{\endgroup\@href {#1}{\urlprefix }}%
\providecommand \urlprefix  [0]{URL }%
\providecommand \Eprint [0]{\href }%
\providecommand \doibase [0]{http://dx.doi.org/}%
\providecommand \selectlanguage [0]{\@gobble}%
\providecommand \bibinfo  [0]{\@secondoftwo}%
\providecommand \bibfield  [0]{\@secondoftwo}%
\providecommand \translation [1]{[#1]}%
\providecommand \BibitemOpen [0]{}%
\providecommand \bibitemStop [0]{}%
\providecommand \bibitemNoStop [0]{.\EOS\space}%
\providecommand \EOS [0]{\spacefactor3000\relax}%
\providecommand \BibitemShut  [1]{\csname bibitem#1\endcsname}%
\let\auto@bib@innerbib\@empty
\bibitem [{\citenamefont {Abbott}\ \emph
  {et~al.}(2016{\natexlab{a}})\citenamefont {Abbott} \emph
  {et~al.}}]{GW150914}%
  \BibitemOpen
  \bibfield  {author} {\bibinfo {author} {\bibfnamefont {B.~P.}\ \bibnamefont
  {Abbott}} \emph {et~al.},\ }\href {\doibase 10.1103/PhysRevLett.116.061102}
  {\bibfield  {journal} {\bibinfo  {journal} {Phys. Rev. Lett.}\ }\textbf
  {\bibinfo {volume} {116}},\ \bibinfo {pages} {061102} (\bibinfo {year}
  {2016}{\natexlab{a}})}\BibitemShut {NoStop}%
\bibitem [{\citenamefont {Abbott}\ \emph
  {et~al.}(2016{\natexlab{b}})\citenamefont {Abbott} \emph
  {et~al.}}]{GW151226}%
  \BibitemOpen
  \bibfield  {author} {\bibinfo {author} {\bibfnamefont {B.~P.}\ \bibnamefont
  {Abbott}} \emph {et~al.},\ }\href {\doibase 10.1103/PhysRevLett.116.241103}
  {\bibfield  {journal} {\bibinfo  {journal} {Phys. Rev. Lett.}\ }\textbf
  {\bibinfo {volume} {116}},\ \bibinfo {pages} {241103} (\bibinfo {year}
  {2016}{\natexlab{b}})}\BibitemShut {NoStop}%
\bibitem [{\citenamefont {Abbott}\ \emph
  {et~al.}(2017{\natexlab{a}})\citenamefont {Abbott} \emph
  {et~al.}}]{GW170104}%
  \BibitemOpen
  \bibfield  {author} {\bibinfo {author} {\bibfnamefont {B.~P.}\ \bibnamefont
  {Abbott}} \emph {et~al.},\ }\href {\doibase 10.1103/PhysRevLett.118.221101}
  {\bibfield  {journal} {\bibinfo  {journal} {Phys. Rev. Lett.}\ }\textbf
  {\bibinfo {volume} {118}},\ \bibinfo {pages} {221101} (\bibinfo {year}
  {2017}{\natexlab{a}})}\BibitemShut {NoStop}%
\bibitem [{\citenamefont {Abbott}\ \emph
  {et~al.}(2017{\natexlab{b}})\citenamefont {Abbott} \emph
  {et~al.}}]{GW170608}%
  \BibitemOpen
  \bibfield  {author} {\bibinfo {author} {\bibfnamefont {B.~P.}\ \bibnamefont
  {Abbott}} \emph {et~al.} (\bibinfo {collaboration} {Virgo, LIGO
  Scientific}),\ }\href {\doibase 10.3847/2041-8213/aa9f0c} {\bibfield
  {journal} {\bibinfo  {journal} {Astrophys. J.}\ }\textbf {\bibinfo {volume}
  {851}},\ \bibinfo {pages} {L35} (\bibinfo {year}
  {2017}{\natexlab{b}})}\BibitemShut {NoStop}%
\bibitem [{\citenamefont {Abbott}\ \emph
  {et~al.}(2017{\natexlab{c}})\citenamefont {Abbott} \emph
  {et~al.}}]{GW170817}%
  \BibitemOpen
  \bibfield  {author} {\bibinfo {author} {\bibfnamefont {B.}~\bibnamefont
  {Abbott}} \emph {et~al.} (\bibinfo {collaboration} {Virgo, LIGO
  Scientific}),\ }\href {\doibase 10.1103/PhysRevLett.119.161101} {\bibfield
  {journal} {\bibinfo  {journal} {Phys. Rev. Lett.}\ }\textbf {\bibinfo
  {volume} {119}},\ \bibinfo {pages} {161101} (\bibinfo {year}
  {2017}{\natexlab{c}})}\BibitemShut {NoStop}%
\bibitem [{\citenamefont {Abbott}\ \emph
  {et~al.}(2017{\natexlab{d}})\citenamefont {Abbott} \emph
  {et~al.}}]{EM170817}%
  \BibitemOpen
  \bibfield  {author} {\bibinfo {author} {\bibfnamefont {B.~P.}\ \bibnamefont
  {Abbott}} \emph {et~al.} (\bibinfo {collaboration} {GROND, SALT Group,
  OzGrav, DFN, INTEGRAL, Virgo, Insight-Hxmt, MAXI Team, Fermi-LAT, J-GEM,
  RATIR, IceCube, CAASTRO, LWA, ePESSTO, GRAWITA, RIMAS, SKA South
  Africa/MeerKAT, H.E.S.S., 1M2H Team, IKI-GW Follow-up, Fermi GBM, Pi of Sky,
  DWF (Deeper Wider Faster Program), Dark Energy Survey, MASTER, AstroSat
  Cadmium Zinc Telluride Imager Team, Swift, Pierre Auger, ASKAP, VINROUGE,
  JAGWAR, Chandra Team at McGill University, TTU-NRAO, GROWTH, AGILE Team, MWA,
  ATCA, AST3, TOROS, Pan-STARRS, NuSTAR, ATLAS Telescopes, BOOTES, CaltechNRAO,
  LIGO Scientific, High Time Resolution Universe Survey, Nordic Optical
  Telescope, Las Cumbres Observatory Group, TZAC Consortium, LOFAR, IPN, DLT40,
  Texas Tech University, HAWC, ANTARES, KU, Dark Energy Camera GW-EM, CALET,
  Euro VLBI Team, ALMA}),\ }\href {\doibase 10.3847/2041-8213/aa91c9}
  {\bibfield  {journal} {\bibinfo  {journal} {Astrophys. J.}\ }\textbf
  {\bibinfo {volume} {848}},\ \bibinfo {pages} {L12} (\bibinfo {year}
  {2017}{\natexlab{d}})}\BibitemShut {NoStop}%
\bibitem [{\citenamefont {Punturo}\ \emph {et~al.}(2010)\citenamefont {Punturo}
  \emph {et~al.}}]{Punturo:2010zza}%
  \BibitemOpen
  \bibfield  {author} {\bibinfo {author} {\bibfnamefont {M.}~\bibnamefont
  {Punturo}} \emph {et~al.},\ }\href {\doibase 10.1088/0264-9381/27/8/084007}
  {\bibfield  {journal} {\bibinfo  {journal} {Class. Quant. Grav.}\ }\textbf
  {\bibinfo {volume} {27}},\ \bibinfo {pages} {084007} (\bibinfo {year}
  {2010})}\BibitemShut {NoStop}%
\bibitem [{\citenamefont {Benhar}\ \emph {et~al.}(2004)\citenamefont {Benhar},
  \citenamefont {Ferrari},\ and\ \citenamefont {Gualtieri}}]{Benhar:2004xg}%
  \BibitemOpen
  \bibfield  {author} {\bibinfo {author} {\bibfnamefont {O.}~\bibnamefont
  {Benhar}}, \bibinfo {author} {\bibfnamefont {V.}~\bibnamefont {Ferrari}}, \
  and\ \bibinfo {author} {\bibfnamefont {L.}~\bibnamefont {Gualtieri}},\ }\href
  {\doibase 10.1103/PhysRevD.70.124015} {\bibfield  {journal} {\bibinfo
  {journal} {Phys. Rev.}\ }\textbf {\bibinfo {volume} {D70}},\ \bibinfo {pages}
  {124015} (\bibinfo {year} {2004})}\BibitemShut {NoStop}%
\bibitem [{\citenamefont {Andersson}\ and\ \citenamefont
  {Kokkotas}(1998)}]{NAKK98}%
  \BibitemOpen
  \bibfield  {author} {\bibinfo {author} {\bibfnamefont {N.}~\bibnamefont
  {Andersson}}\ and\ \bibinfo {author} {\bibfnamefont {K.~D.}\ \bibnamefont
  {Kokkotas}},\ }\href {\doibase 10.1046/j.1365-8711.1998.01840.x} {\bibfield
  {journal} {\bibinfo  {journal} {Mon. Not. R. Astron. Soc.}\ }\textbf
  {\bibinfo {volume} {299}},\ \bibinfo {pages} {1059} (\bibinfo {year}
  {1998})}\BibitemShut {NoStop}%
\bibitem [{\citenamefont {Gaertig}\ and\ \citenamefont
  {Kokkotas}(2011)}]{gaertig_KK2011}%
  \BibitemOpen
  \bibfield  {author} {\bibinfo {author} {\bibfnamefont {E.}~\bibnamefont
  {Gaertig}}\ and\ \bibinfo {author} {\bibfnamefont {K.~D.}\ \bibnamefont
  {Kokkotas}},\ }\href {\doibase 10.1103/PhysRevD.83.064031} {\bibfield
  {journal} {\bibinfo  {journal} {Phys. Rev. D}\ }\textbf {\bibinfo {volume}
  {83}},\ \bibinfo {pages} {064031} (\bibinfo {year} {2011})}\BibitemShut
  {NoStop}%
\bibitem [{\citenamefont {Doneva}\ \emph {et~al.}(2013)\citenamefont {Doneva},
  \citenamefont {Gaertig}, \citenamefont {Kokkotas},\ and\ \citenamefont
  {Kr\"uger}}]{doneva_etal2013}%
  \BibitemOpen
  \bibfield  {author} {\bibinfo {author} {\bibfnamefont {D.~D.}\ \bibnamefont
  {Doneva}}, \bibinfo {author} {\bibfnamefont {E.}~\bibnamefont {Gaertig}},
  \bibinfo {author} {\bibfnamefont {K.~D.}\ \bibnamefont {Kokkotas}}, \ and\
  \bibinfo {author} {\bibfnamefont {C.}~\bibnamefont {Kr\"uger}},\ }\href
  {\doibase 10.1103/PhysRevD.88.044052} {\bibfield  {journal} {\bibinfo
  {journal} {Phys. Rev. D}\ }\textbf {\bibinfo {volume} {88}},\ \bibinfo
  {pages} {044052} (\bibinfo {year} {2013})}\BibitemShut {NoStop}%
\bibitem [{\citenamefont {Detweiler}(1975)}]{detweiler1975}%
  \BibitemOpen
  \bibfield  {author} {\bibinfo {author} {\bibfnamefont {S.~L.}\ \bibnamefont
  {Detweiler}},\ }\href {\doibase 10.1086/153504} {\bibfield  {journal}
  {\bibinfo  {journal} {Astrophys. J.}\ }\textbf {\bibinfo {volume} {197}},\
  \bibinfo {pages} {203} (\bibinfo {year} {1975})}\BibitemShut {NoStop}%
\bibitem [{\citenamefont {Lau}\ \emph {et~al.}(2010)\citenamefont {Lau},
  \citenamefont {Leung},\ and\ \citenamefont {Lin}}]{lau_etal2010}%
  \BibitemOpen
  \bibfield  {author} {\bibinfo {author} {\bibfnamefont {H.~K.}\ \bibnamefont
  {Lau}}, \bibinfo {author} {\bibfnamefont {P.~T.}\ \bibnamefont {Leung}}, \
  and\ \bibinfo {author} {\bibfnamefont {L.~M.}\ \bibnamefont {Lin}},\ }\href
  {\doibase 10.1088/0004-637X/714/2/1234} {\bibfield  {journal} {\bibinfo
  {journal} {Astrophys. J.}\ }\textbf {\bibinfo {volume} {714}},\ \bibinfo
  {pages} {1234} (\bibinfo {year} {2010})}\BibitemShut {NoStop}%
\bibitem [{\citenamefont {Doneva}\ and\ \citenamefont
  {Kokkotas}(2015)}]{doneva_KK2015}%
  \BibitemOpen
  \bibfield  {author} {\bibinfo {author} {\bibfnamefont {D.~D.}\ \bibnamefont
  {Doneva}}\ and\ \bibinfo {author} {\bibfnamefont {K.~D.}\ \bibnamefont
  {Kokkotas}},\ }\href {\doibase 10.1103/PhysRevD.92.124004} {\bibfield
  {journal} {\bibinfo  {journal} {Phys. Rev. D}\ }\textbf {\bibinfo {volume}
  {92}},\ \bibinfo {pages} {124004} (\bibinfo {year} {2015})}\BibitemShut
  {NoStop}%
\bibitem [{\citenamefont {Zink}\ \emph {et~al.}(2010)\citenamefont {Zink},
  \citenamefont {Korobkin}, \citenamefont {Schnetter},\ and\ \citenamefont
  {Stergioulas}}]{zink_etal2010}%
  \BibitemOpen
  \bibfield  {author} {\bibinfo {author} {\bibfnamefont {B.}~\bibnamefont
  {Zink}}, \bibinfo {author} {\bibfnamefont {O.}~\bibnamefont {Korobkin}},
  \bibinfo {author} {\bibfnamefont {E.}~\bibnamefont {Schnetter}}, \ and\
  \bibinfo {author} {\bibfnamefont {N.}~\bibnamefont {Stergioulas}},\ }\href
  {\doibase 10.1103/PhysRevD.81.084055} {\bibfield  {journal} {\bibinfo
  {journal} {Phys. Rev. D}\ }\textbf {\bibinfo {volume} {81}},\ \bibinfo
  {pages} {084055} (\bibinfo {year} {2010})}\BibitemShut {NoStop}%
\bibitem [{\citenamefont {Gualtieri}\ \emph {et~al.}(2014)\citenamefont
  {Gualtieri}, \citenamefont {Kantor}, \citenamefont {Gusakov},\ and\
  \citenamefont {Chugunov}}]{Gualtieri:2014lsa}%
  \BibitemOpen
  \bibfield  {author} {\bibinfo {author} {\bibfnamefont {L.}~\bibnamefont
  {Gualtieri}}, \bibinfo {author} {\bibfnamefont {E.~M.}\ \bibnamefont
  {Kantor}}, \bibinfo {author} {\bibfnamefont {M.~E.}\ \bibnamefont {Gusakov}},
  \ and\ \bibinfo {author} {\bibfnamefont {A.~I.}\ \bibnamefont {Chugunov}},\
  }\href {\doibase 10.1103/PhysRevD.90.024010} {\bibfield  {journal} {\bibinfo
  {journal} {Phys. Rev.}\ }\textbf {\bibinfo {volume} {D90}},\ \bibinfo {pages}
  {024010} (\bibinfo {year} {2014})}\BibitemShut {NoStop}%
\bibitem [{\citenamefont {Gusakov}\ and\ \citenamefont
  {Kantor}(2013)}]{Gusakov:2013eoa}%
  \BibitemOpen
  \bibfield  {author} {\bibinfo {author} {\bibfnamefont {M.~E.}\ \bibnamefont
  {Gusakov}}\ and\ \bibinfo {author} {\bibfnamefont {E.~M.}\ \bibnamefont
  {Kantor}},\ }\href {\doibase 10.1103/PhysRevD.88.101302} {\bibfield
  {journal} {\bibinfo  {journal} {Phys. Rev.}\ }\textbf {\bibinfo {volume}
  {D88}},\ \bibinfo {pages} {101302} (\bibinfo {year} {2013})}\BibitemShut
  {NoStop}%
\bibitem [{\citenamefont {Kantor}\ and\ \citenamefont
  {Gusakov}(2014)}]{Kantor:2014lja}%
  \BibitemOpen
  \bibfield  {author} {\bibinfo {author} {\bibfnamefont {E.~M.}\ \bibnamefont
  {Kantor}}\ and\ \bibinfo {author} {\bibfnamefont {M.~E.}\ \bibnamefont
  {Gusakov}},\ }\href {\doibase 10.1093/mnrasl/slu061} {\bibfield  {journal}
  {\bibinfo  {journal} {Mon. Not. R. Astron. Soc.}\ }\textbf {\bibinfo {volume}
  {442}},\ \bibinfo {pages} {90} (\bibinfo {year} {2014})}\BibitemShut
  {NoStop}%
\bibitem [{\citenamefont {Passamonti}\ \emph {et~al.}(2016)\citenamefont
  {Passamonti}, \citenamefont {Andersson},\ and\ \citenamefont
  {Ho}}]{Passamonti:2015oia}%
  \BibitemOpen
  \bibfield  {author} {\bibinfo {author} {\bibfnamefont {A.}~\bibnamefont
  {Passamonti}}, \bibinfo {author} {\bibfnamefont {N.}~\bibnamefont
  {Andersson}}, \ and\ \bibinfo {author} {\bibfnamefont {W.~C.~G.}\
  \bibnamefont {Ho}},\ }\href {\doibase 10.1093/mnras/stv2149} {\bibfield
  {journal} {\bibinfo  {journal} {Mon. Not. R. Astron. Soc.}\ }\textbf
  {\bibinfo {volume} {455}},\ \bibinfo {pages} {1489} (\bibinfo {year}
  {2016})}\BibitemShut {NoStop}%
\bibitem [{\citenamefont {Dommes}\ and\ \citenamefont
  {Gusakov}(2016)}]{Dommes:2015wul}%
  \BibitemOpen
  \bibfield  {author} {\bibinfo {author} {\bibfnamefont {V.~A.}\ \bibnamefont
  {Dommes}}\ and\ \bibinfo {author} {\bibfnamefont {M.~E.}\ \bibnamefont
  {Gusakov}},\ }\href {\doibase 10.1093/mnras/stv2408} {\bibfield  {journal}
  {\bibinfo  {journal} {Mon. Not. R. Astron. Soc.}\ }\textbf {\bibinfo {volume}
  {455}},\ \bibinfo {pages} {2852} (\bibinfo {year} {2016})}\BibitemShut
  {NoStop}%
\bibitem [{\citenamefont {Gusakov}\ \emph
  {et~al.}(2014{\natexlab{a}})\citenamefont {Gusakov}, \citenamefont
  {Chugunov},\ and\ \citenamefont {Kantor}}]{gusakov_etal2014a}%
  \BibitemOpen
  \bibfield  {author} {\bibinfo {author} {\bibfnamefont {M.~E.}\ \bibnamefont
  {Gusakov}}, \bibinfo {author} {\bibfnamefont {A.~I.}\ \bibnamefont
  {Chugunov}}, \ and\ \bibinfo {author} {\bibfnamefont {E.~M.}\ \bibnamefont
  {Kantor}},\ }\href {\doibase 10.1103/PhysRevLett.112.151101} {\bibfield
  {journal} {\bibinfo  {journal} {Phys. Rev. Lett.}\ }\textbf {\bibinfo
  {volume} {112}},\ \bibinfo {pages} {151101} (\bibinfo {year}
  {2014}{\natexlab{a}})}\BibitemShut {NoStop}%
\bibitem [{\citenamefont {Gusakov}\ \emph
  {et~al.}(2014{\natexlab{b}})\citenamefont {Gusakov}, \citenamefont
  {Chugunov},\ and\ \citenamefont {Kantor}}]{gusakov_etal2014b}%
  \BibitemOpen
  \bibfield  {author} {\bibinfo {author} {\bibfnamefont {M.~E.}\ \bibnamefont
  {Gusakov}}, \bibinfo {author} {\bibfnamefont {A.~I.}\ \bibnamefont
  {Chugunov}}, \ and\ \bibinfo {author} {\bibfnamefont {E.~M.}\ \bibnamefont
  {Kantor}},\ }\href {\doibase 10.1103/PhysRevD.90.063001} {\bibfield
  {journal} {\bibinfo  {journal} {Phys. Rev. D}\ }\textbf {\bibinfo {volume}
  {90}},\ \bibinfo {pages} {063001} (\bibinfo {year}
  {2014}{\natexlab{b}})}\BibitemShut {NoStop}%
\bibitem [{\citenamefont {Chandrasekhar}(1970)}]{chandra1970}%
  \BibitemOpen
  \bibfield  {author} {\bibinfo {author} {\bibfnamefont {S.}~\bibnamefont
  {Chandrasekhar}},\ }\href@noop {} {\bibfield  {journal} {\bibinfo  {journal}
  {Phys. Rev. Lett.}\ }\textbf {\bibinfo {volume} {24}},\ \bibinfo {pages}
  {611} (\bibinfo {year} {1970})}\BibitemShut {NoStop}%
\bibitem [{\citenamefont {Friedman}\ and\ \citenamefont
  {Schutz}(1978{\natexlab{a}})}]{friedman_schutz1978a}%
  \BibitemOpen
  \bibfield  {author} {\bibinfo {author} {\bibfnamefont {J.~L.}\ \bibnamefont
  {Friedman}}\ and\ \bibinfo {author} {\bibfnamefont {B.~F.}\ \bibnamefont
  {Schutz}},\ }\href@noop {} {\bibfield  {journal} {\bibinfo  {journal}
  {Astrophys. J.}\ }\textbf {\bibinfo {volume} {221}},\ \bibinfo {pages} {937}
  (\bibinfo {year} {1978}{\natexlab{a}})}\BibitemShut {NoStop}%
\bibitem [{\citenamefont {Friedman}\ and\ \citenamefont
  {Schutz}(1978{\natexlab{b}})}]{friedman_schutz1978b}%
  \BibitemOpen
  \bibfield  {author} {\bibinfo {author} {\bibfnamefont {J.~L.}\ \bibnamefont
  {Friedman}}\ and\ \bibinfo {author} {\bibfnamefont {B.~F.}\ \bibnamefont
  {Schutz}},\ }\href@noop {} {\bibfield  {journal} {\bibinfo  {journal}
  {Astrophys. J.}\ }\textbf {\bibinfo {volume} {222}},\ \bibinfo {pages} {281}
  (\bibinfo {year} {1978}{\natexlab{b}})}\BibitemShut {NoStop}%
\bibitem [{\citenamefont {Friedman}(1978)}]{friedman1978}%
  \BibitemOpen
  \bibfield  {author} {\bibinfo {author} {\bibfnamefont {J.~L.}\ \bibnamefont
  {Friedman}},\ }\href@noop {} {\bibfield  {journal} {\bibinfo  {journal}
  {Comm. Math. Phys.}\ }\textbf {\bibinfo {volume} {62}},\ \bibinfo {pages}
  {247} (\bibinfo {year} {1978})}\BibitemShut {NoStop}%
\bibitem [{\citenamefont {Chandrasekhar}(1969)}]{chandra_book}%
  \BibitemOpen
  \bibfield  {author} {\bibinfo {author} {\bibfnamefont {S.}~\bibnamefont
  {Chandrasekhar}},\ }\href@noop {} {\emph {\bibinfo {title} {{Ellipsoidal
  figures of equilibrium}}}}\ (\bibinfo  {publisher} {Yale University Press},\
  \bibinfo {year} {1969})\BibitemShut {NoStop}%
\bibitem [{\citenamefont {Andersson}\ and\ \citenamefont
  {Kokkotas}(2001)}]{NAKK01}%
  \BibitemOpen
  \bibfield  {author} {\bibinfo {author} {\bibfnamefont {N.}~\bibnamefont
  {Andersson}}\ and\ \bibinfo {author} {\bibfnamefont {K.~D.}\ \bibnamefont
  {Kokkotas}},\ }\href {\doibase 10.1142/S0218271801001062} {\bibfield
  {journal} {\bibinfo  {journal} {Int. J. Mod. Phys. D}\ }\textbf {\bibinfo
  {volume} {10}},\ \bibinfo {pages} {381} (\bibinfo {year} {2001})}\BibitemShut
  {NoStop}%
\bibitem [{\citenamefont {Andersson}(2003)}]{NA03}%
  \BibitemOpen
  \bibfield  {author} {\bibinfo {author} {\bibfnamefont {N.}~\bibnamefont
  {Andersson}},\ }\href {\doibase 10.1088/0264-9381/20/7/201} {\bibfield
  {journal} {\bibinfo  {journal} {Class. Quant. Grav.}\ }\textbf {\bibinfo
  {volume} {20}},\ \bibinfo {pages} {R105} (\bibinfo {year}
  {2003})}\BibitemShut {NoStop}%
\bibitem [{\citenamefont {L.}\ and\ \citenamefont
  {Stergioulas}(2013)}]{FSbook}%
  \BibitemOpen
  \bibfield  {author} {\bibinfo {author} {\bibfnamefont {F.~J.}\ \bibnamefont
  {L.}}\ and\ \bibinfo {author} {\bibfnamefont {N.}~\bibnamefont
  {Stergioulas}},\ }\href@noop {} {\emph {\bibinfo {title} {{Rotating
  Relativistic Stars}}}}\ (\bibinfo  {publisher} {Cambridge University Press},\
  \bibinfo {year} {2013})\BibitemShut {NoStop}%
\bibitem [{\citenamefont {Andersson}\ and\ \citenamefont
  {Comer}(2007)}]{NAGC2007}%
  \BibitemOpen
  \bibfield  {author} {\bibinfo {author} {\bibfnamefont {N.}~\bibnamefont
  {Andersson}}\ and\ \bibinfo {author} {\bibfnamefont {G.~L.}\ \bibnamefont
  {Comer}},\ }\href {\doibase 10.12942/lrr-2007-1} {\bibfield  {journal}
  {\bibinfo  {journal} {Living Rev. Rel.}\ }\textbf {\bibinfo {volume} {10}},\
  \bibinfo {pages} {83} (\bibinfo {year} {2007})}\BibitemShut {NoStop}%
\bibitem [{\citenamefont {Thorne}(1980)}]{Thorne80}%
  \BibitemOpen
  \bibfield  {author} {\bibinfo {author} {\bibfnamefont {K.~S.}\ \bibnamefont
  {Thorne}},\ }\href {\doibase 10.1103/RevModPhys.52.299} {\bibfield  {journal}
  {\bibinfo  {journal} {Rev. Mod. Phys.}\ }\textbf {\bibinfo {volume} {52}},\
  \bibinfo {pages} {299} (\bibinfo {year} {1980})}\BibitemShut {NoStop}%
\bibitem [{\citenamefont {Ipser}\ and\ \citenamefont
  {Lindblom}(1991)}]{IL1991}%
  \BibitemOpen
  \bibfield  {author} {\bibinfo {author} {\bibfnamefont {J.~R.}\ \bibnamefont
  {Ipser}}\ and\ \bibinfo {author} {\bibfnamefont {L.}~\bibnamefont
  {Lindblom}},\ }\href {\doibase 10.1086/170039} {\bibfield  {journal}
  {\bibinfo  {journal} {Astrophys. J.}\ }\textbf {\bibinfo {volume} {373}},\
  \bibinfo {pages} {213} (\bibinfo {year} {1991})}\BibitemShut {NoStop}%
\bibitem [{\citenamefont {Lamb}(1908)}]{lamb1908}%
  \BibitemOpen
  \bibfield  {author} {\bibinfo {author} {\bibfnamefont {H.}~\bibnamefont
  {Lamb}},\ }\href {http://www.jstor.org/stable/92794} {\bibfield  {journal}
  {\bibinfo  {journal} {Proc. R. Soc. Lond. A}\ }\textbf {\bibinfo {volume}
  {80}},\ \bibinfo {pages} {168} (\bibinfo {year} {1908})}\BibitemShut
  {NoStop}%
\bibitem [{\citenamefont {Pierce}(1974)}]{pierce_book}%
  \BibitemOpen
  \bibfield  {author} {\bibinfo {author} {\bibfnamefont {J.~R.}\ \bibnamefont
  {Pierce}},\ }\href@noop {} {\emph {\bibinfo {title} {Almost All about
  Waves}}}\ (\bibinfo  {publisher} {MIT Press},\ \bibinfo {year}
  {1974})\BibitemShut {NoStop}%
\bibitem [{\citenamefont {Sawyer}(1989)}]{sawyer1989}%
  \BibitemOpen
  \bibfield  {author} {\bibinfo {author} {\bibfnamefont {R.~F.}\ \bibnamefont
  {Sawyer}},\ }\href {\doibase 10.1103/PhysRevD.39.3804} {\bibfield  {journal}
  {\bibinfo  {journal} {Phys. Rev. D}\ }\textbf {\bibinfo {volume} {39}},\
  \bibinfo {pages} {3804} (\bibinfo {year} {1989})}\BibitemShut {NoStop}%
\bibitem [{\citenamefont {Andersson}\ \emph
  {et~al.}(2005{\natexlab{a}})\citenamefont {Andersson}, \citenamefont
  {Comer},\ and\ \citenamefont {Glampedakis}}]{andersson_etal2005b}%
  \BibitemOpen
  \bibfield  {author} {\bibinfo {author} {\bibfnamefont {N.}~\bibnamefont
  {Andersson}}, \bibinfo {author} {\bibfnamefont {G.~L.}\ \bibnamefont
  {Comer}}, \ and\ \bibinfo {author} {\bibfnamefont {K.}~\bibnamefont
  {Glampedakis}},\ }\href {\doibase 10.1016/j.nuclphysa.2005.08.012} {\bibfield
   {journal} {\bibinfo  {journal} {Nucl. Phys. A}\ }\textbf {\bibinfo {volume}
  {763}},\ \bibinfo {pages} {212} (\bibinfo {year}
  {2005}{\natexlab{a}})}\BibitemShut {NoStop}%
\bibitem [{\citenamefont {Brown}\ and\ \citenamefont
  {Scriven}(1980)}]{brown_scriven1980}%
  \BibitemOpen
  \bibfield  {author} {\bibinfo {author} {\bibfnamefont {R.~A.}\ \bibnamefont
  {Brown}}\ and\ \bibinfo {author} {\bibfnamefont {L.~E.}\ \bibnamefont
  {Scriven}},\ }\href@noop {} {\bibfield  {journal} {\bibinfo  {journal} {Proc.
  R. Soc. Lond. A}\ }\textbf {\bibinfo {volume} {371}},\ \bibinfo {pages} {331}
  (\bibinfo {year} {1980})}\BibitemShut {NoStop}%
\bibitem [{\citenamefont {Hill}\ and\ \citenamefont
  {Eaves}(2008)}]{hill_eaves2008}%
  \BibitemOpen
  \bibfield  {author} {\bibinfo {author} {\bibfnamefont {J.~A.}\ \bibnamefont
  {Hill}}\ and\ \bibinfo {author} {\bibfnamefont {L.}~\bibnamefont {Eaves}},\
  }\href@noop {} {\bibfield  {journal} {\bibinfo  {journal} {Phys. Rev. Lett.}\
  }\textbf {\bibinfo {volume} {101}},\ \bibinfo {pages} {234501} (\bibinfo
  {year} {2008})}\BibitemShut {NoStop}%
\bibitem [{\citenamefont {Baiotti}\ \emph {et~al.}(2007)\citenamefont
  {Baiotti}, \citenamefont {De~Pietri}, \citenamefont {Manca},\ and\
  \citenamefont {Rezzolla}}]{baiotti_etal2007}%
  \BibitemOpen
  \bibfield  {author} {\bibinfo {author} {\bibfnamefont {L.}~\bibnamefont
  {Baiotti}}, \bibinfo {author} {\bibfnamefont {R.}~\bibnamefont {De~Pietri}},
  \bibinfo {author} {\bibfnamefont {G.~M.}\ \bibnamefont {Manca}}, \ and\
  \bibinfo {author} {\bibfnamefont {L.}~\bibnamefont {Rezzolla}},\ }\href
  {\doibase 10.1103/PhysRevD.75.044023} {\bibfield  {journal} {\bibinfo
  {journal} {Phys. Rev. D}\ }\textbf {\bibinfo {volume} {75}},\ \bibinfo
  {pages} {044023} (\bibinfo {year} {2007})}\BibitemShut {NoStop}%
\bibitem [{\citenamefont {Manca}\ \emph {et~al.}(2007)\citenamefont {Manca},
  \citenamefont {Baiotti}, \citenamefont {De~Pietri},\ and\ \citenamefont
  {Rezzolla}}]{manca_etal2007}%
  \BibitemOpen
  \bibfield  {author} {\bibinfo {author} {\bibfnamefont {G.~M.}\ \bibnamefont
  {Manca}}, \bibinfo {author} {\bibfnamefont {L.}~\bibnamefont {Baiotti}},
  \bibinfo {author} {\bibfnamefont {R.}~\bibnamefont {De~Pietri}}, \ and\
  \bibinfo {author} {\bibfnamefont {L.}~\bibnamefont {Rezzolla}},\ }\href
  {\doibase 10.1088/0264-9381/24/12/S12} {\bibfield  {journal} {\bibinfo
  {journal} {Class. Quant. Grav.}\ }\textbf {\bibinfo {volume} {24}},\ \bibinfo
  {pages} {S171} (\bibinfo {year} {2007})}\BibitemShut {NoStop}%
\bibitem [{\citenamefont {Franci}\ \emph {et~al.}(2013)\citenamefont {Franci},
  \citenamefont {De~Pietri}, \citenamefont {Dionysopoulou},\ and\ \citenamefont
  {Rezzolla}}]{franci_etal2013}%
  \BibitemOpen
  \bibfield  {author} {\bibinfo {author} {\bibfnamefont {L.}~\bibnamefont
  {Franci}}, \bibinfo {author} {\bibfnamefont {R.}~\bibnamefont {De~Pietri}},
  \bibinfo {author} {\bibfnamefont {K.}~\bibnamefont {Dionysopoulou}}, \ and\
  \bibinfo {author} {\bibfnamefont {L.}~\bibnamefont {Rezzolla}},\ }\href
  {\doibase 10.1103/PhysRevD.88.104028} {\bibfield  {journal} {\bibinfo
  {journal} {Phys. Rev.}\ }\textbf {\bibinfo {volume} {D88}},\ \bibinfo {pages}
  {104028} (\bibinfo {year} {2013})}\BibitemShut {NoStop}%
\bibitem [{\citenamefont {Shapiro}(2000)}]{shapiro2000}%
  \BibitemOpen
  \bibfield  {author} {\bibinfo {author} {\bibfnamefont {S.~L.}\ \bibnamefont
  {Shapiro}},\ }\href {\doibase 10.1086/317209} {\bibfield  {journal} {\bibinfo
   {journal} {Astrophys. J.}\ }\textbf {\bibinfo {volume} {544}},\ \bibinfo
  {pages} {397} (\bibinfo {year} {2000})}\BibitemShut {NoStop}%
\bibitem [{\citenamefont {Shibata}\ \emph {et~al.}(2002)\citenamefont
  {Shibata}, \citenamefont {Karino},\ and\ \citenamefont
  {Eriguchi}}]{shibata_etal2002}%
  \BibitemOpen
  \bibfield  {author} {\bibinfo {author} {\bibfnamefont {M.}~\bibnamefont
  {Shibata}}, \bibinfo {author} {\bibfnamefont {S.}~\bibnamefont {Karino}}, \
  and\ \bibinfo {author} {\bibfnamefont {Y.}~\bibnamefont {Eriguchi}},\ }\href
  {\doibase 10.1046/j.1365-8711.2002.05724.x} {\bibfield  {journal} {\bibinfo
  {journal} {Mon. Not. R. Astron. Soc.}\ }\textbf {\bibinfo {volume} {334}},\
  \bibinfo {pages} {L27} (\bibinfo {year} {2002})}\BibitemShut {NoStop}%
\bibitem [{\citenamefont {Kuroda}\ and\ \citenamefont
  {Umeda}(2010)}]{kuroda_umeda2010}%
  \BibitemOpen
  \bibfield  {author} {\bibinfo {author} {\bibfnamefont {T.}~\bibnamefont
  {Kuroda}}\ and\ \bibinfo {author} {\bibfnamefont {H.}~\bibnamefont {Umeda}},\
  }\href {\doibase 10.1088/0067-0049/191/2/439} {\bibfield  {journal} {\bibinfo
   {journal} {Astrophys. J. Suppl.}\ }\textbf {\bibinfo {volume} {191}},\
  \bibinfo {pages} {439} (\bibinfo {year} {2010})}\BibitemShut {NoStop}%
\bibitem [{\citenamefont {Watts}\ \emph {et~al.}(2005)\citenamefont {Watts},
  \citenamefont {Andersson},\ and\ \citenamefont {Jones}}]{watts_etal2005}%
  \BibitemOpen
  \bibfield  {author} {\bibinfo {author} {\bibfnamefont {A.~L.}\ \bibnamefont
  {Watts}}, \bibinfo {author} {\bibfnamefont {N.}~\bibnamefont {Andersson}}, \
  and\ \bibinfo {author} {\bibfnamefont {D.~I.}\ \bibnamefont {Jones}},\ }\href
  {\doibase 10.1086/427653} {\bibfield  {journal} {\bibinfo  {journal}
  {Astrophys. J.}\ }\textbf {\bibinfo {volume} {618}},\ \bibinfo {pages} {L37}
  (\bibinfo {year} {2005})}\BibitemShut {NoStop}%
\bibitem [{\citenamefont {Passamonti}\ and\ \citenamefont
  {Andersson}(2015)}]{passamonti_andersson2015}%
  \BibitemOpen
  \bibfield  {author} {\bibinfo {author} {\bibfnamefont {A.}~\bibnamefont
  {Passamonti}}\ and\ \bibinfo {author} {\bibfnamefont {N.}~\bibnamefont
  {Andersson}},\ }\href {\doibase 10.1093/mnras/stu2062} {\bibfield  {journal}
  {\bibinfo  {journal} {Mon. Not. R. Astron. Soc.}\ }\textbf {\bibinfo {volume}
  {446}},\ \bibinfo {pages} {555} (\bibinfo {year} {2015})}\BibitemShut
  {NoStop}%
\bibitem [{\citenamefont {Glampedakis}\ and\ \citenamefont
  {Andersson}(2009)}]{KGNA2009}%
  \BibitemOpen
  \bibfield  {author} {\bibinfo {author} {\bibfnamefont {K.}~\bibnamefont
  {Glampedakis}}\ and\ \bibinfo {author} {\bibfnamefont {N.}~\bibnamefont
  {Andersson}},\ }\href {\doibase 10.1103/PhysRevLett.102.141101} {\bibfield
  {journal} {\bibinfo  {journal} {Phys. Rev. Lett.}\ }\textbf {\bibinfo
  {volume} {102}},\ \bibinfo {pages} {141101} (\bibinfo {year}
  {2009})}\BibitemShut {NoStop}%
\bibitem [{\citenamefont {Andersson}\ \emph {et~al.}(2013)\citenamefont
  {Andersson}, \citenamefont {Glampedakis},\ and\ \citenamefont
  {Hogg}}]{andersson_etal2013}%
  \BibitemOpen
  \bibfield  {author} {\bibinfo {author} {\bibfnamefont {N.}~\bibnamefont
  {Andersson}}, \bibinfo {author} {\bibfnamefont {K.}~\bibnamefont
  {Glampedakis}}, \ and\ \bibinfo {author} {\bibfnamefont {M.}~\bibnamefont
  {Hogg}},\ }\href {\doibase 10.1103/PhysRevD.87.063007} {\bibfield  {journal}
  {\bibinfo  {journal} {Phys. Rev. D}\ }\textbf {\bibinfo {volume} {87}},\
  \bibinfo {pages} {063007} (\bibinfo {year} {2013})}\BibitemShut {NoStop}%
\bibitem [{\citenamefont {Friedman}(1983)}]{friedman1983}%
  \BibitemOpen
  \bibfield  {author} {\bibinfo {author} {\bibfnamefont {J.~L.}\ \bibnamefont
  {Friedman}},\ }\href {\doibase 10.1103/PhysRevLett.51.11} {\bibfield
  {journal} {\bibinfo  {journal} {Phys. Rev. Lett.}\ }\textbf {\bibinfo
  {volume} {51}},\ \bibinfo {pages} {11} (\bibinfo {year} {1983})},\ \bibinfo
  {note} {[Erratum: Phys. Rev. Lett.51,718(1983)]}\BibitemShut {NoStop}%
\bibitem [{\citenamefont {Ipser}\ and\ \citenamefont
  {Lindblom}(1989)}]{IL1989}%
  \BibitemOpen
  \bibfield  {author} {\bibinfo {author} {\bibfnamefont {J.~R.}\ \bibnamefont
  {Ipser}}\ and\ \bibinfo {author} {\bibfnamefont {L.}~\bibnamefont
  {Lindblom}},\ }\href {\doibase 10.1103/PhysRevLett.62.2777} {\bibfield
  {journal} {\bibinfo  {journal} {Phys. Rev. Lett.}\ }\textbf {\bibinfo
  {volume} {62}},\ \bibinfo {pages} {2777} (\bibinfo {year}
  {1989})}\BibitemShut {NoStop}%
\bibitem [{\citenamefont {Ipser}\ and\ \citenamefont
  {Lindblom}(1990)}]{IL1990}%
  \BibitemOpen
  \bibfield  {author} {\bibinfo {author} {\bibfnamefont {J.~R.}\ \bibnamefont
  {Ipser}}\ and\ \bibinfo {author} {\bibfnamefont {L.}~\bibnamefont
  {Lindblom}},\ }\href {\doibase 10.1086/168757} {\bibfield  {journal}
  {\bibinfo  {journal} {Astrophys. J.}\ }\textbf {\bibinfo {volume} {355}},\
  \bibinfo {pages} {226} (\bibinfo {year} {1990})}\BibitemShut {NoStop}%
\bibitem [{\citenamefont {Lindblom}\ and\ \citenamefont
  {Mendell}(1995)}]{lindblom_mendell1995}%
  \BibitemOpen
  \bibfield  {author} {\bibinfo {author} {\bibfnamefont {L.}~\bibnamefont
  {Lindblom}}\ and\ \bibinfo {author} {\bibfnamefont {G.}~\bibnamefont
  {Mendell}},\ }\href {\doibase 10.1086/175653} {\bibfield  {journal} {\bibinfo
   {journal} {Astrophys. J.}\ }\textbf {\bibinfo {volume} {444}},\ \bibinfo
  {pages} {804} (\bibinfo {year} {1995})}\BibitemShut {NoStop}%
\bibitem [{\citenamefont {Alpar}\ \emph {et~al.}(1984)\citenamefont {Alpar},
  \citenamefont {Langer},\ and\ \citenamefont {Sauls}}]{als1984}%
  \BibitemOpen
  \bibfield  {author} {\bibinfo {author} {\bibfnamefont {M.~A.}\ \bibnamefont
  {Alpar}}, \bibinfo {author} {\bibfnamefont {S.~A.}\ \bibnamefont {Langer}}, \
  and\ \bibinfo {author} {\bibfnamefont {J.~A.}\ \bibnamefont {Sauls}},\ }\href
  {\doibase 10.1086/162232} {\bibfield  {journal} {\bibinfo  {journal}
  {Astrophys. J.}\ }\textbf {\bibinfo {volume} {282}},\ \bibinfo {pages} {533}
  (\bibinfo {year} {1984})}\BibitemShut {NoStop}%
\bibitem [{\citenamefont {Andersson}\ \emph {et~al.}(2006)\citenamefont
  {Andersson}, \citenamefont {Sidery},\ and\ \citenamefont
  {Comer}}]{andersson_etal2006}%
  \BibitemOpen
  \bibfield  {author} {\bibinfo {author} {\bibfnamefont {N.}~\bibnamefont
  {Andersson}}, \bibinfo {author} {\bibfnamefont {T.}~\bibnamefont {Sidery}}, \
  and\ \bibinfo {author} {\bibfnamefont {G.~L.}\ \bibnamefont {Comer}},\ }\href
  {\doibase 10.1111/j.1365-2966.2006.10147.x} {\bibfield  {journal} {\bibinfo
  {journal} {Mon. Not. R. Astron. Soc.}\ }\textbf {\bibinfo {volume} {368}},\
  \bibinfo {pages} {162} (\bibinfo {year} {2006})}\BibitemShut {NoStop}%
\bibitem [{\citenamefont {Lai}\ and\ \citenamefont
  {Shapiro}(1994)}]{lai_shapiro1994}%
  \BibitemOpen
  \bibfield  {author} {\bibinfo {author} {\bibfnamefont {D.}~\bibnamefont
  {Lai}}\ and\ \bibinfo {author} {\bibfnamefont {S.~L.}\ \bibnamefont
  {Shapiro}},\ }\href {\doibase 10.1086/175438} {\bibfield  {journal} {\bibinfo
   {journal} {Astrophys. J.}\ }\textbf {\bibinfo {volume} {442}},\ \bibinfo
  {pages} {259} (\bibinfo {year} {1994})}\BibitemShut {NoStop}%
\bibitem [{\citenamefont {Stergioulas}(2003)}]{stergioulas2003}%
  \BibitemOpen
  \bibfield  {author} {\bibinfo {author} {\bibfnamefont {N.}~\bibnamefont
  {Stergioulas}},\ }\href {\doibase 10.12942/lrr-2003-3} {\bibfield  {journal}
  {\bibinfo  {journal} {Living Rev. Rel.}\ }\textbf {\bibinfo {volume} {6}},\
  \bibinfo {pages} {109} (\bibinfo {year} {2003})}\BibitemShut {NoStop}%
\bibitem [{\citenamefont {Passamonti}\ \emph {et~al.}(2013)\citenamefont
  {Passamonti}, \citenamefont {Gaertig}, \citenamefont {Kokkotas},\ and\
  \citenamefont {Doneva}}]{passamonti_etal2013}%
  \BibitemOpen
  \bibfield  {author} {\bibinfo {author} {\bibfnamefont {A.}~\bibnamefont
  {Passamonti}}, \bibinfo {author} {\bibfnamefont {E.}~\bibnamefont {Gaertig}},
  \bibinfo {author} {\bibfnamefont {K.~D.}\ \bibnamefont {Kokkotas}}, \ and\
  \bibinfo {author} {\bibfnamefont {D.}~\bibnamefont {Doneva}},\ }\href
  {\doibase 10.1103/PhysRevD.87.084010} {\bibfield  {journal} {\bibinfo
  {journal} {Phys. Rev. D}\ }\textbf {\bibinfo {volume} {87}},\ \bibinfo
  {pages} {084010} (\bibinfo {year} {2013})}\BibitemShut {NoStop}%
\bibitem [{\citenamefont {Owen}\ \emph {et~al.}(1998)\citenamefont {Owen},
  \citenamefont {Lindblom}, \citenamefont {Cutler}, \citenamefont {Schutz},\
  and\ \citenamefont {Andersson}}]{owen_etal1998}%
  \BibitemOpen
  \bibfield  {author} {\bibinfo {author} {\bibfnamefont {B.~J.}\ \bibnamefont
  {Owen}}, \bibinfo {author} {\bibfnamefont {L.}~\bibnamefont {Lindblom}},
  \bibinfo {author} {\bibfnamefont {C.}~\bibnamefont {Cutler}}, \bibinfo
  {author} {\bibfnamefont {B.~F.}\ \bibnamefont {Schutz}}, \ and\ \bibinfo
  {author} {\bibfnamefont {N.}~\bibnamefont {Andersson}},\ }\href {\doibase
  10.1103/PhysRevD.58.084020} {\bibfield  {journal} {\bibinfo  {journal} {Phys.
  Rev. D}\ }\textbf {\bibinfo {volume} {58}},\ \bibinfo {pages} {084020}
  (\bibinfo {year} {1998})}\BibitemShut {NoStop}%
\bibitem [{\citenamefont {Pnigouras}\ and\ \citenamefont
  {Kokkotas}(2015)}]{pnigouras_KK2015}%
  \BibitemOpen
  \bibfield  {author} {\bibinfo {author} {\bibfnamefont {P.}~\bibnamefont
  {Pnigouras}}\ and\ \bibinfo {author} {\bibfnamefont {K.~D.}\ \bibnamefont
  {Kokkotas}},\ }\href {\doibase 10.1103/PhysRevD.92.084018} {\bibfield
  {journal} {\bibinfo  {journal} {Phys. Rev. D}\ }\textbf {\bibinfo {volume}
  {92}},\ \bibinfo {pages} {084018} (\bibinfo {year} {2015})}\BibitemShut
  {NoStop}%
\bibitem [{\citenamefont {Pnigouras}\ and\ \citenamefont
  {Kokkotas}(2016)}]{pnigouras_KK2016}%
  \BibitemOpen
  \bibfield  {author} {\bibinfo {author} {\bibfnamefont {P.}~\bibnamefont
  {Pnigouras}}\ and\ \bibinfo {author} {\bibfnamefont {K.~D.}\ \bibnamefont
  {Kokkotas}},\ }\href {\doibase 10.1103/PhysRevD.94.024053} {\bibfield
  {journal} {\bibinfo  {journal} {Phys. Rev. D}\ }\textbf {\bibinfo {volume}
  {94}},\ \bibinfo {pages} {024053} (\bibinfo {year} {2016})}\BibitemShut
  {NoStop}%
\bibitem [{\citenamefont {Gaertig}\ \emph {et~al.}(2011)\citenamefont
  {Gaertig}, \citenamefont {Glampedakis}, \citenamefont {Kokkotas},\ and\
  \citenamefont {Zink}}]{gaertig_etal2011}%
  \BibitemOpen
  \bibfield  {author} {\bibinfo {author} {\bibfnamefont {E.}~\bibnamefont
  {Gaertig}}, \bibinfo {author} {\bibfnamefont {K.}~\bibnamefont
  {Glampedakis}}, \bibinfo {author} {\bibfnamefont {K.~D.}\ \bibnamefont
  {Kokkotas}}, \ and\ \bibinfo {author} {\bibfnamefont {B.}~\bibnamefont
  {Zink}},\ }\href {\doibase 10.1103/PhysRevLett.107.101102} {\bibfield
  {journal} {\bibinfo  {journal} {Phys. Rev. Lett.}\ }\textbf {\bibinfo
  {volume} {107}},\ \bibinfo {pages} {101102} (\bibinfo {year}
  {2011})}\BibitemShut {NoStop}%
\bibitem [{\citenamefont {Doneva}\ \emph {et~al.}(2015)\citenamefont {Doneva},
  \citenamefont {Kokkotas},\ and\ \citenamefont {Pnigouras}}]{doneva_etal2015}%
  \BibitemOpen
  \bibfield  {author} {\bibinfo {author} {\bibfnamefont {D.~D.}\ \bibnamefont
  {Doneva}}, \bibinfo {author} {\bibfnamefont {K.~D.}\ \bibnamefont
  {Kokkotas}}, \ and\ \bibinfo {author} {\bibfnamefont {P.}~\bibnamefont
  {Pnigouras}},\ }\href {\doibase 10.1103/PhysRevD.92.104040} {\bibfield
  {journal} {\bibinfo  {journal} {Phys. Rev. D}\ }\textbf {\bibinfo {volume}
  {92}},\ \bibinfo {pages} {104040} (\bibinfo {year} {2015})}\BibitemShut
  {NoStop}%
\bibitem [{\citenamefont {Dai}\ and\ \citenamefont {Lu}(1998)}]{dai_lu1998}%
  \BibitemOpen
  \bibfield  {author} {\bibinfo {author} {\bibfnamefont {Z.~G.}\ \bibnamefont
  {Dai}}\ and\ \bibinfo {author} {\bibfnamefont {T.}~\bibnamefont {Lu}},\
  }\href {\doibase 10.1103/PhysRevLett.81.4301} {\bibfield  {journal} {\bibinfo
   {journal} {Phys. Rev. Lett.}\ }\textbf {\bibinfo {volume} {81}},\ \bibinfo
  {pages} {4301} (\bibinfo {year} {1998})}\BibitemShut {NoStop}%
\bibitem [{\citenamefont {Zhang}\ and\ \citenamefont
  {M\'esz\'aros}(2001)}]{zhang_meszaros2001}%
  \BibitemOpen
  \bibfield  {author} {\bibinfo {author} {\bibfnamefont {B.}~\bibnamefont
  {Zhang}}\ and\ \bibinfo {author} {\bibfnamefont {P.}~\bibnamefont
  {M\'esz\'aros}},\ }\href {\doibase 10.1086/320255} {\bibfield  {journal}
  {\bibinfo  {journal} {Astrophys. J.}\ }\textbf {\bibinfo {volume} {552}},\
  \bibinfo {pages} {L35} (\bibinfo {year} {2001})}\BibitemShut {NoStop}%
\bibitem [{\citenamefont {Rowlinson}\ \emph {et~al.}(2013)\citenamefont
  {Rowlinson}, \citenamefont {O'Brien}, \citenamefont {Metzger}, \citenamefont
  {Tanvir},\ and\ \citenamefont {Levan}}]{rowlinson_etal2013}%
  \BibitemOpen
  \bibfield  {author} {\bibinfo {author} {\bibfnamefont {A.}~\bibnamefont
  {Rowlinson}}, \bibinfo {author} {\bibfnamefont {P.~T.}\ \bibnamefont
  {O'Brien}}, \bibinfo {author} {\bibfnamefont {B.~D.}\ \bibnamefont
  {Metzger}}, \bibinfo {author} {\bibfnamefont {N.~R.}\ \bibnamefont {Tanvir}},
  \ and\ \bibinfo {author} {\bibfnamefont {A.~J.}\ \bibnamefont {Levan}},\
  }\href {\doibase 10.1093/mnras/sts683} {\bibfield  {journal} {\bibinfo
  {journal} {Mon. Not. R. Astron. Soc.}\ }\textbf {\bibinfo {volume} {430}},\
  \bibinfo {pages} {1061} (\bibinfo {year} {2013})}\BibitemShut {NoStop}%
\bibitem [{\citenamefont {Rezzolla}\ \emph {et~al.}(2011)\citenamefont
  {Rezzolla}, \citenamefont {Giacomazzo}, \citenamefont {Baiotti},
  \citenamefont {Granot}, \citenamefont {Kouveliotou},\ and\ \citenamefont
  {Aloy}}]{rezzolla_etal2011}%
  \BibitemOpen
  \bibfield  {author} {\bibinfo {author} {\bibfnamefont {L.}~\bibnamefont
  {Rezzolla}}, \bibinfo {author} {\bibfnamefont {B.}~\bibnamefont
  {Giacomazzo}}, \bibinfo {author} {\bibfnamefont {L.}~\bibnamefont {Baiotti}},
  \bibinfo {author} {\bibfnamefont {J.}~\bibnamefont {Granot}}, \bibinfo
  {author} {\bibfnamefont {C.}~\bibnamefont {Kouveliotou}}, \ and\ \bibinfo
  {author} {\bibfnamefont {M.~A.}\ \bibnamefont {Aloy}},\ }\href {\doibase
  10.1088/2041-8205/732/1/L6} {\bibfield  {journal} {\bibinfo  {journal}
  {Astrophys. J. Lett.}\ }\textbf {\bibinfo {volume} {732}},\ \bibinfo {pages}
  {L6} (\bibinfo {year} {2011})}\BibitemShut {NoStop}%
\bibitem [{\citenamefont {Kiuchi}\ \emph {et~al.}(2014)\citenamefont {Kiuchi},
  \citenamefont {Kyutoku}, \citenamefont {Sekiguchi}, \citenamefont {Shibata},\
  and\ \citenamefont {Wada}}]{kiuchi_etal2014}%
  \BibitemOpen
  \bibfield  {author} {\bibinfo {author} {\bibfnamefont {K.}~\bibnamefont
  {Kiuchi}}, \bibinfo {author} {\bibfnamefont {K.}~\bibnamefont {Kyutoku}},
  \bibinfo {author} {\bibfnamefont {Y.}~\bibnamefont {Sekiguchi}}, \bibinfo
  {author} {\bibfnamefont {M.}~\bibnamefont {Shibata}}, \ and\ \bibinfo
  {author} {\bibfnamefont {T.}~\bibnamefont {Wada}},\ }\href {\doibase
  10.1103/PhysRevD.90.041502} {\bibfield  {journal} {\bibinfo  {journal} {Phys.
  Rev. D}\ }\textbf {\bibinfo {volume} {90}},\ \bibinfo {pages} {041502}
  (\bibinfo {year} {2014})}\BibitemShut {NoStop}%
\bibitem [{\citenamefont {Giacomazzo}\ \emph {et~al.}(2015)\citenamefont
  {Giacomazzo}, \citenamefont {Zrake}, \citenamefont {Duffell}, \citenamefont
  {MacFadyen},\ and\ \citenamefont {Perna}}]{giacomazzo_etal2015}%
  \BibitemOpen
  \bibfield  {author} {\bibinfo {author} {\bibfnamefont {B.}~\bibnamefont
  {Giacomazzo}}, \bibinfo {author} {\bibfnamefont {J.}~\bibnamefont {Zrake}},
  \bibinfo {author} {\bibfnamefont {P.~C.}\ \bibnamefont {Duffell}}, \bibinfo
  {author} {\bibfnamefont {A.~I.}\ \bibnamefont {MacFadyen}}, \ and\ \bibinfo
  {author} {\bibfnamefont {R.}~\bibnamefont {Perna}},\ }\href {\doibase
  10.1088/0004-637X/809/1/39} {\bibfield  {journal} {\bibinfo  {journal}
  {Astrophys. J.}\ }\textbf {\bibinfo {volume} {809}},\ \bibinfo {pages} {5}
  (\bibinfo {year} {2015})}\BibitemShut {NoStop}%
\bibitem [{\citenamefont {Metzger}\ and\ \citenamefont
  {Piro}(2014)}]{metzger_piro2014}%
  \BibitemOpen
  \bibfield  {author} {\bibinfo {author} {\bibfnamefont {B.~D.}\ \bibnamefont
  {Metzger}}\ and\ \bibinfo {author} {\bibfnamefont {A.~L.}\ \bibnamefont
  {Piro}},\ }\href {\doibase 10.1093/mnras/stu247} {\bibfield  {journal}
  {\bibinfo  {journal} {Mon. Not. R. Astron. Soc.}\ }\textbf {\bibinfo {volume}
  {439}},\ \bibinfo {pages} {3916} (\bibinfo {year} {2014})}\BibitemShut
  {NoStop}%
\bibitem [{\citenamefont {Lasky}\ and\ \citenamefont
  {Glampedakis}(2016)}]{PLKG2016}%
  \BibitemOpen
  \bibfield  {author} {\bibinfo {author} {\bibfnamefont {P.~D.}\ \bibnamefont
  {Lasky}}\ and\ \bibinfo {author} {\bibfnamefont {K.}~\bibnamefont
  {Glampedakis}},\ }\href {\doibase 10.1093/mnras/stw435} {\bibfield  {journal}
  {\bibinfo  {journal} {Mon. Not. R. Astron. Soc.}\ }\textbf {\bibinfo {volume}
  {458}},\ \bibinfo {pages} {1660} (\bibinfo {year} {2016})}\BibitemShut
  {NoStop}%
\bibitem [{\citenamefont {Yoshida}(2012)}]{yoshida2012}%
  \BibitemOpen
  \bibfield  {author} {\bibinfo {author} {\bibfnamefont {S.}~\bibnamefont
  {Yoshida}},\ }\href {\doibase 10.1103/PhysRevD.86.104055} {\bibfield
  {journal} {\bibinfo  {journal} {Phys. Rev. D}\ }\textbf {\bibinfo {volume}
  {86}},\ \bibinfo {pages} {104055} (\bibinfo {year} {2012})}\BibitemShut
  {NoStop}%
\bibitem [{\citenamefont {Andersson}(1998)}]{NA98}%
  \BibitemOpen
  \bibfield  {author} {\bibinfo {author} {\bibfnamefont {N.}~\bibnamefont
  {Andersson}},\ }\href {\doibase 10.1086/305919} {\bibfield  {journal}
  {\bibinfo  {journal} {Astrophys. J.}\ }\textbf {\bibinfo {volume} {502}},\
  \bibinfo {pages} {708} (\bibinfo {year} {1998})}\BibitemShut {NoStop}%
\bibitem [{\citenamefont {Friedman}\ and\ \citenamefont
  {Morsink}(1998)}]{FM98}%
  \BibitemOpen
  \bibfield  {author} {\bibinfo {author} {\bibfnamefont {J.~L.}\ \bibnamefont
  {Friedman}}\ and\ \bibinfo {author} {\bibfnamefont {S.~M.}\ \bibnamefont
  {Morsink}},\ }\href {\doibase 10.1086/305920} {\bibfield  {journal} {\bibinfo
   {journal} {Astrophys. J.}\ }\textbf {\bibinfo {volume} {502}},\ \bibinfo
  {pages} {714} (\bibinfo {year} {1998})}\BibitemShut {NoStop}%
\bibitem [{\citenamefont {Haskell}(2015)}]{haskell15}%
  \BibitemOpen
  \bibfield  {author} {\bibinfo {author} {\bibfnamefont {B.}~\bibnamefont
  {Haskell}},\ }\href {\doibase 10.1142/S0218301315410074} {\bibfield
  {journal} {\bibinfo  {journal} {Int. Journal of Mod. Phys. E}\ }\textbf
  {\bibinfo {volume} {24}},\ \bibinfo {pages} {1541007} (\bibinfo {year}
  {2015})}\BibitemShut {NoStop}%
\bibitem [{\citenamefont {Lindblom}\ \emph {et~al.}(1998)\citenamefont
  {Lindblom}, \citenamefont {Owen},\ and\ \citenamefont
  {Morsink}}]{lindblom_etal1998}%
  \BibitemOpen
  \bibfield  {author} {\bibinfo {author} {\bibfnamefont {L.}~\bibnamefont
  {Lindblom}}, \bibinfo {author} {\bibfnamefont {B.~J.}\ \bibnamefont {Owen}},
  \ and\ \bibinfo {author} {\bibfnamefont {S.~M.}\ \bibnamefont {Morsink}},\
  }\href {\doibase 10.1103/PhysRevLett.80.4843} {\bibfield  {journal} {\bibinfo
   {journal} {Phys. Rev. Lett.}\ }\textbf {\bibinfo {volume} {80}},\ \bibinfo
  {pages} {4843} (\bibinfo {year} {1998})}\BibitemShut {NoStop}%
\bibitem [{\citenamefont {Lockitch}\ \emph {et~al.}(2001)\citenamefont
  {Lockitch}, \citenamefont {Andersson},\ and\ \citenamefont
  {Friedman}}]{lockitch_etal2001}%
  \BibitemOpen
  \bibfield  {author} {\bibinfo {author} {\bibfnamefont {K.~H.}\ \bibnamefont
  {Lockitch}}, \bibinfo {author} {\bibfnamefont {N.}~\bibnamefont {Andersson}},
  \ and\ \bibinfo {author} {\bibfnamefont {J.~L.}\ \bibnamefont {Friedman}},\
  }\href {\doibase 10.1103/PhysRevD.63.024019} {\bibfield  {journal} {\bibinfo
  {journal} {Phys. Rev. D}\ }\textbf {\bibinfo {volume} {63}},\ \bibinfo
  {pages} {024019} (\bibinfo {year} {2001})}\BibitemShut {NoStop}%
\bibitem [{\citenamefont {Ruoff}\ and\ \citenamefont
  {Kokkotas}(2001)}]{ruoff_KK2001}%
  \BibitemOpen
  \bibfield  {author} {\bibinfo {author} {\bibfnamefont {J.}~\bibnamefont
  {Ruoff}}\ and\ \bibinfo {author} {\bibfnamefont {K.~D.}\ \bibnamefont
  {Kokkotas}},\ }\href {\doibase 10.1046/j.1365-8711.2001.04909.x} {\bibfield
  {journal} {\bibinfo  {journal} {Mon. Not. R. Astron. Soc.}\ }\textbf
  {\bibinfo {volume} {328}},\ \bibinfo {pages} {678} (\bibinfo {year}
  {2001})}\BibitemShut {NoStop}%
\bibitem [{\citenamefont {Yoshida}\ and\ \citenamefont
  {Lee}(2002)}]{yoshida_lee2002}%
  \BibitemOpen
  \bibfield  {author} {\bibinfo {author} {\bibfnamefont {S.}~\bibnamefont
  {Yoshida}}\ and\ \bibinfo {author} {\bibfnamefont {U.}~\bibnamefont {Lee}},\
  }\href {\doibase 10.1086/338663} {\bibfield  {journal} {\bibinfo  {journal}
  {Astrophys. J.}\ }\textbf {\bibinfo {volume} {567}},\ \bibinfo {pages} {1112}
  (\bibinfo {year} {2002})}\BibitemShut {NoStop}%
\bibitem [{\citenamefont {Lockitch}\ \emph {et~al.}(2003)\citenamefont
  {Lockitch}, \citenamefont {Andersson},\ and\ \citenamefont
  {Friedman}}]{lockitch_etal2003}%
  \BibitemOpen
  \bibfield  {author} {\bibinfo {author} {\bibfnamefont {K.~H.}\ \bibnamefont
  {Lockitch}}, \bibinfo {author} {\bibfnamefont {N.}~\bibnamefont {Andersson}},
  \ and\ \bibinfo {author} {\bibfnamefont {J.~L.}\ \bibnamefont {Friedman}},\
  }\href {\doibase 10.1103/PhysRevD.68.124010} {\bibfield  {journal} {\bibinfo
  {journal} {Phys. Rev. D}\ }\textbf {\bibinfo {volume} {68}},\ \bibinfo
  {pages} {124010} (\bibinfo {year} {2003})}\BibitemShut {NoStop}%
\bibitem [{\citenamefont {Idrisy}\ \emph {et~al.}(2015)\citenamefont {Idrisy},
  \citenamefont {Owen},\ and\ \citenamefont {Jones}}]{idrisy_etal2015}%
  \BibitemOpen
  \bibfield  {author} {\bibinfo {author} {\bibfnamefont {A.}~\bibnamefont
  {Idrisy}}, \bibinfo {author} {\bibfnamefont {B.~J.}\ \bibnamefont {Owen}}, \
  and\ \bibinfo {author} {\bibfnamefont {D.~I.}\ \bibnamefont {Jones}},\ }\href
  {\doibase 10.1103/PhysRevD.91.024001} {\bibfield  {journal} {\bibinfo
  {journal} {Phys. Rev. D}\ }\textbf {\bibinfo {volume} {91}},\ \bibinfo
  {pages} {024001} (\bibinfo {year} {2015})}\BibitemShut {NoStop}%
\bibitem [{\citenamefont {Owen}(2010)}]{owen2010}%
  \BibitemOpen
  \bibfield  {author} {\bibinfo {author} {\bibfnamefont {B.~J.}\ \bibnamefont
  {Owen}},\ }\href {\doibase 10.1103/PhysRevD.82.104002} {\bibfield  {journal}
  {\bibinfo  {journal} {Phys. Rev. D}\ }\textbf {\bibinfo {volume} {82}},\
  \bibinfo {pages} {104002} (\bibinfo {year} {2010})}\BibitemShut {NoStop}%
\bibitem [{\citenamefont {Mahmoodifar}\ and\ \citenamefont
  {Strohmayer}(2013)}]{MS2013}%
  \BibitemOpen
  \bibfield  {author} {\bibinfo {author} {\bibfnamefont {S.}~\bibnamefont
  {Mahmoodifar}}\ and\ \bibinfo {author} {\bibfnamefont {T.}~\bibnamefont
  {Strohmayer}},\ }\href {\doibase 10.1088/0004-637X/773/2/140} {\bibfield
  {journal} {\bibinfo  {journal} {Astrophys. J.}\ }\textbf {\bibinfo {volume}
  {773}},\ \bibinfo {pages} {10} (\bibinfo {year} {2013})}\BibitemShut
  {NoStop}%
\bibitem [{\citenamefont {Ho}\ \emph {et~al.}(2011)\citenamefont {Ho},
  \citenamefont {Andersson},\ and\ \citenamefont {Haskell}}]{Ho_etal2011}%
  \BibitemOpen
  \bibfield  {author} {\bibinfo {author} {\bibfnamefont {W.~C.~G.}\
  \bibnamefont {Ho}}, \bibinfo {author} {\bibfnamefont {N.}~\bibnamefont
  {Andersson}}, \ and\ \bibinfo {author} {\bibfnamefont {B.}~\bibnamefont
  {Haskell}},\ }\href {\doibase 10.1103/PhysRevLett.107.101101} {\bibfield
  {journal} {\bibinfo  {journal} {Phys. Rev. Lett.}\ }\textbf {\bibinfo
  {volume} {107}},\ \bibinfo {pages} {101101} (\bibinfo {year}
  {2011})}\BibitemShut {NoStop}%
\bibitem [{\citenamefont {Haskell}\ \emph {et~al.}(2012)\citenamefont
  {Haskell}, \citenamefont {Degenaar},\ and\ \citenamefont
  {Ho}}]{Haskell:2012vg}%
  \BibitemOpen
  \bibfield  {author} {\bibinfo {author} {\bibfnamefont {B.}~\bibnamefont
  {Haskell}}, \bibinfo {author} {\bibfnamefont {N.}~\bibnamefont {Degenaar}}, \
  and\ \bibinfo {author} {\bibfnamefont {W.~C.~G.}\ \bibnamefont {Ho}},\ }\href
  {\doibase 10.1111/j.1365-2966.2012.21171.x} {\bibfield  {journal} {\bibinfo
  {journal} {Mon. Not. Roy. Astron. Soc.}\ }\textbf {\bibinfo {volume} {424}},\
  \bibinfo {pages} {93} (\bibinfo {year} {2012})}\BibitemShut {NoStop}%
\bibitem [{\citenamefont {Papaloizou}\ and\ \citenamefont
  {Pringle}(1978)}]{PP78}%
  \BibitemOpen
  \bibfield  {author} {\bibinfo {author} {\bibfnamefont {J.}~\bibnamefont
  {Papaloizou}}\ and\ \bibinfo {author} {\bibfnamefont {J.~E.}\ \bibnamefont
  {Pringle}},\ }\href {\doibase 10.1093/mnras/184.3.501} {\bibfield  {journal}
  {\bibinfo  {journal} {Mon. Not. R. Astron. Soc.}\ }\textbf {\bibinfo {volume}
  {184}},\ \bibinfo {pages} {501} (\bibinfo {year} {1978})}\BibitemShut
  {NoStop}%
\bibitem [{\citenamefont {Bildsten}(1998{\natexlab{a}})}]{bildsten1998}%
  \BibitemOpen
  \bibfield  {author} {\bibinfo {author} {\bibfnamefont {L.}~\bibnamefont
  {Bildsten}},\ }\href {\doibase 10.1086/311440} {\bibfield  {journal}
  {\bibinfo  {journal} {Astrophys. J.}\ }\textbf {\bibinfo {volume} {501}},\
  \bibinfo {pages} {L89} (\bibinfo {year} {1998}{\natexlab{a}})}\BibitemShut
  {NoStop}%
\bibitem [{\citenamefont {Andersson}\ \emph {et~al.}(1999)\citenamefont
  {Andersson}, \citenamefont {Kokkotas},\ and\ \citenamefont
  {Stergioulas}}]{andersson_etal1999}%
  \BibitemOpen
  \bibfield  {author} {\bibinfo {author} {\bibfnamefont {N.}~\bibnamefont
  {Andersson}}, \bibinfo {author} {\bibfnamefont {K.~D.}\ \bibnamefont
  {Kokkotas}}, \ and\ \bibinfo {author} {\bibfnamefont {N.}~\bibnamefont
  {Stergioulas}},\ }\href {\doibase 10.1086/307082} {\bibfield  {journal}
  {\bibinfo  {journal} {Astrophys. J.}\ }\textbf {\bibinfo {volume} {516}},\
  \bibinfo {pages} {307} (\bibinfo {year} {1999})}\BibitemShut {NoStop}%
\bibitem [{\citenamefont {Ghosh}\ and\ \citenamefont
  {Lamb}(1979)}]{ghosh_lamb1979}%
  \BibitemOpen
  \bibfield  {author} {\bibinfo {author} {\bibfnamefont {P.}~\bibnamefont
  {Ghosh}}\ and\ \bibinfo {author} {\bibfnamefont {F.~K.}\ \bibnamefont
  {Lamb}},\ }\href {\doibase 10.1086/157498} {\bibfield  {journal} {\bibinfo
  {journal} {Astrophys. J.}\ }\textbf {\bibinfo {volume} {234}},\ \bibinfo
  {pages} {296} (\bibinfo {year} {1979})}\BibitemShut {NoStop}%
\bibitem [{\citenamefont {Wang}(1995)}]{wang1995}%
  \BibitemOpen
  \bibfield  {author} {\bibinfo {author} {\bibfnamefont {Y.-M.}\ \bibnamefont
  {Wang}},\ }\href {\doibase 10.1086/309649} {\bibfield  {journal} {\bibinfo
  {journal} {Astrophys. J. Lett.}\ }\textbf {\bibinfo {volume} {449}},\
  \bibinfo {pages} {L153} (\bibinfo {year} {1995})}\BibitemShut {NoStop}%
\bibitem [{\citenamefont {Rappaport}\ \emph {et~al.}(2004)\citenamefont
  {Rappaport}, \citenamefont {Fregeau},\ and\ \citenamefont
  {Spruit}}]{rappaport_etal2004}%
  \BibitemOpen
  \bibfield  {author} {\bibinfo {author} {\bibfnamefont {S.~A.}\ \bibnamefont
  {Rappaport}}, \bibinfo {author} {\bibfnamefont {J.~M.}\ \bibnamefont
  {Fregeau}}, \ and\ \bibinfo {author} {\bibfnamefont {H.}~\bibnamefont
  {Spruit}},\ }\href {\doibase 10.1086/382863} {\bibfield  {journal} {\bibinfo
  {journal} {Astrophys. J.}\ }\textbf {\bibinfo {volume} {606}},\ \bibinfo
  {pages} {436} (\bibinfo {year} {2004})}\BibitemShut {NoStop}%
\bibitem [{\citenamefont {Andersson}\ \emph
  {et~al.}(2005{\natexlab{b}})\citenamefont {Andersson}, \citenamefont
  {Glampedakis}, \citenamefont {Haskell},\ and\ \citenamefont
  {Watts}}]{andersson_etal2005}%
  \BibitemOpen
  \bibfield  {author} {\bibinfo {author} {\bibfnamefont {N.}~\bibnamefont
  {Andersson}}, \bibinfo {author} {\bibfnamefont {K.}~\bibnamefont
  {Glampedakis}}, \bibinfo {author} {\bibfnamefont {B.}~\bibnamefont
  {Haskell}}, \ and\ \bibinfo {author} {\bibfnamefont {A.~L.}\ \bibnamefont
  {Watts}},\ }\href {\doibase 10.1111/j.1365-2966.2005.09167.x} {\bibfield
  {journal} {\bibinfo  {journal} {Mon. Not. R. Astron. Soc.}\ }\textbf
  {\bibinfo {volume} {361}},\ \bibinfo {pages} {1153} (\bibinfo {year}
  {2005}{\natexlab{b}})}\BibitemShut {NoStop}%
\bibitem [{\citenamefont {Haskell}\ and\ \citenamefont
  {Patruno}(2011)}]{haskell_patruno2011}%
  \BibitemOpen
  \bibfield  {author} {\bibinfo {author} {\bibfnamefont {B.}~\bibnamefont
  {Haskell}}\ and\ \bibinfo {author} {\bibfnamefont {A.}~\bibnamefont
  {Patruno}},\ }\href {\doibase 10.1088/2041-8205/738/1/L14} {\bibfield
  {journal} {\bibinfo  {journal} {Astrophys. J. Lett.}\ }\textbf {\bibinfo
  {volume} {738}},\ \bibinfo {pages} {L14} (\bibinfo {year}
  {2011})}\BibitemShut {NoStop}%
\bibitem [{\citenamefont {Patruno}\ and\ \citenamefont
  {Watts}(2012)}]{patruno_watts2012}%
  \BibitemOpen
  \bibfield  {author} {\bibinfo {author} {\bibfnamefont {A.}~\bibnamefont
  {Patruno}}\ and\ \bibinfo {author} {\bibfnamefont {A.~L.}\ \bibnamefont
  {Watts}},\ }in\ \href@noop {} {\emph {\bibinfo {booktitle} {{Timing neutron
  stars: pulsations, oscillations and explosions}}}},\ \bibinfo {editor}
  {edited by\ \bibinfo {editor} {\bibnamefont {{T. Belloni, M. Mendez, C.M.
  Zhang}}}}\ (\bibinfo  {publisher} {{Springer}},\ \bibinfo {year}
  {2012})\BibitemShut {NoStop}%
\bibitem [{\citenamefont {Bhattacharyya}\ and\ \citenamefont
  {Chakrabarty}(2017)}]{bhattacharyya_chakrabarty2017}%
  \BibitemOpen
  \bibfield  {author} {\bibinfo {author} {\bibfnamefont {S.}~\bibnamefont
  {Bhattacharyya}}\ and\ \bibinfo {author} {\bibfnamefont {D.}~\bibnamefont
  {Chakrabarty}},\ }\href {\doibase 10.3847/1538-4357/835/1/4} {\bibfield
  {journal} {\bibinfo  {journal} {Astrophys. J.}\ }\textbf {\bibinfo {volume}
  {835}},\ \bibinfo {pages} {4} (\bibinfo {year} {2017})}\BibitemShut {NoStop}%
\bibitem [{\citenamefont {Bhattacharyya}(2017)}]{bhattacharyya2017}%
  \BibitemOpen
  \bibfield  {author} {\bibinfo {author} {\bibfnamefont {S.}~\bibnamefont
  {Bhattacharyya}},\ }\href {\doibase 10.3847/1538-4357/aa8169} {\bibfield
  {journal} {\bibinfo  {journal} {Astrophys. J.}\ }\textbf {\bibinfo {volume}
  {847}},\ \bibinfo {pages} {8} (\bibinfo {year} {2017})}\BibitemShut {NoStop}%
\bibitem [{\citenamefont {Levin}(1999)}]{levin99}%
  \BibitemOpen
  \bibfield  {author} {\bibinfo {author} {\bibfnamefont {Y.}~\bibnamefont
  {Levin}},\ }\href {\doibase 10.1086/307196} {\bibfield  {journal} {\bibinfo
  {journal} {Astrophys. J.}\ }\textbf {\bibinfo {volume} {517}},\ \bibinfo
  {pages} {328} (\bibinfo {year} {1999})}\BibitemShut {NoStop}%
\bibitem [{\citenamefont {Andersson}\ \emph {et~al.}(2000)\citenamefont
  {Andersson}, \citenamefont {Jones}, \citenamefont {Kokkotas},\ and\
  \citenamefont {Stergioulas}}]{andersson_etal2000}%
  \BibitemOpen
  \bibfield  {author} {\bibinfo {author} {\bibfnamefont {N.}~\bibnamefont
  {Andersson}}, \bibinfo {author} {\bibfnamefont {D.~I.}\ \bibnamefont
  {Jones}}, \bibinfo {author} {\bibfnamefont {K.~D.}\ \bibnamefont {Kokkotas}},
  \ and\ \bibinfo {author} {\bibfnamefont {N.}~\bibnamefont {Stergioulas}},\
  }\href {\doibase 10.1086/312643} {\bibfield  {journal} {\bibinfo  {journal}
  {Astrophys. J.}\ }\textbf {\bibinfo {volume} {534}},\ \bibinfo {pages} {L75}
  (\bibinfo {year} {2000})}\BibitemShut {NoStop}%
\bibitem [{\citenamefont {Bondarescu}\ \emph {et~al.}(2007)\citenamefont
  {Bondarescu}, \citenamefont {Teukolsky},\ and\ \citenamefont
  {Wasserman}}]{bondarescu_etal2007}%
  \BibitemOpen
  \bibfield  {author} {\bibinfo {author} {\bibfnamefont {R.}~\bibnamefont
  {Bondarescu}}, \bibinfo {author} {\bibfnamefont {S.~A.}\ \bibnamefont
  {Teukolsky}}, \ and\ \bibinfo {author} {\bibfnamefont {I.}~\bibnamefont
  {Wasserman}},\ }\href {\doibase 10.1103/PhysRevD.76.064019} {\bibfield
  {journal} {\bibinfo  {journal} {Phys. Rev. D}\ }\textbf {\bibinfo {volume}
  {76}},\ \bibinfo {pages} {064019} (\bibinfo {year} {2007})}\BibitemShut
  {NoStop}%
\bibitem [{\citenamefont {Shapiro}\ and\ \citenamefont
  {Teukolsky}(1986)}]{STbook}%
  \BibitemOpen
  \bibfield  {author} {\bibinfo {author} {\bibfnamefont {S.~L.}\ \bibnamefont
  {Shapiro}}\ and\ \bibinfo {author} {\bibfnamefont {S.~A.}\ \bibnamefont
  {Teukolsky}},\ }\href@noop {} {\emph {\bibinfo {title} {{Black Holes, White
  Dwarfs and Neutron Stars: The Physics of Compact Objects}}}}\ (\bibinfo
  {publisher} {Wiley},\ \bibinfo {year} {1986})\BibitemShut {NoStop}%
\bibitem [{\citenamefont {Heyl}(2002)}]{heyl2002}%
  \BibitemOpen
  \bibfield  {author} {\bibinfo {author} {\bibfnamefont {J.~S.}\ \bibnamefont
  {Heyl}},\ }\href {\doibase 10.1086/342263} {\bibfield  {journal} {\bibinfo
  {journal} {Astrophys. J.}\ }\textbf {\bibinfo {volume} {574}},\ \bibinfo
  {pages} {L57} (\bibinfo {year} {2002})}\BibitemShut {NoStop}%
\bibitem [{\citenamefont {Andersson}\ \emph {et~al.}(2002)\citenamefont
  {Andersson}, \citenamefont {Jones},\ and\ \citenamefont
  {Kokkotas}}]{andersson_etal2002}%
  \BibitemOpen
  \bibfield  {author} {\bibinfo {author} {\bibfnamefont {N.}~\bibnamefont
  {Andersson}}, \bibinfo {author} {\bibfnamefont {D.~I.}\ \bibnamefont
  {Jones}}, \ and\ \bibinfo {author} {\bibfnamefont {K.~D.}\ \bibnamefont
  {Kokkotas}},\ }\href {\doibase 10.1046/j.1365-8711.2002.05837.x} {\bibfield
  {journal} {\bibinfo  {journal} {Mon. Not. R. Astron. Soc.}\ }\textbf
  {\bibinfo {volume} {337}},\ \bibinfo {pages} {1224} (\bibinfo {year}
  {2002})}\BibitemShut {NoStop}%
\bibitem [{\citenamefont {Nayyar}\ and\ \citenamefont
  {Owen}(2006)}]{nayyar_owen2006}%
  \BibitemOpen
  \bibfield  {author} {\bibinfo {author} {\bibfnamefont {M.}~\bibnamefont
  {Nayyar}}\ and\ \bibinfo {author} {\bibfnamefont {B.~J.}\ \bibnamefont
  {Owen}},\ }\href {\doibase 10.1103/PhysRevD.73.084001} {\bibfield  {journal}
  {\bibinfo  {journal} {Phys. Rev. D}\ }\textbf {\bibinfo {volume} {73}},\
  \bibinfo {pages} {084001} (\bibinfo {year} {2006})}\BibitemShut {NoStop}%
\bibitem [{\citenamefont {Haskell}\ and\ \citenamefont
  {Andersson}(2010)}]{haskell_andersson2010}%
  \BibitemOpen
  \bibfield  {author} {\bibinfo {author} {\bibfnamefont {B.}~\bibnamefont
  {Haskell}}\ and\ \bibinfo {author} {\bibfnamefont {N.}~\bibnamefont
  {Andersson}},\ }\href {\doibase 10.1111/j.1365-2966.2010.17255.x} {\bibfield
  {journal} {\bibinfo  {journal} {Mon. Not. R. Astron. Soc.}\ }\textbf
  {\bibinfo {volume} {408}},\ \bibinfo {pages} {1897} (\bibinfo {year}
  {2010})}\BibitemShut {NoStop}%
\bibitem [{\citenamefont {Alford}\ and\ \citenamefont
  {Schwenzer}(2015)}]{alford_schwenzer2015}%
  \BibitemOpen
  \bibfield  {author} {\bibinfo {author} {\bibfnamefont {M.~G.}\ \bibnamefont
  {Alford}}\ and\ \bibinfo {author} {\bibfnamefont {K.}~\bibnamefont
  {Schwenzer}},\ }\href {\doibase 10.1093/mnras/stu2361} {\bibfield  {journal}
  {\bibinfo  {journal} {Mon. Not. R. Astron. Soc.}\ }\textbf {\bibinfo {volume}
  {446}},\ \bibinfo {pages} {3631} (\bibinfo {year} {2015})}\BibitemShut
  {NoStop}%
\bibitem [{\citenamefont {Aasi}\ \emph {et~al.}(2015)\citenamefont {Aasi} \emph
  {et~al.}}]{aasi_etal2015}%
  \BibitemOpen
  \bibfield  {author} {\bibinfo {author} {\bibfnamefont {J.}~\bibnamefont
  {Aasi}} \emph {et~al.},\ }\href@noop {} {\bibfield  {journal} {\bibinfo
  {journal} {Astrophys. J.}\ }\textbf {\bibinfo {volume} {813}},\ \bibinfo
  {pages} {16} (\bibinfo {year} {2015})}\BibitemShut {NoStop}%
\bibitem [{\citenamefont {Pappito}\ \emph {et~al.}(2011)\citenamefont
  {Pappito}, \citenamefont {Riggio}, \citenamefont {Bunderi}, \citenamefont
  {di~Salvo}, \citenamefont {D'A\'i},\ and\ \citenamefont
  {Iaria}}]{papitto_etal2011}%
  \BibitemOpen
  \bibfield  {author} {\bibinfo {author} {\bibfnamefont {A.}~\bibnamefont
  {Pappito}}, \bibinfo {author} {\bibfnamefont {A.}~\bibnamefont {Riggio}},
  \bibinfo {author} {\bibfnamefont {L.}~\bibnamefont {Bunderi}}, \bibinfo
  {author} {\bibfnamefont {T.}~\bibnamefont {di~Salvo}}, \bibinfo {author}
  {\bibfnamefont {A.}~\bibnamefont {D'A\'i}}, \ and\ \bibinfo {author}
  {\bibfnamefont {R.}~\bibnamefont {Iaria}},\ }\href {\doibase
  10.1051/0004-6361/201014837} {\bibfield  {journal} {\bibinfo  {journal}
  {Astron. Astrophys.}\ }\textbf {\bibinfo {volume} {528}},\ \bibinfo {pages}
  {6} (\bibinfo {year} {2011})}\BibitemShut {NoStop}%
\bibitem [{\citenamefont {Schenk}\ \emph {et~al.}(2002)\citenamefont {Schenk},
  \citenamefont {Arras}, \citenamefont {Flanagan}, \citenamefont {Teukolsky},\
  and\ \citenamefont {Wasserman}}]{schenk_etal2002}%
  \BibitemOpen
  \bibfield  {author} {\bibinfo {author} {\bibfnamefont {A.~K.}\ \bibnamefont
  {Schenk}}, \bibinfo {author} {\bibfnamefont {P.}~\bibnamefont {Arras}},
  \bibinfo {author} {\bibfnamefont {E.~E.}\ \bibnamefont {Flanagan}}, \bibinfo
  {author} {\bibfnamefont {S.~A.}\ \bibnamefont {Teukolsky}}, \ and\ \bibinfo
  {author} {\bibfnamefont {I.}~\bibnamefont {Wasserman}},\ }\href {\doibase
  10.1103/PhysRevD.65.024001} {\bibfield  {journal} {\bibinfo  {journal} {Phys.
  Rev. D}\ }\textbf {\bibinfo {volume} {65}},\ \bibinfo {pages} {024001}
  (\bibinfo {year} {2002})}\BibitemShut {NoStop}%
\bibitem [{\citenamefont {Arras}\ \emph {et~al.}(2003)\citenamefont {Arras},
  \citenamefont {Flanagan}, \citenamefont {Morsink}, \citenamefont {Schenk},
  \citenamefont {A.},\ and\ \citenamefont {Wasserman}}]{arras_etal2003}%
  \BibitemOpen
  \bibfield  {author} {\bibinfo {author} {\bibfnamefont {P.}~\bibnamefont
  {Arras}}, \bibinfo {author} {\bibfnamefont {E.~E.}\ \bibnamefont {Flanagan}},
  \bibinfo {author} {\bibfnamefont {S.~M.}\ \bibnamefont {Morsink}}, \bibinfo
  {author} {\bibfnamefont {A.~K.}\ \bibnamefont {Schenk}}, \bibinfo {author}
  {\bibfnamefont {T.~S.}\ \bibnamefont {A.}}, \ and\ \bibinfo {author}
  {\bibfnamefont {I.}~\bibnamefont {Wasserman}},\ }\href {\doibase
  10.1086/374657} {\bibfield  {journal} {\bibinfo  {journal} {Astrophys. J.}\
  }\textbf {\bibinfo {volume} {591}},\ \bibinfo {pages} {1129} (\bibinfo {year}
  {2003})}\BibitemShut {NoStop}%
\bibitem [{\citenamefont {Bondarescu}\ \emph {et~al.}(2009)\citenamefont
  {Bondarescu}, \citenamefont {A.},\ and\ \citenamefont
  {Wasserman}}]{bondarescu_etal2009}%
  \BibitemOpen
  \bibfield  {author} {\bibinfo {author} {\bibfnamefont {R.}~\bibnamefont
  {Bondarescu}}, \bibinfo {author} {\bibfnamefont {T.~S.}\ \bibnamefont {A.}},
  \ and\ \bibinfo {author} {\bibfnamefont {I.}~\bibnamefont {Wasserman}},\
  }\href {\doibase 10.1103/PhysRevD.79.104003} {\bibfield  {journal} {\bibinfo
  {journal} {Phys. Rev. D}\ }\textbf {\bibinfo {volume} {79}},\ \bibinfo
  {pages} {104003} (\bibinfo {year} {2009})}\BibitemShut {NoStop}%
\bibitem [{\citenamefont {Alford}\ and\ \citenamefont
  {Schwenzer}(2014{\natexlab{a}})}]{alford_schwenzer2014b}%
  \BibitemOpen
  \bibfield  {author} {\bibinfo {author} {\bibfnamefont {M.~G.}\ \bibnamefont
  {Alford}}\ and\ \bibinfo {author} {\bibfnamefont {K.}~\bibnamefont
  {Schwenzer}},\ }\href {\doibase 10.1103/PhysRevLett.113.251102} {\bibfield
  {journal} {\bibinfo  {journal} {Phys. Rev. Lett.}\ }\textbf {\bibinfo
  {volume} {113}},\ \bibinfo {pages} {251102} (\bibinfo {year}
  {2014}{\natexlab{a}})}\BibitemShut {NoStop}%
\bibitem [{\citenamefont {Brown}\ and\ \citenamefont
  {Ushomirsky}(2000)}]{BU2000}%
  \BibitemOpen
  \bibfield  {author} {\bibinfo {author} {\bibfnamefont {E.~F.}\ \bibnamefont
  {Brown}}\ and\ \bibinfo {author} {\bibfnamefont {G.}~\bibnamefont
  {Ushomirsky}},\ }\href {\doibase 10.1086/308969} {\bibfield  {journal}
  {\bibinfo  {journal} {Astrophys. J.}\ }\textbf {\bibinfo {volume} {536}},\
  \bibinfo {pages} {915} (\bibinfo {year} {2000})}\BibitemShut {NoStop}%
\bibitem [{\citenamefont {Haskell}\ \emph {et~al.}(2014)\citenamefont
  {Haskell}, \citenamefont {Glampedakis},\ and\ \citenamefont
  {Andersson}}]{haskell_etal2014}%
  \BibitemOpen
  \bibfield  {author} {\bibinfo {author} {\bibfnamefont {B.}~\bibnamefont
  {Haskell}}, \bibinfo {author} {\bibfnamefont {K.}~\bibnamefont
  {Glampedakis}}, \ and\ \bibinfo {author} {\bibfnamefont {N.}~\bibnamefont
  {Andersson}},\ }\href {\doibase 10.1093/mnras/stu535} {\bibfield  {journal}
  {\bibinfo  {journal} {Mon. Not. R. Astron. Soc.}\ }\textbf {\bibinfo {volume}
  {441}},\ \bibinfo {pages} {1662} (\bibinfo {year} {2014})}\BibitemShut
  {NoStop}%
\bibitem [{\citenamefont {Alford}\ and\ \citenamefont
  {Schwenzer}(2014{\natexlab{b}})}]{alford_schwenzer2014a}%
  \BibitemOpen
  \bibfield  {author} {\bibinfo {author} {\bibfnamefont {M.~G.}\ \bibnamefont
  {Alford}}\ and\ \bibinfo {author} {\bibfnamefont {K.}~\bibnamefont
  {Schwenzer}},\ }\href {\doibase 10.1088/0004-637X/781/1/26} {\bibfield
  {journal} {\bibinfo  {journal} {Astrophys. J.}\ }\textbf {\bibinfo {volume}
  {781}},\ \bibinfo {pages} {22} (\bibinfo {year}
  {2014}{\natexlab{b}})}\BibitemShut {NoStop}%
\bibitem [{\citenamefont {Kokkotas}\ and\ \citenamefont
  {Schwenzer}(2016)}]{KK_schwenzer2016}%
  \BibitemOpen
  \bibfield  {author} {\bibinfo {author} {\bibfnamefont {K.~D.}\ \bibnamefont
  {Kokkotas}}\ and\ \bibinfo {author} {\bibfnamefont {K.}~\bibnamefont
  {Schwenzer}},\ }\href {\doibase 10.1140/epja/i2016-16038-9} {\bibfield
  {journal} {\bibinfo  {journal} {Eur. Phys. J. A}\ }\textbf {\bibinfo {volume}
  {52}},\ \bibinfo {pages} {15} (\bibinfo {year} {2016})}\BibitemShut {NoStop}%
\bibitem [{\citenamefont {Strohmayer}\ and\ \citenamefont
  {Mahmoodifar}(2014)}]{sm2014}%
  \BibitemOpen
  \bibfield  {author} {\bibinfo {author} {\bibfnamefont {T.}~\bibnamefont
  {Strohmayer}}\ and\ \bibinfo {author} {\bibfnamefont {S.}~\bibnamefont
  {Mahmoodifar}},\ }\href {\doibase 10.1088/2041-8205/793/2/L38} {\bibfield
  {journal} {\bibinfo  {journal} {Astrophys. J. Lett.}\ }\textbf {\bibinfo
  {volume} {793}},\ \bibinfo {pages} {L38} (\bibinfo {year}
  {2014})}\BibitemShut {NoStop}%
\bibitem [{\citenamefont {Andersson}\ \emph {et~al.}(2014)\citenamefont
  {Andersson}, \citenamefont {Jones},\ and\ \citenamefont
  {Ho}}]{andersson_etal2014}%
  \BibitemOpen
  \bibfield  {author} {\bibinfo {author} {\bibfnamefont {N.}~\bibnamefont
  {Andersson}}, \bibinfo {author} {\bibfnamefont {D.~I.}\ \bibnamefont
  {Jones}}, \ and\ \bibinfo {author} {\bibfnamefont {W.~C.~G.}\ \bibnamefont
  {Ho}},\ }\href {\doibase 10.1093/mnras/stu870} {\bibfield  {journal}
  {\bibinfo  {journal} {Mon. Not. R. Astron. Soc.}\ }\textbf {\bibinfo {volume}
  {442}},\ \bibinfo {pages} {1786} (\bibinfo {year} {2014})}\BibitemShut
  {NoStop}%
\bibitem [{LIG()}]{LIGOcurve}%
  \BibitemOpen
  \href@noop {} {}\bibinfo {howpublished} {{\tt
  https://dcc.ligo.org/cgi-bin/DocDB/ShowDocument?docid=T0900288}},\ \bibinfo
  {note} {sensitivity curve of Advanced LIGO}\BibitemShut {NoStop}%
\bibitem [{ETc()}]{ETcurve}%
  \BibitemOpen
  \href@noop {} {}\bibinfo {howpublished} {{\tt
  https://workarea.et-gw.eu/et/WG4-Astrophysics}},\ \bibinfo {note}
  {sensitivity curve of ET}\BibitemShut {NoStop}%
\bibitem [{\citenamefont {Jones}(2001)}]{jones2001}%
  \BibitemOpen
  \bibfield  {author} {\bibinfo {author} {\bibfnamefont {P.~B.}\ \bibnamefont
  {Jones}},\ }\href {\doibase 10.1103/PhysRevLett.86.1384} {\bibfield
  {journal} {\bibinfo  {journal} {Phys. Rev. Lett.}\ }\textbf {\bibinfo
  {volume} {86}},\ \bibinfo {pages} {1384} (\bibinfo {year}
  {2001})}\BibitemShut {NoStop}%
\bibitem [{\citenamefont {Lindblom}\ and\ \citenamefont
  {Owen}(2002)}]{lindblom_owen2002}%
  \BibitemOpen
  \bibfield  {author} {\bibinfo {author} {\bibfnamefont {L.}~\bibnamefont
  {Lindblom}}\ and\ \bibinfo {author} {\bibfnamefont {B.~J.}\ \bibnamefont
  {Owen}},\ }\href {\doibase 10.1103/PhysRevD.65.063006} {\bibfield  {journal}
  {\bibinfo  {journal} {Phys. Rev. D}\ }\textbf {\bibinfo {volume} {65}},\
  \bibinfo {pages} {063006} (\bibinfo {year} {2002})}\BibitemShut {NoStop}%
\bibitem [{\citenamefont {Madsen}(1998)}]{madsen1998}%
  \BibitemOpen
  \bibfield  {author} {\bibinfo {author} {\bibfnamefont {J.}~\bibnamefont
  {Madsen}},\ }\href {\doibase 10.1103/PhysRevLett.81.3311} {\bibfield
  {journal} {\bibinfo  {journal} {Phys. Rev. Lett.}\ }\textbf {\bibinfo
  {volume} {81}},\ \bibinfo {pages} {3311} (\bibinfo {year}
  {1998})}\BibitemShut {NoStop}%
\bibitem [{\citenamefont {Madsen}(2000)}]{madsen2000}%
  \BibitemOpen
  \bibfield  {author} {\bibinfo {author} {\bibfnamefont {J.}~\bibnamefont
  {Madsen}},\ }\href {\doibase 10.1103/PhysRevLett.85.10} {\bibfield  {journal}
  {\bibinfo  {journal} {Phys. Rev. Lett.}\ }\textbf {\bibinfo {volume} {85}},\
  \bibinfo {pages} {10} (\bibinfo {year} {2000})}\BibitemShut {NoStop}%
\bibitem [{\citenamefont {Mannarelli}\ \emph {et~al.}(2008)\citenamefont
  {Mannarelli}, \citenamefont {Manuel},\ and\ \citenamefont
  {Sa'D}}]{mannarelli_etal2008}%
  \BibitemOpen
  \bibfield  {author} {\bibinfo {author} {\bibfnamefont {M.}~\bibnamefont
  {Mannarelli}}, \bibinfo {author} {\bibfnamefont {C.}~\bibnamefont {Manuel}},
  \ and\ \bibinfo {author} {\bibfnamefont {B.~A.}\ \bibnamefont {Sa'D}},\
  }\href {\doibase 10.1103/PhysRevLett.101.241101} {\bibfield  {journal}
  {\bibinfo  {journal} {Phys. Rev. Lett.}\ }\textbf {\bibinfo {volume} {101}},\
  \bibinfo {pages} {241101} (\bibinfo {year} {2008})}\BibitemShut {NoStop}%
\bibitem [{\citenamefont {Andersson}\ \emph {et~al.}(2010)\citenamefont
  {Andersson}, \citenamefont {Haskell},\ and\ \citenamefont
  {Comer}}]{andersson_etal2010}%
  \BibitemOpen
  \bibfield  {author} {\bibinfo {author} {\bibfnamefont {N.}~\bibnamefont
  {Andersson}}, \bibinfo {author} {\bibfnamefont {B.}~\bibnamefont {Haskell}},
  \ and\ \bibinfo {author} {\bibfnamefont {G.~L.}\ \bibnamefont {Comer}},\
  }\href {\doibase 10.1103/PhysRevD.82.023007} {\bibfield  {journal} {\bibinfo
  {journal} {Phys. Rev. D}\ }\textbf {\bibinfo {volume} {82}},\ \bibinfo
  {pages} {023007} (\bibinfo {year} {2010})}\BibitemShut {NoStop}%
\bibitem [{\citenamefont {Lindblom}\ and\ \citenamefont
  {Mendell}(2000)}]{lindblom_mendell2000}%
  \BibitemOpen
  \bibfield  {author} {\bibinfo {author} {\bibfnamefont {L.}~\bibnamefont
  {Lindblom}}\ and\ \bibinfo {author} {\bibfnamefont {G.}~\bibnamefont
  {Mendell}},\ }\href {\doibase 10.1103/PhysRevD.61.104003} {\bibfield
  {journal} {\bibinfo  {journal} {Phys. Rev. D}\ }\textbf {\bibinfo {volume}
  {61}},\ \bibinfo {pages} {104003} (\bibinfo {year} {2000})}\BibitemShut
  {NoStop}%
\bibitem [{\citenamefont {Ruderman}\ \emph {et~al.}()\citenamefont {Ruderman},
  \citenamefont {Zhu},\ and\ \citenamefont {Chen}}]{ruderman_etal1998}%
  \BibitemOpen
  \bibfield  {author} {\bibinfo {author} {\bibfnamefont {M.}~\bibnamefont
  {Ruderman}}, \bibinfo {author} {\bibfnamefont {T.}~\bibnamefont {Zhu}}, \
  and\ \bibinfo {author} {\bibfnamefont {K.}~\bibnamefont {Chen}},\ }\href
  {\doibase 10.1086/305026} {\bibfield  {journal} {\bibinfo  {journal}
  {Astrophys. J.}\ }\textbf {\bibinfo {volume} {492}},\ \bibinfo {pages}
  {267}}\BibitemShut {NoStop}%
\bibitem [{\citenamefont {Link}(2003)}]{link2003}%
  \BibitemOpen
  \bibfield  {author} {\bibinfo {author} {\bibfnamefont {B.}~\bibnamefont
  {Link}},\ }\href {\doibase 10.1103/PhysRevLett.91.101101} {\bibfield
  {journal} {\bibinfo  {journal} {Phys. Rev. Lett.}\ }\textbf {\bibinfo
  {volume} {91}},\ \bibinfo {pages} {101101} (\bibinfo {year}
  {2003})}\BibitemShut {NoStop}%
\bibitem [{\citenamefont {Haskell}\ \emph {et~al.}(2009)\citenamefont
  {Haskell}, \citenamefont {Andersson},\ and\ \citenamefont
  {Passamonti}}]{haskell_etal2009}%
  \BibitemOpen
  \bibfield  {author} {\bibinfo {author} {\bibfnamefont {B.}~\bibnamefont
  {Haskell}}, \bibinfo {author} {\bibfnamefont {N.}~\bibnamefont {Andersson}},
  \ and\ \bibinfo {author} {\bibfnamefont {A.}~\bibnamefont {Passamonti}},\
  }\href {\doibase 10.1111/j.1365-2966.2009.14963.x} {\bibfield  {journal}
  {\bibinfo  {journal} {Mon. Not. R. Astron. Soc.}\ }\textbf {\bibinfo {volume}
  {397}},\ \bibinfo {pages} {1464} (\bibinfo {year} {2009})}\BibitemShut
  {NoStop}%
\bibitem [{\citenamefont {Chugunov}\ \emph {et~al.}(2017)\citenamefont
  {Chugunov}, \citenamefont {Kantor},\ and\ \citenamefont
  {Gusakov}}]{Gusakov:2016drh}%
  \BibitemOpen
  \bibfield  {author} {\bibinfo {author} {\bibfnamefont {A.~I.}\ \bibnamefont
  {Chugunov}}, \bibinfo {author} {\bibfnamefont {E.~M.}\ \bibnamefont
  {Kantor}}, \ and\ \bibinfo {author} {\bibfnamefont {M.~E.}\ \bibnamefont
  {Gusakov}},\ }\href {\doibase 10.1093/mnras/stx391} {\bibfield  {journal}
  {\bibinfo  {journal} {Mon. Not. Roy. Astron. Soc.}\ }\textbf {\bibinfo
  {volume} {468}},\ \bibinfo {pages} {291} (\bibinfo {year}
  {2017})}\BibitemShut {NoStop}%
\bibitem [{\citenamefont {Kantor}\ and\ \citenamefont
  {Gusakov}(2017)}]{Kantor:2017xuo}%
  \BibitemOpen
  \bibfield  {author} {\bibinfo {author} {\bibfnamefont {E.~M.}\ \bibnamefont
  {Kantor}}\ and\ \bibinfo {author} {\bibfnamefont {M.~E.}\ \bibnamefont
  {Gusakov}},\ }\href {\doibase 10.1093/mnras/stx1075} {\bibfield  {journal}
  {\bibinfo  {journal} {Mon. Not. Roy. Astron. Soc.}\ }\textbf {\bibinfo
  {volume} {469}},\ \bibinfo {pages} {3928} (\bibinfo {year}
  {2017})}\BibitemShut {NoStop}%
\bibitem [{\citenamefont {Rezzolla}\ \emph {et~al.}(2000)\citenamefont
  {Rezzolla}, \citenamefont {Lamb},\ and\ \citenamefont
  {Shapiro}}]{rezzolla_etal2000}%
  \BibitemOpen
  \bibfield  {author} {\bibinfo {author} {\bibfnamefont {L.}~\bibnamefont
  {Rezzolla}}, \bibinfo {author} {\bibfnamefont {F.~K.}\ \bibnamefont {Lamb}},
  \ and\ \bibinfo {author} {\bibfnamefont {S.~L.}\ \bibnamefont {Shapiro}},\
  }\href {\doibase 10.1086/312539} {\bibfield  {journal} {\bibinfo  {journal}
  {Astrophys. J.}\ }\textbf {\bibinfo {volume} {531}},\ \bibinfo {pages} {L139}
  (\bibinfo {year} {2000})}\BibitemShut {NoStop}%
\bibitem [{\citenamefont {Rezzolla}\ \emph
  {et~al.}(2001{\natexlab{a}})\citenamefont {Rezzolla}, \citenamefont {Lamb},
  \citenamefont {Markovic},\ and\ \citenamefont
  {Shapiro}}]{rezzolla_etal2001a}%
  \BibitemOpen
  \bibfield  {author} {\bibinfo {author} {\bibfnamefont {L.}~\bibnamefont
  {Rezzolla}}, \bibinfo {author} {\bibfnamefont {F.~K.}\ \bibnamefont {Lamb}},
  \bibinfo {author} {\bibfnamefont {D.}~\bibnamefont {Markovic}}, \ and\
  \bibinfo {author} {\bibfnamefont {S.~L.}\ \bibnamefont {Shapiro}},\ }\href
  {\doibase 10.1103/PhysRevD.64.104013} {\bibfield  {journal} {\bibinfo
  {journal} {Phys. Rev. D}\ }\textbf {\bibinfo {volume} {64}},\ \bibinfo
  {pages} {104013} (\bibinfo {year} {2001}{\natexlab{a}})}\BibitemShut
  {NoStop}%
\bibitem [{\citenamefont {Rezzolla}\ \emph
  {et~al.}(2001{\natexlab{b}})\citenamefont {Rezzolla}, \citenamefont {Lamb},
  \citenamefont {Markovic},\ and\ \citenamefont
  {Shapiro}}]{rezzolla_etal2001b}%
  \BibitemOpen
  \bibfield  {author} {\bibinfo {author} {\bibfnamefont {L.}~\bibnamefont
  {Rezzolla}}, \bibinfo {author} {\bibfnamefont {F.~K.}\ \bibnamefont {Lamb}},
  \bibinfo {author} {\bibfnamefont {D.}~\bibnamefont {Markovic}}, \ and\
  \bibinfo {author} {\bibfnamefont {S.~L.}\ \bibnamefont {Shapiro}},\ }\href
  {\doibase 10.1103/PhysRevD.64.104014} {\bibfield  {journal} {\bibinfo
  {journal} {Phys. Rev. D}\ }\textbf {\bibinfo {volume} {64}},\ \bibinfo
  {pages} {104014} (\bibinfo {year} {2001}{\natexlab{b}})}\BibitemShut
  {NoStop}%
\bibitem [{\citenamefont {Cuofano}\ \emph {et~al.}(2012)\citenamefont
  {Cuofano}, \citenamefont {Dall'Osso}, \citenamefont {Drago},\ and\
  \citenamefont {Stella}}]{cuofano_etal2012}%
  \BibitemOpen
  \bibfield  {author} {\bibinfo {author} {\bibfnamefont {C.}~\bibnamefont
  {Cuofano}}, \bibinfo {author} {\bibfnamefont {S.}~\bibnamefont {Dall'Osso}},
  \bibinfo {author} {\bibfnamefont {A.}~\bibnamefont {Drago}}, \ and\ \bibinfo
  {author} {\bibfnamefont {L.}~\bibnamefont {Stella}},\ }\href {\doibase
  10.1103/PhysRevD.86.044004} {\bibfield  {journal} {\bibinfo  {journal} {Phys.
  Rev. D}\ }\textbf {\bibinfo {volume} {86}},\ \bibinfo {pages} {044004}
  (\bibinfo {year} {2012})}\BibitemShut {NoStop}%
\bibitem [{\citenamefont {Chugunov}(2015)}]{chugunov2015}%
  \BibitemOpen
  \bibfield  {author} {\bibinfo {author} {\bibfnamefont {A.~I.}\ \bibnamefont
  {Chugunov}},\ }\href {\doibase 10.1093/mnras/stv1092} {\bibfield  {journal}
  {\bibinfo  {journal} {Mon. Not. R. Astron. Soc.}\ }\textbf {\bibinfo {volume}
  {451}},\ \bibinfo {pages} {2772} (\bibinfo {year} {2015})}\BibitemShut
  {NoStop}%
\bibitem [{\citenamefont {Friedman}\ \emph {et~al.}(2016)\citenamefont
  {Friedman}, \citenamefont {Lindblom},\ and\ \citenamefont
  {Lockitch}}]{friedman_etal2016}%
  \BibitemOpen
  \bibfield  {author} {\bibinfo {author} {\bibfnamefont {J.~L.}\ \bibnamefont
  {Friedman}}, \bibinfo {author} {\bibfnamefont {L.}~\bibnamefont {Lindblom}},
  \ and\ \bibinfo {author} {\bibfnamefont {K.~H.}\ \bibnamefont {Lockitch}},\
  }\href {\doibase 10.1103/PhysRevD.93.024023} {\bibfield  {journal} {\bibinfo
  {journal} {Phys. Rev. D}\ }\textbf {\bibinfo {volume} {93}},\ \bibinfo
  {pages} {024023} (\bibinfo {year} {2016})}\BibitemShut {NoStop}%
\bibitem [{\citenamefont {Bildsten}\ and\ \citenamefont
  {Ushomirsky}(2000)}]{BU2000b}%
  \BibitemOpen
  \bibfield  {author} {\bibinfo {author} {\bibfnamefont {L.}~\bibnamefont
  {Bildsten}}\ and\ \bibinfo {author} {\bibfnamefont {G.}~\bibnamefont
  {Ushomirsky}},\ }\href {\doibase 10.1086/312454} {\bibfield  {journal}
  {\bibinfo  {journal} {Astrophys. J.}\ }\textbf {\bibinfo {volume} {529}},\
  \bibinfo {pages} {L33} (\bibinfo {year} {2000})}\BibitemShut {NoStop}%
\bibitem [{\citenamefont {Lindblom}\ \emph {et~al.}(2000)\citenamefont
  {Lindblom}, \citenamefont {Owen},\ and\ \citenamefont
  {Ushomirsky}}]{lindblom_etal2000}%
  \BibitemOpen
  \bibfield  {author} {\bibinfo {author} {\bibfnamefont {L.}~\bibnamefont
  {Lindblom}}, \bibinfo {author} {\bibfnamefont {B.~J.}\ \bibnamefont {Owen}},
  \ and\ \bibinfo {author} {\bibfnamefont {G.}~\bibnamefont {Ushomirsky}},\
  }\href {\doibase 10.1103/PhysRevD.62.084030} {\bibfield  {journal} {\bibinfo
  {journal} {Phys. Rev. D}\ }\textbf {\bibinfo {volume} {62}},\ \bibinfo
  {pages} {084030} (\bibinfo {year} {2000})}\BibitemShut {NoStop}%
\bibitem [{\citenamefont {Rieutord}(2001)}]{rieutord2001}%
  \BibitemOpen
  \bibfield  {author} {\bibinfo {author} {\bibfnamefont {M.}~\bibnamefont
  {Rieutord}},\ }\href {\doibase 10.1086/319705} {\bibfield  {journal}
  {\bibinfo  {journal} {Astrophys. J.}\ }\textbf {\bibinfo {volume} {550}},\
  \bibinfo {pages} {443} (\bibinfo {year} {2001})}\BibitemShut {NoStop}%
\bibitem [{\citenamefont {Glampedakis}\ and\ \citenamefont
  {Andersson}(2006{\natexlab{a}})}]{KG_NA2006b}%
  \BibitemOpen
  \bibfield  {author} {\bibinfo {author} {\bibfnamefont {K.}~\bibnamefont
  {Glampedakis}}\ and\ \bibinfo {author} {\bibfnamefont {N.}~\bibnamefont
  {Andersson}},\ }\href {\doibase 10.1111/j.1365-2966.2006.10749.x} {\bibfield
  {journal} {\bibinfo  {journal} {Mon. Not. R. Astron. Soc.}\ }\textbf
  {\bibinfo {volume} {371}},\ \bibinfo {pages} {1311} (\bibinfo {year}
  {2006}{\natexlab{a}})}\BibitemShut {NoStop}%
\bibitem [{\citenamefont {Levin}\ and\ \citenamefont
  {Ushomirsky}(2001)}]{LU2001}%
  \BibitemOpen
  \bibfield  {author} {\bibinfo {author} {\bibfnamefont {Y.}~\bibnamefont
  {Levin}}\ and\ \bibinfo {author} {\bibfnamefont {G.}~\bibnamefont
  {Ushomirsky}},\ }\href {\doibase 10.1046/j.1365-8711.2001.04323.x} {\bibfield
   {journal} {\bibinfo  {journal} {Mon. Not. R. Astron. Soc.}\ }\textbf
  {\bibinfo {volume} {324}},\ \bibinfo {pages} {917} (\bibinfo {year}
  {2001})}\BibitemShut {NoStop}%
\bibitem [{\citenamefont {Glampedakis}\ and\ \citenamefont
  {Andersson}(2006{\natexlab{b}})}]{KG_NA2006a}%
  \BibitemOpen
  \bibfield  {author} {\bibinfo {author} {\bibfnamefont {K.}~\bibnamefont
  {Glampedakis}}\ and\ \bibinfo {author} {\bibfnamefont {N.}~\bibnamefont
  {Andersson}},\ }\href {\doibase 10.1103/PhysRevD.74.044040} {\bibfield
  {journal} {\bibinfo  {journal} {Phys. Rev. D}\ }\textbf {\bibinfo {volume}
  {74}},\ \bibinfo {pages} {044040} (\bibinfo {year}
  {2006}{\natexlab{b}})}\BibitemShut {NoStop}%
\bibitem [{\citenamefont {Shternin}\ and\ \citenamefont
  {Yakovlev}(2008)}]{shternin_yakovlev2008}%
  \BibitemOpen
  \bibfield  {author} {\bibinfo {author} {\bibfnamefont {P.~S.}\ \bibnamefont
  {Shternin}}\ and\ \bibinfo {author} {\bibfnamefont {D.~G.}\ \bibnamefont
  {Yakovlev}},\ }\href {\doibase 10.1103/PhysRevD.78.063006} {\bibfield
  {journal} {\bibinfo  {journal} {Phys. Rev. D}\ }\textbf {\bibinfo {volume}
  {78}},\ \bibinfo {pages} {063006} (\bibinfo {year} {2008})}\BibitemShut
  {NoStop}%
\bibitem [{\citenamefont {Wen}\ \emph {et~al.}(2012)\citenamefont {Wen},
  \citenamefont {Newton},\ and\ \citenamefont {Li}}]{wen_etal2012}%
  \BibitemOpen
  \bibfield  {author} {\bibinfo {author} {\bibfnamefont {D.-H.}\ \bibnamefont
  {Wen}}, \bibinfo {author} {\bibfnamefont {W.~G.}\ \bibnamefont {Newton}}, \
  and\ \bibinfo {author} {\bibfnamefont {B.-A.}\ \bibnamefont {Li}},\ }\href
  {\doibase 10.1103/PhysRevC.85.025801} {\bibfield  {journal} {\bibinfo
  {journal} {Phys. Rev. D}\ }\textbf {\bibinfo {volume} {85}},\ \bibinfo
  {pages} {025801} (\bibinfo {year} {2012})}\BibitemShut {NoStop}%
\bibitem [{\citenamefont {Mendell}(2001)}]{mendell2001}%
  \BibitemOpen
  \bibfield  {author} {\bibinfo {author} {\bibfnamefont {G.}~\bibnamefont
  {Mendell}},\ }\href {\doibase 10.1103/PhysRevD.64.044009} {\bibfield
  {journal} {\bibinfo  {journal} {Phys. Rev. D}\ }\textbf {\bibinfo {volume}
  {64}},\ \bibinfo {pages} {044009} (\bibinfo {year} {2001})}\BibitemShut
  {NoStop}%
\bibitem [{\citenamefont {Glampedakis}\ \emph
  {et~al.}(2011{\natexlab{a}})\citenamefont {Glampedakis}, \citenamefont
  {Andersson},\ and\ \citenamefont {Samuelsson}}]{KG_etal2011}%
  \BibitemOpen
  \bibfield  {author} {\bibinfo {author} {\bibfnamefont {K.}~\bibnamefont
  {Glampedakis}}, \bibinfo {author} {\bibfnamefont {N.}~\bibnamefont
  {Andersson}}, \ and\ \bibinfo {author} {\bibfnamefont {L.}~\bibnamefont
  {Samuelsson}},\ }\href {\doibase 10.1111/j.1365-2966.2010.17484.x} {\bibfield
   {journal} {\bibinfo  {journal} {Mon. Not. R. Astron. Soc.}\ }\textbf
  {\bibinfo {volume} {410}},\ \bibinfo {pages} {805} (\bibinfo {year}
  {2011}{\natexlab{a}})}\BibitemShut {NoStop}%
\bibitem [{\citenamefont {Glampedakis}\ \emph {et~al.}(tion)\citenamefont
  {Glampedakis}, \citenamefont {Andersson},\ and\ \citenamefont
  {Jones}}]{KGetal_inprep}%
  \BibitemOpen
  \bibfield  {author} {\bibinfo {author} {\bibfnamefont {K.}~\bibnamefont
  {Glampedakis}}, \bibinfo {author} {\bibfnamefont {N.}~\bibnamefont
  {Andersson}}, \ and\ \bibinfo {author} {\bibfnamefont {D.~I.}\ \bibnamefont
  {Jones}},\ }\href@noop {} {\  (\bibinfo {year} {2017, in
  preparation})}\BibitemShut {NoStop}%
\bibitem [{\citenamefont {Glampedakis}\ \emph {et~al.}(2006)\citenamefont
  {Glampedakis}, \citenamefont {Samuelsson},\ and\ \citenamefont
  {Andersson}}]{Glampedakis:2006apa}%
  \BibitemOpen
  \bibfield  {author} {\bibinfo {author} {\bibfnamefont {K.}~\bibnamefont
  {Glampedakis}}, \bibinfo {author} {\bibfnamefont {L.}~\bibnamefont
  {Samuelsson}}, \ and\ \bibinfo {author} {\bibfnamefont {N.}~\bibnamefont
  {Andersson}},\ }\href {\doibase 10.1111/j.1745-3933.2006.00211.x} {\bibfield
  {journal} {\bibinfo  {journal} {Mon. Not. Roy. Astron. Soc.}\ }\textbf
  {\bibinfo {volume} {371}},\ \bibinfo {pages} {L74} (\bibinfo {year}
  {2006})}\BibitemShut {NoStop}%
\bibitem [{\citenamefont {Gabler}\ \emph {et~al.}(2011)\citenamefont {Gabler},
  \citenamefont {Cerda-Duran}, \citenamefont {Font}, \citenamefont {Muller},\
  and\ \citenamefont {Stergioulas}}]{Gabler:2010rp}%
  \BibitemOpen
  \bibfield  {author} {\bibinfo {author} {\bibfnamefont {M.}~\bibnamefont
  {Gabler}}, \bibinfo {author} {\bibfnamefont {P.}~\bibnamefont {Cerda-Duran}},
  \bibinfo {author} {\bibfnamefont {J.~A.}\ \bibnamefont {Font}}, \bibinfo
  {author} {\bibfnamefont {E.}~\bibnamefont {Muller}}, \ and\ \bibinfo {author}
  {\bibfnamefont {N.}~\bibnamefont {Stergioulas}},\ }\href {\doibase
  10.1111/j.1745-3933.2010.00974.x} {\bibfield  {journal} {\bibinfo  {journal}
  {Mon. Not. Roy. Astron. Soc.}\ }\textbf {\bibinfo {volume} {410}},\ \bibinfo
  {pages} {37} (\bibinfo {year} {2011})}\BibitemShut {NoStop}%
\bibitem [{\citenamefont {Colaiuda}\ and\ \citenamefont
  {Kokkotas}(2011)}]{Colaiuda:2010pc}%
  \BibitemOpen
  \bibfield  {author} {\bibinfo {author} {\bibfnamefont {A.}~\bibnamefont
  {Colaiuda}}\ and\ \bibinfo {author} {\bibfnamefont {K.~D.}\ \bibnamefont
  {Kokkotas}},\ }\href {\doibase 10.1111/j.1365-2966.2011.18602.x} {\bibfield
  {journal} {\bibinfo  {journal} {Mon. Not. Roy. Astron. Soc.}\ }\textbf
  {\bibinfo {volume} {414}},\ \bibinfo {pages} {3014} (\bibinfo {year}
  {2011})}\BibitemShut {NoStop}%
\bibitem [{\citenamefont {Gabler}\ \emph {et~al.}(2012)\citenamefont {Gabler},
  \citenamefont {Duran}, \citenamefont {Stergioulas}, \citenamefont {Font},\
  and\ \citenamefont {Muller}}]{Gabler:2011am}%
  \BibitemOpen
  \bibfield  {author} {\bibinfo {author} {\bibfnamefont {M.}~\bibnamefont
  {Gabler}}, \bibinfo {author} {\bibfnamefont {P.~C.}\ \bibnamefont {Duran}},
  \bibinfo {author} {\bibfnamefont {N.}~\bibnamefont {Stergioulas}}, \bibinfo
  {author} {\bibfnamefont {J.~A.}\ \bibnamefont {Font}}, \ and\ \bibinfo
  {author} {\bibfnamefont {E.}~\bibnamefont {Muller}},\ }\href {\doibase
  10.1111/j.1365-2966.2012.20454.x} {\bibfield  {journal} {\bibinfo  {journal}
  {Mon. Not. R. Astron. Soc.}\ }\textbf {\bibinfo {volume} {421}},\ \bibinfo
  {pages} {2054} (\bibinfo {year} {2012})}\BibitemShut {NoStop}%
\bibitem [{\citenamefont {Gabler}\ \emph {et~al.}(2016)\citenamefont {Gabler},
  \citenamefont {Cerd{\'a}-Dur{\'a}n}, \citenamefont {Stergioulas},
  \citenamefont {Font},\ and\ \citenamefont {M{\"u}ller}}]{Gabler:2016rth}%
  \BibitemOpen
  \bibfield  {author} {\bibinfo {author} {\bibfnamefont {M.}~\bibnamefont
  {Gabler}}, \bibinfo {author} {\bibfnamefont {P.}~\bibnamefont
  {Cerd{\'a}-Dur{\'a}n}}, \bibinfo {author} {\bibfnamefont {N.}~\bibnamefont
  {Stergioulas}}, \bibinfo {author} {\bibfnamefont {J.~A.}\ \bibnamefont
  {Font}}, \ and\ \bibinfo {author} {\bibfnamefont {E.}~\bibnamefont
  {M{\"u}ller}},\ }\href {\doibase 10.1093/mnras/stw1272} {\bibfield  {journal}
  {\bibinfo  {journal} {Mon. Not. R. Astron. Soc.}\ }\textbf {\bibinfo {volume}
  {460}},\ \bibinfo {pages} {4242} (\bibinfo {year} {2016})}\BibitemShut
  {NoStop}%
\bibitem [{\citenamefont {Kinney}\ and\ \citenamefont
  {Mendell}(2003)}]{kinney_mendell2003}%
  \BibitemOpen
  \bibfield  {author} {\bibinfo {author} {\bibfnamefont {J.~B.}\ \bibnamefont
  {Kinney}}\ and\ \bibinfo {author} {\bibfnamefont {G.}~\bibnamefont
  {Mendell}},\ }\href {\doibase 10.1103/PhysRevD.67.024032} {\bibfield
  {journal} {\bibinfo  {journal} {Phys. Rev. D}\ }\textbf {\bibinfo {volume}
  {67}},\ \bibinfo {pages} {024032} (\bibinfo {year} {2003})}\BibitemShut
  {NoStop}%
\bibitem [{\citenamefont {Gearheart}\ \emph {et~al.}(2011)\citenamefont
  {Gearheart}, \citenamefont {Newton}, \citenamefont {Hooker},\ and\
  \citenamefont {Li}}]{gearheart_etal2011}%
  \BibitemOpen
  \bibfield  {author} {\bibinfo {author} {\bibfnamefont {M.}~\bibnamefont
  {Gearheart}}, \bibinfo {author} {\bibfnamefont {W.~G.}\ \bibnamefont
  {Newton}}, \bibinfo {author} {\bibfnamefont {J.}~\bibnamefont {Hooker}}, \
  and\ \bibinfo {author} {\bibfnamefont {B.-A.}\ \bibnamefont {Li}},\ }\href
  {\doibase 10.1111/j.1365-2966.2011.19628.x} {\bibfield  {journal} {\bibinfo
  {journal} {Mon. Not. R. Astron. Soc.}\ }\textbf {\bibinfo {volume} {418}},\
  \bibinfo {pages} {2343} (\bibinfo {year} {2011})}\BibitemShut {NoStop}%
\bibitem [{\citenamefont {{Duncan}}\ and\ \citenamefont
  {{Thompson}}(1992)}]{Duncan:1992dt}%
  \BibitemOpen
  \bibfield  {author} {\bibinfo {author} {\bibfnamefont {R.~C.}\ \bibnamefont
  {{Duncan}}}\ and\ \bibinfo {author} {\bibfnamefont {C.}~\bibnamefont
  {{Thompson}}},\ }\href {\doibase 10.1086/186413} {\bibfield  {journal}
  {\bibinfo  {journal} {Astrophys. J. Lett.}\ }\textbf {\bibinfo {volume}
  {392}},\ \bibinfo {pages} {L9} (\bibinfo {year} {1992})}\BibitemShut
  {NoStop}%
\bibitem [{\citenamefont {{Woods}}\ and\ \citenamefont
  {{Thompson}}(2006)}]{Woods:2004kb}%
  \BibitemOpen
  \bibfield  {author} {\bibinfo {author} {\bibfnamefont {P.~M.}\ \bibnamefont
  {{Woods}}}\ and\ \bibinfo {author} {\bibfnamefont {C.}~\bibnamefont
  {{Thompson}}},\ }\bibfield  {booktitle} {\emph {\bibinfo {booktitle} {Compact
  stellar X-ray sources}},\ }\href@noop {} {\ ,\ \bibinfo {pages} {547}
  (\bibinfo {year} {2006})}\BibitemShut {NoStop}%
\bibitem [{\citenamefont {Turolla}\ \emph {et~al.}(2015)\citenamefont
  {Turolla}, \citenamefont {Zane},\ and\ \citenamefont
  {Watts}}]{Turolla:2015mwa}%
  \BibitemOpen
  \bibfield  {author} {\bibinfo {author} {\bibfnamefont {R.}~\bibnamefont
  {Turolla}}, \bibinfo {author} {\bibfnamefont {S.}~\bibnamefont {Zane}}, \
  and\ \bibinfo {author} {\bibfnamefont {A.}~\bibnamefont {Watts}},\ }\href
  {\doibase 10.1088/0034-4885/78/11/116901} {\bibfield  {journal} {\bibinfo
  {journal} {Rept. Prog. Phys.}\ }\textbf {\bibinfo {volume} {78}},\ \bibinfo
  {pages} {116901} (\bibinfo {year} {2015})}\BibitemShut {NoStop}%
\bibitem [{\citenamefont {Israel}\ \emph {et~al.}(2005)\citenamefont {Israel},
  \citenamefont {Belloni}, \citenamefont {Stella}, \citenamefont {Rephaeli},
  \citenamefont {Gruber}, \citenamefont {Casella}, \citenamefont {Dall'Osso},
  \citenamefont {Rea}, \citenamefont {Persic},\ and\ \citenamefont
  {Rothschild}}]{Israel:2005av}%
  \BibitemOpen
  \bibfield  {author} {\bibinfo {author} {\bibfnamefont {G.}~\bibnamefont
  {Israel}}, \bibinfo {author} {\bibfnamefont {T.}~\bibnamefont {Belloni}},
  \bibinfo {author} {\bibfnamefont {L.}~\bibnamefont {Stella}}, \bibinfo
  {author} {\bibfnamefont {Y.}~\bibnamefont {Rephaeli}}, \bibinfo {author}
  {\bibfnamefont {D.}~\bibnamefont {Gruber}}, \bibinfo {author} {\bibfnamefont
  {P.~G.}\ \bibnamefont {Casella}}, \bibinfo {author} {\bibfnamefont
  {S.}~\bibnamefont {Dall'Osso}}, \bibinfo {author} {\bibfnamefont
  {N.}~\bibnamefont {Rea}}, \bibinfo {author} {\bibfnamefont {M.}~\bibnamefont
  {Persic}}, \ and\ \bibinfo {author} {\bibfnamefont {R.}~\bibnamefont
  {Rothschild}},\ }\href {\doibase 10.1086/432615} {\bibfield  {journal}
  {\bibinfo  {journal} {Astrophys. J.}\ }\textbf {\bibinfo {volume} {628}},\
  \bibinfo {pages} {L53} (\bibinfo {year} {2005})}\BibitemShut {NoStop}%
\bibitem [{\citenamefont {Strohmayer}\ and\ \citenamefont
  {Watts}(2006)}]{Strohmayer:2006py}%
  \BibitemOpen
  \bibfield  {author} {\bibinfo {author} {\bibfnamefont {T.~E.}\ \bibnamefont
  {Strohmayer}}\ and\ \bibinfo {author} {\bibfnamefont {A.~L.}\ \bibnamefont
  {Watts}},\ }\href {\doibase 10.1086/508703} {\bibfield  {journal} {\bibinfo
  {journal} {Astrophys. J.}\ }\textbf {\bibinfo {volume} {653}},\ \bibinfo
  {pages} {593} (\bibinfo {year} {2006})}\BibitemShut {NoStop}%
\bibitem [{\citenamefont {Strohmayer}\ and\ \citenamefont
  {Watts}(2005)}]{Strohmayer:2005ks}%
  \BibitemOpen
  \bibfield  {author} {\bibinfo {author} {\bibfnamefont {T.~E.}\ \bibnamefont
  {Strohmayer}}\ and\ \bibinfo {author} {\bibfnamefont {A.~L.}\ \bibnamefont
  {Watts}},\ }\href {\doibase 10.1086/497911} {\bibfield  {journal} {\bibinfo
  {journal} {Astrophys. J.}\ }\textbf {\bibinfo {volume} {632}},\ \bibinfo
  {pages} {L111} (\bibinfo {year} {2005})}\BibitemShut {NoStop}%
\bibitem [{\citenamefont {Huppenkothen}\ \emph
  {et~al.}(2014{\natexlab{a}})\citenamefont {Huppenkothen}, \citenamefont
  {Heil}, \citenamefont {Watts},\ and\ \citenamefont {G{\"o}{\u g}{\"u}{\c
  s}}}]{Huppenkothen:2014cla}%
  \BibitemOpen
  \bibfield  {author} {\bibinfo {author} {\bibfnamefont {D.}~\bibnamefont
  {Huppenkothen}}, \bibinfo {author} {\bibfnamefont {L.~M.}\ \bibnamefont
  {Heil}}, \bibinfo {author} {\bibfnamefont {A.~L.}\ \bibnamefont {Watts}}, \
  and\ \bibinfo {author} {\bibfnamefont {E.}~\bibnamefont {G{\"o}{\u g}{\"u}{\c
  s}}},\ }\href {\doibase 10.1088/0004-637X/795/2/114} {\bibfield  {journal}
  {\bibinfo  {journal} {Astrophys. J.}\ }\textbf {\bibinfo {volume} {795}},\
  \bibinfo {pages} {114} (\bibinfo {year} {2014}{\natexlab{a}})}\BibitemShut
  {NoStop}%
\bibitem [{\citenamefont {Huppenkothen}\ \emph
  {et~al.}(2014{\natexlab{b}})\citenamefont {Huppenkothen} \emph
  {et~al.}}]{Huppenkothen:2014pba}%
  \BibitemOpen
  \bibfield  {author} {\bibinfo {author} {\bibfnamefont {D.}~\bibnamefont
  {Huppenkothen}} \emph {et~al.},\ }\href {\doibase
  10.1088/0004-637X/787/2/128} {\bibfield  {journal} {\bibinfo  {journal}
  {Astrophys. J.}\ }\textbf {\bibinfo {volume} {787}},\ \bibinfo {pages} {128}
  (\bibinfo {year} {2014}{\natexlab{b}})}\BibitemShut {NoStop}%
\bibitem [{\citenamefont {Samuelsson}\ and\ \citenamefont
  {Andersson}(2007)}]{Samuelsson:2006tt}%
  \BibitemOpen
  \bibfield  {author} {\bibinfo {author} {\bibfnamefont {L.}~\bibnamefont
  {Samuelsson}}\ and\ \bibinfo {author} {\bibfnamefont {N.}~\bibnamefont
  {Andersson}},\ }\href {\doibase 10.1111/j.1365-2966.2006.11147.x} {\bibfield
  {journal} {\bibinfo  {journal} {Mon. Not. R. Astron. Soc.}\ }\textbf
  {\bibinfo {volume} {374}},\ \bibinfo {pages} {256} (\bibinfo {year}
  {2007})}\BibitemShut {NoStop}%
\bibitem [{\citenamefont {Watts}\ and\ \citenamefont
  {Reddy}(2007)}]{Watts:2006hk}%
  \BibitemOpen
  \bibfield  {author} {\bibinfo {author} {\bibfnamefont {A.~L.}\ \bibnamefont
  {Watts}}\ and\ \bibinfo {author} {\bibfnamefont {S.}~\bibnamefont {Reddy}},\
  }\href {\doibase 10.1111/j.1745-3933.2007.00336.x} {\bibfield  {journal}
  {\bibinfo  {journal} {Mon. Not. R. Astron. Soc.}\ }\textbf {\bibinfo {volume}
  {379}},\ \bibinfo {pages} {L63} (\bibinfo {year} {2007})}\BibitemShut
  {NoStop}%
\bibitem [{\citenamefont {Sotani}\ \emph {et~al.}(2012)\citenamefont {Sotani},
  \citenamefont {Nakazato}, \citenamefont {Iida},\ and\ \citenamefont
  {Oyamatsu}}]{Sotani:2012qc}%
  \BibitemOpen
  \bibfield  {author} {\bibinfo {author} {\bibfnamefont {H.}~\bibnamefont
  {Sotani}}, \bibinfo {author} {\bibfnamefont {K.}~\bibnamefont {Nakazato}},
  \bibinfo {author} {\bibfnamefont {K.}~\bibnamefont {Iida}}, \ and\ \bibinfo
  {author} {\bibfnamefont {K.}~\bibnamefont {Oyamatsu}},\ }\href {\doibase
  10.1103/PhysRevLett.108.201101} {\bibfield  {journal} {\bibinfo  {journal}
  {Phys. Rev. Lett.}\ }\textbf {\bibinfo {volume} {108}},\ \bibinfo {pages}
  {201101} (\bibinfo {year} {2012})}\BibitemShut {NoStop}%
\bibitem [{\citenamefont {Sotani}\ \emph
  {et~al.}(2013{\natexlab{a}})\citenamefont {Sotani}, \citenamefont {Nakazato},
  \citenamefont {Iida},\ and\ \citenamefont {Oyamatsu}}]{Sotani:2013jya}%
  \BibitemOpen
  \bibfield  {author} {\bibinfo {author} {\bibfnamefont {H.}~\bibnamefont
  {Sotani}}, \bibinfo {author} {\bibfnamefont {K.}~\bibnamefont {Nakazato}},
  \bibinfo {author} {\bibfnamefont {K.}~\bibnamefont {Iida}}, \ and\ \bibinfo
  {author} {\bibfnamefont {K.}~\bibnamefont {Oyamatsu}},\ }\href {\doibase
  10.1093/mnras/stt1152} {\bibfield  {journal} {\bibinfo  {journal} {Mon. Not.
  R. Astron. Soc.}\ }\textbf {\bibinfo {volume} {434}},\ \bibinfo {pages}
  {2060} (\bibinfo {year} {2013}{\natexlab{a}})}\BibitemShut {NoStop}%
\bibitem [{\citenamefont {Watts}\ \emph {et~al.}(2016)\citenamefont {Watts}
  \emph {et~al.}}]{Watts:2016uzu}%
  \BibitemOpen
  \bibfield  {author} {\bibinfo {author} {\bibfnamefont {A.~L.}\ \bibnamefont
  {Watts}} \emph {et~al.},\ }\href {\doibase 10.1103/RevModPhys.88.021001}
  {\bibfield  {journal} {\bibinfo  {journal} {Rev. Mod. Phys.}\ }\textbf
  {\bibinfo {volume} {88}},\ \bibinfo {pages} {021001} (\bibinfo {year}
  {2016})}\BibitemShut {NoStop}%
\bibitem [{\citenamefont {Levin}(2006)}]{Levin:2006ck}%
  \BibitemOpen
  \bibfield  {author} {\bibinfo {author} {\bibfnamefont {Y.}~\bibnamefont
  {Levin}},\ }\href {\doibase 10.1111/j.1745-3933.2006.00155.x} {\bibfield
  {journal} {\bibinfo  {journal} {Mon. Not. Roy. Astron. Soc.}\ }\textbf
  {\bibinfo {volume} {368}},\ \bibinfo {pages} {L35} (\bibinfo {year}
  {2006})}\BibitemShut {NoStop}%
\bibitem [{\citenamefont {Levin}(2007)}]{Levin:2006qd}%
  \BibitemOpen
  \bibfield  {author} {\bibinfo {author} {\bibfnamefont {Y.}~\bibnamefont
  {Levin}},\ }\href {\doibase 10.1111/j.1365-2966.2007.11582.x} {\bibfield
  {journal} {\bibinfo  {journal} {Mon. Not. R. Astron. Soc.}\ }\textbf
  {\bibinfo {volume} {377}},\ \bibinfo {pages} {159} (\bibinfo {year}
  {2007})}\BibitemShut {NoStop}%
\bibitem [{\citenamefont {Sotani}\ \emph {et~al.}(2008)\citenamefont {Sotani},
  \citenamefont {Kokkotas},\ and\ \citenamefont {Stergioulas}}]{Sotani:2007pp}%
  \BibitemOpen
  \bibfield  {author} {\bibinfo {author} {\bibfnamefont {H.}~\bibnamefont
  {Sotani}}, \bibinfo {author} {\bibfnamefont {K.~D.}\ \bibnamefont
  {Kokkotas}}, \ and\ \bibinfo {author} {\bibfnamefont {N.}~\bibnamefont
  {Stergioulas}},\ }\href {\doibase 10.1111/j.1745-3933.2007.00420.x}
  {\bibfield  {journal} {\bibinfo  {journal} {Mon. Not. R. Astron. Soc.}\
  }\textbf {\bibinfo {volume} {385}},\ \bibinfo {pages} {L5} (\bibinfo {year}
  {2008})}\BibitemShut {NoStop}%
\bibitem [{\citenamefont {van Hoven}\ and\ \citenamefont
  {Levin}(2011)}]{vanHoven:2010gy}%
  \BibitemOpen
  \bibfield  {author} {\bibinfo {author} {\bibfnamefont {M.}~\bibnamefont {van
  Hoven}}\ and\ \bibinfo {author} {\bibfnamefont {Y.}~\bibnamefont {Levin}},\
  }\href {\doibase 10.1111/j.1365-2966.2010.17499.x} {\bibfield  {journal}
  {\bibinfo  {journal} {Mon. Not. R. Astron. Soc.}\ }\textbf {\bibinfo {volume}
  {410}},\ \bibinfo {pages} {1036} (\bibinfo {year} {2011})}\BibitemShut
  {NoStop}%
\bibitem [{\citenamefont {van Hoven}\ and\ \citenamefont
  {Levin}(2012)}]{vanHoven:2011it}%
  \BibitemOpen
  \bibfield  {author} {\bibinfo {author} {\bibfnamefont {M.}~\bibnamefont {van
  Hoven}}\ and\ \bibinfo {author} {\bibfnamefont {Y.}~\bibnamefont {Levin}},\
  }\href {\doibase 10.1111/j.1365-2966.2011.20177.x} {\bibfield  {journal}
  {\bibinfo  {journal} {Mon. Not. R. Astron. Soc.}\ }\textbf {\bibinfo {volume}
  {420}},\ \bibinfo {pages} {3035} (\bibinfo {year} {2012})}\BibitemShut
  {NoStop}%
\bibitem [{\citenamefont {{Asai}}\ and\ \citenamefont
  {{Lee}}(2014)}]{Asai:2014al}%
  \BibitemOpen
  \bibfield  {author} {\bibinfo {author} {\bibfnamefont {H.}~\bibnamefont
  {{Asai}}}\ and\ \bibinfo {author} {\bibfnamefont {U.}~\bibnamefont {{Lee}}},\
  }\href {\doibase 10.1088/0004-637X/790/1/66} {\bibfield  {journal} {\bibinfo
  {journal} {Astrophys. J.}\ }\textbf {\bibinfo {volume} {790}},\ \bibinfo
  {eid} {66} (\bibinfo {year} {2014})}\BibitemShut {NoStop}%
\bibitem [{\citenamefont {Asai}\ \emph {et~al.}(2016)\citenamefont {Asai},
  \citenamefont {Lee},\ and\ \citenamefont {Yoshida}}]{Asai:2015qha}%
  \BibitemOpen
  \bibfield  {author} {\bibinfo {author} {\bibfnamefont {H.}~\bibnamefont
  {Asai}}, \bibinfo {author} {\bibfnamefont {U.}~\bibnamefont {Lee}}, \ and\
  \bibinfo {author} {\bibfnamefont {S.}~\bibnamefont {Yoshida}},\ }\href
  {\doibase 10.1093/mnras/stv2368} {\bibfield  {journal} {\bibinfo  {journal}
  {Mon. Not. Roy. Astron. Soc.}\ }\textbf {\bibinfo {volume} {455}},\ \bibinfo
  {pages} {2228} (\bibinfo {year} {2016})}\BibitemShut {NoStop}%
\bibitem [{\citenamefont {Passamonti}\ and\ \citenamefont
  {Lander}(2013)}]{Passamonti:2012ma}%
  \BibitemOpen
  \bibfield  {author} {\bibinfo {author} {\bibfnamefont {A.}~\bibnamefont
  {Passamonti}}\ and\ \bibinfo {author} {\bibfnamefont {S.~K.}\ \bibnamefont
  {Lander}},\ }\href {\doibase 10.1093/mnras/sts372} {\bibfield  {journal}
  {\bibinfo  {journal} {Mon. Not. R. Astron. Soc.}\ }\textbf {\bibinfo {volume}
  {429}},\ \bibinfo {pages} {767} (\bibinfo {year} {2013})}\BibitemShut
  {NoStop}%
\bibitem [{\citenamefont {Sotani}\ \emph
  {et~al.}(2013{\natexlab{b}})\citenamefont {Sotani}, \citenamefont {Nakazato},
  \citenamefont {Iida},\ and\ \citenamefont {Oyamatsu}}]{Sotani:2012xd}%
  \BibitemOpen
  \bibfield  {author} {\bibinfo {author} {\bibfnamefont {H.}~\bibnamefont
  {Sotani}}, \bibinfo {author} {\bibfnamefont {K.}~\bibnamefont {Nakazato}},
  \bibinfo {author} {\bibfnamefont {K.}~\bibnamefont {Iida}}, \ and\ \bibinfo
  {author} {\bibfnamefont {K.}~\bibnamefont {Oyamatsu}},\ }\href {\doibase
  10.1093/mnrasl/sls006} {\bibfield  {journal} {\bibinfo  {journal} {Mon. Not.
  R. Astron. Soc.}\ }\textbf {\bibinfo {volume} {428}},\ \bibinfo {pages} {L21}
  (\bibinfo {year} {2013}{\natexlab{b}})}\BibitemShut {NoStop}%
\bibitem [{\citenamefont {Passamonti}\ and\ \citenamefont
  {Lander}(2014)}]{Passamonti:2013zra}%
  \BibitemOpen
  \bibfield  {author} {\bibinfo {author} {\bibfnamefont {A.}~\bibnamefont
  {Passamonti}}\ and\ \bibinfo {author} {\bibfnamefont {S.~K.}\ \bibnamefont
  {Lander}},\ }\href {\doibase 10.1093/mnras/stt2134} {\bibfield  {journal}
  {\bibinfo  {journal} {Mon. Not. R. Astron. Soc.}\ }\textbf {\bibinfo {volume}
  {438}},\ \bibinfo {pages} {156} (\bibinfo {year} {2014})}\BibitemShut
  {NoStop}%
\bibitem [{\citenamefont {Glampedakis}\ and\ \citenamefont
  {Jones}(2014)}]{Jones:2013yea}%
  \BibitemOpen
  \bibfield  {author} {\bibinfo {author} {\bibfnamefont {K.}~\bibnamefont
  {Glampedakis}}\ and\ \bibinfo {author} {\bibfnamefont {D.~I.}\ \bibnamefont
  {Jones}},\ }\href {\doibase 10.1093/mnras/stu017} {\bibfield  {journal}
  {\bibinfo  {journal} {Mon. Not. R. Astron. Soc.}\ }\textbf {\bibinfo {volume}
  {439}},\ \bibinfo {pages} {1522} (\bibinfo {year} {2014})}\BibitemShut
  {NoStop}%
\bibitem [{\citenamefont {Gabler}\ \emph {et~al.}(2013)\citenamefont {Gabler},
  \citenamefont {Cerd{\'a}-Dur{\'a}n}, \citenamefont {Stergioulas},
  \citenamefont {Font},\ and\ \citenamefont {M{\"u}ller}}]{Gabler:2013ova}%
  \BibitemOpen
  \bibfield  {author} {\bibinfo {author} {\bibfnamefont {M.}~\bibnamefont
  {Gabler}}, \bibinfo {author} {\bibfnamefont {P.}~\bibnamefont
  {Cerd{\'a}-Dur{\'a}n}}, \bibinfo {author} {\bibfnamefont {N.}~\bibnamefont
  {Stergioulas}}, \bibinfo {author} {\bibfnamefont {J.~A.}\ \bibnamefont
  {Font}}, \ and\ \bibinfo {author} {\bibfnamefont {E.}~\bibnamefont
  {M{\"u}ller}},\ }\href {\doibase 10.1103/PhysRevLett.111.211102} {\bibfield
  {journal} {\bibinfo  {journal} {Phys. Rev. Lett.}\ }\textbf {\bibinfo
  {volume} {111}},\ \bibinfo {pages} {211102} (\bibinfo {year}
  {2013})}\BibitemShut {NoStop}%
\bibitem [{\citenamefont {Gabler}\ \emph {et~al.}(2014)\citenamefont {Gabler},
  \citenamefont {Cerd{\'a}-Dur{\'a}n}, \citenamefont {Stergioulas},
  \citenamefont {Font},\ and\ \citenamefont {M{\"u}ller}}]{Gabler:2014bza}%
  \BibitemOpen
  \bibfield  {author} {\bibinfo {author} {\bibfnamefont {M.}~\bibnamefont
  {Gabler}}, \bibinfo {author} {\bibfnamefont {P.}~\bibnamefont
  {Cerd{\'a}-Dur{\'a}n}}, \bibinfo {author} {\bibfnamefont {N.}~\bibnamefont
  {Stergioulas}}, \bibinfo {author} {\bibfnamefont {J.~A.}\ \bibnamefont
  {Font}}, \ and\ \bibinfo {author} {\bibfnamefont {E.}~\bibnamefont
  {M{\"u}ller}},\ }\href {\doibase 10.1093/mnras/stu1263} {\bibfield  {journal}
  {\bibinfo  {journal} {Mon. Not. R. Astron. Soc.}\ }\textbf {\bibinfo {volume}
  {443}},\ \bibinfo {pages} {1416} (\bibinfo {year} {2014})}\BibitemShut
  {NoStop}%
\bibitem [{\citenamefont {Passamonti}\ and\ \citenamefont
  {Pons}(2016)}]{Passamonti:2016jfo}%
  \BibitemOpen
  \bibfield  {author} {\bibinfo {author} {\bibfnamefont {A.}~\bibnamefont
  {Passamonti}}\ and\ \bibinfo {author} {\bibfnamefont {J.~A.}\ \bibnamefont
  {Pons}},\ }\href {\doibase 10.1093/mnras/stw1880} {\bibfield  {journal}
  {\bibinfo  {journal} {Mon. Not. R. Astron. Soc.}\ }\textbf {\bibinfo {volume}
  {463}},\ \bibinfo {pages} {1173} (\bibinfo {year} {2016})}\BibitemShut
  {NoStop}%
\bibitem [{\citenamefont {Colaiuda}\ and\ \citenamefont
  {Kokkotas}(2012)}]{Colaiuda:2011aa}%
  \BibitemOpen
  \bibfield  {author} {\bibinfo {author} {\bibfnamefont {A.}~\bibnamefont
  {Colaiuda}}\ and\ \bibinfo {author} {\bibfnamefont {K.~D.}\ \bibnamefont
  {Kokkotas}},\ }\href {\doibase 10.1111/j.1365-2966.2012.20919.x} {\bibfield
  {journal} {\bibinfo  {journal} {Mon. Not. R. Astron. Soc.}\ }\textbf
  {\bibinfo {volume} {423}},\ \bibinfo {pages} {811} (\bibinfo {year}
  {2012})}\BibitemShut {NoStop}%
\bibitem [{\citenamefont {Sotani}(2015)}]{Sotani:2015rla}%
  \BibitemOpen
  \bibfield  {author} {\bibinfo {author} {\bibfnamefont {H.}~\bibnamefont
  {Sotani}},\ }\href {\doibase 10.1103/PhysRevD.92.104024} {\bibfield
  {journal} {\bibinfo  {journal} {Phys. Rev.}\ }\textbf {\bibinfo {volume}
  {D92}},\ \bibinfo {pages} {104024} (\bibinfo {year} {2015})}\BibitemShut
  {NoStop}%
\bibitem [{\citenamefont {Link}\ and\ \citenamefont {van
  Eysden}(2015)}]{Link:2015cza}%
  \BibitemOpen
  \bibfield  {author} {\bibinfo {author} {\bibfnamefont {B.}~\bibnamefont
  {Link}}\ and\ \bibinfo {author} {\bibfnamefont {C.~A.}\ \bibnamefont {van
  Eysden}},\ }\href@noop {} {\  (\bibinfo {year} {2015})}\BibitemShut {NoStop}%
\bibitem [{\citenamefont {Link}\ and\ \citenamefont {van
  Eysden}(2016)}]{Link:2016dxt}%
  \BibitemOpen
  \bibfield  {author} {\bibinfo {author} {\bibfnamefont {B.}~\bibnamefont
  {Link}}\ and\ \bibinfo {author} {\bibfnamefont {C.~A.}\ \bibnamefont {van
  Eysden}},\ }\href {\doibase 10.3847/2041-8205/823/1/L1,
  10.3847/0067-0049/224/1/6} {\bibfield  {journal} {\bibinfo  {journal}
  {Astrophys. J.}\ }\textbf {\bibinfo {volume} {823}},\ \bibinfo {pages} {L1}
  (\bibinfo {year} {2016})},\ \bibinfo {note} {[Erratum: Astrophys. J.
  Suppl.224,no.1,6(2016)]}\BibitemShut {NoStop}%
\bibitem [{\citenamefont {Huppenkothen}\ \emph
  {et~al.}(2014{\natexlab{c}})\citenamefont {Huppenkothen}, \citenamefont
  {Watts},\ and\ \citenamefont {Levin}}]{Huppenkothen:2014ufa}%
  \BibitemOpen
  \bibfield  {author} {\bibinfo {author} {\bibfnamefont {D.}~\bibnamefont
  {Huppenkothen}}, \bibinfo {author} {\bibfnamefont {A.~L.}\ \bibnamefont
  {Watts}}, \ and\ \bibinfo {author} {\bibfnamefont {Y.}~\bibnamefont
  {Levin}},\ }\href {\doibase 10.1088/0004-637X/793/2/129} {\bibfield
  {journal} {\bibinfo  {journal} {Astrophys. J.}\ }\textbf {\bibinfo {volume}
  {793}},\ \bibinfo {pages} {129} (\bibinfo {year}
  {2014}{\natexlab{c}})}\BibitemShut {NoStop}%
\bibitem [{\citenamefont {Arzoumanian}\ \emph {et~al.}(2014)\citenamefont
  {Arzoumanian}, \citenamefont {Gendreau}, \citenamefont {Baker}, \citenamefont
  {Cazeau}, \citenamefont {Hestnes}, \citenamefont {Kellogg}, \citenamefont
  {Kenyon}, \citenamefont {Kozon}, \citenamefont {Liu}, \citenamefont
  {Manthripragada} \emph {et~al.}}]{arzoumanian2014neutron}%
  \BibitemOpen
  \bibfield  {author} {\bibinfo {author} {\bibfnamefont {Z.}~\bibnamefont
  {Arzoumanian}}, \bibinfo {author} {\bibfnamefont {K.}~\bibnamefont
  {Gendreau}}, \bibinfo {author} {\bibfnamefont {C.}~\bibnamefont {Baker}},
  \bibinfo {author} {\bibfnamefont {T.}~\bibnamefont {Cazeau}}, \bibinfo
  {author} {\bibfnamefont {P.}~\bibnamefont {Hestnes}}, \bibinfo {author}
  {\bibfnamefont {J.}~\bibnamefont {Kellogg}}, \bibinfo {author} {\bibfnamefont
  {S.}~\bibnamefont {Kenyon}}, \bibinfo {author} {\bibfnamefont
  {R.}~\bibnamefont {Kozon}}, \bibinfo {author} {\bibfnamefont {K.-C.}\
  \bibnamefont {Liu}}, \bibinfo {author} {\bibfnamefont {S.}~\bibnamefont
  {Manthripragada}},  \emph {et~al.},\ }in\ \href@noop {} {\emph {\bibinfo
  {booktitle} {SPIE Astronomical Telescopes+ Instrumentation}}}\ (\bibinfo
  {organization} {International Society for Optics and Photonics},\ \bibinfo
  {year} {2014})\ pp.\ \bibinfo {pages} {914420--914420}\BibitemShut {NoStop}%
\bibitem [{\citenamefont {Feroci}\ \emph {et~al.}(2012)\citenamefont {Feroci},
  \citenamefont {Stella}, \citenamefont {Van~der Klis}, \citenamefont
  {Courvoisier}, \citenamefont {Hernanz}, \citenamefont {Hudec}, \citenamefont
  {Santangelo}, \citenamefont {Walton}, \citenamefont {Zdziarski},
  \citenamefont {Barret} \emph {et~al.}}]{feroci2012large}%
  \BibitemOpen
  \bibfield  {author} {\bibinfo {author} {\bibfnamefont {M.}~\bibnamefont
  {Feroci}}, \bibinfo {author} {\bibfnamefont {L.}~\bibnamefont {Stella}},
  \bibinfo {author} {\bibfnamefont {M.}~\bibnamefont {Van~der Klis}}, \bibinfo
  {author} {\bibfnamefont {T.-L.}\ \bibnamefont {Courvoisier}}, \bibinfo
  {author} {\bibfnamefont {M.}~\bibnamefont {Hernanz}}, \bibinfo {author}
  {\bibfnamefont {R.}~\bibnamefont {Hudec}}, \bibinfo {author} {\bibfnamefont
  {A.}~\bibnamefont {Santangelo}}, \bibinfo {author} {\bibfnamefont
  {D.}~\bibnamefont {Walton}}, \bibinfo {author} {\bibfnamefont
  {A.}~\bibnamefont {Zdziarski}}, \bibinfo {author} {\bibfnamefont
  {D.}~\bibnamefont {Barret}},  \emph {et~al.},\ }\href@noop {} {\bibfield
  {journal} {\bibinfo  {journal} {Exper. Astron.}\ }\textbf {\bibinfo {volume}
  {34}},\ \bibinfo {pages} {415} (\bibinfo {year} {2012})}\BibitemShut
  {NoStop}%
\bibitem [{eXT()}]{eXTP}%
  \BibitemOpen
  \href@noop {} {}\bibinfo {howpublished} {{\tt
  http://www.isdc.unige.ch/extp/}{http://www.isdc.unige.ch/extp/}}\BibitemShut
  {NoStop}%
\bibitem [{\citenamefont {Zink}\ \emph {et~al.}(2012)\citenamefont {Zink},
  \citenamefont {Lasky},\ and\ \citenamefont {Kokkotas}}]{Zink:2011kq}%
  \BibitemOpen
  \bibfield  {author} {\bibinfo {author} {\bibfnamefont {B.}~\bibnamefont
  {Zink}}, \bibinfo {author} {\bibfnamefont {P.~D.}\ \bibnamefont {Lasky}}, \
  and\ \bibinfo {author} {\bibfnamefont {K.~D.}\ \bibnamefont {Kokkotas}},\
  }\href {\doibase 10.1103/PhysRevD.85.024030} {\bibfield  {journal} {\bibinfo
  {journal} {Phys. Rev.}\ }\textbf {\bibinfo {volume} {D85}},\ \bibinfo {pages}
  {024030} (\bibinfo {year} {2012})}\BibitemShut {NoStop}%
\bibitem [{\citenamefont {Thompson}\ and\ \citenamefont
  {Duncan}(1995)}]{Thompson:1995gw}%
  \BibitemOpen
  \bibfield  {author} {\bibinfo {author} {\bibfnamefont {C.}~\bibnamefont
  {Thompson}}\ and\ \bibinfo {author} {\bibfnamefont {R.~C.}\ \bibnamefont
  {Duncan}},\ }\href@noop {} {\bibfield  {journal} {\bibinfo  {journal} {Mon.
  Not. Roy. Astron. Soc.}\ }\textbf {\bibinfo {volume} {275}},\ \bibinfo
  {pages} {255} (\bibinfo {year} {1995})}\BibitemShut {NoStop}%
\bibitem [{\citenamefont {Thompson}\ and\ \citenamefont
  {Duncan}(2001)}]{Thompson:2001ie}%
  \BibitemOpen
  \bibfield  {author} {\bibinfo {author} {\bibfnamefont {C.}~\bibnamefont
  {Thompson}}\ and\ \bibinfo {author} {\bibfnamefont {R.~C.}\ \bibnamefont
  {Duncan}},\ }\href {\doibase 10.1086/323256} {\bibfield  {journal} {\bibinfo
  {journal} {Astrophys. J.}\ }\textbf {\bibinfo {volume} {561}},\ \bibinfo
  {pages} {980} (\bibinfo {year} {2001})}\BibitemShut {NoStop}%
\bibitem [{\citenamefont {Lyutikov}(2006)}]{Lyutikov:2005un}%
  \BibitemOpen
  \bibfield  {author} {\bibinfo {author} {\bibfnamefont {M.}~\bibnamefont
  {Lyutikov}},\ }\href {\doibase 10.1111/j.1365-2966.2006.10069.x} {\bibfield
  {journal} {\bibinfo  {journal} {Mon. Not. Roy. Astron. Soc.}\ }\textbf
  {\bibinfo {volume} {367}},\ \bibinfo {pages} {1594} (\bibinfo {year}
  {2006})}\BibitemShut {NoStop}%
\bibitem [{\citenamefont {Gill}\ and\ \citenamefont
  {Heyl}(2010)}]{Gill:2010qx}%
  \BibitemOpen
  \bibfield  {author} {\bibinfo {author} {\bibfnamefont {R.}~\bibnamefont
  {Gill}}\ and\ \bibinfo {author} {\bibfnamefont {J.~S.}\ \bibnamefont
  {Heyl}},\ }\href {\doibase 10.1111/j.1365-2966.2010.17038.x} {\bibfield
  {journal} {\bibinfo  {journal} {Mon. Not. Roy. Astron. Soc.}\ }\textbf
  {\bibinfo {volume} {407}},\ \bibinfo {pages} {1926} (\bibinfo {year}
  {2010})}\BibitemShut {NoStop}%
\bibitem [{\citenamefont {Corsi}\ and\ \citenamefont
  {Owen}(2011)}]{Corsi:2011zi}%
  \BibitemOpen
  \bibfield  {author} {\bibinfo {author} {\bibfnamefont {A.}~\bibnamefont
  {Corsi}}\ and\ \bibinfo {author} {\bibfnamefont {B.~J.}\ \bibnamefont
  {Owen}},\ }\href {\doibase 10.1103/PhysRevD.83.104014} {\bibfield  {journal}
  {\bibinfo  {journal} {Phys. Rev.}\ }\textbf {\bibinfo {volume} {D83}},\
  \bibinfo {pages} {104014} (\bibinfo {year} {2011})}\BibitemShut {NoStop}%
\bibitem [{\citenamefont {Levin}\ and\ \citenamefont {van
  Hoven}(2011)}]{Levin:2011vh}%
  \BibitemOpen
  \bibfield  {author} {\bibinfo {author} {\bibfnamefont {Y.}~\bibnamefont
  {Levin}}\ and\ \bibinfo {author} {\bibfnamefont {M.}~\bibnamefont {van
  Hoven}},\ }\href {\doibase 10.1111/j.1365-2966.2011.19515.x} {\bibfield
  {journal} {\bibinfo  {journal} {Mon. Not. R. Astron. Soc.}\ }\textbf
  {\bibinfo {volume} {418}},\ \bibinfo {pages} {659} (\bibinfo {year}
  {2011})}\BibitemShut {NoStop}%
\bibitem [{\citenamefont {Ciolfi}\ and\ \citenamefont
  {Rezzolla}(2012)}]{Ciolfi:2012en}%
  \BibitemOpen
  \bibfield  {author} {\bibinfo {author} {\bibfnamefont {R.}~\bibnamefont
  {Ciolfi}}\ and\ \bibinfo {author} {\bibfnamefont {L.}~\bibnamefont
  {Rezzolla}},\ }\href {\doibase 10.1088/0004-637X/760/1/1} {\bibfield
  {journal} {\bibinfo  {journal} {Astrophys. J.}\ }\textbf {\bibinfo {volume}
  {760}},\ \bibinfo {pages} {1} (\bibinfo {year} {2012})}\BibitemShut {NoStop}%
\bibitem [{\citenamefont {Zimmermann}\ and\ \citenamefont
  {Szedenits}(1979)}]{Zimmermann:1979ip}%
  \BibitemOpen
  \bibfield  {author} {\bibinfo {author} {\bibfnamefont {M.}~\bibnamefont
  {Zimmermann}}\ and\ \bibinfo {author} {\bibfnamefont {E.}~\bibnamefont
  {Szedenits}},\ }\href {\doibase 10.1103/PhysRevD.20.351} {\bibfield
  {journal} {\bibinfo  {journal} {Phys. Rev.}\ }\textbf {\bibinfo {volume}
  {D20}},\ \bibinfo {pages} {351} (\bibinfo {year} {1979})}\BibitemShut
  {NoStop}%
\bibitem [{\citenamefont {Bonazzola}\ and\ \citenamefont
  {Gourgoulhon}(1996)}]{Bonazzola:1995rb}%
  \BibitemOpen
  \bibfield  {author} {\bibinfo {author} {\bibfnamefont {S.}~\bibnamefont
  {Bonazzola}}\ and\ \bibinfo {author} {\bibfnamefont {E.}~\bibnamefont
  {Gourgoulhon}},\ }\href@noop {} {\bibfield  {journal} {\bibinfo  {journal}
  {Astron. Astrophys.}\ }\textbf {\bibinfo {volume} {312}},\ \bibinfo {pages}
  {675} (\bibinfo {year} {1996})}\BibitemShut {NoStop}%
\bibitem [{\citenamefont {Haskell}\ \emph {et~al.}(2008)\citenamefont
  {Haskell}, \citenamefont {Samuelsson}, \citenamefont {Glampedakis},\ and\
  \citenamefont {Andersson}}]{Haskell:2007bh}%
  \BibitemOpen
  \bibfield  {author} {\bibinfo {author} {\bibfnamefont {B.}~\bibnamefont
  {Haskell}}, \bibinfo {author} {\bibfnamefont {L.}~\bibnamefont {Samuelsson}},
  \bibinfo {author} {\bibfnamefont {K.}~\bibnamefont {Glampedakis}}, \ and\
  \bibinfo {author} {\bibfnamefont {N.}~\bibnamefont {Andersson}},\ }\href
  {\doibase 10.1111/j.1365-2966.2008.12861.x} {\bibfield  {journal} {\bibinfo
  {journal} {Mon. Not. R. Astron. Soc.}\ }\textbf {\bibinfo {volume} {385}},\
  \bibinfo {pages} {531} (\bibinfo {year} {2008})}\BibitemShut {NoStop}%
\bibitem [{\citenamefont {Ciolfi}\ \emph {et~al.}(2010)\citenamefont {Ciolfi},
  \citenamefont {Ferrari},\ and\ \citenamefont {Gualtieri}}]{Ciolfi:2010td}%
  \BibitemOpen
  \bibfield  {author} {\bibinfo {author} {\bibfnamefont {R.}~\bibnamefont
  {Ciolfi}}, \bibinfo {author} {\bibfnamefont {V.}~\bibnamefont {Ferrari}}, \
  and\ \bibinfo {author} {\bibfnamefont {L.}~\bibnamefont {Gualtieri}},\ }\href
  {\doibase 10.1111/j.1365-2966.2010.16847.x} {\bibfield  {journal} {\bibinfo
  {journal} {Mon. Not. R. Astron. Soc.}\ }\textbf {\bibinfo {volume} {406}},\
  \bibinfo {pages} {2540} (\bibinfo {year} {2010})}\BibitemShut {NoStop}%
\bibitem [{\citenamefont {Bildsten}(1998{\natexlab{b}})}]{Bildsten:1998ey}%
  \BibitemOpen
  \bibfield  {author} {\bibinfo {author} {\bibfnamefont {L.}~\bibnamefont
  {Bildsten}},\ }\href {\doibase 10.1086/311440} {\bibfield  {journal}
  {\bibinfo  {journal} {Astrophys. J.}\ }\textbf {\bibinfo {volume} {501}},\
  \bibinfo {pages} {L89} (\bibinfo {year} {1998}{\natexlab{b}})}\BibitemShut
  {NoStop}%
\bibitem [{\citenamefont {Ushomirsky}\ \emph {et~al.}(2000)\citenamefont
  {Ushomirsky}, \citenamefont {Cutler},\ and\ \citenamefont
  {Bildsten}}]{Ushomirsky:2000ax}%
  \BibitemOpen
  \bibfield  {author} {\bibinfo {author} {\bibfnamefont {G.}~\bibnamefont
  {Ushomirsky}}, \bibinfo {author} {\bibfnamefont {C.}~\bibnamefont {Cutler}},
  \ and\ \bibinfo {author} {\bibfnamefont {L.}~\bibnamefont {Bildsten}},\
  }\href {\doibase 10.1046/j.1365-8711.2000.03938.x} {\bibfield  {journal}
  {\bibinfo  {journal} {Mon. Not. R. Astron. Soc.}\ }\textbf {\bibinfo {volume}
  {319}},\ \bibinfo {pages} {902} (\bibinfo {year} {2000})}\BibitemShut
  {NoStop}%
\bibitem [{\citenamefont {Aasi}\ \emph {et~al.}(2014)\citenamefont {Aasi} \emph
  {et~al.}}]{Aasi:2013sia}%
  \BibitemOpen
  \bibfield  {author} {\bibinfo {author} {\bibfnamefont {J.}~\bibnamefont
  {Aasi}} \emph {et~al.} (\bibinfo {collaboration} {LIGO Scientific}),\ }\href
  {\doibase 10.1088/0004-637X/785/2/119} {\bibfield  {journal} {\bibinfo
  {journal} {Astrophys. J.}\ }\textbf {\bibinfo {volume} {785}},\ \bibinfo
  {pages} {119} (\bibinfo {year} {2014})}\BibitemShut {NoStop}%
\bibitem [{\citenamefont {{Chandrasekhar}}\ and\ \citenamefont
  {{Fermi}}(1953)}]{Chandrasekhar:1953ef}%
  \BibitemOpen
  \bibfield  {author} {\bibinfo {author} {\bibfnamefont {S.}~\bibnamefont
  {{Chandrasekhar}}}\ and\ \bibinfo {author} {\bibfnamefont {E.}~\bibnamefont
  {{Fermi}}},\ }\href {\doibase 10.1086/145732} {\bibfield  {journal} {\bibinfo
   {journal} {Astrophys. J.}\ }\textbf {\bibinfo {volume} {118}},\ \bibinfo
  {pages} {116} (\bibinfo {year} {1953})}\BibitemShut {NoStop}%
\bibitem [{\citenamefont {{Ferraro}}(1954)}]{Ferraro:1954ff}%
  \BibitemOpen
  \bibfield  {author} {\bibinfo {author} {\bibfnamefont {V.~C.~A.}\
  \bibnamefont {{Ferraro}}},\ }\href {\doibase 10.1086/145838} {\bibfield
  {journal} {\bibinfo  {journal} {Astrophys. J.}\ }\textbf {\bibinfo {volume}
  {119}},\ \bibinfo {pages} {407} (\bibinfo {year} {1954})}\BibitemShut
  {NoStop}%
\bibitem [{\citenamefont {{Wentzel}}(1960)}]{Wentzel:1960ww}%
  \BibitemOpen
  \bibfield  {author} {\bibinfo {author} {\bibfnamefont {D.~G.}\ \bibnamefont
  {{Wentzel}}},\ }\href {\doibase 10.1086/190055} {\bibfield  {journal}
  {\bibinfo  {journal} {Astrophys. J. Suppl.}\ }\textbf {\bibinfo {volume}
  {5}},\ \bibinfo {pages} {187} (\bibinfo {year} {1960})}\BibitemShut {NoStop}%
\bibitem [{\citenamefont {{Ostriker}}\ and\ \citenamefont
  {{Gunn}}(1969)}]{Ostriker:1969og}%
  \BibitemOpen
  \bibfield  {author} {\bibinfo {author} {\bibfnamefont {J.~P.}\ \bibnamefont
  {{Ostriker}}}\ and\ \bibinfo {author} {\bibfnamefont {J.~E.}\ \bibnamefont
  {{Gunn}}},\ }\href {\doibase 10.1086/150160} {\bibfield  {journal} {\bibinfo
  {journal} {Astrophys. J.}\ }\textbf {\bibinfo {volume} {157}},\ \bibinfo
  {pages} {1395} (\bibinfo {year} {1969})}\BibitemShut {NoStop}%
\bibitem [{\citenamefont {Cutler}(2002)}]{Cutler:2002nw}%
  \BibitemOpen
  \bibfield  {author} {\bibinfo {author} {\bibfnamefont {C.}~\bibnamefont
  {Cutler}},\ }\href {\doibase 10.1103/PhysRevD.66.084025} {\bibfield
  {journal} {\bibinfo  {journal} {Phys. Rev.}\ }\textbf {\bibinfo {volume}
  {D66}},\ \bibinfo {pages} {084025} (\bibinfo {year} {2002})}\BibitemShut
  {NoStop}%
\bibitem [{\citenamefont {Glampedakis}\ \emph
  {et~al.}(2011{\natexlab{b}})\citenamefont {Glampedakis}, \citenamefont
  {Andersson},\ and\ \citenamefont {Samuelsson}}]{Glampedakis:2010sk}%
  \BibitemOpen
  \bibfield  {author} {\bibinfo {author} {\bibfnamefont {K.}~\bibnamefont
  {Glampedakis}}, \bibinfo {author} {\bibfnamefont {N.}~\bibnamefont
  {Andersson}}, \ and\ \bibinfo {author} {\bibfnamefont {L.}~\bibnamefont
  {Samuelsson}},\ }\href {\doibase 10.1111/j.1365-2966.2010.17484.x} {\bibfield
   {journal} {\bibinfo  {journal} {Mon. Not. Roy. Astron. Soc.}\ }\textbf
  {\bibinfo {volume} {410}},\ \bibinfo {pages} {805} (\bibinfo {year}
  {2011}{\natexlab{b}})}\BibitemShut {NoStop}%
\bibitem [{\citenamefont {{Jones}}(1975)}]{Jones:1975jj}%
  \BibitemOpen
  \bibfield  {author} {\bibinfo {author} {\bibfnamefont {P.~B.}\ \bibnamefont
  {{Jones}}},\ }\href {\doibase 10.1007/BF00646019} {\bibfield  {journal}
  {\bibinfo  {journal} {Astrophys. \& Space Sc.}\ }\textbf {\bibinfo {volume}
  {33}},\ \bibinfo {pages} {215} (\bibinfo {year} {1975})}\BibitemShut
  {NoStop}%
\bibitem [{\citenamefont {{Prendergast}}(1956)}]{Prendergast:1956kh}%
  \BibitemOpen
  \bibfield  {author} {\bibinfo {author} {\bibfnamefont {K.~H.}\ \bibnamefont
  {{Prendergast}}},\ }\href {\doibase 10.1086/146186} {\bibfield  {journal}
  {\bibinfo  {journal} {Astrophys. J.}\ }\textbf {\bibinfo {volume} {123}},\
  \bibinfo {pages} {498} (\bibinfo {year} {1956})}\BibitemShut {NoStop}%
\bibitem [{\citenamefont {{Tayler}}(1973)}]{Tayler:1973ty}%
  \BibitemOpen
  \bibfield  {author} {\bibinfo {author} {\bibfnamefont {R.~J.}\ \bibnamefont
  {{Tayler}}},\ }\href {\doibase 10.1093/mnras/161.4.365} {\bibfield  {journal}
  {\bibinfo  {journal} {Mon. Not. R. Astron. Soc.}\ }\textbf {\bibinfo {volume}
  {161}},\ \bibinfo {pages} {365} (\bibinfo {year} {1973})}\BibitemShut
  {NoStop}%
\bibitem [{\citenamefont {Braithwaite}(2006)}]{Braithwaite:2005su}%
  \BibitemOpen
  \bibfield  {author} {\bibinfo {author} {\bibfnamefont {J.}~\bibnamefont
  {Braithwaite}},\ }\href {\doibase 10.1051/0004-6361:20041282} {\bibfield
  {journal} {\bibinfo  {journal} {Astron. Astrophys.}\ }\textbf {\bibinfo
  {volume} {453}},\ \bibinfo {pages} {687} (\bibinfo {year}
  {2006})}\BibitemShut {NoStop}%
\bibitem [{\citenamefont {Braithwaite}(2007)}]{Braithwaite:2007fy}%
  \BibitemOpen
  \bibfield  {author} {\bibinfo {author} {\bibfnamefont {J.}~\bibnamefont
  {Braithwaite}},\ }\href {\doibase 10.1051/0004-6361:20065903} {\bibfield
  {journal} {\bibinfo  {journal} {Astron. Astrophys.}\ }\textbf {\bibinfo
  {volume} {469}},\ \bibinfo {pages} {275} (\bibinfo {year}
  {2007})}\BibitemShut {NoStop}%
\bibitem [{\citenamefont {Akgun}\ and\ \citenamefont
  {Wasserman}(2008)}]{Akgun:2007ph}%
  \BibitemOpen
  \bibfield  {author} {\bibinfo {author} {\bibfnamefont {T.}~\bibnamefont
  {Akgun}}\ and\ \bibinfo {author} {\bibfnamefont {I.}~\bibnamefont
  {Wasserman}},\ }\href {\doibase 10.1111/j.1365-2966.2007.12660.x} {\bibfield
  {journal} {\bibinfo  {journal} {Mon. Not. Roy. Astron. Soc.}\ }\textbf
  {\bibinfo {volume} {383}},\ \bibinfo {pages} {1551} (\bibinfo {year}
  {2008})}\BibitemShut {NoStop}%
\bibitem [{\citenamefont {Braithwaite}\ and\ \citenamefont
  {Spruit}(2004)}]{Braithwaite:2005ps}%
  \BibitemOpen
  \bibfield  {author} {\bibinfo {author} {\bibfnamefont {J.}~\bibnamefont
  {Braithwaite}}\ and\ \bibinfo {author} {\bibfnamefont {H.~C.}\ \bibnamefont
  {Spruit}},\ }\href {\doibase 10.1038/nature02934} {\bibfield  {journal}
  {\bibinfo  {journal} {Nature}\ }\textbf {\bibinfo {volume} {431}},\ \bibinfo
  {pages} {819} (\bibinfo {year} {2004})}\BibitemShut {NoStop}%
\bibitem [{\citenamefont {Braithwaite}\ and\ \citenamefont
  {Nordlund}(2006)}]{Braithwaite:2005xi}%
  \BibitemOpen
  \bibfield  {author} {\bibinfo {author} {\bibfnamefont {J.}~\bibnamefont
  {Braithwaite}}\ and\ \bibinfo {author} {\bibfnamefont {A.}~\bibnamefont
  {Nordlund}},\ }\href {\doibase 10.1051/0004-6361:20041980} {\bibfield
  {journal} {\bibinfo  {journal} {Astron. Astrophys.}\ }\textbf {\bibinfo
  {volume} {450}},\ \bibinfo {pages} {1077} (\bibinfo {year}
  {2006})}\BibitemShut {NoStop}%
\bibitem [{\citenamefont {{Lander}}\ and\ \citenamefont
  {{Jones}}(2009)}]{Lander:2010lj}%
  \BibitemOpen
  \bibfield  {author} {\bibinfo {author} {\bibfnamefont {S.~K.}\ \bibnamefont
  {{Lander}}}\ and\ \bibinfo {author} {\bibfnamefont {D.~I.}\ \bibnamefont
  {{Jones}}},\ }\href {\doibase 10.1111/j.1365-2966.2009.14667.x} {\bibfield
  {journal} {\bibinfo  {journal} {Mon. Not. R. Astron. Soc.}\ }\textbf
  {\bibinfo {volume} {395}},\ \bibinfo {pages} {2162} (\bibinfo {year}
  {2009})}\BibitemShut {NoStop}%
\bibitem [{\citenamefont {Pili}\ \emph {et~al.}(2014)\citenamefont {Pili},
  \citenamefont {Bucciantini},\ and\ \citenamefont {Del~Zanna}}]{Pili:2014npa}%
  \BibitemOpen
  \bibfield  {author} {\bibinfo {author} {\bibfnamefont {A.~G.}\ \bibnamefont
  {Pili}}, \bibinfo {author} {\bibfnamefont {N.}~\bibnamefont {Bucciantini}}, \
  and\ \bibinfo {author} {\bibfnamefont {L.}~\bibnamefont {Del~Zanna}},\ }\href
  {\doibase 10.1093/mnras/stu215} {\bibfield  {journal} {\bibinfo  {journal}
  {Mon. Not. Roy. Astron. Soc.}\ }\textbf {\bibinfo {volume} {439}},\ \bibinfo
  {pages} {3541} (\bibinfo {year} {2014})}\BibitemShut {NoStop}%
\bibitem [{\citenamefont {Uryu}\ \emph {et~al.}(2014)\citenamefont {Uryu},
  \citenamefont {Gourgoulhon}, \citenamefont {Markakis}, \citenamefont
  {Fujisawa}, \citenamefont {Tsokaros},\ and\ \citenamefont
  {Eriguchi}}]{Uryu:2014tda}%
  \BibitemOpen
  \bibfield  {author} {\bibinfo {author} {\bibfnamefont {K.}~\bibnamefont
  {Uryu}}, \bibinfo {author} {\bibfnamefont {E.}~\bibnamefont {Gourgoulhon}},
  \bibinfo {author} {\bibfnamefont {C.}~\bibnamefont {Markakis}}, \bibinfo
  {author} {\bibfnamefont {K.}~\bibnamefont {Fujisawa}}, \bibinfo {author}
  {\bibfnamefont {A.}~\bibnamefont {Tsokaros}}, \ and\ \bibinfo {author}
  {\bibfnamefont {Y.}~\bibnamefont {Eriguchi}},\ }\href {\doibase
  10.1103/PhysRevD.90.101501} {\bibfield  {journal} {\bibinfo  {journal} {Phys.
  Rev.}\ }\textbf {\bibinfo {volume} {D90}},\ \bibinfo {pages} {101501}
  (\bibinfo {year} {2014})}\BibitemShut {NoStop}%
\bibitem [{\citenamefont {Ciolfi}\ and\ \citenamefont
  {Rezzolla}(2013)}]{Ciolfi:2013dta}%
  \BibitemOpen
  \bibfield  {author} {\bibinfo {author} {\bibfnamefont {R.}~\bibnamefont
  {Ciolfi}}\ and\ \bibinfo {author} {\bibfnamefont {L.}~\bibnamefont
  {Rezzolla}},\ }\href {\doibase 10.1093/mnrasl/slt092} {\bibfield  {journal}
  {\bibinfo  {journal} {Mon. Not. R. Astron. Soc.}\ }\textbf {\bibinfo {volume}
  {435}},\ \bibinfo {pages} {L43} (\bibinfo {year} {2013})}\BibitemShut
  {NoStop}%
\bibitem [{\citenamefont {Mastrano}\ \emph {et~al.}(2011)\citenamefont
  {Mastrano}, \citenamefont {Melatos}, \citenamefont {Reisenegger},\ and\
  \citenamefont {Akgun}}]{Mastrano:2011tf}%
  \BibitemOpen
  \bibfield  {author} {\bibinfo {author} {\bibfnamefont {A.}~\bibnamefont
  {Mastrano}}, \bibinfo {author} {\bibfnamefont {A.}~\bibnamefont {Melatos}},
  \bibinfo {author} {\bibfnamefont {A.}~\bibnamefont {Reisenegger}}, \ and\
  \bibinfo {author} {\bibfnamefont {T.}~\bibnamefont {Akgun}},\ }\href
  {\doibase 10.1111/j.1365-2966.2011.19410.x} {\bibfield  {journal} {\bibinfo
  {journal} {Mon. Not. R. Astron. Soc.}\ }\textbf {\bibinfo {volume} {417}},\
  \bibinfo {pages} {2288} (\bibinfo {year} {2011})}\BibitemShut {NoStop}%
\bibitem [{\citenamefont {{Glampedakis}}\ and\ \citenamefont
  {{Lasky}}(2016)}]{Glampedakis:2016gl}%
  \BibitemOpen
  \bibfield  {author} {\bibinfo {author} {\bibfnamefont {K.}~\bibnamefont
  {{Glampedakis}}}\ and\ \bibinfo {author} {\bibfnamefont {P.~D.}\ \bibnamefont
  {{Lasky}}},\ }\href {\doibase 10.1093/mnras/stw2115} {\bibfield  {journal}
  {\bibinfo  {journal} {Mon. Not. R. Astron. Soc.}\ }\textbf {\bibinfo {volume}
  {463}},\ \bibinfo {pages} {2542} (\bibinfo {year} {2016})}\BibitemShut
  {NoStop}%
\bibitem [{\citenamefont {Mastrano}\ \emph {et~al.}(2015)\citenamefont
  {Mastrano}, \citenamefont {Suvorov},\ and\ \citenamefont
  {Melatos}}]{Mastrano:2015rfa}%
  \BibitemOpen
  \bibfield  {author} {\bibinfo {author} {\bibfnamefont {A.}~\bibnamefont
  {Mastrano}}, \bibinfo {author} {\bibfnamefont {A.~G.}\ \bibnamefont
  {Suvorov}}, \ and\ \bibinfo {author} {\bibfnamefont {A.}~\bibnamefont
  {Melatos}},\ }\href {\doibase 10.1093/mnras/stu2671} {\bibfield  {journal}
  {\bibinfo  {journal} {Mon. Not. R. Astron. Soc.}\ }\textbf {\bibinfo {volume}
  {447}},\ \bibinfo {pages} {3475} (\bibinfo {year} {2015})}\BibitemShut
  {NoStop}%
\bibitem [{\citenamefont {Lasky}\ and\ \citenamefont
  {Melatos}(2013)}]{Lasky:2013bpa}%
  \BibitemOpen
  \bibfield  {author} {\bibinfo {author} {\bibfnamefont {P.~D.}\ \bibnamefont
  {Lasky}}\ and\ \bibinfo {author} {\bibfnamefont {A.}~\bibnamefont
  {Melatos}},\ }\href {\doibase 10.1103/PhysRevD.88.103005} {\bibfield
  {journal} {\bibinfo  {journal} {Phys. Rev.}\ }\textbf {\bibinfo {volume}
  {D88}},\ \bibinfo {pages} {103005} (\bibinfo {year} {2013})}\BibitemShut
  {NoStop}%
\bibitem [{\citenamefont {{Mendell}}(1991)}]{Mendell:1991mn}%
  \BibitemOpen
  \bibfield  {author} {\bibinfo {author} {\bibfnamefont {G.}~\bibnamefont
  {{Mendell}}},\ }\href {\doibase 10.1086/170609} {\bibfield  {journal}
  {\bibinfo  {journal} {Astrophys. J.}\ }\textbf {\bibinfo {volume} {380}},\
  \bibinfo {pages} {515} (\bibinfo {year} {1991})}\BibitemShut {NoStop}%
\bibitem [{\citenamefont {Lander}\ \emph {et~al.}(2012)\citenamefont {Lander},
  \citenamefont {Andersson},\ and\ \citenamefont
  {Glampedakis}}]{Lander:2011yr}%
  \BibitemOpen
  \bibfield  {author} {\bibinfo {author} {\bibfnamefont {S.~K.}\ \bibnamefont
  {Lander}}, \bibinfo {author} {\bibfnamefont {N.}~\bibnamefont {Andersson}}, \
  and\ \bibinfo {author} {\bibfnamefont {K.}~\bibnamefont {Glampedakis}},\
  }\href {\doibase 10.1111/j.1365-2966.2011.19720.x} {\bibfield  {journal}
  {\bibinfo  {journal} {Mon. Not. R. Astron. Soc.}\ }\textbf {\bibinfo {volume}
  {419}},\ \bibinfo {pages} {732} (\bibinfo {year} {2012})}\BibitemShut
  {NoStop}%
\bibitem [{\citenamefont {{Henriksson}}\ and\ \citenamefont
  {{Wasserman}}(2013)}]{Henriksson:2012qz}%
  \BibitemOpen
  \bibfield  {author} {\bibinfo {author} {\bibfnamefont {K.~T.}\ \bibnamefont
  {{Henriksson}}}\ and\ \bibinfo {author} {\bibfnamefont {I.}~\bibnamefont
  {{Wasserman}}},\ }\href {\doibase 10.1093/mnras/stt338} {\bibfield  {journal}
  {\bibinfo  {journal} {Mon. Not. Roy. Astron. Soc.}\ }\textbf {\bibinfo
  {volume} {431}},\ \bibinfo {pages} {2986} (\bibinfo {year}
  {2013})}\BibitemShut {NoStop}%
\bibitem [{\citenamefont {Lander}(2013)}]{Lander:2012zz}%
  \BibitemOpen
  \bibfield  {author} {\bibinfo {author} {\bibfnamefont {S.~K.}\ \bibnamefont
  {Lander}},\ }\href {\doibase 10.1103/PhysRevLett.110.071101} {\bibfield
  {journal} {\bibinfo  {journal} {Phys. Rev. Lett.}\ }\textbf {\bibinfo
  {volume} {110}},\ \bibinfo {pages} {071101} (\bibinfo {year}
  {2013})}\BibitemShut {NoStop}%
\bibitem [{ATN()}]{ATNF}%
  \BibitemOpen
  \href@noop {} {}\bibinfo {howpublished} {{\tt
  http://www.atnf.csiro.au/people/pulsar/psrcat/.}}\BibitemShut {Stop}%
\bibitem [{\citenamefont {{Goldreich}}\ and\ \citenamefont
  {{Reisenegger}}(1992)}]{Goldreich:1992gr}%
  \BibitemOpen
  \bibfield  {author} {\bibinfo {author} {\bibfnamefont {P.}~\bibnamefont
  {{Goldreich}}}\ and\ \bibinfo {author} {\bibfnamefont {A.}~\bibnamefont
  {{Reisenegger}}},\ }\href {\doibase 10.1086/171646} {\bibfield  {journal}
  {\bibinfo  {journal} {Astrophys. J.}\ }\textbf {\bibinfo {volume} {395}},\
  \bibinfo {pages} {250} (\bibinfo {year} {1992})}\BibitemShut {NoStop}%
\bibitem [{\citenamefont {Pons}\ and\ \citenamefont
  {Geppert}(2007)}]{Pons:2007vf}%
  \BibitemOpen
  \bibfield  {author} {\bibinfo {author} {\bibfnamefont {J.~A.}\ \bibnamefont
  {Pons}}\ and\ \bibinfo {author} {\bibfnamefont {U.}~\bibnamefont {Geppert}},\
  }\href {\doibase 10.1051/0004-6361:20077456} {\bibfield  {journal} {\bibinfo
  {journal} {Astron. Astrophys.}\ }\textbf {\bibinfo {volume} {470}},\ \bibinfo
  {pages} {303} (\bibinfo {year} {2007})}\BibitemShut {NoStop}%
\bibitem [{\citenamefont {Kouveliotou}\ \emph {et~al.}(1998)\citenamefont
  {Kouveliotou} \emph {et~al.}}]{Kouveliotou:1998ze}%
  \BibitemOpen
  \bibfield  {author} {\bibinfo {author} {\bibfnamefont {C.}~\bibnamefont
  {Kouveliotou}} \emph {et~al.},\ }\href {\doibase 10.1038/30410} {\bibfield
  {journal} {\bibinfo  {journal} {Nature}\ }\textbf {\bibinfo {volume} {393}},\
  \bibinfo {pages} {235} (\bibinfo {year} {1998})}\BibitemShut {NoStop}%
\bibitem [{\citenamefont {Dall'Osso}\ \emph {et~al.}(2009)\citenamefont
  {Dall'Osso}, \citenamefont {Shore},\ and\ \citenamefont
  {Stella}}]{DallOsso:2008kll}%
  \BibitemOpen
  \bibfield  {author} {\bibinfo {author} {\bibfnamefont {S.}~\bibnamefont
  {Dall'Osso}}, \bibinfo {author} {\bibfnamefont {S.~N.}\ \bibnamefont
  {Shore}}, \ and\ \bibinfo {author} {\bibfnamefont {L.}~\bibnamefont
  {Stella}},\ }\href {\doibase 10.1111/j.1365-2966.2008.14054.x} {\bibfield
  {journal} {\bibinfo  {journal} {Mon. Not. Roy. Astron. Soc.}\ }\textbf
  {\bibinfo {volume} {398}},\ \bibinfo {pages} {1869} (\bibinfo {year}
  {2009})}\BibitemShut {NoStop}%
\bibitem [{\citenamefont {Camelio}\ \emph {et~al.}(2016)\citenamefont
  {Camelio}, \citenamefont {Gualtieri}, \citenamefont {Pons},\ and\
  \citenamefont {Ferrari}}]{Camelio:2016fan}%
  \BibitemOpen
  \bibfield  {author} {\bibinfo {author} {\bibfnamefont {G.}~\bibnamefont
  {Camelio}}, \bibinfo {author} {\bibfnamefont {L.}~\bibnamefont {Gualtieri}},
  \bibinfo {author} {\bibfnamefont {J.~A.}\ \bibnamefont {Pons}}, \ and\
  \bibinfo {author} {\bibfnamefont {V.}~\bibnamefont {Ferrari}},\ }\href
  {\doibase 10.1103/PhysRevD.94.024008} {\bibfield  {journal} {\bibinfo
  {journal} {Phys. Rev.}\ }\textbf {\bibinfo {volume} {D94}},\ \bibinfo {pages}
  {024008} (\bibinfo {year} {2016})}\BibitemShut {NoStop}%
\bibitem [{\citenamefont {Andersson}\ \emph {et~al.}(2011)\citenamefont
  {Andersson}, \citenamefont {Ferrari}, \citenamefont {Jones}, \citenamefont
  {Kokkotas}, \citenamefont {Krishnan}, \citenamefont {Read}, \citenamefont
  {Rezzolla},\ and\ \citenamefont {Zink}}]{Andersson:2009yt}%
  \BibitemOpen
  \bibfield  {author} {\bibinfo {author} {\bibfnamefont {N.}~\bibnamefont
  {Andersson}}, \bibinfo {author} {\bibfnamefont {V.}~\bibnamefont {Ferrari}},
  \bibinfo {author} {\bibfnamefont {D.~I.}\ \bibnamefont {Jones}}, \bibinfo
  {author} {\bibfnamefont {K.~D.}\ \bibnamefont {Kokkotas}}, \bibinfo {author}
  {\bibfnamefont {B.}~\bibnamefont {Krishnan}}, \bibinfo {author}
  {\bibfnamefont {J.~S.}\ \bibnamefont {Read}}, \bibinfo {author}
  {\bibfnamefont {L.}~\bibnamefont {Rezzolla}}, \ and\ \bibinfo {author}
  {\bibfnamefont {B.}~\bibnamefont {Zink}},\ }\href {\doibase
  10.1007/s10714-010-1059-4} {\bibfield  {journal} {\bibinfo  {journal} {Gen.
  Rel. Grav.}\ }\textbf {\bibinfo {volume} {43}},\ \bibinfo {pages} {409}
  (\bibinfo {year} {2011})}\BibitemShut {NoStop}%
\bibitem [{\citenamefont {Zhang}(2013)}]{Zhang:2012wt}%
  \BibitemOpen
  \bibfield  {author} {\bibinfo {author} {\bibfnamefont {B.}~\bibnamefont
  {Zhang}},\ }\href {\doibase 10.1088/2041-8205/763/1/L22} {\bibfield
  {journal} {\bibinfo  {journal} {Astrophys. J.}\ }\textbf {\bibinfo {volume}
  {763}},\ \bibinfo {pages} {L22} (\bibinfo {year} {2013})}\BibitemShut
  {NoStop}%
\bibitem [{\citenamefont {Giacomazzo}\ and\ \citenamefont
  {Perna}(2013)}]{Giacomazzo:2013uua}%
  \BibitemOpen
  \bibfield  {author} {\bibinfo {author} {\bibfnamefont {B.}~\bibnamefont
  {Giacomazzo}}\ and\ \bibinfo {author} {\bibfnamefont {R.}~\bibnamefont
  {Perna}},\ }\href {\doibase 10.1088/2041-8205/771/2/L26} {\bibfield
  {journal} {\bibinfo  {journal} {Astrophys. J.}\ }\textbf {\bibinfo {volume}
  {771}},\ \bibinfo {pages} {L26} (\bibinfo {year} {2013})}\BibitemShut
  {NoStop}%
\bibitem [{\citenamefont {Ravi}\ and\ \citenamefont
  {Lasky}(2014)}]{Ravi:2014gxa}%
  \BibitemOpen
  \bibfield  {author} {\bibinfo {author} {\bibfnamefont {V.}~\bibnamefont
  {Ravi}}\ and\ \bibinfo {author} {\bibfnamefont {P.~D.}\ \bibnamefont
  {Lasky}},\ }\href {\doibase 10.1093/mnras/stu720} {\bibfield  {journal}
  {\bibinfo  {journal} {Mon. Not. Roy. Astron. Soc.}\ }\textbf {\bibinfo
  {volume} {441}},\ \bibinfo {pages} {2433} (\bibinfo {year}
  {2014})}\BibitemShut {NoStop}%
\bibitem [{\citenamefont {Dall'Osso}\ \emph {et~al.}(2015)\citenamefont
  {Dall'Osso}, \citenamefont {Giacomazzo}, \citenamefont {Perna},\ and\
  \citenamefont {Stella}}]{DallOsso:2014hpa}%
  \BibitemOpen
  \bibfield  {author} {\bibinfo {author} {\bibfnamefont {S.}~\bibnamefont
  {Dall'Osso}}, \bibinfo {author} {\bibfnamefont {B.}~\bibnamefont
  {Giacomazzo}}, \bibinfo {author} {\bibfnamefont {R.}~\bibnamefont {Perna}}, \
  and\ \bibinfo {author} {\bibfnamefont {L.}~\bibnamefont {Stella}},\ }\href
  {\doibase 10.1088/0004-637X/798/1/25} {\bibfield  {journal} {\bibinfo
  {journal} {Astrophys. J.}\ }\textbf {\bibinfo {volume} {798}},\ \bibinfo
  {pages} {25} (\bibinfo {year} {2015})}\BibitemShut {NoStop}%
\bibitem [{\citenamefont {Corsi}\ and\ \citenamefont
  {Meszaros}(2009)}]{Corsi:2009jt}%
  \BibitemOpen
  \bibfield  {author} {\bibinfo {author} {\bibfnamefont {A.}~\bibnamefont
  {Corsi}}\ and\ \bibinfo {author} {\bibfnamefont {P.}~\bibnamefont
  {Meszaros}},\ }\href {\doibase 10.1088/0004-637X/702/2/1171} {\bibfield
  {journal} {\bibinfo  {journal} {Astrophys. J.}\ }\textbf {\bibinfo {volume}
  {702}},\ \bibinfo {pages} {1171} (\bibinfo {year} {2009})}\BibitemShut
  {NoStop}%
\bibitem [{\citenamefont {Payne}\ and\ \citenamefont
  {Melatos}(2004)}]{Payne:2004vt}%
  \BibitemOpen
  \bibfield  {author} {\bibinfo {author} {\bibfnamefont {D.~J.~B.}\
  \bibnamefont {Payne}}\ and\ \bibinfo {author} {\bibfnamefont
  {A.}~\bibnamefont {Melatos}},\ }\href {\doibase
  10.1111/j.1365-2966.2004.07798.x} {\bibfield  {journal} {\bibinfo  {journal}
  {Mon. Not. R. Astron. Soc.}\ }\textbf {\bibinfo {volume} {351}},\ \bibinfo
  {pages} {569} (\bibinfo {year} {2004})}\BibitemShut {NoStop}%
\bibitem [{\citenamefont {Melatos}\ and\ \citenamefont
  {Payne}(2005)}]{Melatos:2005ez}%
  \BibitemOpen
  \bibfield  {author} {\bibinfo {author} {\bibfnamefont {A.}~\bibnamefont
  {Melatos}}\ and\ \bibinfo {author} {\bibfnamefont {D.~J.~B.}\ \bibnamefont
  {Payne}},\ }\href {\doibase 10.1086/428600} {\bibfield  {journal} {\bibinfo
  {journal} {Astrophys. J.}\ }\textbf {\bibinfo {volume} {623}},\ \bibinfo
  {pages} {1044} (\bibinfo {year} {2005})}\BibitemShut {NoStop}%
\bibitem [{\citenamefont {Vigelius}\ and\ \citenamefont
  {Melatos}(2008)}]{Vigelius:2008fv}%
  \BibitemOpen
  \bibfield  {author} {\bibinfo {author} {\bibfnamefont {M.}~\bibnamefont
  {Vigelius}}\ and\ \bibinfo {author} {\bibfnamefont {A.}~\bibnamefont
  {Melatos}},\ }\href {\doibase 10.1111/j.1365-2966.2008.13139.x} {\bibfield
  {journal} {\bibinfo  {journal} {Mon. Not. Roy. Astron. Soc.}\ }\textbf
  {\bibinfo {volume} {386}},\ \bibinfo {pages} {1294} (\bibinfo {year}
  {2008})}\BibitemShut {NoStop}%
\bibitem [{\citenamefont {Shibazaki}\ \emph {et~al.}(1989)\citenamefont
  {Shibazaki}, \citenamefont {Murakami}, \citenamefont {Shaham},\ and\
  \citenamefont {Nomoto}}]{Shibazaki:1989sm}%
  \BibitemOpen
  \bibfield  {author} {\bibinfo {author} {\bibfnamefont {N.}~\bibnamefont
  {Shibazaki}}, \bibinfo {author} {\bibfnamefont {T.}~\bibnamefont {Murakami}},
  \bibinfo {author} {\bibfnamefont {J.}~\bibnamefont {Shaham}}, \ and\ \bibinfo
  {author} {\bibfnamefont {K.}~\bibnamefont {Nomoto}},\ }\href {\doibase
  10.1038/342656a0} {\bibfield  {journal} {\bibinfo  {journal} {Nature}\
  }\textbf {\bibinfo {volume} {342}},\ \bibinfo {pages} {656} (\bibinfo {year}
  {1989})}\BibitemShut {NoStop}%
\bibitem [{\citenamefont {Priymak}\ \emph {et~al.}(2011)\citenamefont
  {Priymak}, \citenamefont {Melatos},\ and\ \citenamefont
  {Payne}}]{Priymak:2011zv}%
  \BibitemOpen
  \bibfield  {author} {\bibinfo {author} {\bibfnamefont {M.}~\bibnamefont
  {Priymak}}, \bibinfo {author} {\bibfnamefont {A.}~\bibnamefont {Melatos}}, \
  and\ \bibinfo {author} {\bibfnamefont {D.~J.~B.}\ \bibnamefont {Payne}},\
  }\href {\doibase 10.1111/j.1365-2966.2011.19431.x} {\bibfield  {journal}
  {\bibinfo  {journal} {Mon. Not. R. Astron. Soc.}\ }\textbf {\bibinfo {volume}
  {417}},\ \bibinfo {pages} {2696} (\bibinfo {year} {2011})}\BibitemShut
  {NoStop}%
\bibitem [{\citenamefont {Haskell}\ \emph {et~al.}(2015)\citenamefont
  {Haskell}, \citenamefont {Priymak}, \citenamefont {Patruno}, \citenamefont
  {Oppenoorth}, \citenamefont {Melatos},\ and\ \citenamefont
  {Lasky}}]{Haskell:2015psa}%
  \BibitemOpen
  \bibfield  {author} {\bibinfo {author} {\bibfnamefont {B.}~\bibnamefont
  {Haskell}}, \bibinfo {author} {\bibfnamefont {M.}~\bibnamefont {Priymak}},
  \bibinfo {author} {\bibfnamefont {A.}~\bibnamefont {Patruno}}, \bibinfo
  {author} {\bibfnamefont {M.}~\bibnamefont {Oppenoorth}}, \bibinfo {author}
  {\bibfnamefont {A.}~\bibnamefont {Melatos}}, \ and\ \bibinfo {author}
  {\bibfnamefont {P.~D.}\ \bibnamefont {Lasky}},\ }\href {\doibase
  10.1093/mnras/stv726} {\bibfield  {journal} {\bibinfo  {journal} {Mon. Not.
  R. Astron. Soc.}\ }\textbf {\bibinfo {volume} {450}},\ \bibinfo {pages}
  {2393} (\bibinfo {year} {2015})}\BibitemShut {NoStop}%
\bibitem [{\citenamefont {Vigelius}\ and\ \citenamefont
  {Melatos}(2009)}]{Vigelius:2009eg}%
  \BibitemOpen
  \bibfield  {author} {\bibinfo {author} {\bibfnamefont {M.}~\bibnamefont
  {Vigelius}}\ and\ \bibinfo {author} {\bibfnamefont {A.}~\bibnamefont
  {Melatos}},\ }\href {\doibase 10.1111/j.1365-2966.2009.14698.x} {\bibfield
  {journal} {\bibinfo  {journal} {Mon. Not. Roy. Astron. Soc.}\ }\textbf
  {\bibinfo {volume} {395}},\ \bibinfo {pages} {1985} (\bibinfo {year}
  {2009})}\BibitemShut {NoStop}%
\bibitem [{\citenamefont {{Haensel}}\ and\ \citenamefont
  {{Zdunik}}(1990)}]{Haensel:1990hz}%
  \BibitemOpen
  \bibfield  {author} {\bibinfo {author} {\bibfnamefont {P.}~\bibnamefont
  {{Haensel}}}\ and\ \bibinfo {author} {\bibfnamefont {J.~L.}\ \bibnamefont
  {{Zdunik}}},\ }\href@noop {} {\bibfield  {journal} {\bibinfo  {journal}
  {Astron. Astrophys.}\ }\textbf {\bibinfo {volume} {227}},\ \bibinfo {pages}
  {431} (\bibinfo {year} {1990})}\BibitemShut {NoStop}%
\bibitem [{\citenamefont {Ushomirsky}\ and\ \citenamefont
  {Rutledge}(2001)}]{Ushomirsky:2001pd}%
  \BibitemOpen
  \bibfield  {author} {\bibinfo {author} {\bibfnamefont {G.}~\bibnamefont
  {Ushomirsky}}\ and\ \bibinfo {author} {\bibfnamefont {R.~E.}\ \bibnamefont
  {Rutledge}},\ }\href {\doibase 10.1046/j.1365-8711.2001.04515.x} {\bibfield
  {journal} {\bibinfo  {journal} {Mon. Not. Roy. Astron. Soc.}\ }\textbf
  {\bibinfo {volume} {325}},\ \bibinfo {pages} {1157} (\bibinfo {year}
  {2001})}\BibitemShut {NoStop}%
\bibitem [{\citenamefont {Archibald}\ \emph {et~al.}(2009)\citenamefont
  {Archibald} \emph {et~al.}}]{Archibald:2009zb}%
  \BibitemOpen
  \bibfield  {author} {\bibinfo {author} {\bibfnamefont {A.~M.}\ \bibnamefont
  {Archibald}} \emph {et~al.},\ }\href {\doibase 10.1126/science.1172740}
  {\bibfield  {journal} {\bibinfo  {journal} {Science}\ }\textbf {\bibinfo
  {volume} {324}},\ \bibinfo {pages} {1411} (\bibinfo {year}
  {2009})}\BibitemShut {NoStop}%
\bibitem [{\citenamefont {Haskell}\ and\ \citenamefont
  {Patruno}(2017)}]{Haskell:2017ajb}%
  \BibitemOpen
  \bibfield  {author} {\bibinfo {author} {\bibfnamefont {B.}~\bibnamefont
  {Haskell}}\ and\ \bibinfo {author} {\bibfnamefont {A.}~\bibnamefont
  {Patruno}},\ }\href@noop {} {\  (\bibinfo {year} {2017})},\ \Eprint
  {http://arxiv.org/abs/1703.08374} {arXiv:1703.08374 [astro-ph.HE]}
  \BibitemShut {NoStop}%
\bibitem [{\citenamefont {Haskell}\ \emph {et~al.}(2006)\citenamefont
  {Haskell}, \citenamefont {Jones},\ and\ \citenamefont
  {Andersson}}]{Haskell:2006sv}%
  \BibitemOpen
  \bibfield  {author} {\bibinfo {author} {\bibfnamefont {B.}~\bibnamefont
  {Haskell}}, \bibinfo {author} {\bibfnamefont {D.~I.}\ \bibnamefont {Jones}},
  \ and\ \bibinfo {author} {\bibfnamefont {N.}~\bibnamefont {Andersson}},\
  }\href {\doibase 10.1111/j.1365-2966.2006.10998.x} {\bibfield  {journal}
  {\bibinfo  {journal} {Mon. Not. Roy. Astron. Soc.}\ }\textbf {\bibinfo
  {volume} {373}},\ \bibinfo {pages} {1423} (\bibinfo {year}
  {2006})}\BibitemShut {NoStop}%
\bibitem [{\citenamefont {Johnson-McDaniel}\ and\ \citenamefont
  {Owen}(2013)}]{JohnsonMcDaniel:2012wg}%
  \BibitemOpen
  \bibfield  {author} {\bibinfo {author} {\bibfnamefont {N.~K.}\ \bibnamefont
  {Johnson-McDaniel}}\ and\ \bibinfo {author} {\bibfnamefont {B.~J.}\
  \bibnamefont {Owen}},\ }\href {\doibase 10.1103/PhysRevD.88.044004}
  {\bibfield  {journal} {\bibinfo  {journal} {Phys. Rev.}\ }\textbf {\bibinfo
  {volume} {D88}},\ \bibinfo {pages} {044004} (\bibinfo {year}
  {2013})}\BibitemShut {NoStop}%
\bibitem [{\citenamefont {Horowitz}\ and\ \citenamefont
  {Kadau}(2009)}]{Horowitz:2009ya}%
  \BibitemOpen
  \bibfield  {author} {\bibinfo {author} {\bibfnamefont {C.~J.}\ \bibnamefont
  {Horowitz}}\ and\ \bibinfo {author} {\bibfnamefont {K.}~\bibnamefont
  {Kadau}},\ }\href {\doibase 10.1103/PhysRevLett.102.191102} {\bibfield
  {journal} {\bibinfo  {journal} {Phys. Rev. Lett.}\ }\textbf {\bibinfo
  {volume} {102}},\ \bibinfo {pages} {191102} (\bibinfo {year}
  {2009})}\BibitemShut {NoStop}%
\bibitem [{\citenamefont {Glampedakis}\ \emph {et~al.}(2012)\citenamefont
  {Glampedakis}, \citenamefont {Jones},\ and\ \citenamefont
  {Samuelsson}}]{KG_etal2012}%
  \BibitemOpen
  \bibfield  {author} {\bibinfo {author} {\bibfnamefont {K.}~\bibnamefont
  {Glampedakis}}, \bibinfo {author} {\bibfnamefont {D.~I.}\ \bibnamefont
  {Jones}}, \ and\ \bibinfo {author} {\bibfnamefont {L.}~\bibnamefont
  {Samuelsson}},\ }\href {\doibase 10.1103/PhysRevLett.109.081103} {\bibfield
  {journal} {\bibinfo  {journal} {Phys. Rev. Lett.}\ }\textbf {\bibinfo
  {volume} {109}},\ \bibinfo {pages} {081103} (\bibinfo {year}
  {2012})}\BibitemShut {NoStop}%
\bibitem [{\citenamefont {Alford}\ \emph {et~al.}(2008)\citenamefont {Alford},
  \citenamefont {Schmitt}, \citenamefont {Rajagopal},\ and\ \citenamefont
  {Sh\"afer}}]{alford_etal2008}%
  \BibitemOpen
  \bibfield  {author} {\bibinfo {author} {\bibfnamefont {M.~G.}\ \bibnamefont
  {Alford}}, \bibinfo {author} {\bibfnamefont {A.}~\bibnamefont {Schmitt}},
  \bibinfo {author} {\bibfnamefont {K.}~\bibnamefont {Rajagopal}}, \ and\
  \bibinfo {author} {\bibfnamefont {T.}~\bibnamefont {Sh\"afer}},\ }\href
  {\doibase 10.1103/RevModPhys.80.1455} {\bibfield  {journal} {\bibinfo
  {journal} {Rev. Mod. Phys.}\ }\textbf {\bibinfo {volume} {80}},\ \bibinfo
  {pages} {1455} (\bibinfo {year} {2008})}\BibitemShut {NoStop}%
\bibitem [{\citenamefont {Iida}\ and\ \citenamefont
  {Baym}(2002)}]{iida_baym2002}%
  \BibitemOpen
  \bibfield  {author} {\bibinfo {author} {\bibfnamefont {K.}~\bibnamefont
  {Iida}}\ and\ \bibinfo {author} {\bibfnamefont {G.}~\bibnamefont {Baym}},\
  }\href {\doibase 10.1103/PhysRevD.66.014015} {\bibfield  {journal} {\bibinfo
  {journal} {Phys. Rev. D}\ }\textbf {\bibinfo {volume} {66}},\ \bibinfo
  {pages} {014015} (\bibinfo {year} {2002})}\BibitemShut {NoStop}%
\bibitem [{\citenamefont {Alford}\ and\ \citenamefont
  {Sedrakian}(2010)}]{alford_sedrakian2010}%
  \BibitemOpen
  \bibfield  {author} {\bibinfo {author} {\bibfnamefont {M.~G.}\ \bibnamefont
  {Alford}}\ and\ \bibinfo {author} {\bibfnamefont {A.}~\bibnamefont
  {Sedrakian}},\ }\href {\doibase 10.1088/0954-3899/37/7/075202} {\bibfield
  {journal} {\bibinfo  {journal} {J. Phys. G}\ }\textbf {\bibinfo {volume}
  {37}},\ \bibinfo {pages} {075202} (\bibinfo {year} {2010})}\BibitemShut
  {NoStop}%
\bibitem [{\citenamefont {Ciolfi}\ \emph {et~al.}(2009)\citenamefont {Ciolfi},
  \citenamefont {Ferrari}, \citenamefont {Gualtieri},\ and\ \citenamefont
  {Pons}}]{ciolfi_etal2009}%
  \BibitemOpen
  \bibfield  {author} {\bibinfo {author} {\bibfnamefont {R.}~\bibnamefont
  {Ciolfi}}, \bibinfo {author} {\bibfnamefont {V.}~\bibnamefont {Ferrari}},
  \bibinfo {author} {\bibfnamefont {L.}~\bibnamefont {Gualtieri}}, \ and\
  \bibinfo {author} {\bibfnamefont {J.~A.}\ \bibnamefont {Pons}},\ }\href
  {\doibase 10.1111/j.1365-2966.2009.14990.x} {\bibfield  {journal} {\bibinfo
  {journal} {Mon. Not. R. Astron. Soc.}\ }\textbf {\bibinfo {volume} {397}},\
  \bibinfo {pages} {913} (\bibinfo {year} {2009})}\BibitemShut {NoStop}%
\bibitem [{\citenamefont {Lander}\ and\ \citenamefont
  {Jones}(2009)}]{lander_jones2009}%
  \BibitemOpen
  \bibfield  {author} {\bibinfo {author} {\bibfnamefont {S.~K.}\ \bibnamefont
  {Lander}}\ and\ \bibinfo {author} {\bibfnamefont {D.~I.}\ \bibnamefont
  {Jones}},\ }\href {\doibase 10.1111/j.1365-2966.2009.14667.x} {\bibfield
  {journal} {\bibinfo  {journal} {Mon. Not. R. Astron. Soc.}\ }\textbf
  {\bibinfo {volume} {395}},\ \bibinfo {pages} {2162} (\bibinfo {year}
  {2009})}\BibitemShut {NoStop}%
\bibitem [{\citenamefont {Rajagopal}\ and\ \citenamefont
  {Sharma}(2006)}]{rajagopal_sharma2006}%
  \BibitemOpen
  \bibfield  {author} {\bibinfo {author} {\bibfnamefont {K.}~\bibnamefont
  {Rajagopal}}\ and\ \bibinfo {author} {\bibfnamefont {R.}~\bibnamefont
  {Sharma}},\ }\href {\doibase 10.1103/PhysRevD.74.094019} {\bibfield
  {journal} {\bibinfo  {journal} {Phys. Rev. D}\ }\textbf {\bibinfo {volume}
  {74}},\ \bibinfo {pages} {094019} (\bibinfo {year} {2006})}\BibitemShut
  {NoStop}%
\bibitem [{\citenamefont {Mannarelli}\ \emph {et~al.}(2007)\citenamefont
  {Mannarelli}, \citenamefont {Rajagopal},\ and\ \citenamefont
  {Sharma}}]{mannarelli_etal2007}%
  \BibitemOpen
  \bibfield  {author} {\bibinfo {author} {\bibfnamefont {M.}~\bibnamefont
  {Mannarelli}}, \bibinfo {author} {\bibfnamefont {K.}~\bibnamefont
  {Rajagopal}}, \ and\ \bibinfo {author} {\bibfnamefont {R.}~\bibnamefont
  {Sharma}},\ }\href {\doibase 10.1103/PhysRevD.76.074026} {\bibfield
  {journal} {\bibinfo  {journal} {Phys. Rev. D}\ }\textbf {\bibinfo {volume}
  {76}},\ \bibinfo {pages} {074026} (\bibinfo {year} {2007})}\BibitemShut
  {NoStop}%
\bibitem [{\citenamefont {Owen}(2005)}]{owen2005}%
  \BibitemOpen
  \bibfield  {author} {\bibinfo {author} {\bibfnamefont {B.~J.}\ \bibnamefont
  {Owen}},\ }\href {\doibase 10.1103/PhysRevLett.95.211101} {\bibfield
  {journal} {\bibinfo  {journal} {Phys. Rev. Lett.}\ }\textbf {\bibinfo
  {volume} {95}},\ \bibinfo {pages} {211101} (\bibinfo {year}
  {2005})}\BibitemShut {NoStop}%
\bibitem [{\citenamefont {{B. Haskell, N. Andersson, D. I. Jones and L.
  Samuelsson}}(2007)}]{haskell_etal2007}%
  \BibitemOpen
  \bibfield  {author} {\bibinfo {author} {\bibnamefont {{B. Haskell, N.
  Andersson, D. I. Jones and L. Samuelsson}}},\ }\href {\doibase
  10.1103/PhysRevLett.99.231101} {\bibfield  {journal} {\bibinfo  {journal}
  {Phys. Rev. Lett.}\ }\textbf {\bibinfo {volume} {99}},\ \bibinfo {pages}
  {231101} (\bibinfo {year} {2007})}\BibitemShut {NoStop}%
\bibitem [{\citenamefont {Abbott}\ \emph
  {et~al.}(2017{\natexlab{e}})\citenamefont {Abbott} \emph
  {et~al.}}]{abbott_etal2017}%
  \BibitemOpen
  \bibfield  {author} {\bibinfo {author} {\bibfnamefont {B.~P.}\ \bibnamefont
  {Abbott}} \emph {et~al.} (\bibinfo {collaboration} {Virgo, LIGO
  Scientific}),\ }\href {\doibase 10.3847/1538-4357/aa677f} {\bibfield
  {journal} {\bibinfo  {journal} {Astrophys. J.}\ }\textbf {\bibinfo {volume}
  {839}},\ \bibinfo {pages} {12} (\bibinfo {year}
  {2017}{\natexlab{e}})}\BibitemShut {NoStop}%
\bibitem [{\citenamefont {Jones}(2010)}]{jones2010}%
  \BibitemOpen
  \bibfield  {author} {\bibinfo {author} {\bibfnamefont {D.~I.}\ \bibnamefont
  {Jones}},\ }\href {\doibase 10.1111/j.1365-2966.2009.16059.x} {\bibfield
  {journal} {\bibinfo  {journal} {Mon. Not. R. Astron. Soc.}\ }\textbf
  {\bibinfo {volume} {402}},\ \bibinfo {pages} {2503} (\bibinfo {year}
  {2010})}\BibitemShut {NoStop}%
\bibitem [{\citenamefont {Glampedakis}\ and\ \citenamefont
  {Jones}(2010)}]{KG_DIJ2010}%
  \BibitemOpen
  \bibfield  {author} {\bibinfo {author} {\bibfnamefont {K.}~\bibnamefont
  {Glampedakis}}\ and\ \bibinfo {author} {\bibfnamefont {D.~I.}\ \bibnamefont
  {Jones}},\ }\href {\doibase 10.1111/j.1745-3933.2010.00846.x} {\bibfield
  {journal} {\bibinfo  {journal} {Mon. Not. R. Astron. Soc.}\ }\textbf
  {\bibinfo {volume} {405}},\ \bibinfo {pages} {L6} (\bibinfo {year}
  {2010})}\BibitemShut {NoStop}%
\bibitem [{\citenamefont {Jones}(2002)}]{jones2002}%
  \BibitemOpen
  \bibfield  {author} {\bibinfo {author} {\bibfnamefont {D.~I.}\ \bibnamefont
  {Jones}},\ }\href {\doibase 10.1088/0264-9381/19/7/304} {\bibfield  {journal}
  {\bibinfo  {journal} {Class. Quant. Grav.}\ }\textbf {\bibinfo {volume}
  {19}},\ \bibinfo {pages} {1255} (\bibinfo {year} {2002})}\BibitemShut
  {NoStop}%
\bibitem [{\citenamefont {Ruderman}(1991{\natexlab{a}})}]{ruderman1991a}%
  \BibitemOpen
  \bibfield  {author} {\bibinfo {author} {\bibfnamefont {M.}~\bibnamefont
  {Ruderman}},\ }\href {\doibase 10.1086/170744} {\bibfield  {journal}
  {\bibinfo  {journal} {Astrophys. J.}\ }\textbf {\bibinfo {volume} {382}},\
  \bibinfo {pages} {576} (\bibinfo {year} {1991}{\natexlab{a}})}\BibitemShut
  {NoStop}%
\bibitem [{\citenamefont {Ruderman}(1991{\natexlab{b}})}]{ruderman1991b}%
  \BibitemOpen
  \bibfield  {author} {\bibinfo {author} {\bibfnamefont {M.}~\bibnamefont
  {Ruderman}},\ }\href {\doibase 10.1086/170745} {\bibfield  {journal}
  {\bibinfo  {journal} {Astrophys. J.}\ }\textbf {\bibinfo {volume} {382}},\
  \bibinfo {pages} {587} (\bibinfo {year} {1991}{\natexlab{b}})}\BibitemShut
  {NoStop}%
\bibitem [{\citenamefont {Link}\ and\ \citenamefont
  {Cutler}(2002)}]{link_cutler2002}%
  \BibitemOpen
  \bibfield  {author} {\bibinfo {author} {\bibfnamefont {B.}~\bibnamefont
  {Link}}\ and\ \bibinfo {author} {\bibfnamefont {C.}~\bibnamefont {Cutler}},\
  }\href {\doibase 10.1046/j.1365-8711.2002.05726.x} {\bibfield  {journal}
  {\bibinfo  {journal} {Mon. Not. R. Astron. Soc.}\ }\textbf {\bibinfo {volume}
  {336}},\ \bibinfo {pages} {211} (\bibinfo {year} {2002})}\BibitemShut
  {NoStop}%
\bibitem [{\citenamefont {{Radhakrishnan}}\ and\ \citenamefont
  {{Manchester}}(1969)}]{Radjakrishnan:1969rm}%
  \BibitemOpen
  \bibfield  {author} {\bibinfo {author} {\bibfnamefont {V.}~\bibnamefont
  {{Radhakrishnan}}}\ and\ \bibinfo {author} {\bibfnamefont {R.~N.}\
  \bibnamefont {{Manchester}}},\ }\href {\doibase 10.1038/222228a0} {\bibfield
  {journal} {\bibinfo  {journal} {\nat}\ }\textbf {\bibinfo {volume} {222}},\
  \bibinfo {pages} {228} (\bibinfo {year} {1969})}\BibitemShut {NoStop}%
\bibitem [{\citenamefont {{Reichley}}\ and\ \citenamefont
  {{Downs}}(1969)}]{Reichley:1969rd}%
  \BibitemOpen
  \bibfield  {author} {\bibinfo {author} {\bibfnamefont {P.~E.}\ \bibnamefont
  {{Reichley}}}\ and\ \bibinfo {author} {\bibfnamefont {G.~S.}\ \bibnamefont
  {{Downs}}},\ }\href {\doibase 10.1038/222229a0} {\bibfield  {journal}
  {\bibinfo  {journal} {\nat}\ }\textbf {\bibinfo {volume} {222}},\ \bibinfo
  {pages} {229} (\bibinfo {year} {1969})}\BibitemShut {NoStop}%
\bibitem [{\citenamefont {Boynton}\ \emph {et~al.}(1969)\citenamefont
  {Boynton}, \citenamefont {Groth~III}, \citenamefont {Partridge},\ and\
  \citenamefont {Wilkinson}}]{boynton:1969pr}%
  \BibitemOpen
  \bibfield  {author} {\bibinfo {author} {\bibfnamefont {P.}~\bibnamefont
  {Boynton}}, \bibinfo {author} {\bibfnamefont {E.}~\bibnamefont {Groth~III}},
  \bibinfo {author} {\bibfnamefont {R.}~\bibnamefont {Partridge}}, \ and\
  \bibinfo {author} {\bibfnamefont {D.~T.}\ \bibnamefont {Wilkinson}},\
  }\href@noop {} {\bibfield  {journal} {\bibinfo  {journal} {Astrophys. J.}\
  }\textbf {\bibinfo {volume} {157}},\ \bibinfo {pages} {L197} (\bibinfo {year}
  {1969})}\BibitemShut {NoStop}%
\bibitem [{\citenamefont {{Richards}}\ \emph {et~al.}(1969)\citenamefont
  {{Richards}}, \citenamefont {{Pettengill}}, \citenamefont {{Roberts}},
  \citenamefont {{Counselman}},\ and\ \citenamefont
  {{Rankin}}}]{richards:1969gl}%
  \BibitemOpen
  \bibfield  {author} {\bibinfo {author} {\bibfnamefont {D.~W.}\ \bibnamefont
  {{Richards}}}, \bibinfo {author} {\bibfnamefont {G.~H.}\ \bibnamefont
  {{Pettengill}}}, \bibinfo {author} {\bibfnamefont {J.~A.}\ \bibnamefont
  {{Roberts}}}, \bibinfo {author} {\bibfnamefont {C.~C.}\ \bibnamefont
  {{Counselman}}}, \ and\ \bibinfo {author} {\bibfnamefont {J.}~\bibnamefont
  {{Rankin}}},\ }\href@noop {} {\bibfield  {journal} {\bibinfo  {journal}
  {\iaucirc}\ }\textbf {\bibinfo {volume} {2181}} (\bibinfo {year}
  {1969})}\BibitemShut {NoStop}%
\bibitem [{\citenamefont {Espinoza}\ \emph {et~al.}(2011)\citenamefont
  {Espinoza}, \citenamefont {Lyne}, \citenamefont {Stappers},\ and\
  \citenamefont {Kramer}}]{Espinosa:2011els}%
  \BibitemOpen
  \bibfield  {author} {\bibinfo {author} {\bibfnamefont {C.~M.}\ \bibnamefont
  {Espinoza}}, \bibinfo {author} {\bibfnamefont {A.~G.}\ \bibnamefont {Lyne}},
  \bibinfo {author} {\bibfnamefont {B.~W.}\ \bibnamefont {Stappers}}, \ and\
  \bibinfo {author} {\bibfnamefont {M.}~\bibnamefont {Kramer}},\ }\href
  {\doibase 10.1111/j.1365-2966.2011.18503.x} {\bibfield  {journal} {\bibinfo
  {journal} {Mon. Not. Roy. Astron. Soc.}\ }\textbf {\bibinfo {volume} {414}},\
  \bibinfo {pages} {1679} (\bibinfo {year} {2011})}\BibitemShut {NoStop}%
\bibitem [{\citenamefont {{Ray}}\ \emph {et~al.}(2011)\citenamefont {{Ray}}
  \emph {et~al.}}]{Ray:2011rk}%
  \BibitemOpen
  \bibfield  {author} {\bibinfo {author} {\bibfnamefont {P.~S.}\ \bibnamefont
  {{Ray}}} \emph {et~al.},\ }\href {\doibase 10.1088/0067-0049/194/2/17}
  {\bibfield  {journal} {\bibinfo  {journal} {Astrophys. J. Suppl.}\ }\textbf
  {\bibinfo {volume} {194}},\ \bibinfo {eid} {17} (\bibinfo {year}
  {2011})}\BibitemShut {NoStop}%
\bibitem [{\citenamefont {Pletsch}\ \emph {et~al.}(2012)\citenamefont {Pletsch}
  \emph {et~al.}}]{Pletsch:2012ct}%
  \BibitemOpen
  \bibfield  {author} {\bibinfo {author} {\bibfnamefont {H.~J.}\ \bibnamefont
  {Pletsch}} \emph {et~al.},\ }\href {\doibase 10.1088/2041-8205/755/1/L20}
  {\bibfield  {journal} {\bibinfo  {journal} {Astrophys. J. Lett.}\ }\textbf
  {\bibinfo {volume} {755}},\ \bibinfo {pages} {L20} (\bibinfo {year}
  {2012})}\BibitemShut {NoStop}%
\bibitem [{\citenamefont {{Dib}}\ \emph {et~al.}(2008)\citenamefont {{Dib}},
  \citenamefont {{Kaspi}},\ and\ \citenamefont {{Gavriil}}}]{Dib:2008dk}%
  \BibitemOpen
  \bibfield  {author} {\bibinfo {author} {\bibfnamefont {R.}~\bibnamefont
  {{Dib}}}, \bibinfo {author} {\bibfnamefont {V.~M.}\ \bibnamefont {{Kaspi}}},
  \ and\ \bibinfo {author} {\bibfnamefont {F.~P.}\ \bibnamefont {{Gavriil}}},\
  }\href {\doibase 10.1086/524653} {\bibfield  {journal} {\bibinfo  {journal}
  {Astrophys. J.}\ }\textbf {\bibinfo {volume} {673}},\ \bibinfo {eid}
  {1044-1061} (\bibinfo {year} {2008})}\BibitemShut {NoStop}%
\bibitem [{\citenamefont {Cognard}\ and\ \citenamefont
  {Backer}(2004)}]{Cognard:2004av}%
  \BibitemOpen
  \bibfield  {author} {\bibinfo {author} {\bibfnamefont {I.}~\bibnamefont
  {Cognard}}\ and\ \bibinfo {author} {\bibfnamefont {D.~C.}\ \bibnamefont
  {Backer}},\ }\href {\doibase 10.1086/424692} {\bibfield  {journal} {\bibinfo
  {journal} {Astrophys. J.}\ }\textbf {\bibinfo {volume} {612}},\ \bibinfo
  {pages} {L125} (\bibinfo {year} {2004})}\BibitemShut {NoStop}%
\bibitem [{\citenamefont {Dodson}\ \emph {et~al.}(2002)\citenamefont {Dodson},
  \citenamefont {McCulloch},\ and\ \citenamefont {Lewis}}]{Dodson:2002gy}%
  \BibitemOpen
  \bibfield  {author} {\bibinfo {author} {\bibfnamefont {R.~G.}\ \bibnamefont
  {Dodson}}, \bibinfo {author} {\bibfnamefont {P.~M.}\ \bibnamefont
  {McCulloch}}, \ and\ \bibinfo {author} {\bibfnamefont {D.~R.}\ \bibnamefont
  {Lewis}},\ }\href {\doibase 10.1086/339068} {\bibfield  {journal} {\bibinfo
  {journal} {Astrophys. J.}\ }\textbf {\bibinfo {volume} {564}},\ \bibinfo
  {pages} {L85} (\bibinfo {year} {2002})}\BibitemShut {NoStop}%
\bibitem [{\citenamefont {Haskell}\ and\ \citenamefont
  {Melatos}(2015)}]{Haskell:2015jra}%
  \BibitemOpen
  \bibfield  {author} {\bibinfo {author} {\bibfnamefont {B.}~\bibnamefont
  {Haskell}}\ and\ \bibinfo {author} {\bibfnamefont {A.}~\bibnamefont
  {Melatos}},\ }\href {\doibase 10.1142/S0218271815300086} {\bibfield
  {journal} {\bibinfo  {journal} {Int. J. Mod. Phys. D}\ }\textbf {\bibinfo
  {volume} {24}},\ \bibinfo {pages} {1530008} (\bibinfo {year}
  {2015})}\BibitemShut {NoStop}%
\bibitem [{\citenamefont {Abadie}\ \emph {et~al.}(2011)\citenamefont {Abadie}
  \emph {et~al.}}]{Abadie:2010sf}%
  \BibitemOpen
  \bibfield  {author} {\bibinfo {author} {\bibfnamefont {J.}~\bibnamefont
  {Abadie}} \emph {et~al.} (\bibinfo {collaboration} {LIGO}),\ }\href {\doibase
  10.1103/PhysRevD.83.042001, 10.1103/PhysRevD.83.069902,
  10.1103/PhysRevD.85.089902} {\bibfield  {journal} {\bibinfo  {journal} {Phys.
  Rev.}\ }\textbf {\bibinfo {volume} {D83}},\ \bibinfo {pages} {042001}
  (\bibinfo {year} {2011})}\BibitemShut {NoStop}%
\bibitem [{\citenamefont {Sidery}\ \emph {et~al.}(2010)\citenamefont {Sidery},
  \citenamefont {Passamonti},\ and\ \citenamefont {Andersson}}]{Sidery:2009at}%
  \BibitemOpen
  \bibfield  {author} {\bibinfo {author} {\bibfnamefont {T.}~\bibnamefont
  {Sidery}}, \bibinfo {author} {\bibfnamefont {A.}~\bibnamefont {Passamonti}},
  \ and\ \bibinfo {author} {\bibfnamefont {N.}~\bibnamefont {Andersson}},\
  }\href {\doibase 10.1111/j.1365-2966.2010.16497.x} {\bibfield  {journal}
  {\bibinfo  {journal} {Mon. Not. Roy. Astron. Soc.}\ }\textbf {\bibinfo
  {volume} {405}},\ \bibinfo {pages} {1061} (\bibinfo {year}
  {2010})}\BibitemShut {NoStop}%
\bibitem [{\citenamefont {Keer}\ and\ \citenamefont
  {Jones}(2015)}]{Keer:2014uva}%
  \BibitemOpen
  \bibfield  {author} {\bibinfo {author} {\bibfnamefont {L.}~\bibnamefont
  {Keer}}\ and\ \bibinfo {author} {\bibfnamefont {D.~I.}\ \bibnamefont
  {Jones}},\ }\href {\doibase 10.1093/mnras/stu2123} {\bibfield  {journal}
  {\bibinfo  {journal} {Mon. Not. Roy. Astron. Soc.}\ }\textbf {\bibinfo
  {volume} {446}},\ \bibinfo {pages} {865} (\bibinfo {year}
  {2015})}\BibitemShut {NoStop}%
\bibitem [{\citenamefont {Sedrakian}\ \emph {et~al.}(2003)\citenamefont
  {Sedrakian}, \citenamefont {Benacquista}, \citenamefont {Shahabassian},
  \citenamefont {Sadoyan},\ and\ \citenamefont
  {Hairapetyan}}]{Sedrakian:2003sb}%
  \BibitemOpen
  \bibfield  {author} {\bibinfo {author} {\bibfnamefont {D.~M.}\ \bibnamefont
  {Sedrakian}}, \bibinfo {author} {\bibfnamefont {M.}~\bibnamefont
  {Benacquista}}, \bibinfo {author} {\bibfnamefont {K.~M.}\ \bibnamefont
  {Shahabassian}}, \bibinfo {author} {\bibfnamefont {A.~A.}\ \bibnamefont
  {Sadoyan}}, \ and\ \bibinfo {author} {\bibfnamefont {M.~V.}\ \bibnamefont
  {Hairapetyan}},\ }\href {\doibase 10.1023/B:ASYS.0000003260.80468.69}
  {\bibfield  {journal} {\bibinfo  {journal} {Astrophysics}\ }\textbf {\bibinfo
  {volume} {46}},\ \bibinfo {pages} {445} (\bibinfo {year} {2003})}\BibitemShut
  {NoStop}%
\bibitem [{\citenamefont {Santiago-Prieto}\ \emph {et~al.}(2012)\citenamefont
  {Santiago-Prieto}, \citenamefont {Heng}, \citenamefont {Jones},\ and\
  \citenamefont {Clark}}]{SantiagoPrieto:2012ew}%
  \BibitemOpen
  \bibfield  {author} {\bibinfo {author} {\bibfnamefont {I.}~\bibnamefont
  {Santiago-Prieto}}, \bibinfo {author} {\bibfnamefont {I.~S.}\ \bibnamefont
  {Heng}}, \bibinfo {author} {\bibfnamefont {D.~I.}\ \bibnamefont {Jones}}, \
  and\ \bibinfo {author} {\bibfnamefont {J.}~\bibnamefont {Clark}},\ }\bibfield
   {booktitle} {\emph {\bibinfo {booktitle} {{Proceedings of the 9th Amaldi
  Conference and NRDA Meeting, Cardiff, UK, 2011}}},\ }\href {\doibase
  10.1088/1742-6596/363/1/012042} {\bibfield  {journal} {\bibinfo  {journal}
  {J. Phys. Conf. Ser.}\ }\textbf {\bibinfo {volume} {363}},\ \bibinfo {pages}
  {012042} (\bibinfo {year} {2012})}\BibitemShut {NoStop}%
\bibitem [{\citenamefont {Kokkotas}\ \emph {et~al.}(2001)\citenamefont
  {Kokkotas}, \citenamefont {Apostolatos},\ and\ \citenamefont
  {Andersson}}]{Kokkotas:1999mn}%
  \BibitemOpen
  \bibfield  {author} {\bibinfo {author} {\bibfnamefont {K.~D.}\ \bibnamefont
  {Kokkotas}}, \bibinfo {author} {\bibfnamefont {T.~A.}\ \bibnamefont
  {Apostolatos}}, \ and\ \bibinfo {author} {\bibfnamefont {N.}~\bibnamefont
  {Andersson}},\ }\href {\doibase 10.1046/j.1365-8711.2001.03945.x} {\bibfield
  {journal} {\bibinfo  {journal} {Mon. Not. R. Astron. Soc.}\ }\textbf
  {\bibinfo {volume} {320}},\ \bibinfo {pages} {307} (\bibinfo {year}
  {2001})}\BibitemShut {NoStop}%
\bibitem [{\citenamefont {Andersson}\ and\ \citenamefont
  {Comer}(2001)}]{Andersson:2001kx}%
  \BibitemOpen
  \bibfield  {author} {\bibinfo {author} {\bibfnamefont {N.}~\bibnamefont
  {Andersson}}\ and\ \bibinfo {author} {\bibfnamefont {G.~L.}\ \bibnamefont
  {Comer}},\ }\href {\doibase 10.1103/PhysRevLett.87.241101} {\bibfield
  {journal} {\bibinfo  {journal} {Phys. Rev. Lett.}\ }\textbf {\bibinfo
  {volume} {87}},\ \bibinfo {pages} {241101} (\bibinfo {year}
  {2001})}\BibitemShut {NoStop}%
\bibitem [{\citenamefont {Warszawski}\ and\ \citenamefont
  {Melatos}(2011)}]{Warszawski:2011vy}%
  \BibitemOpen
  \bibfield  {author} {\bibinfo {author} {\bibfnamefont {L.}~\bibnamefont
  {Warszawski}}\ and\ \bibinfo {author} {\bibfnamefont {A.}~\bibnamefont
  {Melatos}},\ }\href {\doibase 10.1111/j.1365-2966.2011.18803.x} {\bibfield
  {journal} {\bibinfo  {journal} {Mon. Not. Roy. Astron. Soc.}\ }\textbf
  {\bibinfo {volume} {415}},\ \bibinfo {pages} {1611} (\bibinfo {year}
  {2011})}\BibitemShut {NoStop}%
\bibitem [{\citenamefont {Warszawski}\ and\ \citenamefont
  {Melatos}(2012)}]{Warszawski:2012zq}%
  \BibitemOpen
  \bibfield  {author} {\bibinfo {author} {\bibfnamefont {L.}~\bibnamefont
  {Warszawski}}\ and\ \bibinfo {author} {\bibfnamefont {A.}~\bibnamefont
  {Melatos}},\ }\href {\doibase 10.1111/j.1365-2966.2012.20977.x} {\bibfield
  {journal} {\bibinfo  {journal} {Mon. Not. Roy. Astron. Soc.}\ }\textbf
  {\bibinfo {volume} {423}},\ \bibinfo {pages} {2058} (\bibinfo {year}
  {2012})}\BibitemShut {NoStop}%
\bibitem [{\citenamefont {{Warszawski}}\ \emph {et~al.}(2012)\citenamefont
  {{Warszawski}}, \citenamefont {{Melatos}},\ and\ \citenamefont
  {{Berloff}}}]{Warszawski:2012mb}%
  \BibitemOpen
  \bibfield  {author} {\bibinfo {author} {\bibfnamefont {L.}~\bibnamefont
  {{Warszawski}}}, \bibinfo {author} {\bibfnamefont {A.}~\bibnamefont
  {{Melatos}}}, \ and\ \bibinfo {author} {\bibfnamefont {N.~G.}\ \bibnamefont
  {{Berloff}}},\ }\href {\doibase 10.1103/PhysRevB.85.104503} {\bibfield
  {journal} {\bibinfo  {journal} {\prb}\ }\textbf {\bibinfo {volume} {85}},\
  \bibinfo {eid} {104503} (\bibinfo {year} {2012})}\BibitemShut {NoStop}%
\bibitem [{\citenamefont {{Warszawski}}\ and\ \citenamefont
  {{Melatos}}(2013)}]{Warszawski:2013wm}%
  \BibitemOpen
  \bibfield  {author} {\bibinfo {author} {\bibfnamefont {L.}~\bibnamefont
  {{Warszawski}}}\ and\ \bibinfo {author} {\bibfnamefont {A.}~\bibnamefont
  {{Melatos}}},\ }\href {\doibase 10.1093/mnras/sts108} {\bibfield  {journal}
  {\bibinfo  {journal} {Mon. Not. R. Astron. Soc.}\ }\textbf {\bibinfo {volume}
  {428}},\ \bibinfo {pages} {1911} (\bibinfo {year} {2013})}\BibitemShut
  {NoStop}%
\bibitem [{\citenamefont {Melatos}\ \emph {et~al.}(2015)\citenamefont
  {Melatos}, \citenamefont {Douglass},\ and\ \citenamefont
  {Simula}}]{Melatos:2015oca}%
  \BibitemOpen
  \bibfield  {author} {\bibinfo {author} {\bibfnamefont {A.}~\bibnamefont
  {Melatos}}, \bibinfo {author} {\bibfnamefont {J.~A.}\ \bibnamefont
  {Douglass}}, \ and\ \bibinfo {author} {\bibfnamefont {T.~P.}\ \bibnamefont
  {Simula}},\ }\href {\doibase 10.1088/0004-637X/807/2/132} {\bibfield
  {journal} {\bibinfo  {journal} {Astrophys. J.}\ }\textbf {\bibinfo {volume}
  {807}},\ \bibinfo {pages} {132} (\bibinfo {year} {2015})}\BibitemShut
  {NoStop}%
\bibitem [{\citenamefont {Benton}\ and\ \citenamefont
  {Clark~Jr}(1974)}]{benton1974spin}%
  \BibitemOpen
  \bibfield  {author} {\bibinfo {author} {\bibfnamefont {E.~R.}\ \bibnamefont
  {Benton}}\ and\ \bibinfo {author} {\bibfnamefont {A.}~\bibnamefont
  {Clark~Jr}},\ }\href@noop {} {\bibfield  {journal} {\bibinfo  {journal}
  {Annual Review of Fluid Mechanics}\ }\textbf {\bibinfo {volume} {6}},\
  \bibinfo {pages} {257} (\bibinfo {year} {1974})}\BibitemShut {NoStop}%
\bibitem [{\citenamefont {van Eysden}\ and\ \citenamefont
  {Melatos}(2008)}]{vanEysden:2008pd}%
  \BibitemOpen
  \bibfield  {author} {\bibinfo {author} {\bibfnamefont {C.~A.}\ \bibnamefont
  {van Eysden}}\ and\ \bibinfo {author} {\bibfnamefont {A.}~\bibnamefont
  {Melatos}},\ }\href {\doibase 10.1088/0264-9381/25/22/225020} {\bibfield
  {journal} {\bibinfo  {journal} {Class. Quant. Grav.}\ }\textbf {\bibinfo
  {volume} {25}},\ \bibinfo {pages} {225020} (\bibinfo {year}
  {2008})}\BibitemShut {NoStop}%
\bibitem [{\citenamefont {Bennett}\ \emph {et~al.}(2010)\citenamefont
  {Bennett}, \citenamefont {van Eysden},\ and\ \citenamefont
  {Melatos}}]{Bennett:2010tm}%
  \BibitemOpen
  \bibfield  {author} {\bibinfo {author} {\bibfnamefont {M.~F.}\ \bibnamefont
  {Bennett}}, \bibinfo {author} {\bibfnamefont {C.~A.}\ \bibnamefont {van
  Eysden}}, \ and\ \bibinfo {author} {\bibfnamefont {A.}~\bibnamefont
  {Melatos}},\ }\href {\doibase 10.1111/j.1365-2966.2010.17416.x} {\bibfield
  {journal} {\bibinfo  {journal} {Mon. Not. Roy. Astron. Soc.}\ }\textbf
  {\bibinfo {volume} {409}},\ \bibinfo {pages} {1705} (\bibinfo {year}
  {2010})}\BibitemShut {NoStop}%
\bibitem [{\citenamefont {Singh}(2017)}]{Singh:2016ilt}%
  \BibitemOpen
  \bibfield  {author} {\bibinfo {author} {\bibfnamefont {A.}~\bibnamefont
  {Singh}},\ }\href {\doibase 10.1103/PhysRevD.95.024022} {\bibfield  {journal}
  {\bibinfo  {journal} {Phys. Rev.}\ }\textbf {\bibinfo {volume} {D95}},\
  \bibinfo {pages} {024022} (\bibinfo {year} {2017})}\BibitemShut {NoStop}%
\bibitem [{\citenamefont {Prix}\ \emph {et~al.}(2011)\citenamefont {Prix},
  \citenamefont {Giampanis},\ and\ \citenamefont {Messenger}}]{Prix:2011qv}%
  \BibitemOpen
  \bibfield  {author} {\bibinfo {author} {\bibfnamefont {R.}~\bibnamefont
  {Prix}}, \bibinfo {author} {\bibfnamefont {S.}~\bibnamefont {Giampanis}}, \
  and\ \bibinfo {author} {\bibfnamefont {C.}~\bibnamefont {Messenger}},\ }\href
  {\doibase 10.1103/PhysRevD.84.023007} {\bibfield  {journal} {\bibinfo
  {journal} {Phys. Rev.}\ }\textbf {\bibinfo {volume} {D84}},\ \bibinfo {pages}
  {023007} (\bibinfo {year} {2011})}\BibitemShut {NoStop}%
\bibitem [{\citenamefont {Stopnitzky}\ and\ \citenamefont
  {Profumo}(2014)}]{Stopnitzky:2013wza}%
  \BibitemOpen
  \bibfield  {author} {\bibinfo {author} {\bibfnamefont {E.}~\bibnamefont
  {Stopnitzky}}\ and\ \bibinfo {author} {\bibfnamefont {S.}~\bibnamefont
  {Profumo}},\ }\href {\doibase 10.1088/0004-637X/787/2/114} {\bibfield
  {journal} {\bibinfo  {journal} {Astrophys. J.}\ }\textbf {\bibinfo {volume}
  {787}},\ \bibinfo {pages} {114} (\bibinfo {year} {2014})}\BibitemShut
  {NoStop}%
\end{thebibliography}%

\end{document}